\documentclass[prd,onecolumn,showpacs,showkeys,floatfix,superscriptaddress,11pt]{revtex4}

\usepackage[utf8]{inputenc} 
\usepackage[T1]{fontenc}    
\usepackage{hyperref}       
\usepackage{url}            
\usepackage{booktabs}       
\usepackage{amsfonts}       
\usepackage{nicefrac}       
\usepackage{microtype}      
\usepackage{lipsum}		
\usepackage{graphicx}
\usepackage{natbib}
\usepackage{doi}
\usepackage{graphicx}
\usepackage{subfigure}
\usepackage{epsf}
\usepackage{bm}
\usepackage{amsmath}
\usepackage{amsfonts}
\usepackage{amssymb}
\usepackage{graphicx}
\usepackage{tabularx}
\usepackage{color}%
\providecommand{\U}[1]{\protect\rule{.1in}{.1in}}

\newcommand{\be}{\begin{equation}}
\newcommand{\ee}{\end{equation}}

\newcommand{\mincir}{\raise
-3.truept\hbox{\rlap{\hbox{$\sim$}}\raise4.truept\hbox{$<$}\ }}
\newcommand{\magcir}{\raise
-3.truept\hbox{\rlap{\hbox{$\sim$}}\raise4.truept\hbox{$>$}\ }}

\DeclareMathOperator{\arccot}{arccot}
\providecommand{\U}[1]{\protect\rule{.1in}{.1in}}

\usepackage{tikz,xcolor,hyperref}

\definecolor{lime}{HTML}{A6CE39}
\DeclareRobustCommand{\orcidicon}{%
	\begin{tikzpicture}
	\draw[lime, fill=lime] (0,0) 
	circle [radius=0.16] 
	node[white] {{\fontfamily{qag}\selectfont \tiny ID}};
	\draw[white, fill=white] (-0.0625,0.095) 
	circle [radius=0.007];
	\end{tikzpicture}
	\hspace{-2mm}
}

\foreach \x in {A, ..., Z}{%
	\expandafter\xdef\csname orcid\x\endcsname{\noexpand\href{https://orcid.org/\csname orcidauthor\x\endcsname}{\noexpand\orcidicon}}
}

\begin{document}

\title{Phase-space analysis of an Einstein-Gauss-Bonnet scalar field cosmology}

\author{Alfredo D. Millano\orcidC{}}
\email{alfredo.millano@alumnos.ucn.cl}
\affiliation{Departamento de Matem\'{a}ticas, Universidad Cat\'{o}lica del Norte, Avda.
Angamos 0610, Casilla 1280 Antofagasta, Chile}

\author{Genly Leon\orcidA{}}
\email{genly.leon@ucn.cl}
\affiliation{Departamento de Matem\'{a}ticas, Universidad Cat\'{o}lica del Norte, Avda.
Angamos 0610, Casilla 1280 Antofagasta, Chile}
\affiliation{Institute of Systems Science, Durban University of Technology, PO Box 1334,
Durban 4000, South Africa}

\author{Andronikos Paliathanasis\orcidB{}}
\email{anpaliat@phys.uoa.gr}
\affiliation{Institute of Systems Science, Durban University of Technology, PO Box 1334,
Durban 4000, South Africa}
\affiliation{Departamento de Matem\'{a}ticas, Universidad Cat\'{o}lica del Norte, Avda.
Angamos 0610, Casilla 1280 Antofagasta, Chile}

\begin{abstract}
We perform a detailed study of the phase-space of the field equations of an Einstein-Gauss-Bonnet scalar field cosmology for a spatially flat Friedmann--Lema\^{\i}tre--Robertson--Walker spacetime. For the scalar field potential, we consider the exponential function. In contrast, for the coupling function of the scalar field with the Gauss-Bonnet term, we assume two cases, the exponential function and the power-law function. We write the field equations in dimensionless variables and study the equilibrium points using Poincare variables. For the exponential coupling function, the asymptotic solutions describe de Sitter universes or spacetimes where the Gauss-Bonnet term dominates. We recovered previous results but found new asymptotic solutions not previously studied. For the power-law coupling function, equilibrium points which describe the scaling solution appear. Finally, the power-law coupling provides a rich cosmological phenomenology.
\end{abstract}
\date{\today}
\maketitle

\section{Introduction}

The analysis of the cosmological observations suggests that our universe, on
large scales, is isotropic and homogeneous, as described by the four-dimensional
Friedmann--Lema\^{\i}tre--Robertson--Walker (FLRW) geometry. The primary theoretical mechanism proposed to explain the observations is
the so-called cosmic inflation \cite{Aref1,guth}. Indeed, a rapid expansion of
the size of the universe provides that the latter effectively loses its memory
on the initial conditions; hence the flatness and homogeneity problems can be
solved by inflation \cite{f1,f2}.

In the context of Einstein's General Relativity, inflation is described by a
scalar field, known as \textquotedblleft inflaton\textquotedblright.
Specifically, the inflationary mechanism introduces a scalar field in the
cosmic fluid, and the cosmic expansion appears when the scalar field potential
dominates to drive the dynamics. \cite{ref1a,ref1,ref2,ref3,newinf}.
The additional degrees of freedom provided by the scalar field can describe
higher-order geometric invariants introduced in the Einstein-Hilbert Action.
Indeed, in the Starobinsky model for inflation \cite{star} inspired by field
theory, a quadratic term of the Ricci scalar has been introduced to modify the
Einstein-Hilbert Action. The higher-order derivatives are attributed by a
scalar field which can provide an inflationary epoch, see also the recent
studies \cite{ib1,ib2}.

Furthermore, at present, the universe is under a second acceleration
phase \cite{sup1}, which is attributed to an exotic matter source with negative
pressure is known as dark energy. The nature of the dark energy is unknown. The
two acceleration phases of the universe challenge the theory of General
Relativity and various modified and alternative theories of gravities have
been proposed by cosmologists in the last decades
\cite{Clifton1,Nojiri:2017ncd,sf8}.

General Relativity's main characteristic is a second-order theory
of gravity. Moreover, according to Lovelock's theorem General Relativity is
the unique second-order gravitational theory in the four dimensions where the
field equations are generated from an Action Integral \cite{lvl}. However, General Relativity is only a case of Lovelock
gravity in higher dimensions.\ The latter is a second-order theory of gravity in higher dimensions
where higher order invariants are introduced in the gravitational Action
Integral \cite{lvl2,lvl3}. The Gauss-Bonnet invariant is the only invariant
derived by the Riemann tensor quadratic products that do not introduce any
terms with higher-order derivatives into the field equations \cite{lvl2}. On
the other hand, in the case of four dimensions, the Gauss-Bonnet invariant is a topological invariant, that is, a total derivative that, when it is
introduced in the gravitational Lagrangian, does not affect the field
equations. The Einstein Gauss-Bonnet theory is the most straightforward extension of
Einstein's General Relativity and belongs to Lovelock's theories. The Einstein
Gauss-Bonnet terms have been widely studied in higher-order theories of gravities,
see for instance \cite{gb1,gb2,gb3,gb4,gb6,gb7,gb8,gb9,gb10} and references
therein. Furthermore, the Gauss-Bonnet term can describe the quantum
corrections to gravity, mainly related to the heterotic string
\cite{gb11}. An essential property of the Einstein Gauss-Bonnet theory is that
it is a ghost-free theory of gravity \cite{hor}.

In the case of four dimensions, because the Gauss-Bonnet is a topological
invariant, it can be introduced in gravitational Action Integral
only with modifications. Indeed, there is a family of theories known as
$f\left(  G\right)  $ theories of gravity where nonlinear functions of the Gauss-Bonnet invariant are introduced in the Gravitational Integral
\cite{gbm01,gbm02,gbm03}. Another attempt is to introduce a scalar field coupled to the Gauss-Bonnet invariant. In that case, a coupling function exists between the Gauss-Bonnet term and the scalar field. That is the
cosmological scenario we deal with in this work, known as Einstein
Gauss-Bonnet scalar field theory \cite{bb1}. The properties of astrophysical objects in this theory were the
subject of study for various studies \cite{bb2,bb2b,bb3,bb4,bb5}.

In cosmological studies, the four-dimensional
Einstein\ Gauss-Bonnet scalar field theory has been applied for the
description of various epochs of cosmological evolution. It has been found
that the Gauss-Bonnet invariant and the coupling function introduce
non-trivial effects on the early inflationary stage of the universe
\cite{in1} and that a small transition exists to Einstein's General
Relativity at the end of the inflationary epoch. In \cite{in2}, some exact
solutions which describe cosmic inflation were derived. On the other hand,
inflationary models with a Gauss-Bonnet term were constrained in the view of
GW170817 event in a series of studies \cite{in3,in4,in5}. In the presence of a
nonzero spatial curvature for the background space, exact solutions in
Einstein Gauss-Bonnet scalar field theory derived before in \cite{in6}. It was
found that the quadratic coupling function of the scalar field to the
Gauss-Bonnet term is of particular importance because the singularity-free theory provides inflationary solutions.

In \cite{dn1}, the dynamics of the cosmological field equations were
investigated for the four-dimensional Einstein Gauss-Bonnet scalar field
theory, where the authors have assumed that the Hubble function is
that of a scaling solution, however in \cite{dn2}, the most general case was
studied, and the equilibrium points of the field equations were
investigated. The analysis in \cite{dn2} shows that the only
equilibrium points where the Gauss-Bonnet term contributes to the
cosmological fluid are that of the de Sitter universe. However, as we shall
show in the following lines, additional equilibrium points exist that
describe scaling solutions to which the Gauss-Bonnet term contributes. In
particular, we perform a detailed analysis of the phase for the cosmological
field equations in the Einstein-Gauss-Bonnet scalar field theory to reconstruct the cosmological history and understand the evolution of the cosmological parameters. Such analysis provides essential information about the significant cosmological eras provided by the theory. Simultaneously, important conclusions about the viability of the theory can be made. 
Section \ref{II} presents the gravitational Action integral for the Einstein-Gauss-Bonnet scalar field theory in a four-dimensional, spatially flat FLRW geometry. We present the field equations where we observe that they depend on two functions, the scalar field potential $V(\phi)$ and the coupling function $f(\phi)$ of the scalar field with the Gauss-Bonnet scalar. Moreover, the scalar field can be a quintessence or a phantom field. We perform a global analysis of the field equations' phase space to reconstruct the cosmological parameters' evolution. In Section \ref{III}, we study the equilibrium points for linear function $f(\phi)=f_{0}\phi$, while in Section \ref{IV}, we perform the same analysis for the exponential function $f(\phi)=f_{0}e^{ \zeta \phi}$. As far as the scalar field potential is concerned, we consider the exponential function $V\left(  \phi\right)  =V_{0} e^{\lambda\phi}$. Finally, Section \ref{V} discusses our results and presents our conclusions.

\section{Einstein-Gauss-Bonnet scalar field 4D Cosmology}
\label{II}
The gravitational Action Integral for the Einstein-Gauss-Bonnet scalar field theory
of gravity in a four-dimensional Riemannian manifold with the metric tensor
$g_{\mu\nu}$ is defined as follows
\begin{equation}
S=\int d^{4}x\sqrt{-g}\left(  \frac{R}{2}-\frac{\varepsilon}{2}g_{\mu\nu}%
\phi^{;\mu}\phi^{;\nu}-V\left(  \phi\right)  -f\left(  \phi\right)  G\right), 
\label{ai.01}%
\end{equation}
where $R$ is the Ricciscalar of the metric tensor, $\phi$ is the scalar field,
$V\left(  \phi\right)  $ the scalar field potential and $G$ is the
Gauss-Bonnet term
\begin{equation}
G=R^{2}-4R_{\mu\nu}R^{\mu\nu}+R_{\mu\nu\kappa\lambda}R^{\mu\nu\kappa\lambda}.
\end{equation}
Function $f\left(  \phi\right)  $ is the coupling function between the scalar
field and the Gauss-Bonnet term and$~\varepsilon=\pm1$ indicates if the scalar
field $\phi$ is quintessence $\left(  \varepsilon=+1\right)  $ or phantom
$\left(  \varepsilon=-1\right)  $. In the case where $f\left(  \phi\right)  $
is a constant function of the gravitational Action Integral (\ref{ai.01}) reduces
to that of General Relativity with a minimally coupled scalar field.

On very large scales, the universe is considered to be isotropic and
homogeneous. The physical space is described by the FLRW metric tensor with
line element%
\begin{equation}
ds^{2}=-dt^{2}+a^{2}\left(  t\right)  \left(  dr^{2}+r^{2}\left(  d\theta
^{2}+\sin^{2}\theta d\varphi^{2}\right)  \right).\label{ww.16}%
\end{equation}
The three-dimensional surface is a maximally symmetric space and admits six
isometries. Moreover, we assume that the scalar field inherits the symmetries
of the background space, which means that $\phi=\phi\left(  t\right)$.

For the line element (\ref{ww.16}) the Ricci scalar is derived%
\begin{equation}
R=6\left(  2H^{2}+\dot{H}\right),
\end{equation}
where a dot means derivative with respect to $t$,  $\dot{H}=\frac{dH}{dt}$ and $H=\frac{d}{dt}\left(  \ln a\right)  $ is
the Hubble function. Moreover, the Gauss-Bonnet term is calculated as%
\begin{equation}
G=24H^{2}\left(  \dot{H}+H^{2}\right).
\end{equation}

By replacing the latter in the Action Integral (\ref{ai.01}) and by integration by parts, we end with the point-like Lagrangian function%
\[
L\left(  a,\dot{a},\phi,\dot{\phi}\right)  =-3a\dot{a}^{2}+\frac{\varepsilon
}{2}a^{3}\dot{\phi}^{2}+8\dot{a}^{3}f_{,\phi}\dot{\phi}-a^{3}V\left(
\phi\right)  ,
\]
while the field equations are%
\begin{align}
& -48 H^3 \dot{\phi} f'(\phi )+6 H^2-2 V(\phi
   )-\epsilon  \dot{\phi}^2=0,\\
    &-16 H \dot{H} \dot{\phi} f'(\phi )-16 H^3
   \dot{\phi} f'(\phi )-V(\phi )+\frac{1}{2}
   \epsilon  \dot{\phi}^2 +H^2 \left(-8 \dot{\phi}^2 f''(\phi )-8 \ddot{\phi} f'(\phi
   )+3\right)+2 \dot{H}=0, \\
& 3 H \left(-8 H \left(\dot{H}+H^2\right)
   f'(\phi )-\epsilon  \dot{\phi}\right)-V'(\phi
   )-\epsilon  \ddot{\phi}=0,
\end{align}
where the comma means derivative with respect the argument of the function.

The effective density and pressure of the scalar field are given by 
\begin{align}
\rho_\phi & =   \frac{1}{2} \dot{\phi} \left(48 H^3 f'(\phi)+\epsilon  \dot{\phi}\right)+V(\phi), \\
p_\phi & =\frac{8 H^2 f'(\phi) V'(\phi)}{-8 \epsilon  H \dot{\phi} f'(\phi)+96 H^4 f'(\phi)^2+\epsilon
   }-\frac{\epsilon  V(\phi)}{-8 \epsilon  H \dot{\phi} f'(\phi)+96 H^4 f'(\phi)^2+\epsilon
   } \nonumber\\  & +\frac{192 H^6 f'(\phi)^2+\epsilon  \dot{\phi} \left(16 H^2 \left(\dot{\phi} f''(\phi)-4 H f'(\phi
  )\right)-\epsilon  \dot{\phi}\right)}{16 \epsilon  H \dot{\phi} f'(\phi)-2 \left(96 H^4 f'(\phi
  )^2+\epsilon \right)}. \label{weff}
\end{align}
And we define the effective equation of state (EoS) 
\begin{align}
    w_\phi = \frac{p_\phi}{\rho_\phi}. 
\end{align}

In the following, we shall perform a detailed analysis of the phase-space for
the exponential scalar field potential~$V\left(  \phi\right)  =V_{0}%
e^{\lambda\phi}$ and for two coupling functions $f\left(  \phi\right)  $, the
linear $f\left(  \phi\right)  =f_{0}\phi$ and the exponential $f_{,\phi\phi
}=\zeta f_{,\phi}$, where $f_0$ and $\zeta$ are constants.

\section{Phase space Analysis for linear $f$: $f(\phi)=f_0 \phi$}
\label{III}

The field equations for the linear coupling function $f(\phi)=f_0 \phi$ read
\begin{align}
    &-48 f_0 H^3 \dot{\phi}+6 H^2-2 V(\phi
   )-\epsilon  \dot{\phi}^2=0\\
   & -16 f_0 H \dot{H} \dot{\phi}+H^2 \left(3-8
   f_0 \ddot{\phi}\right)-16 f_0 H^3
   \dot{\phi}+2 \dot{H}-V(\phi )+\frac{1}{2}
   \epsilon  \dot{\phi}^2=0\\
   & -3 H \left(8 f_0 H
   \left(\dot{H}+H^2\right)+\epsilon  \dot{\phi} \right)-V'(\phi )-\epsilon  \ddot{\phi}=0
\end{align}

In order to study the phase space, we introduce the following normalized variables, 
\begin{equation}
    {x}=\frac{\dot{\phi}}{\sqrt{1+H^2}}, \; {y}=  \frac{\sqrt{V(\phi)}}{\sqrt{1+H^2}}, \; \eta=\frac{H}{\sqrt{1+H^2}}.
\end{equation}

With these definitions, the first modified Friedmann equation is written in the algebraic form
\begin{equation}\label{Friedmann-new-var}
  -48 f_0\eta  ^3 x +\epsilon  \left(\eta  ^2-1\right) x ^2+2 \left(\eta  ^2-1\right) \left(y ^2-3 \eta  ^2\right)=0. 
\end{equation}

Moreover, the rest of the field equations are described by the following system of first-order ordinary differential equations
\begin{align}
    \frac{dx}{d\tau} &=\frac{1}{K(x,y,\eta,f_0,\epsilon)}\Big[6 x \eta \left(\left(64 f_0^2+\epsilon \right) \eta ^6-4 \epsilon  \eta ^4+5 \epsilon  \eta
   ^2-2 \epsilon \right) \nonumber \\ & -48 f_0 \eta ^4 \left(\eta ^2-1\right)-8 f_0 \epsilon  \eta^2  \left(2
   \eta ^4+13 \eta ^2-15\right) x^2+\left(\eta ^2-1\right)^2 x^3 \eta \nonumber\\&-2 \left(\eta ^2-1\right)
   y^2 \left(2 (\lambda -12 f_0) \eta ^2+\eta  x \left(16 f_0 \lambda +\eta ^2 (8
   f_0 \lambda +\epsilon )-\epsilon \right)-2 \lambda \right)\Big], \label{linearf3D-1}\\
   \frac{dy}{d\tau} & =\frac{y}{K(x,y,\eta,f_0,\epsilon)} \Big[6 \eta ^3 \left(\left(64 f_0^2+\epsilon \right) \eta ^4-2 \epsilon  \eta ^2+\epsilon
   \right)-16 f_0 \epsilon  \left(\eta ^2-1\right) \eta ^4 x \nonumber \\&-2 \eta \left(\eta ^2-1\right) y^2
   \left(\eta ^2 (8 f_0 \lambda +\epsilon )-\epsilon \right)+\eta \left(\eta ^2-1\right)^2 x^2+2
   \lambda  x\Big], \label{linearf3D-2}\\
         \frac{d\eta}{d\tau}   &=\frac{1}{K(x,y,\eta,f_0,\epsilon)}\Big[ 6 \eta ^2\left(\eta ^2-1\right) \left(\left(64 f_0^2+\epsilon \right) \eta ^4-2 \epsilon  \eta ^2+\epsilon \right) \nonumber \\&-16 f_0 \epsilon  \left(\eta ^2-1\right)^2 \eta ^3 x-2 \left(\eta ^2-1\right)^2 y^2 \left(\eta ^2 (8 f_0 \lambda +\epsilon )-\epsilon \right)+\left(\eta
   ^2-1\right)^3 x^2\Big]. \label{linearf3D-3}
\end{align}
We define the function $K(x,y,\eta,f_0,\epsilon)=4 \left(\left(96 f_0^2+\epsilon
   \right) \eta ^4+8 f_0 \epsilon  \left(\eta ^2-1\right) \eta  x-2 \epsilon  \eta ^2+\epsilon
   \right),$ and introduce the time derivative 
$d f/d\tau =1/\sqrt{1+H^2}df/dt$. 
   
Since $y>0$, we can solve equation \eqref{Friedmann-new-var} for $y$ and reduce the dimension of the system; the expression for $y$ is
\begin{equation}
    \label{eq-for-y}
   y= \sqrt{\frac{\left(\eta ^2-1\right) x^2 \epsilon -6 \eta ^2 \left(8 f_0 \eta  x+\eta ^2-1\right)}{2(1-\eta ^2)}},
\end{equation}
the dynamics of the model with linear $f$ and $\epsilon= \pm 1$ is given by
\begin{align}
    \frac{dx}{d\tau} = \frac{1}{K(x,y,\eta,f_0,\epsilon)}&\Big[384 f_0^2 \left(\eta ^2+3\right) \eta ^5 x+96 f_0 \left(\eta ^2-1\right) \eta ^4  \nonumber \\
   &-16 f_0 \left(4 \eta ^4+5 \eta ^2-9\right) \eta^2  x^2 \epsilon \nonumber\\
   &+\left(\eta ^2-1\right)^2 x^3 \left(\epsilon
   ^2+1\right)\eta -12 \left(\eta ^2-1\right)^2 x \eta \epsilon \nonumber\\
   &-2 \lambda  \left(4 f_0 \left(\eta ^2+2\right) \eta
    x+\eta ^2-1\right) \left(48 f_0 \eta ^3 x+6 \eta ^4-\eta ^2 \left(x^2 \epsilon +6\right)+x^2 \epsilon
   \right)\Big], \label{linearf2D-1} \\
         \frac{d\eta}{d\tau}= \frac{1}{K(x,y,\eta,f_0,\epsilon)}&\Big[48 f_0 \left(\eta ^2-1\right) \eta ^4 \left(\eta ^2 (8 f_0-\lambda )+\lambda \right)  \nonumber \\&+\left(\eta
   ^2-1\right)^2 x^2 \left(8 f_0 \eta ^2 \lambda  \epsilon +\left(\eta ^2-1\right) \left(\epsilon
   ^2+1\right)\right) \nonumber \\&-64 f_0 \left(\eta ^2-1\right) \eta ^3 x \left(6 f_0 \eta ^2 \lambda +\left(\eta
   ^2-1\right) \epsilon \right)\Big]. \label{linearf2D-2}
\end{align}
The effective equation of state parameter \eqref{weff} can be expressed in term of $x$ and $\eta$ as 
\begin{align}
w_\phi= & -\frac{\eta ^4 (8  f_0 (4  f_0+\lambda )+\epsilon )}{\left(96  f_0^2+\epsilon \right) \eta ^4+8  f_0 \epsilon  \left(\eta
   ^2-1\right) \eta  x-2 \epsilon  \eta ^2+\epsilon } \nonumber \\
   & +\frac{-8  f_0 \eta  x \left(\eta ^2 (24  f_0 \lambda +7
   \epsilon )-7 \epsilon \right)+6 \eta ^2 (4  f_0 \lambda +\epsilon )-3 \epsilon }{3 \left(\left(96  f_0^2+\epsilon \right) \eta
   ^4+8  f_0 \epsilon  \left(\eta ^2-1\right) \eta  x-2 \epsilon  \eta ^2+\epsilon \right)} \nonumber \\
   & +\frac{\epsilon  \left(\eta ^2-1\right) x^2 \left(\eta ^2 (4  f_0 \lambda +\epsilon )-\epsilon \right)}{3 \eta ^2 \left(\left(96  f_0^2+\epsilon
   \right) \eta ^4+8  f_0 \epsilon  \left(\eta ^2-1\right) \eta  x-2 \epsilon  \eta ^2+\epsilon \right)}, 
\end{align}
whereas the deceleration parameter 
$q= -1- \dot{H}/H^2$, can be expressed as 
\begin{align}
    q= & \frac{\left(\epsilon  \left(\eta ^2-1\right) x^2-48 f_0 \eta ^3 x\right) \left(\eta ^2 (4 f_0 \lambda +\epsilon
   )-\epsilon \right)}{2 \eta ^2 \left(\left(96 f_0^2+\epsilon \right) \eta ^4+8 f_0 \epsilon  \left(\eta ^2-1\right) \eta 
   x-2 \epsilon  \eta ^2+\epsilon \right)} \nonumber \\
   & -\frac{\left(\eta ^2-1\right) \left(\eta ^2 (12 f_0 \lambda +\epsilon )-\epsilon
   \right)}{\left(96 f_0^2+\epsilon \right) \eta ^4+8 f_0 \epsilon  \left(\eta ^2-1\right) \eta  x-2 \epsilon  \eta
   ^2+\epsilon }.
\end{align}
\subsection{Dynamical system analysis of 2D system for $\epsilon=1$}
\begin{figure}[ht!]
    \centering
    \includegraphics[scale=0.45]{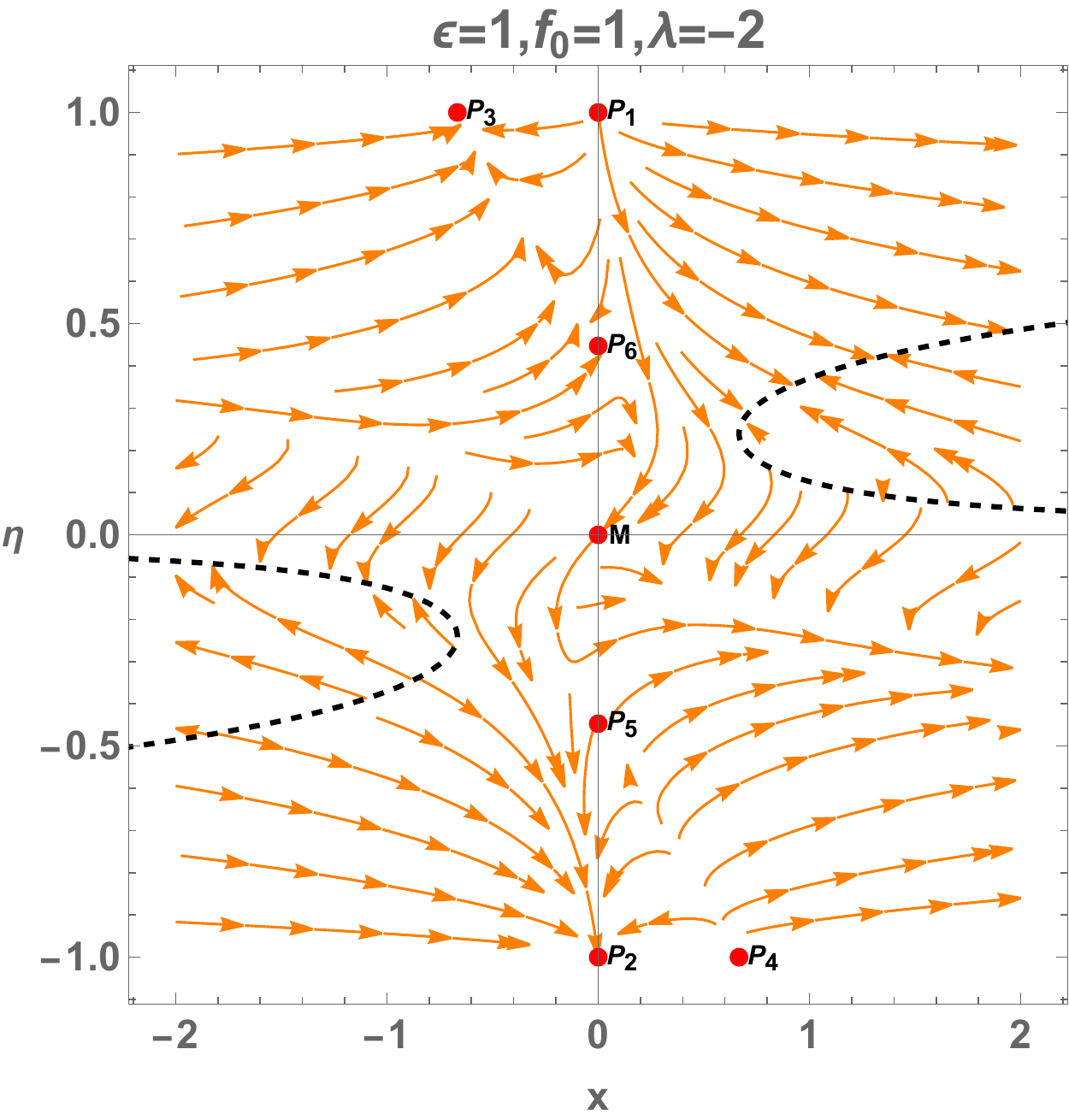}
    \includegraphics[scale=0.45]{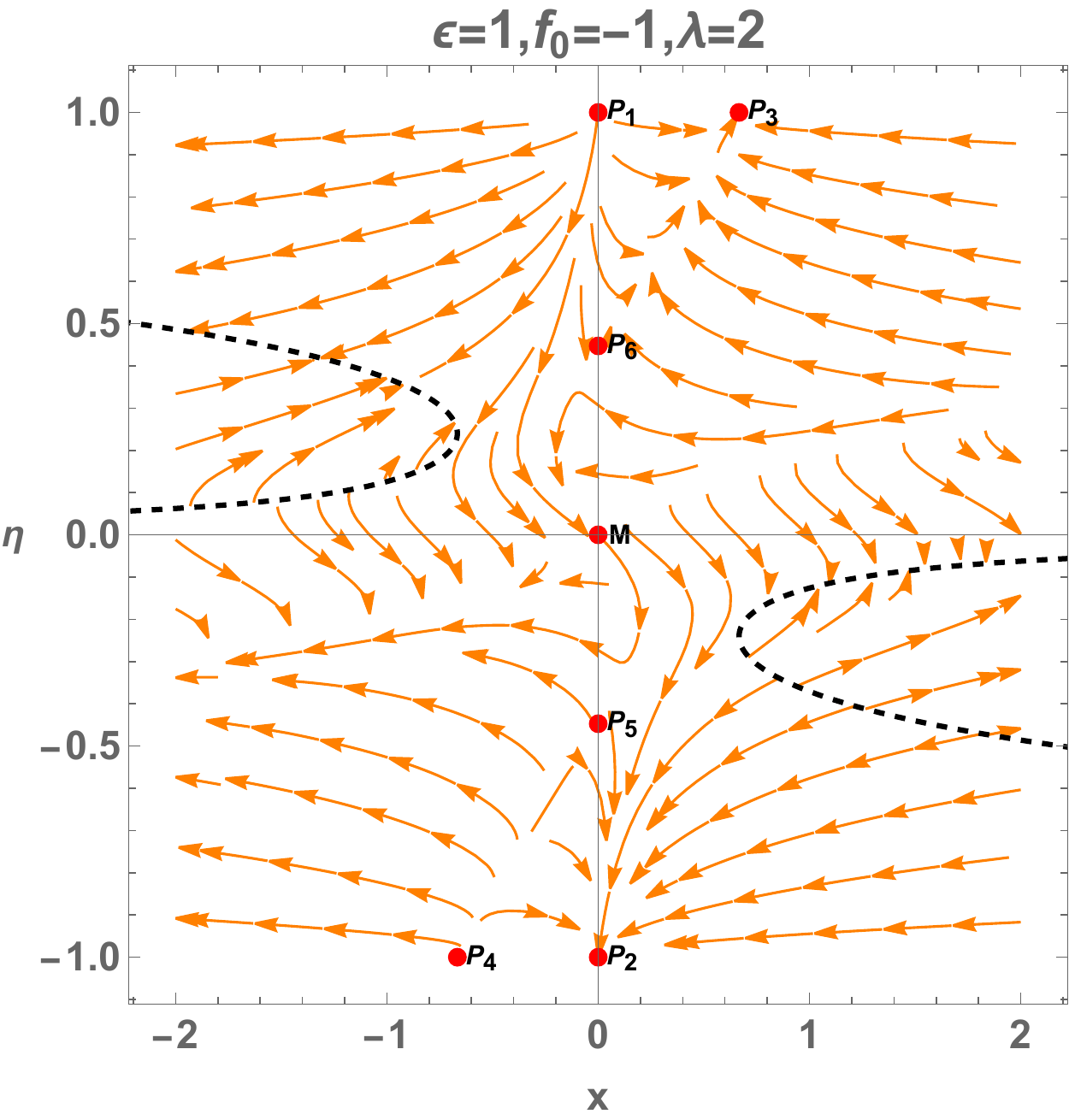}
    \includegraphics[scale=0.45]{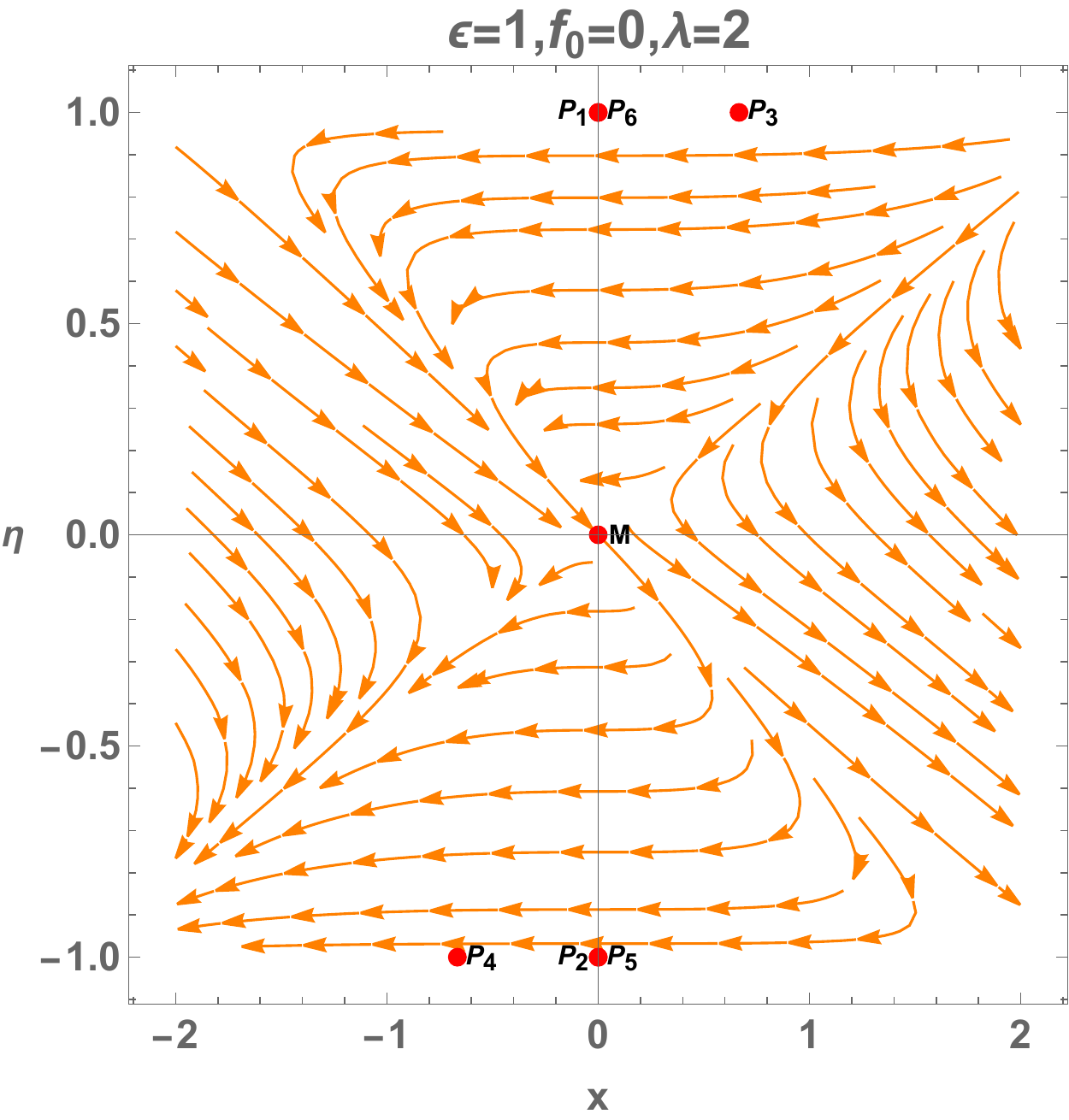}
    \includegraphics[scale=0.45]{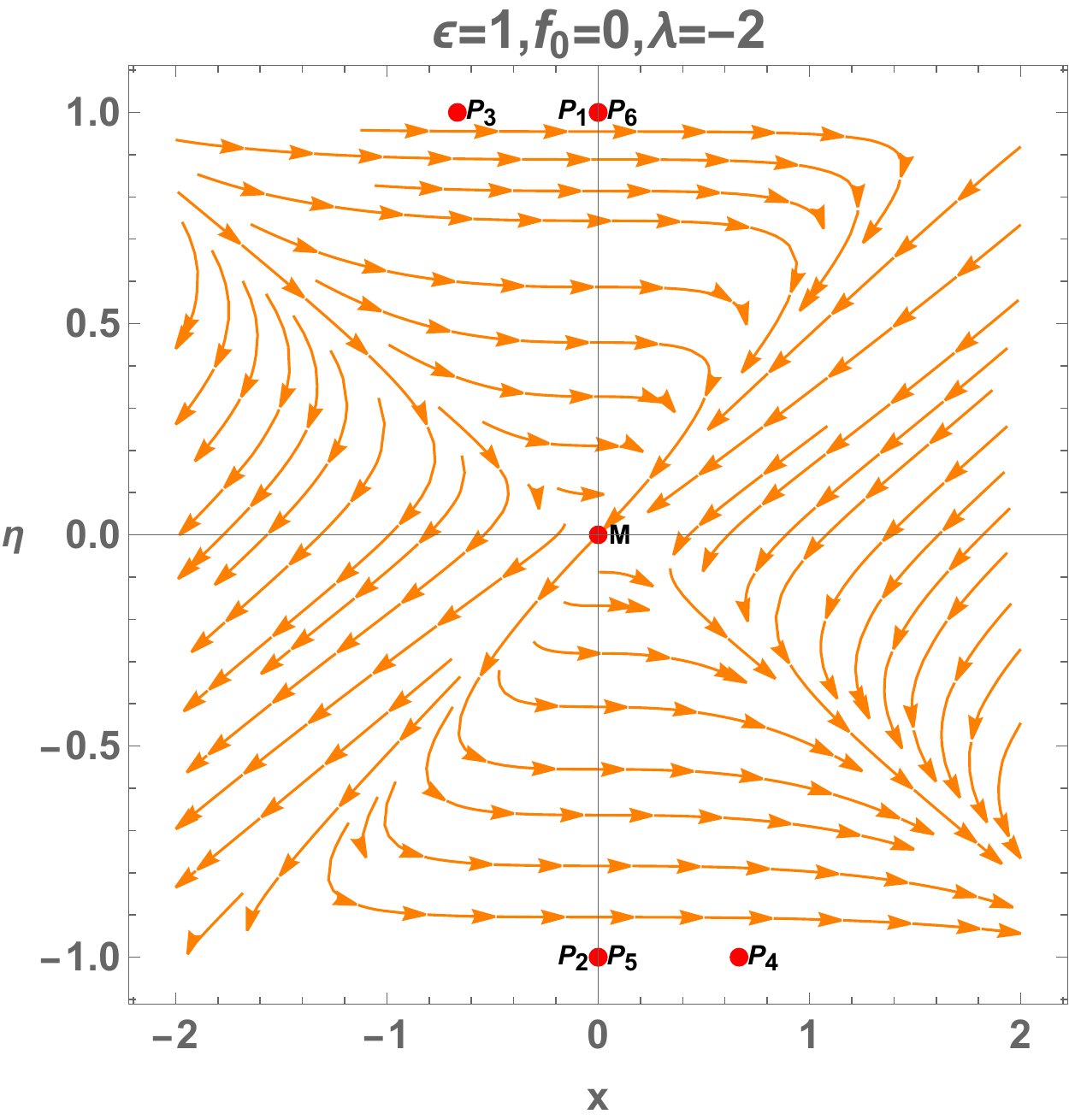}
    \caption{Phase plots for \eqref{linearf2D-1}, \eqref{linearf2D-2} for $\epsilon=1$ and different values of $f_0$ and $\lambda.$ The dashed black lines in the plot correspond to the values of $x$ and $y$ for which $K=0,$ which corresponds to singular curves where the flow direction and the stability changes. }
    \label{fig:1}
\end{figure}

In this section, we perform the stability analysis for the equilibrium points of system \eqref{linearf2D-1}, \eqref{linearf2D-2} taking $\epsilon=1$. The equilibrium points in the coordinates $(x,\eta)$ are the following:

\begin{enumerate}
    \item $M=(0,0),$ with eigenvalues $\{0,0\}$. The asymptotic solution is that of the Minkowski spacetime. 
    \item $P_1=(0,1),$ with eigenvalues $\{2,4\}.$ The asymptotic solution describes a universe dominated by the Gauss-Bonnet term with deceleration parameter $q(P_1)=0$. This point is a source.
    \item $P_2=(0,-1),$ with eigenvalues $\{-2,-4\}.$ This point is a sink. The asymptotic solution is similar to that of point $P_1$.
    \item $P_3=(\frac{4}{3\lambda},1),$ with eigenvalues $\{-4,-\frac{2}{3}\}.$ This point is a sink. We derive that $q(P_3)=0$. The asymptotic solution is similar to that of point $P_1$.
    \item $P_4=(-\frac{4}{3\lambda},-1),$ with eigenvalues $\{4,\frac{2}{3}\}.$ This point is a source. Moreover, for the deceleration parameter, it follows $q(P_4)=0$. The asymptotic solution is similar to that of point $P_1$.
    \item $P_5=(0,-\sqrt{\frac{\lambda }{\lambda -8
   f_0}}),$ with eigenvalues $\left\{\frac{\sqrt{\lambda } \left(3 \sqrt{3 \lambda ^2+2}-\sqrt{51 \lambda ^2+18}\right)}{2 \sqrt{3 \lambda ^2+2}
   \sqrt{\lambda -8 f_0}},\frac{\sqrt{\lambda } \left(3 \sqrt{3 \lambda ^2+2}+\sqrt{51 \lambda ^2+18}\right)}{2
   \sqrt{3 \lambda ^2+2} \sqrt{\lambda -8 f_0}}\right\}.$  Moreover, the deceleration parameter is calculated $q(P_5)=-1$; hence, the asymptotic solution describes the de Sitter universe.  This point is a saddle that exists for
   \begin{enumerate}       
       \item $\lambda <0, \frac{\lambda }{8}<f_0$
       \item $\lambda >0, f_0<\frac{\lambda }{8}.$
   \end{enumerate}
    \item $P_6=(0,\sqrt{\frac{\lambda }{\lambda -8
   f_0}}),$ with eigenvalues $\left\{-\frac{\sqrt{\lambda } \left(3 \sqrt{3 \lambda ^2+2}+\sqrt{51 \lambda ^2+18}\right)}{2 \sqrt{3 \lambda ^2+2}
   \sqrt{\lambda -8 f_0}},\frac{\sqrt{\lambda } \left(\sqrt{51 \lambda ^2+18}-3 \sqrt{3 \lambda ^2+2}\right)}{2
   \sqrt{3 \lambda ^2+2} \sqrt{\lambda -8 f_0}}\right\}.$   Point  $P_6$ describes a de Sitter universe, i.e. $q(P_6)=-1$. This point is a saddle that exists for
   \begin{enumerate}
       \item $\lambda <0, \frac{\lambda }{8}<f_0$
       \item $\lambda >0, 0<f_0<\frac{\lambda }{8}.$
   \end{enumerate}
\end{enumerate}
The above results are summarized in Table \ref{tab:1}. Phase-space diagrams for the dynamical system \eqref{linearf2D-1}, \eqref{linearf2D-2} where the scalar field is a quintessence are presented in Fig. \ref{fig:1}.
\begin{table}[ht!]
    \caption{Equilibrium points of system \eqref{linearf2D-1}, \eqref{linearf2D-2}  for $\epsilon=+1$ with their stability conditions. Also includes the value of $\omega_{\phi}$ and $q.$}
    \label{tab:1}
\newcolumntype{C}{>{\centering\arraybackslash}X}
\centering
    \setlength{\tabcolsep}{4.6mm}
\begin{tabularx}{\textwidth}{cccccc}
\toprule 
  \text{Label}  & \; $x$& $\eta$ & \text{Stability}& $\omega_{\phi}$&$q$\\
  \midrule  
  $M$ & $0$ & $0$  & nonhyperbolic & \text{indeterminate} & \text{indeterminate}\\  \midrule 
  $P_{1}$ & $0$ & $1$ & $\text{source}$ & $-\frac{1}{3}$& $0$\\  \midrule 
  $P_{2}$ & $0$ & $-1$ & $\text{sink}$ & $-\frac{1}{3}$& $0$\\  \midrule 
  $P_{3}$ & $\frac{4}{3\lambda}$& $1$ & $\text{sink}$& $-\frac{1}{3}$ & $0$\\  \midrule 
  $P_{4}$ & $-\frac{4}{3\lambda}$& $-1$ & $\text{source}$& $-\frac{1}{3}$ & $0$\\  \midrule 
  $P_{5}$ & $0$ & $-\sqrt{\frac{\lambda}{\lambda-8f_0}}$  & $\text{saddle}$& $-1$& $-1$\\  \midrule 
  $P_{6}$ & $0$ & $\sqrt{\frac{\lambda}{\lambda-8f_0}}$ & $\text{saddle}$& $-1$ & $-1$\\
\bottomrule
    \end{tabularx}
\end{table}
\subsection{Dynamical system analysis of 2D system for $\epsilon=-1$}
\begin{figure}[ht!]
    \centering
    \includegraphics[scale=0.6]{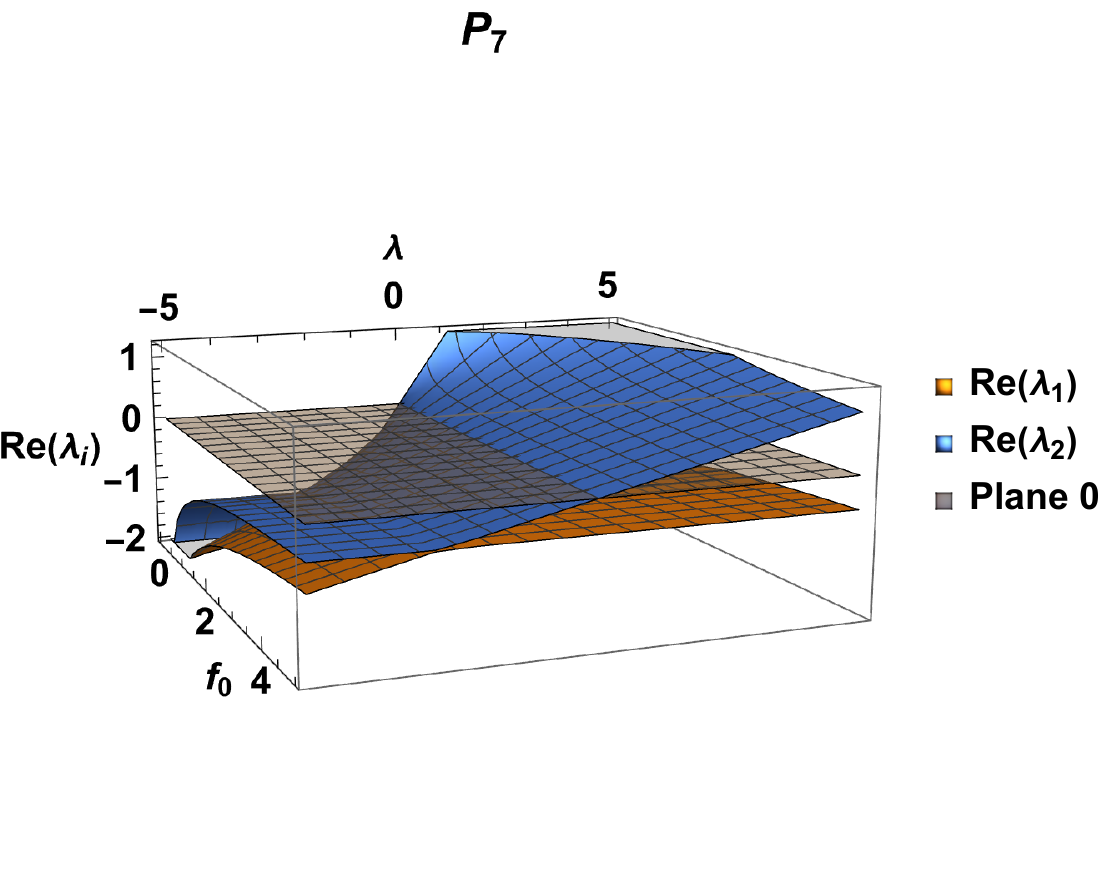}
    \includegraphics[scale=0.6]{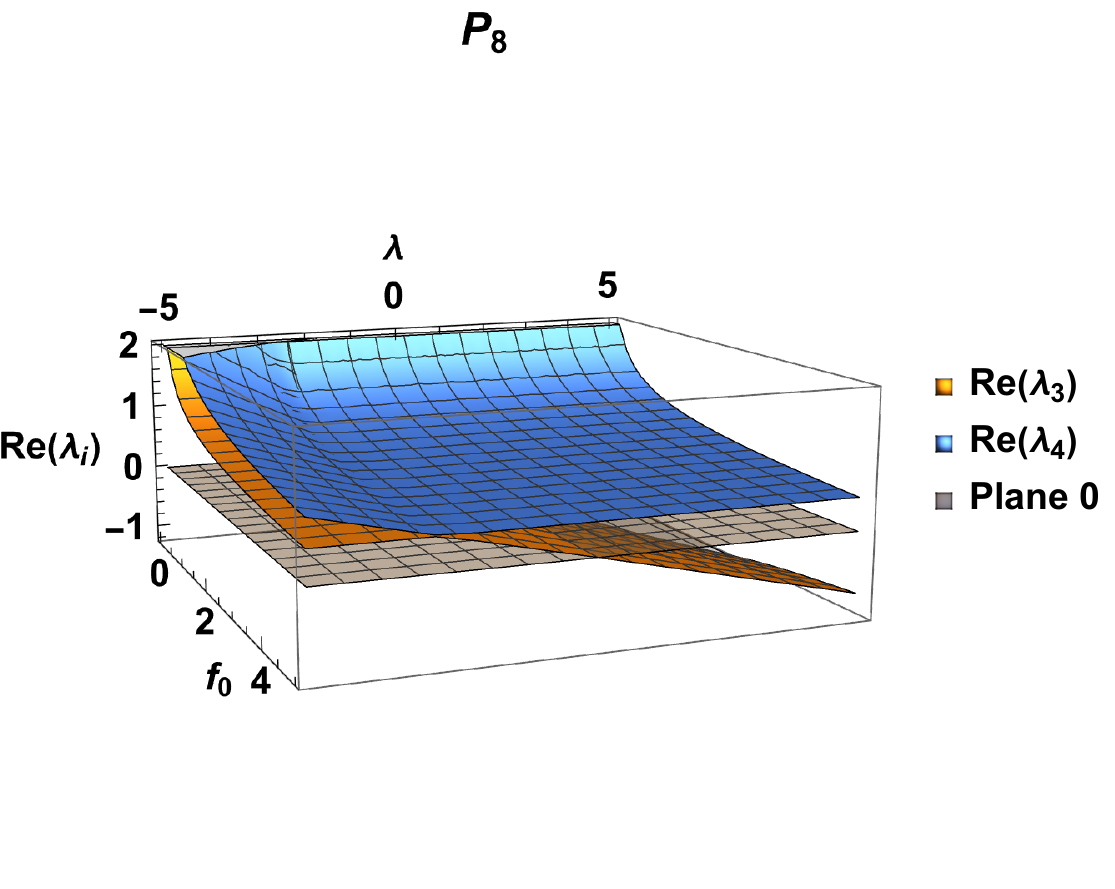}
    \includegraphics[scale=0.6]{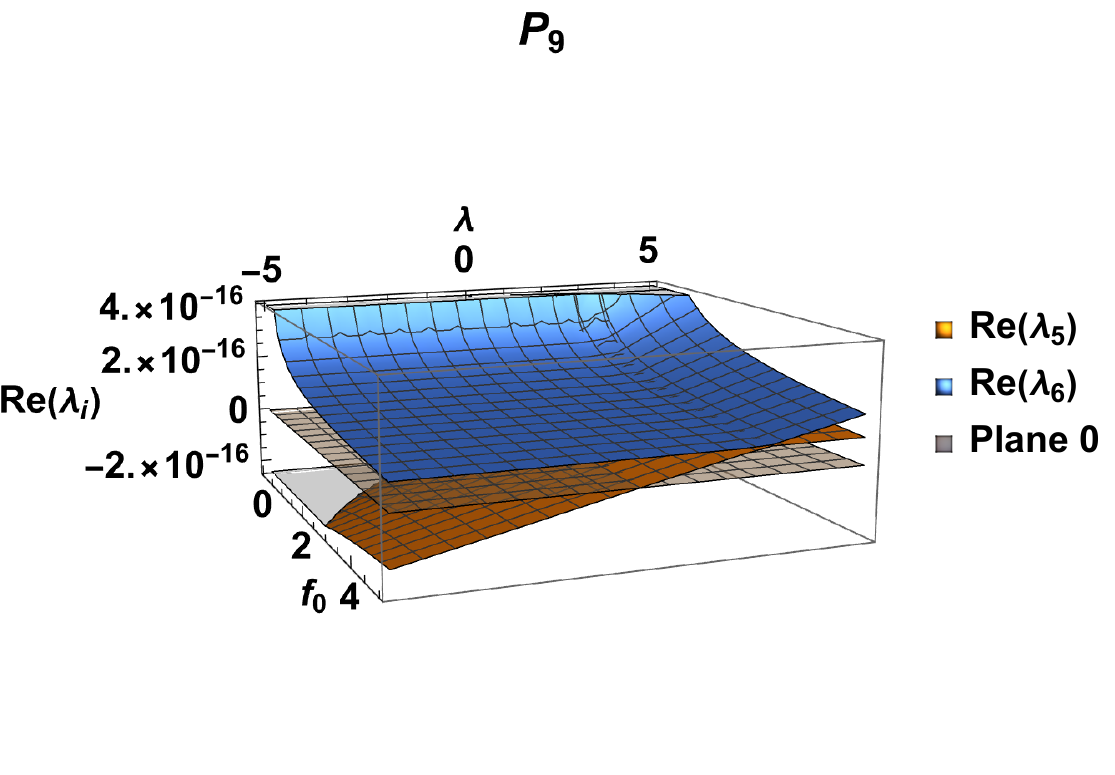}
    \includegraphics[scale=0.6]{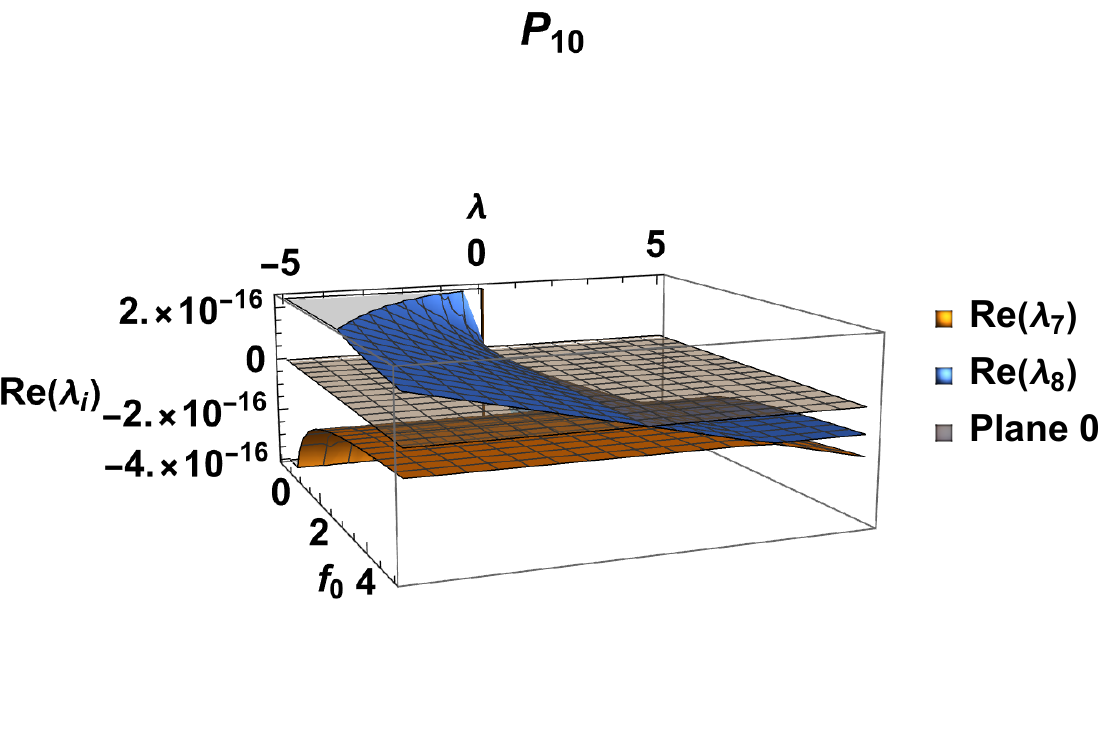}
    \caption{Real part of the eigenvalues $\lambda_i$ where $i=1,\ldots 8.$ for points $P_7, P_8, P_9, P_{10}.$}
    \label{fig:2}
\end{figure}

In this section, we perform the stability analysis for the equilibrium points of system \eqref{linearf2D-1}, \eqref{linearf2D-2} taking $\epsilon=-1$. The equilibrium points for in the coordinates $(x,\eta)$ are the following.
\begin{enumerate}
    \item $M=(0,0),$ with eigenvalues $\{0,0\}$. The asymptotic solution corresponds to the Minkowski spacetime.
    \item $P_1=(0,1),$ with eigenvalues $\{2,4\}.$ The deceleration parameter is $q(P_1)=0$. That means the asymptotic solution describes a universe dominated by the Gauss-Bonnet term. This point is a source.
    \item $P_2=(0,-1),$ with eigenvalues $\{-2,-4\}$ with $q(P_2)=0$. This point is a sink. The asymptotic solution is similar to that of point $P_1$.
    \item $P_3=\left(\frac{4}{3\lambda},1\right),$ with eigenvalues $\{-4,-\frac{2}{3}\}.$ This point is a sink. Moreover, $q(P_3)=0$ means that the asymptotic behaviour is similar to that of $P_1.$
    \item $P_4=\left(-\frac{4}{3\lambda},-1\right),$ with eigenvalues $\{4,\frac{2}{3}\},$ while the deceleration parameter is calculated $q(P_4)=0$. This point is a source. As before, The asymptotic solution is similar to point $P_1$.
      \item $P_5=\left(0,-\sqrt{\frac{\lambda }{\lambda -8
   f_0}}\right),$ with eigenvalues  $\left\{\frac{\sqrt{\lambda } \left(3 \sqrt{3 \lambda ^2-2}-\sqrt{51 \lambda ^2-18}\right)}{2 \sqrt{3 \lambda ^2-2}
   \sqrt{\lambda -8 f_0}},\frac{\sqrt{\lambda } \left(3 \sqrt{3 \lambda ^2-2}+\sqrt{51 \lambda ^2-18}\right)}{2
   \sqrt{3 \lambda ^2-2} \sqrt{\lambda -8 f_0}}\right\}.$ This point corresponds to a de Sitter solution, i.e. $q(P_5)=-1$.  This point exists for
   \begin{enumerate}
       \item $\lambda  <0, \frac{\lambda }{8}<f_0,$
       \item $\lambda >0, f_0<\frac{\lambda }{8}$
   \end{enumerate}
   and is a saddle. 
   
        \item $P_6=\left(0,\sqrt{\frac{\lambda }{\lambda -8
   f_0}}\right),$ is a de Sitter point that is $q(P_6)=-1,$ with eigenvalues \newline $\left\{-\frac{\sqrt{\lambda } \left(3 \sqrt{3 \lambda ^2-2}+\sqrt{51 \lambda ^2-18}\right)}{2 \sqrt{3 \lambda ^2-2}
   \sqrt{\lambda -8 f_0}},\frac{\sqrt{\lambda } \left(\sqrt{51 \lambda ^2-18}-3 \sqrt{3 \lambda ^2-2}\right)}{2
   \sqrt{3 \lambda ^2-2} \sqrt{\lambda -8 f_0}}\right\}.$ This point exists for
   \begin{enumerate}
       \item $\lambda   <0, \frac{\lambda }{8}<f_0,$
       \item $\lambda >0,
  f_0<\frac{\lambda }{8}$
   \end{enumerate}
   and is a saddle. 
   
          \item $P_7=\left(\frac{3 \sqrt{\frac{2}{5}}}{\sqrt{4 \sqrt{30} f_0+3}},\frac{\sqrt{3}}{\sqrt{4 \sqrt{30} f_0+3}}\right).$ This point exist for $f_0\geq 0,$ has eigenvalues \newline  $\{\lambda_1(\lambda,f_0),\lambda_2(\lambda,f_0)\}$ and is a 
   \begin{enumerate}
       \item sink for $\lambda<0,$
              \item saddle for $\lambda>0,$
              \item nonhyperbolic for $\lambda=0.$
   \end{enumerate} 
   Moreover, $q(P_7)=-1$ from where we infer that the asymptotic solution is that of the de Sitter universe. The numerical analysis of the real part of the eigenvalues for $P_7$ is presented in Fig.  \ref{fig:2}. 
            \item $P_8=\left(-\frac{3 \sqrt{\frac{2}{5}}}{\sqrt{4
   \sqrt{30} f_0+3}},-\frac{1}{\sqrt{4
   \sqrt{\frac{10}{3}} f_0+1}}\right)$ describes a de Sitter solution because $q(P_8)=-1$. This point exist for $f_0\geq 0,$ has eigenvalues $\{\lambda_3(\lambda,f_0),\lambda_4(\lambda,f_0)\}$ and is a 
   \begin{enumerate}
       \item source  for $\lambda<0,$
              \item saddle for $\lambda>0,$
              \item nonhyperbolic for $\lambda=0.$
              As before, the numerical analysis of the real part of the eigenvalues for $P_8$ is presented in Fig.  \ref{fig:2}.  
   \end{enumerate}
              \item $P_9=\left(\frac{3 \sqrt{\frac{2}{5}}}{\sqrt{3-4
   \sqrt{30} f_0}},-\frac{1}{\sqrt{1-4
   \sqrt{\frac{10}{3}} f_0}}\right).$
   This point exist for $f_0\leq 0,$  it describes a de Sitter solution because $q(P_9)=-1,$ has eigenvalues $\{\lambda_5(\lambda,f_0),\lambda_6(\lambda,f_0)\}$ and is a 
   \begin{enumerate}
       \item source for  $\lambda>0,$
              \item saddle for $\lambda<0,$
              \item nonhyperbolic for $\lambda=0.$ The numerical analysis of the real part of $\lambda_5$  and $\lambda_6$ for $P_9$ is presented in Fig. \ref{fig:2}.
   \end{enumerate}
                \item Finally, the de Sitter point $P_{10}=\left(-\frac{3 \sqrt{\frac{2}{5}}}{\sqrt{3-4 \sqrt{30} f_0}},\frac{1}{\sqrt{1-4 \sqrt{\frac{10}{3}}
   f_0}}\right).$ This point exist for $f_0\leq 0,$ has eigenvalues $\{\lambda_7(\lambda,f_0),\lambda_8(\lambda,f_0)\}$ and is a
   \begin{enumerate}
       \item sink for $\lambda>0,$
              \item saddle for $\lambda<0,$
              \item nonhyperbolic for $\lambda=0.$
              As before, the numerical analysis of the real part of $\lambda_7$  and $\lambda_8$ for $P_{10}$ is presented in Fig. \ref{fig:2}.
   \end{enumerate}
\end{enumerate}

\begin{figure}[ht!]
    \centering
    \includegraphics[scale=0.45]{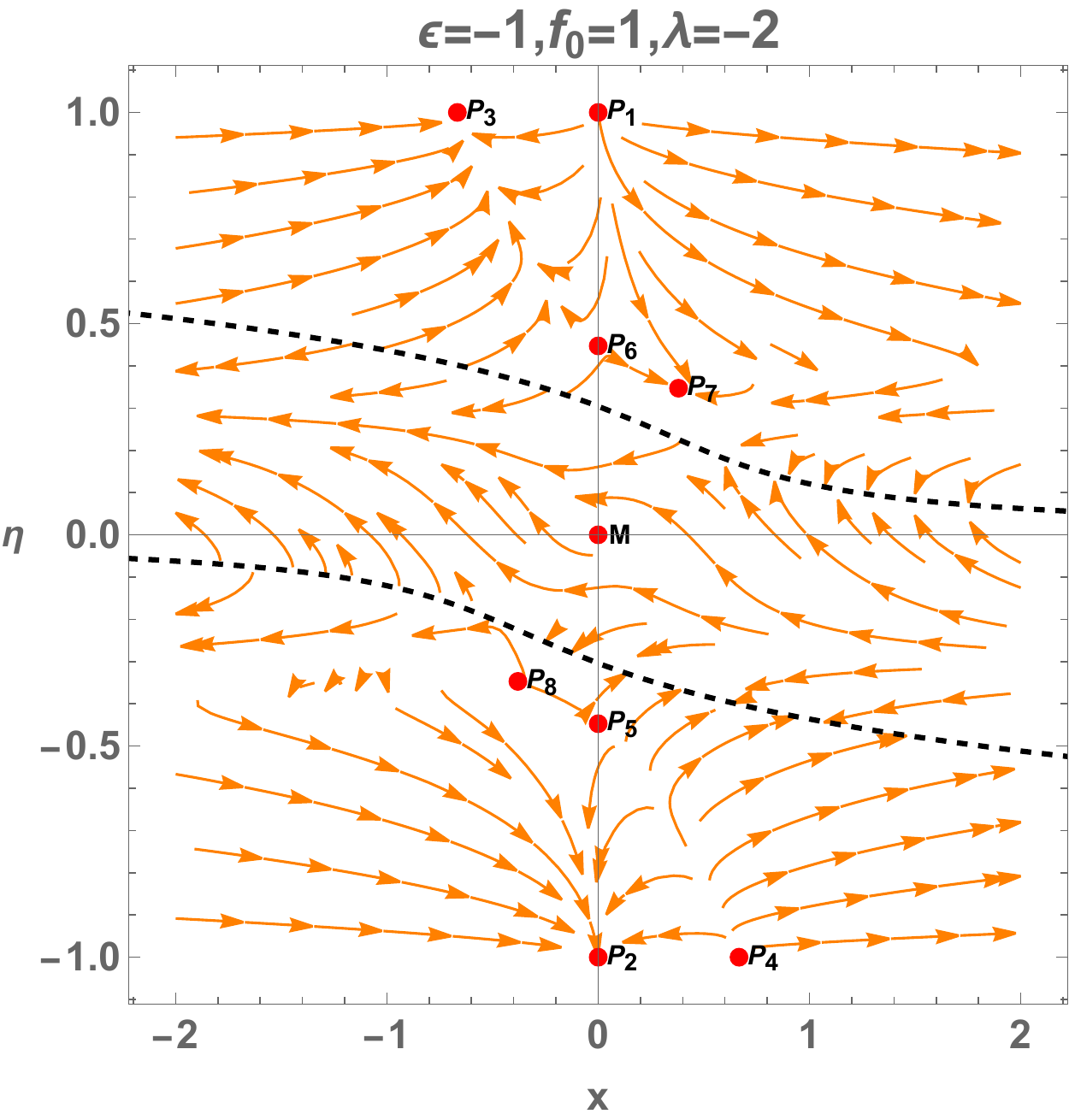}
    \includegraphics[scale=0.45]{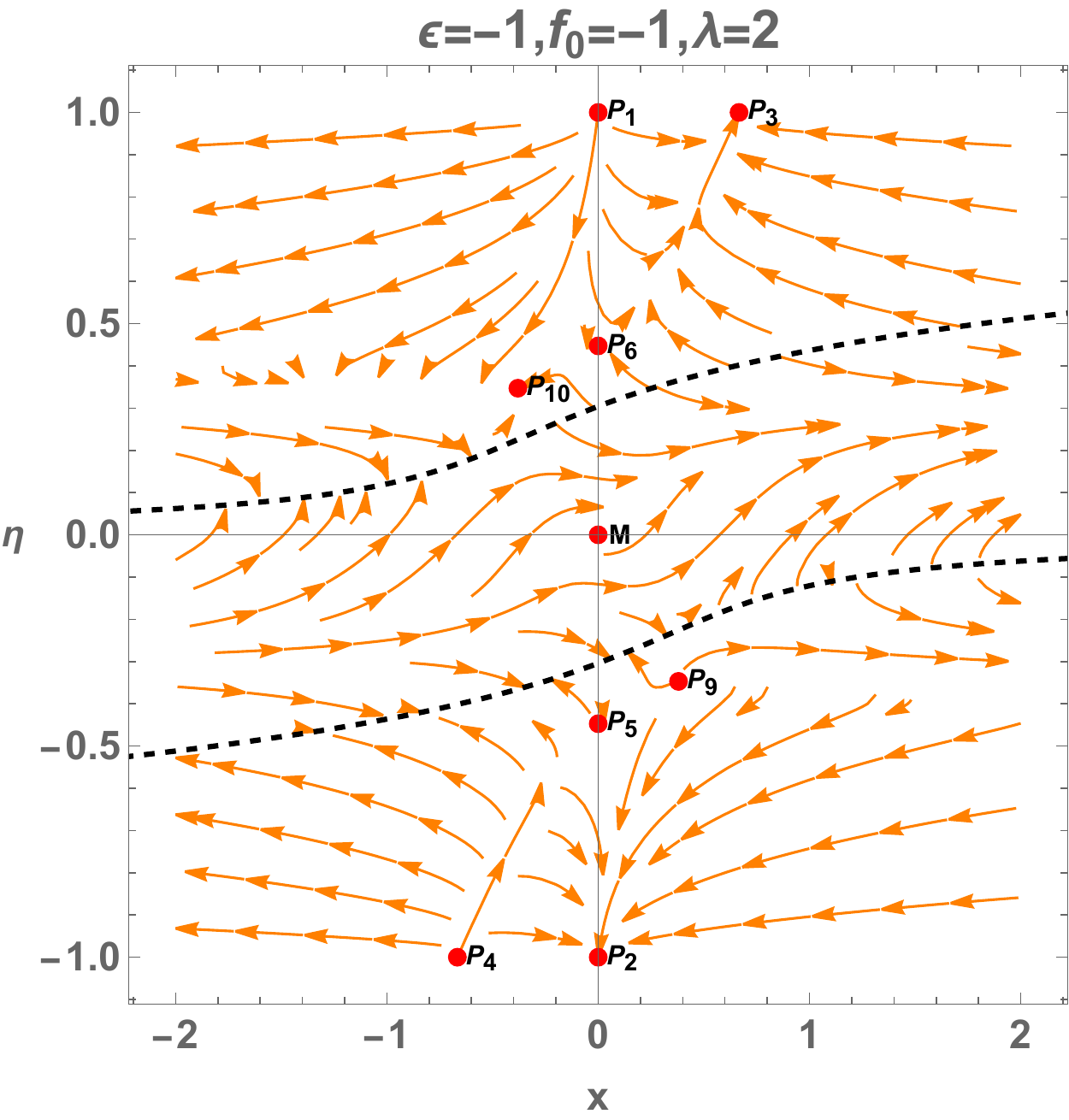}
    \includegraphics[scale=0.45]{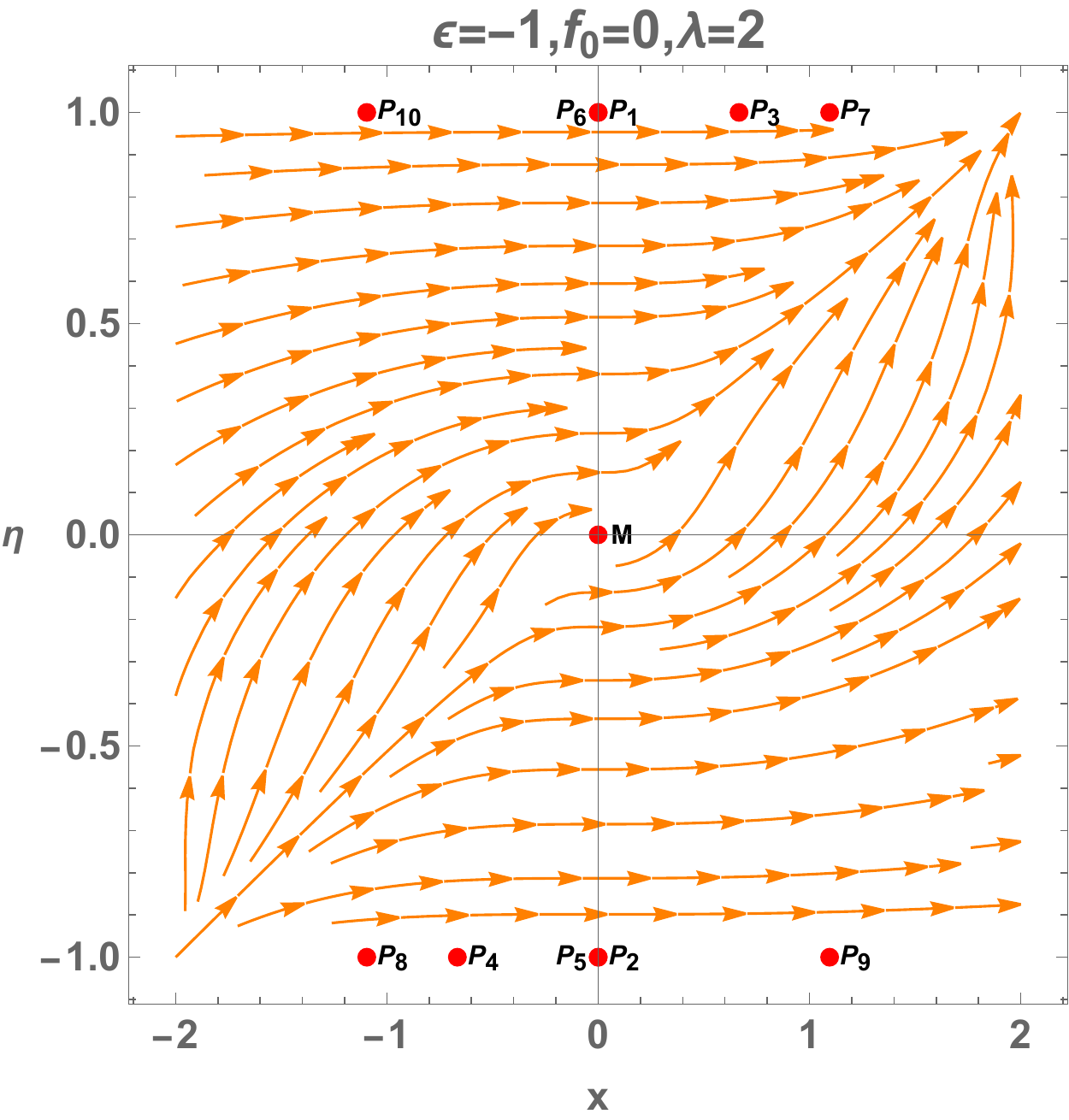}
    \includegraphics[scale=0.45]{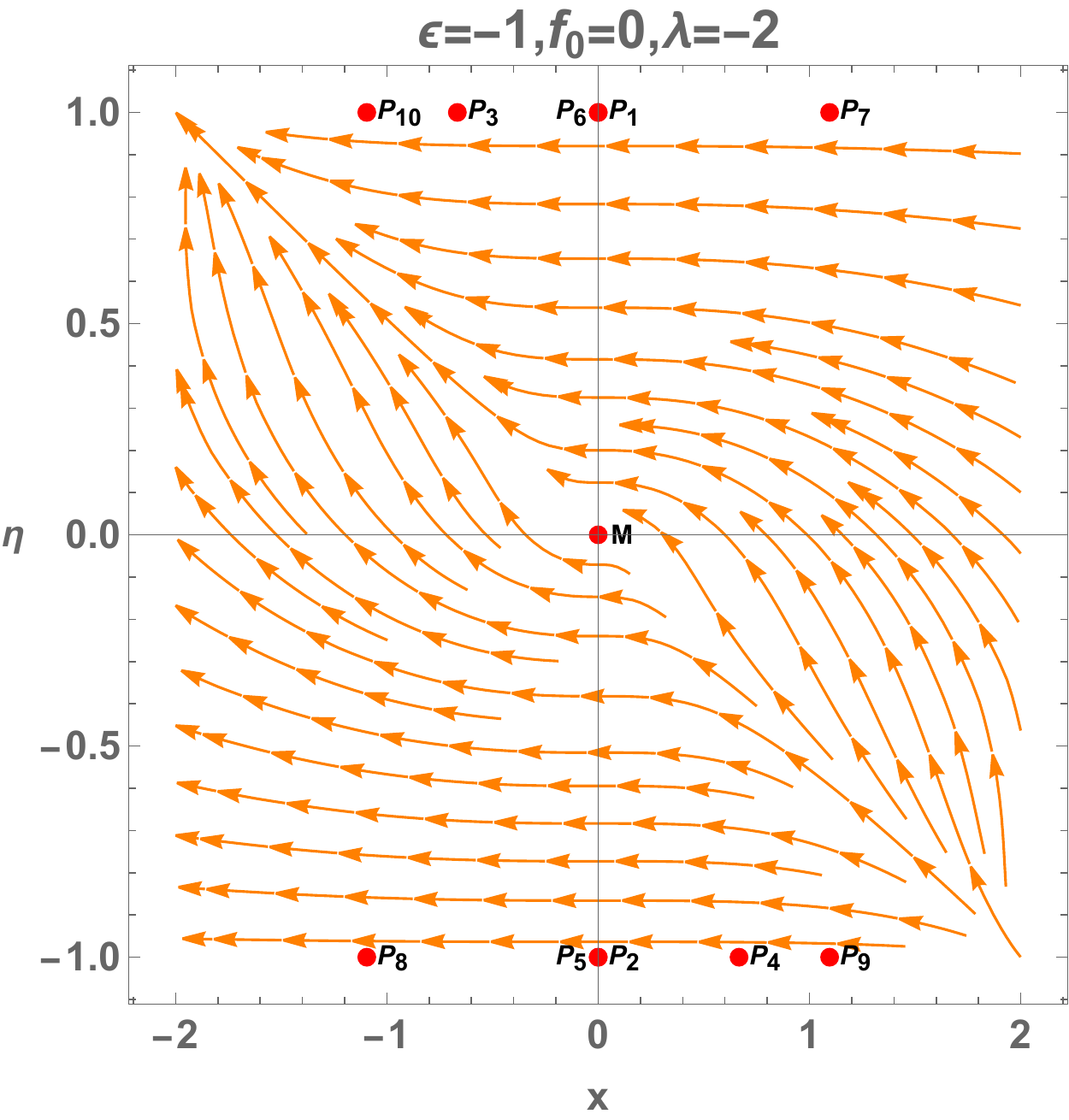}
    \caption{Phase plots for \eqref{linearf2D-1}, \eqref{linearf2D-2} for $\epsilon=-1$ and different values of $f_0$ and $\lambda.$ The dashed black lines in the plot correspond to the values of $x$ and $y$ for which $K=0,$ which corresponds to singular curves where the flow direction and the stability changes.}
    \label{fig:3}
\end{figure}
\begin{table}[ht!]
    \caption{Equilibrium points of system \eqref{linearf2D-1}, \eqref{linearf2D-2}  for $\epsilon=-1$ with their stability conditions. It also includes the value of $\omega_{\phi}$ and $q.$}
    \label{tab:2}
\newcolumntype{C}{>{\centering\arraybackslash}X}
\centering
    \setlength{\tabcolsep}{2.6mm}
\begin{tabularx}{\textwidth}{cccccc}
\toprule 
  \text{Label}  &  $x$ & $\eta$ & \text{Stability}& $\omega_{\phi}$ & $q$\\\midrule  
  $M$ & $0$ & $0$ & nonhyperbolic & \text{indeterminate}&\text{indeterminate}\\\midrule
  $P_{1}$ & $0$ & $1$  & $\text{source}$& $-\frac{1}{3}$& $0$\\\midrule
  $P_{2}$ & $0$ & $-1$ & $\text{sink}$& $-\frac{1}{3}$& $0$\\\midrule
  $P_{3}$ & $\frac{4}{3\lambda}$ & $1$ & $\text{sink}$& $-\frac{1}{3}$& $0$\\\midrule
  $P_{4}$ & $-\frac{4}{3\lambda}$ & $-1$  & $\text{source}$& $-\frac{1}{3}$& $0$\\\midrule
  $P_{5}$ & $0$ & $\sqrt{\frac{\lambda}{\lambda-8f_0}}$ & $\text{saddle}$& $-1$& $-1$\\\midrule
  $P_{6}$ & $0$ & $-\sqrt{\frac{\lambda}{\lambda-8f_0}}$ & $\text{saddle}$& $-1$& $-1$\\\midrule
  $P_{7}$ & $\frac{3 \sqrt{\frac{2}{5}}}{\sqrt{4\sqrt{30} f_0+3}}$ & $\frac{1}{\sqrt{4\sqrt{\frac{10}{3}} f_0+1}}$   
  &            sink for $\lambda<0$ & & \\
           &&& saddle for $\lambda>0$ && \\
           &&& nonhyperbolic for $\lambda=0$& $-1$& $-1$\\\midrule
  $P_{8}$ & $-\frac{3 \sqrt{\frac{2}{5}}}{\sqrt{4\sqrt{30} f_0+3}}$ & $-\frac{1}{\sqrt{4\sqrt{\frac{10}{3}} f_0+1}}$  
    &   source  for $\lambda<0$ && \\
    &&& saddle for $\lambda>0$ && \\
    &&& nonhyperbolic for $\lambda=0$& $-1$& $-1$\\\midrule
  $P_{9}$ & $\frac{3 \sqrt{\frac{2}{5}}}{\sqrt{3-4 \sqrt{30} f_0}}$ & $-\frac{1}{\sqrt{1-4 \sqrt{\frac{10}{3}} f_0}}$  
           &  source for  $\lambda>0$ && \\
           &&& saddle for $\lambda<0$ && \\
           &&& nonhyperbolic for $\lambda=0$ & $-1$& $-1$\\\midrule
  $P_{10}$ & $-\frac{3 \sqrt{\frac{2}{5}}}{\sqrt{3-4 \sqrt{30} f_0}}$ & $\frac{1}{\sqrt{1-4 \sqrt{\frac{10}{3}} f_0}}$  
          &      sink for $\lambda>0$ && \\
          &&& saddle for $\lambda<0$ && \\
          &&& nonhyperbolic for $\lambda=0$ & $-1$& $-1$\\\bottomrule
    \end{tabularx}
\end{table}

The above results are summarized in Table \ref{tab:2}. Phase-space diagrams for the dynamical system \eqref{linearf2D-1}, \eqref{linearf2D-2} where the scalar field is a phantom field, that is, $\epsilon=-1$ are presented in Fig. \ref{fig:3} for various values of the free parameters. 
\subsection{Analysis of system \eqref{linearf2D-1}-\eqref{linearf2D-2} at infinity}

The numerical results presented in Figure  \ref{fig:2} and \ref{fig:3} suggest that there are non-trivial dynamics when $x \rightarrow \pm \infty$. For that reason, we introduce the compactified variable 
\begin{equation}
    u= \frac{x}{\sqrt{1 + x^2}}, 
\end{equation}
and the new time variable 
\begin{equation}
  {f}^{\prime} =  \sqrt{1 - u^2} \frac{df}{d \tau},
\end{equation}
we obtain the compactified dynamical system 
\begin{align}
  u^{\prime} =   \frac{1}{L} \left(1-u^2\right) &\Bigg(u^3 \Big(\eta ^5 \left(-576 f_0^2+4 f_0 \lambda  (\epsilon +18)+\epsilon  (\epsilon +6)\right)+24 f_0 (\lambda -8 f_0)
   \eta ^7 \nonumber \\
   & -2 \eta ^3 (\epsilon  (\epsilon +6)-2 f_0 \lambda  (\epsilon -24))+\epsilon  \eta  (-8 f_0 \lambda +\epsilon +6)\Big) \nonumber \\
   & +\sqrt{1-u^2} u^2 \big(\eta ^4
   \left(-384 f_0^2 \lambda +8 f_0 (6-5 \epsilon )+\lambda  (\epsilon -12)\right) \nonumber \\
   & -2 \eta ^6 (8 f_0 (12 f_0 \lambda +2 \epsilon +3)-3 \lambda )+\eta ^2 (72 f_0 \epsilon -2 \lambda  (\epsilon -3))+\lambda  \epsilon \big) \nonumber \\
   & -6 \eta ^2 \left(\eta ^2-1\right) \sqrt{1-u^2} \left((\lambda -8 f_0) \eta ^2-\lambda \right) \nonumber \\
   & -6 \eta  u \left(4 f_0 (\lambda -8 f_0) \eta ^6+\eta ^4 (12 f_0 (\lambda -8 f_0)+\epsilon )   -2 \eta ^2 (8 f_0 \lambda +\epsilon )+\epsilon
   \right)\Bigg), \label{compact_1-1}\\
  \eta^{\prime} = \frac{1}{L} \left(\eta ^2-1\right) &\Bigg(24 f_0 \eta ^4 \left((8 f_0-\lambda ) \eta ^2+\lambda \right) -32 f_0 \eta ^3 u \sqrt{1-u^2} \left(\eta ^2 (6 f_0 \lambda +\epsilon )-\epsilon \right) \nonumber \\
  & +u^2 \Big(24 f_0 (\lambda -8 f_0) \eta ^6+\eta ^4 (4 f_0 \lambda  (\epsilon -6)+1)  -2 \epsilon \eta ^2 (2 f_0 \lambda +\epsilon )+1\Big)\Bigg), \label{compact_1-2}
\end{align}
where 
$L= 2 \left(\sqrt{1-u^2} \left(\left(96 f_0^2+\epsilon \right) \eta ^4-2 \epsilon  \eta ^2+\epsilon \right)+8 f_0 \epsilon  \eta  \left(\eta ^2-1\right) u\right).$
The limits $u\rightarrow \pm 1$ corresponds to $x \rightarrow \pm \infty$. 

The equilibrium points of system  \eqref{compact_1-1} and \eqref{compact_1-2} at the finite region as the same of  \eqref{linearf2D-1}, \eqref{linearf2D-2} by the rescaling 
$x \mapsto x/\sqrt{1+x^2}$. The points at infinity are those satisfying $u=\pm 1$, say 

\begin{enumerate}
  \item $Q_1=(1,1)$, with eigenvalues $\{-2 \lambda,6 \lambda \}.$ This point is a saddle or nonhyperbolic for $\lambda=0.$ The value of the deceleration parameter is $q(Q_1)=0$. That means the asymptotic solution describes a universe dominated by the Gauss-Bonnet term.
    \item $Q_2=(1,-1)$, with eigenvalues $\{-2 \lambda,6 \lambda \}.$  This point is a saddle or nonhyperbolic for $\lambda=0.$ The value of the deceleration parameter is $q(Q_2)=0.$ The asymptotic behaviour is the same as $Q_1.$
    \item $Q_3=(-1,1)$, with eigenvalues $\{2 \lambda, -6 \lambda \}.$  This point is a saddle or nonhyperbolic for $\lambda=0.$ The value of the deceleration parameter is $q(Q_3)=0.$ The asymptotic behaviour is the same as $Q_1.$
    \item $Q_4=(-1,-1)$, with eigenvalues $\{2 \lambda, -6 \lambda \}.$  This point is a saddle or nonhyperbolic for $\lambda=0.$ The value of the deceleration parameter is $q(Q_4)=0.$ The asymptotic behaviour is the same as $Q_1.$
    \item $Q_5= (1,  \frac{1}{\sqrt{4 f_0\lambda  \epsilon +1}})$, with eigenvalues $\left\{-\lambda,-\frac{\lambda }{2}\right\}$. This point is 
    \begin{enumerate}
        \item a source for $\lambda \geq 0,$
        \item a sink for $\lambda\leq 0,$
        \item nonhyperbolic for $\lambda=0.$ 
    \end{enumerate} 
    Note that for 
    \begin{enumerate}
        \item $\epsilon=1,$ the point has $\eta=\frac{1}{\sqrt{4 f_0\lambda   +1}}.$ This point exists for $\lambda <0, f_0\leq 0$ or $ \lambda =0$ or $ \lambda >0,  f_0\geq 0.$
        \item $\epsilon=-1,$ the point has $\eta=\frac{1}{\sqrt{-4 f_0\lambda   +1}}.$ This point exists for $\lambda <0, f_0\geq 0$ or $ \lambda =0$ or $ \lambda >0,  f_0\leq 0.$
    \end{enumerate}
    The value of the deceleration parameter is $q(Q_5)=-1.$ The asymptotic solution is a de Sitter universe.
    \item $Q_6= (-1,  \frac{1}{\sqrt{4 f_0\lambda  \epsilon +1}})$, with eigenvalues $\left\{\lambda, \frac{\lambda }{2}\right\}$. This point is 
    \begin{enumerate}
        \item a sink for $\lambda \geq 0,$
        \item a source for $\lambda\leq 0,$
        \item nonhyperbolic for $\lambda=0.$  
        \end{enumerate}
        The existence conditions for $\epsilon=\pm 1$ are the same as $Q_5.$ The value of the deceleration parameter is $q(Q_6)=-1.$ The asymptotic behaviour is the same as $Q_5.$
    \item $Q_7= (1,  -\frac{1}{\sqrt{4 f_0\lambda  \epsilon +1}})$, with eigenvalues $\left\{-\lambda,-\frac{\lambda }{2}\right\}$. This point is 
    \begin{enumerate}
        \item a source for $\lambda \geq 0,$
        \item a sink for $\lambda\leq 0,$
        \item nonhyperbolic for $\lambda=0.$ 
        The existence conditions for $\epsilon=\pm 1$ are the same as $Q_5.$ The value of the deceleration parameter is $q(Q_7)=-1.$ The asymptotic behaviour is the same as $Q_5.$
        \end{enumerate}
    \item $Q_8= (-1,  -\frac{1}{\sqrt{4 f_0\lambda  \epsilon +1}})$, with eigenvalues $\left\{\lambda, \frac{\lambda }{2}\right\}$. \begin{enumerate}
        \item a sink for $\lambda \geq 0,$
        \item a source for $\lambda\leq 0,$
        \item nonhyperbolic for $\lambda=0.$  
        \end{enumerate}
  The existence conditions for $\epsilon=\pm 1$ are the same as $Q_5.$ The value of the deceleration parameter is $q(Q_8)=-1.$ The asymptotic behaviour is the same as $Q_5.$
\end{enumerate}

\begin{figure}[ht!]
    \centering
    \includegraphics[scale=0.45]{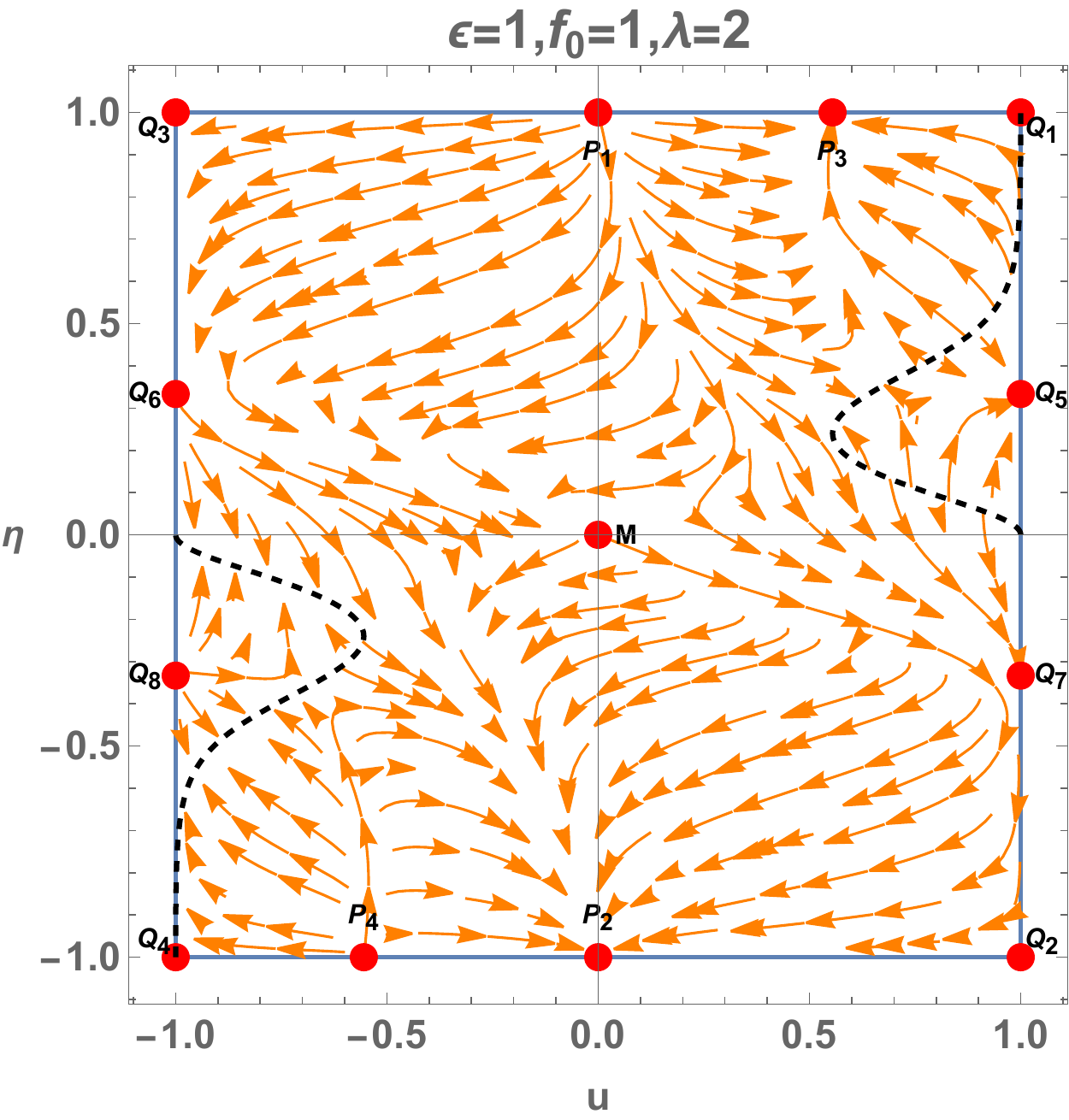}
    \includegraphics[scale=0.45]{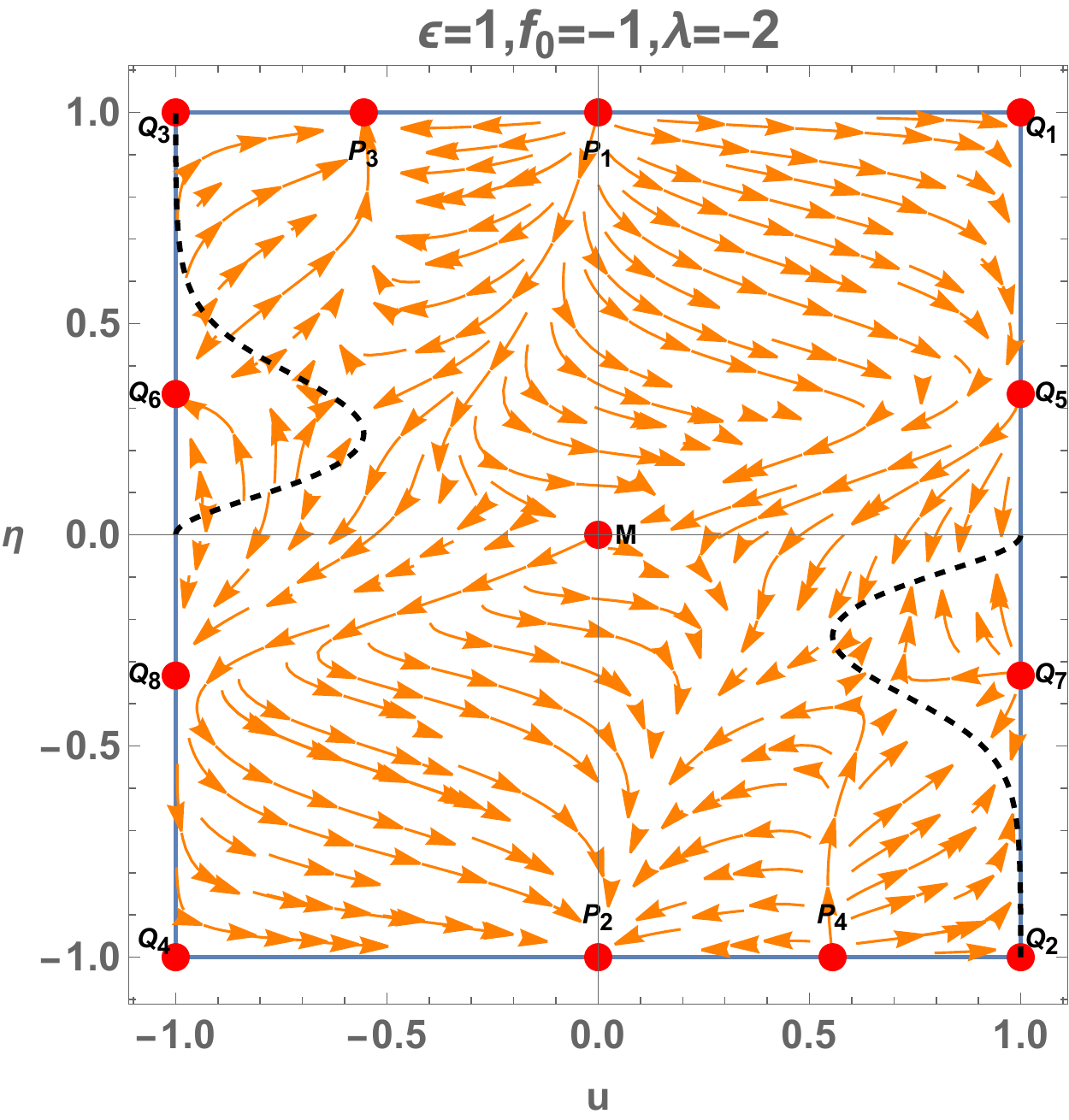}
    \includegraphics[scale=0.45]{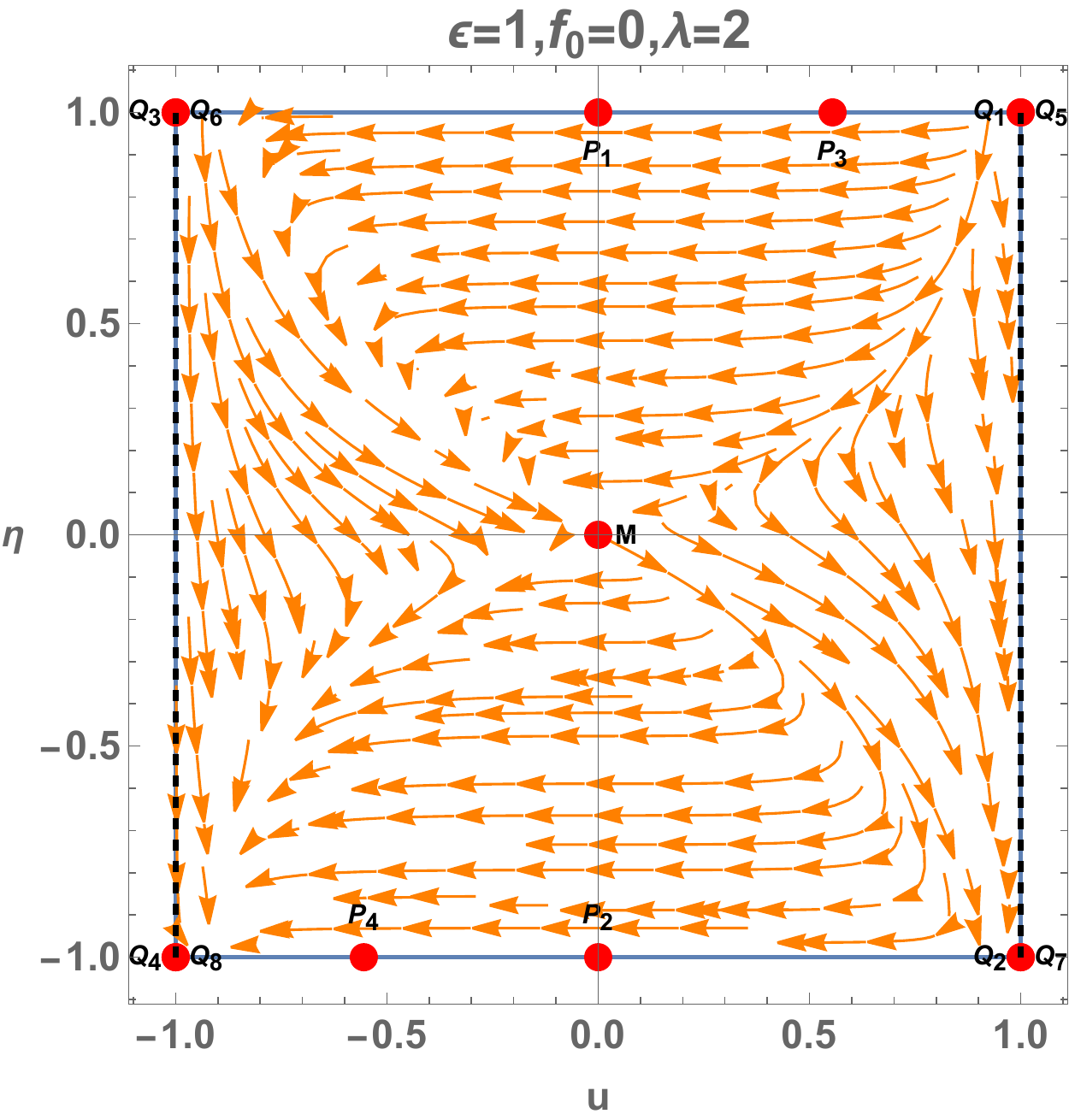}
    \includegraphics[scale=0.45]{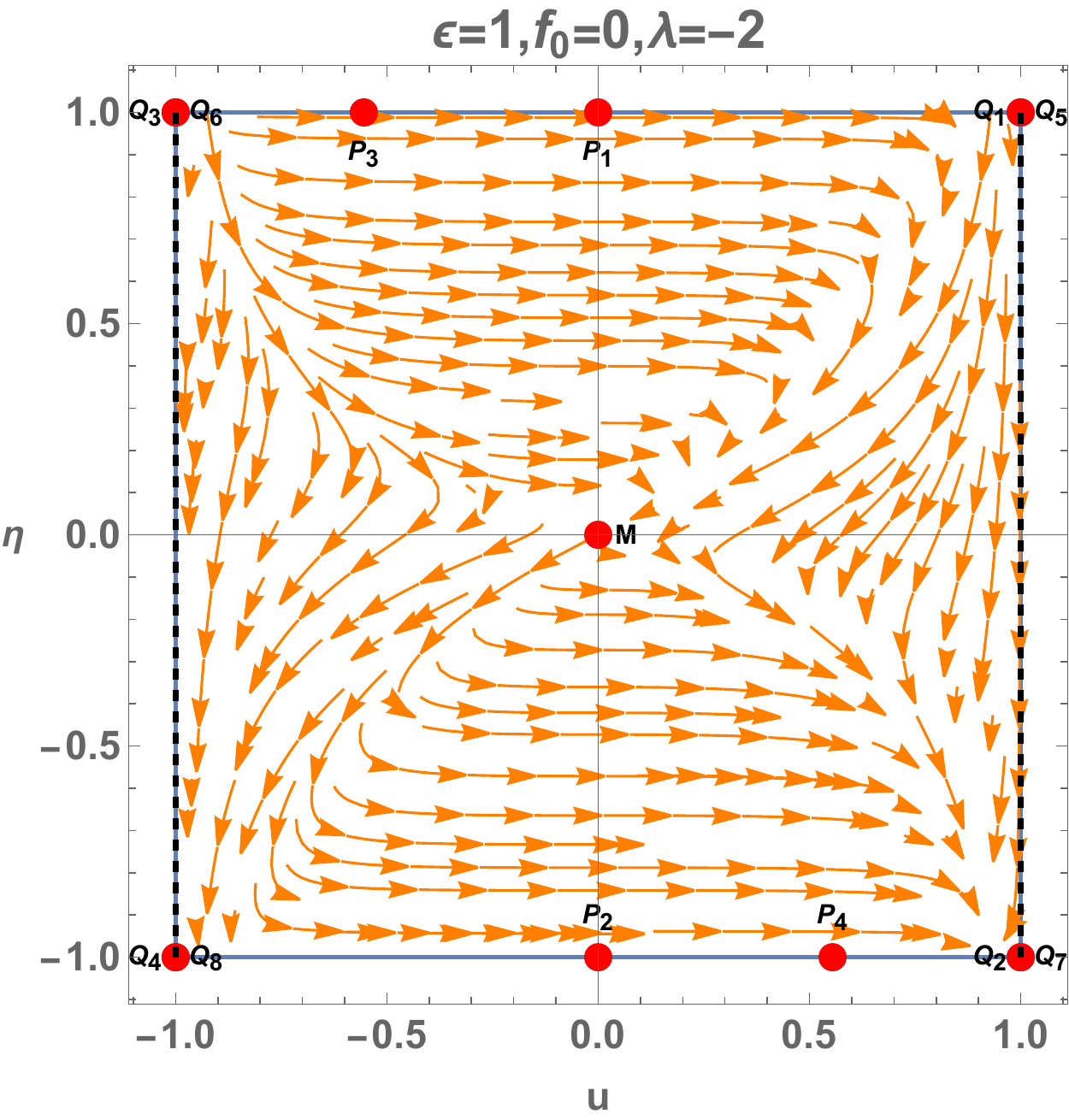}
    \caption{Phase plots for system \eqref{compact_1-1}, \eqref{compact_1-2} for $\epsilon=1$ and different values of $f_0$ and $\lambda.$ The dashed black lines in the plot correspond to the values of $u$ and $\eta$ for which $L=0,$ which corresponds to singular curves where the flow direction and the stability changes.}
    \label{fig:4}
\end{figure}

\begin{figure}[ht!]
    \centering
    \includegraphics[scale=0.45]{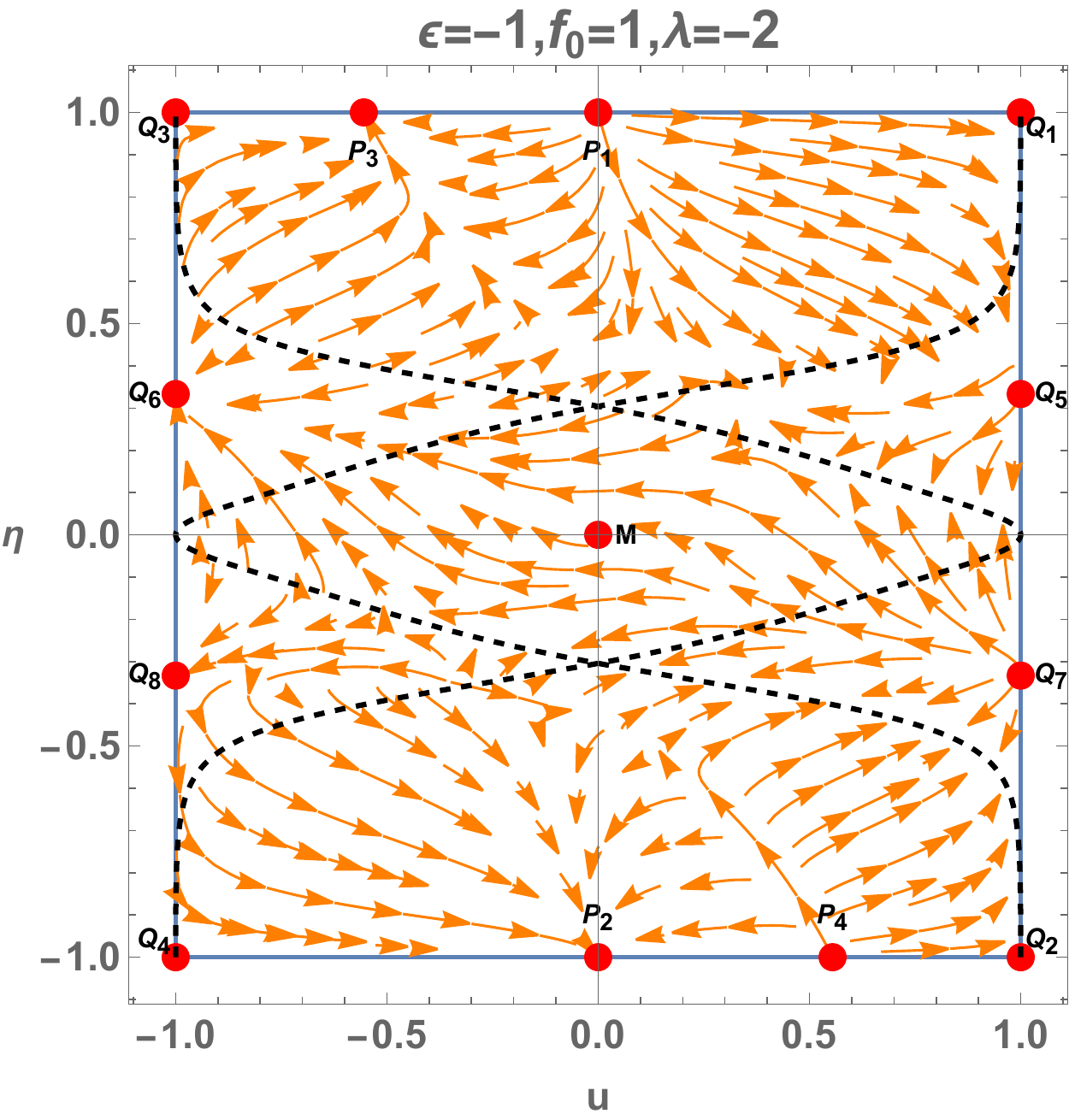}
    \includegraphics[scale=0.45]{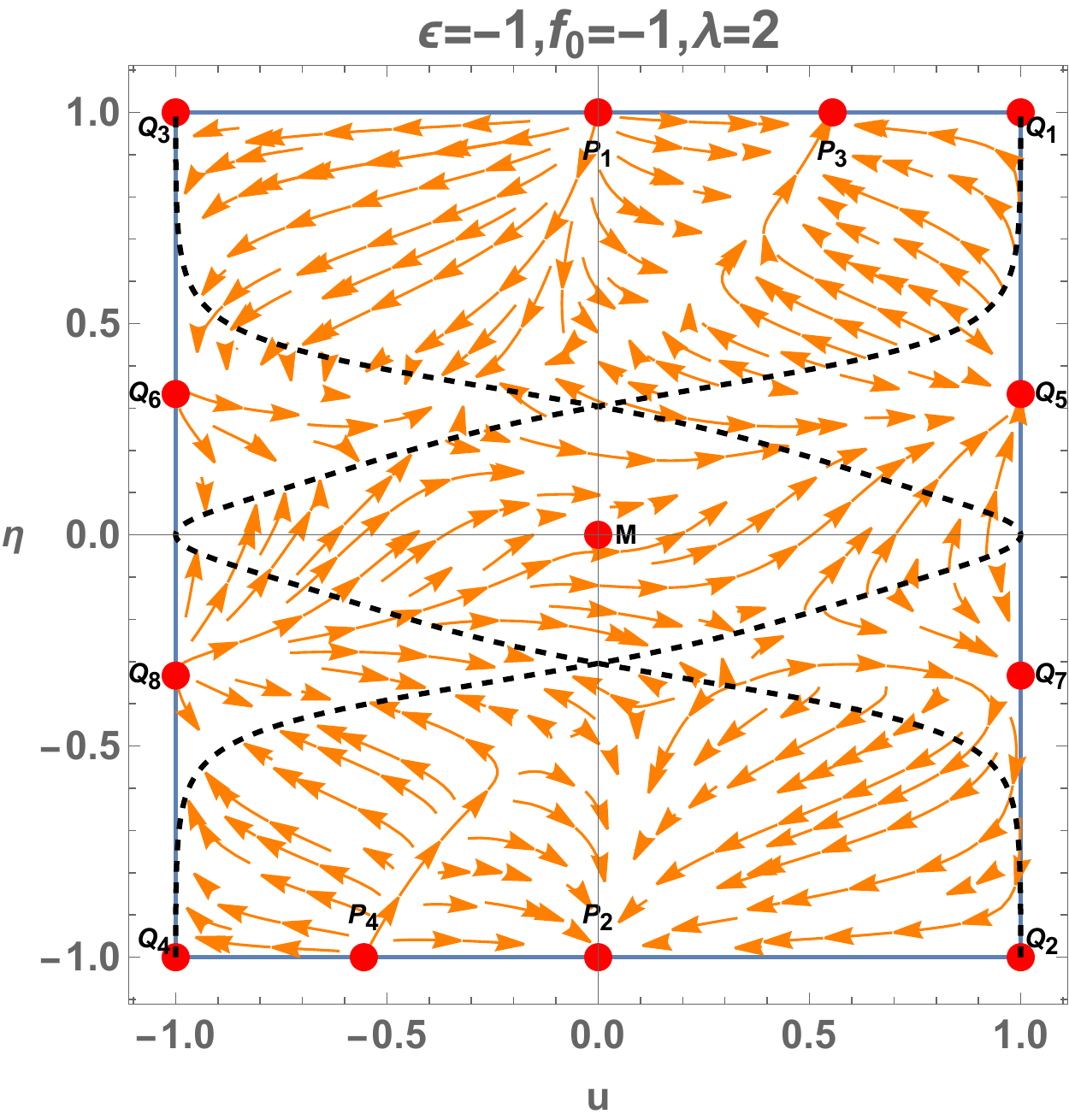}
    \includegraphics[scale=0.45]{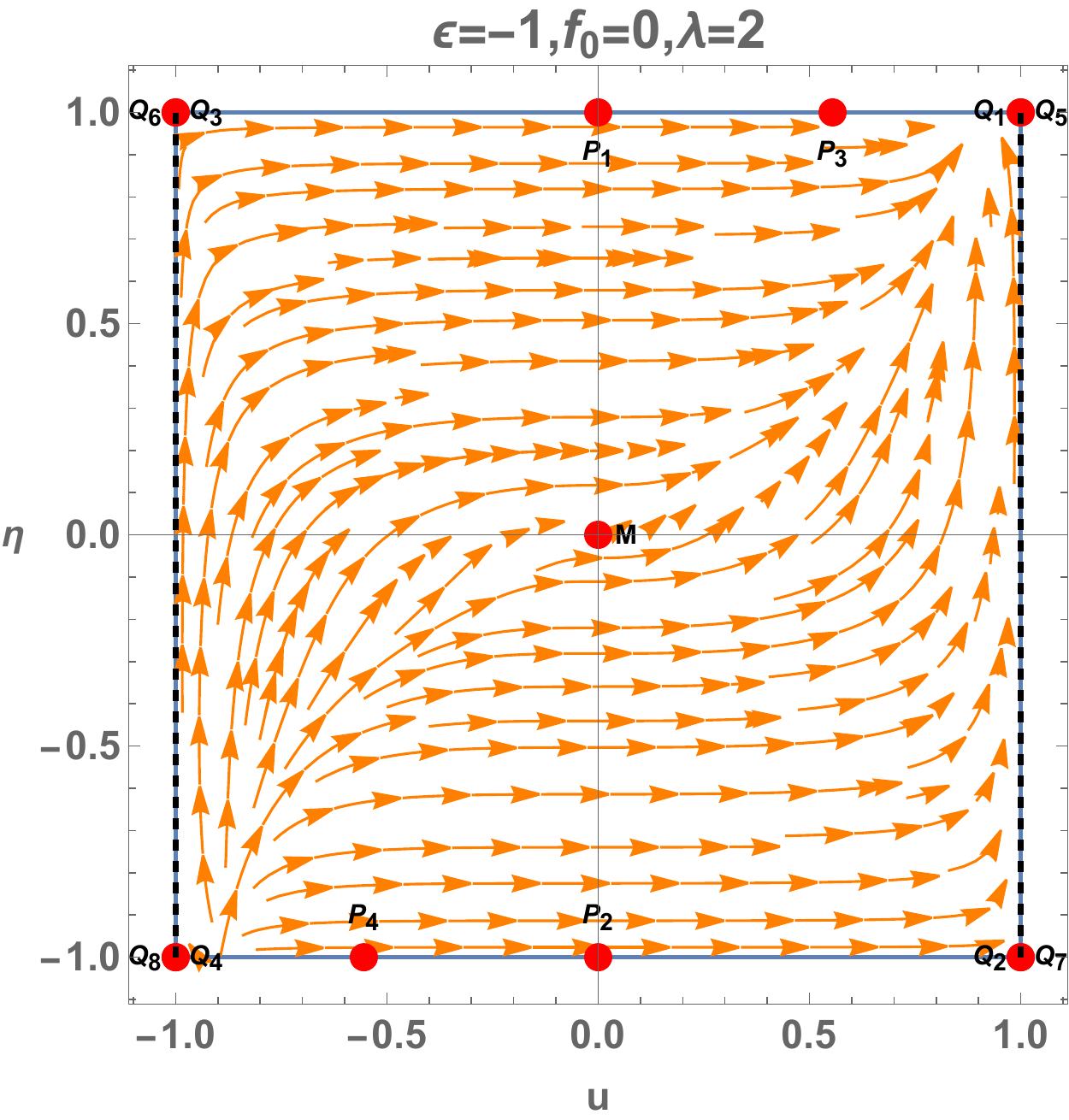}
    \includegraphics[scale=0.45]{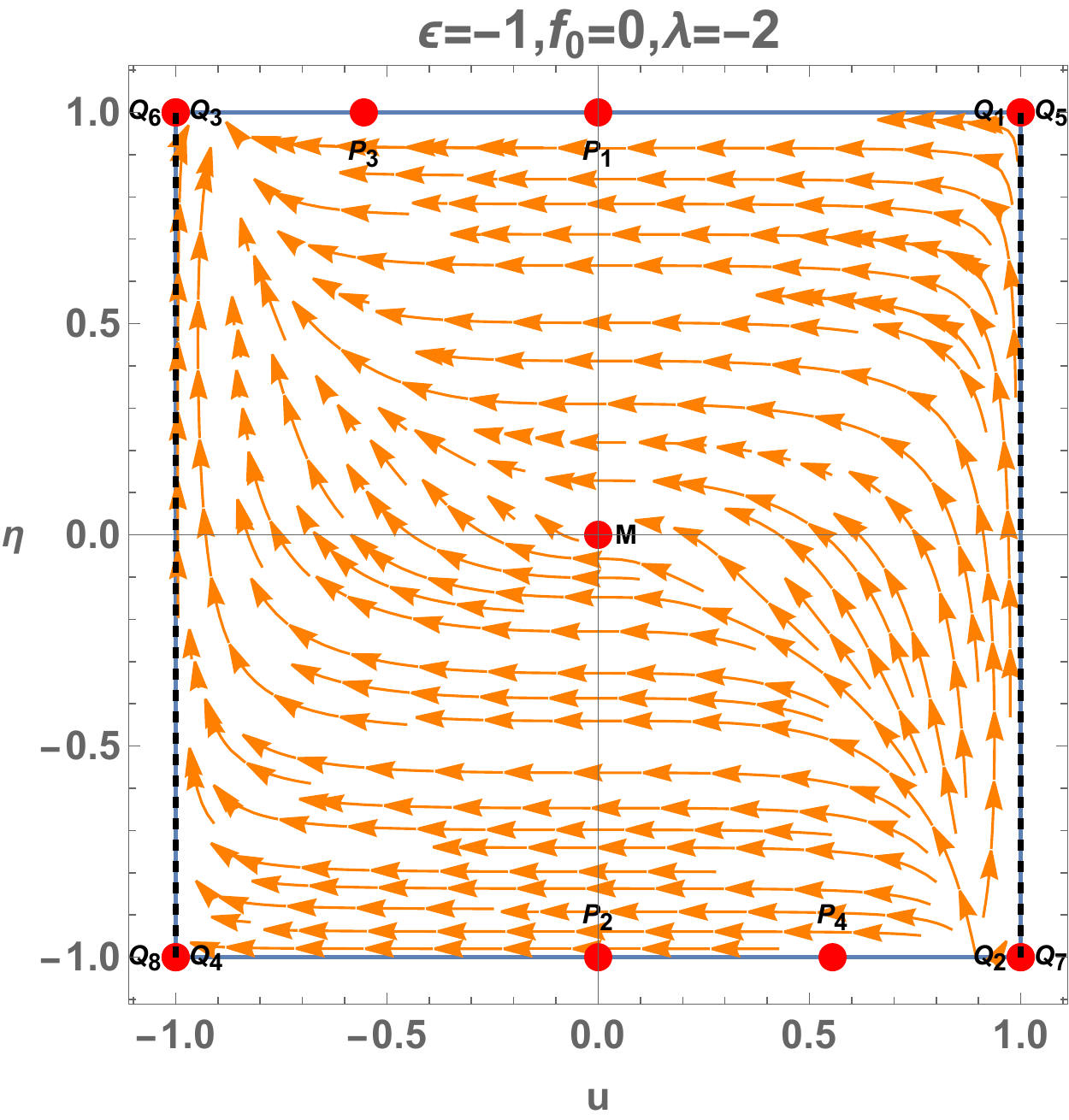}
    \caption{Phase plots for system \eqref{compact_1-1}, \eqref{compact_1-2} for $\epsilon=-1$ and different values of $f_0$ and $\lambda.$ The dashed black lines in the plot correspond to the values of $u$ and $\eta$ for which $L=0,$ which corresponds to singular curves where the flow direction and the stability changes.}
    \label{fig:5}
\end{figure}

The phase-space of the field equations at the new Poincare variables is presented in Figs. \ref{fig:4} and \ref{fig:5} for different values of the free parameters. 
As far as the physical properties of the asymptotic solutions are concerned, we find that $Q_1$, $Q_2$, $Q_3$ and $Q_4$ are Gauss-Bonnet points with deceleration parameter $q=0$, while points $Q_5$, $Q_6$, $Q_7$ and $Q_8$ are de Sitter points with $q=-1$.

\begin{table}[ht!]
    \caption{Equilibrium points of system (\ref{compact_1-1}), (\ref{compact_1-2})  for $\epsilon=\pm 1$ with their  stability conditions.}
    \label{tab:3}
    \centering
\newcolumntype{C}{>{\centering\arraybackslash}X}
\centering
    \setlength{\tabcolsep}{2.6mm}
\begin{tabularx}{\textwidth}{cccccc}
\toprule 
\text{Label}  &  $u$ & $\eta$ & \text{Stability}& $\omega_{\phi}$ & $q$\\\midrule  
$Q_1$ &$1$& $1$& saddle for $\lambda\neq 0$, nonhyperbolic for $\lambda=0$ & $-\frac{1}{3}$& $0$\\\midrule
$Q_{2}$&$1$& $-1$& saddle for $\lambda \neq 0$, nonhyperbolic for $\lambda=0$ & $-\frac{1}{3}$& $0$\\\midrule$Q_{3}$&$-1$&$1$& saddle for $\lambda \neq 0$, nonhyperbolic for $\lambda=0$ & $-\frac{1}{3}$& $0$\\\midrule$Q_{4}$&$-1$& $-1$& saddle for $\lambda \neq 0$, nonhyperbolic for $\lambda=0$ & $-\frac{1}{3}$& $0$\\\midrule$Q_{5}$ & $1$ & $\frac{1}{\sqrt{4 f_0\lambda  \epsilon +1}}$ 
& sink for $\lambda>0$, source for $\lambda<0$, nonhyperbolic for $\lambda=0$  & $-1$& $-1$\\\midrule
  $Q_{6}$ & $-1$ & $\frac{1}{\sqrt{4 f_0\lambda  \epsilon +1}}$  
   & source for $\lambda>0$, sink for $\lambda<0$ , nonhyperbolic for $\lambda=0$ & $-1$& $-1$\\\midrule
   $Q_{7}$ & $1$ & $-\frac{1}{\sqrt{4 f_0\lambda  \epsilon +1}}$ 
   & sink for $\lambda>0$, source for $\lambda<0$, nonhyperbolic for $\lambda=0$ & $-1$& $-1$\\\midrule 
   $Q_{8}$ & $-1$ & $-\frac{1}{\sqrt{4 f_0\lambda  \epsilon +1}}$  
   & source for $\lambda>0$, sink for $\lambda<0$, nonhyperbolic for $\lambda=0$ & $-1$& $-1$\\\bottomrule
    \end{tabularx}
\end{table}
\section{Phase space Analysis for exponential $f$: $f(\phi)=f_0e^{\zeta\phi}$}
\label{IV}

The field equations for the exponential coupling $f(\phi)=f_0e^{\zeta\phi}$ are given by the following expressions
\begin{align}
& -48 H^3 \dot{\phi} f'(\phi )+6 H^2-2 V(\phi )-\epsilon  \dot{\phi}^2=0, \\ 
& -16 H \dot{H} \dot{\phi} f'(\phi )-16 H^3 \dot{\phi} f'(\phi )+H^2 \left(-8 \dot{\phi}^2 f''(\phi )-8 \ddot{\phi} f'(\phi
   )+3\right)+2 \dot{H}-V(\phi )+\frac{1}{2} \epsilon  \dot{\phi}^2=0,\\
& 3 H \left(-8 H \left(\dot{H}+H^2\right)
   f'(\phi )-\epsilon  \dot{\phi}\right)-V'(\phi )-\epsilon  \ddot{\phi}=0,
\end{align}
where the dot means derivative with respect to $t$ and the comma means derivative with respect the argument of the function. 

Defining the normalized variables 
\begin{equation}
    {x}=\frac{\dot{\phi}}{\sqrt{1+H^2}}, \; {y}=  \frac{\sqrt{V(\phi)}}{\sqrt{1+H^2}}, \; \eta=\frac{H}{\sqrt{1+H^2}},\; z=\frac{H^3 f'(\phi )}{\sqrt{H^2+1}},
\end{equation}
We can write the Friedmann equation as
\begin{equation}\label{Friedmann-new-var2}
  6 \eta ^2-x (x+48 z)-2 y^2=0. 
\end{equation}
Using equation \eqref{Friedmann-new-var2} we define $z$ as
\begin{equation}
    z=\frac{6 \eta ^2+x^2-2 y^2}{48 x},
\end{equation}
and the dynamical system is given by 
\begin{align}
     \frac{dx}{d\tau}= &\frac{2 x }{\Tilde{K}(x,y,\eta,\zeta,\epsilon)} \Bigg\{\zeta  \left(2 \eta ^2-1\right) x^5+5 \eta  \left(\eta ^2-3\right) x^4 \nonumber \\
     & -4 x \left(9 \zeta  \eta ^4+y^4
   \left(\zeta +2 \lambda +\lambda  \eta ^2\right)-3 y^2 \left(2 \zeta  \eta ^2+\lambda  \eta ^4\right)\right) \nonumber \\
   & +2 \epsilon  x^3
   \left(-6 \zeta  \eta ^4+6 \zeta  \eta ^2+y^2 \left((2 \zeta -\lambda ) \eta ^2-2 (\zeta +\lambda )\right)\right) \nonumber \\
   & +12 \epsilon  \eta
    x^2 \left(3 \eta ^4+\eta ^2-\left(\eta ^2+2\right) y^2\right)+4 \eta  \left(y^2-3 \eta ^2\right) \left(\left(\eta
   ^2+3\right) y^2-3 \left(\eta ^4+\eta ^2\right)\right)\Bigg\}, \label{exponentialf-1}
\\
  \frac{dy}{d\tau}=  & \frac{y}{\Tilde{K}(x,y,\eta,\zeta,\epsilon)}\Bigg\{72 \eta ^7+12 \eta ^4 x \left(\lambda  \left(2
   y^2+3\right)-2 \zeta  \epsilon  x^2\right) \nonumber \\
   & +4 \eta ^2 x \left(\epsilon  x^2+2 y^2\right) \left(\zeta  \epsilon  x^2-\lambda 
   \left(y^2+3\right)\right)  +24 \eta ^5 \left(3 \epsilon  x^2-2 y^2\right) \nonumber\\
   & +2 \eta ^3 \left(-12 \epsilon  x^2 y^2+5 x^4+4
   y^4\right)+\lambda  x \left(\epsilon  x^2+2 y^2\right) \left(5 \epsilon  x^2+2 y^2\right)\Bigg\}, \label{exponentialf-2}
\\
   \frac{d\eta}{d\tau}=  &\frac{2 \eta  \left(\eta
   ^2-1\right)}{\Tilde{K}(x,y,\eta,\zeta,\epsilon)}\Bigg\{2 \zeta  x^5+5 \eta  x^4-4 \lambda  x y^2 \left(y^2-3 \eta ^2\right) \nonumber\\
   & +2 \epsilon  x^3 \left((2 \zeta
   -\lambda ) y^2-6 \zeta  \eta ^2\right)-12 \epsilon  \eta  x^2 \left(y^2-3 \eta ^2\right)+4 \eta  \left(y^2-3 \eta
   ^2\right)^2\Bigg\}. \label{exponentialf-3}
\end{align}
where $\Tilde{K}=24 \epsilon  x^2 \left(y^2-\eta ^2\right)+10 x^4+8 \left(y^2-3 \eta ^2\right)^2$. Also, the deceleration and EoS parameters are given by

\begin{align}
        q=&\frac{2 x \left(\epsilon  x^2+2 y^2\right) \left(\zeta  \epsilon  x^2-\lambda  y^2\right)}{\eta  \left(12 \epsilon  x^2
   \left(y^2-\eta ^2\right)+5 \epsilon ^2 x^4+4 \left(y^2-3 \eta ^2\right)^2\right)}\nonumber\\ 
   &+\frac{12 \eta  x \left(\lambda 
   y^2-\zeta  \epsilon  x^2\right)}{12 \epsilon  x^2 \left(y^2-\eta ^2\right)+5 \epsilon ^2 x^4+4 \left(y^2-3 \eta
   ^2\right)^2}\nonumber\\ 
   &+\frac{2 x \left(24 \epsilon  \eta ^3 x-12 \epsilon  \eta  x y^2\right)}{\eta  \left(12 \epsilon  x^2
   \left(y^2-\eta ^2\right)+5 \epsilon ^2 x^4+4 \left(y^2-3 \eta ^2\right)^2\right)},
\end{align}
and 
\begin{small}
\begin{align}
        \omega_{\phi}=&-\frac{4 \epsilon  x^3 \left(6 \zeta  \eta ^2+(\lambda -2 \zeta ) y^2\right)}{3 \eta  \left(12 \epsilon  x^2 \left(y^2-\eta ^2\right)+5 \epsilon ^2 x^4+4 \left(y^2-3 \eta ^2\right)^2\right)}  -\frac{-4 \zeta  \epsilon ^2 x^5+5 \epsilon ^2 \eta 
   x^4+4 \eta  \left(y^2-3 \eta ^2\right)^2}{3 \eta  \left(12 \epsilon  x^2 \left(y^2-\eta ^2\right)+5 \epsilon ^2 x^4+4
   \left(y^2-3 \eta ^2\right)^2\right)} \nonumber\\&-\frac{8 \lambda  x y^2 \left(y^2-3 \eta ^2\right)}{3 \eta  \left(12 \epsilon  x^2
   \left(y^2-\eta ^2\right)+5 \epsilon ^2 x^4+4 \left(y^2-3 \eta ^2\right)^2\right)}  -\frac{4 \epsilon  x^2 \left(5 y^2-9 \eta
   ^2\right)}{12 \epsilon  x^2 \left(y^2-\eta ^2\right)+5 \epsilon ^2 x^4+4 \left(y^2-3 \eta ^2\right)^2}.
\end{align}
\end{small}
\subsection{Dynamical system analysis of 3D system for $\epsilon=1$}

The equilibrium points in the coordinates $(x,y,\eta)$ for system \eqref{exponentialf-1}, \eqref{exponentialf-2}, \eqref{exponentialf-3} and $\epsilon=1$ are the following.

\begin{enumerate}
    \item $Z_1=(0,y,0),$ with eigenvalues $\{0,0,0\}.$ This set of equilibrium points exist for $y>0$ and is nonhyperbolic. The asymptotic solution at the point describes the Minkowski spacetime.
    \item $Z_2=(0,0,1),$ with eigenvalues $\{2,2,1\}.$ This point is a source. For the deceleration parameter, we derive $q(Z_2)=0$. The asymptotic solution describes a universe dominated by the Gauss-Bonnet term.
    \item $Z_3=(0,0,-1),$ with eigenvalues $\{-2,-2,-1\}.$ This point is a sink. Since $q(Z_3)=0$, the physical properties are similar to point $Z_2.$
    \item $Z_4=(\sqrt{6},0,1)$ with eigenvalues $\left\{6,\sqrt{6} \zeta -6,\sqrt{\frac{3}{2}} \lambda +3\right\}.$ Moreover, $q(Z_4)=2$ means that the asymptotic solution describes a stiff fluid solution. This point is a
    \begin{enumerate}
        \item source for $\lambda >-\sqrt{6}$, $ \zeta >\sqrt{6},$
        \item saddle for $\lambda <-\sqrt{6}$  or $ \zeta <\sqrt{6},$
        \item nonhyperbolic for $\lambda =-\sqrt{6}$  or $ \zeta =\sqrt{6}.$
    \end{enumerate}
    \item $Z_5=(\sqrt{6},0,-1),$ with eigenvalues $\left\{-6,\sqrt{6} \zeta +6,\sqrt{\frac{3}{2}} \lambda -3\right\}$ and  $q(Z_5)=2,$ it represents a stiff fluid solution. This point is a
    \begin{enumerate}
        \item sink for $\lambda <\sqrt{6}$, $ \zeta <-\sqrt{6},$
        \item saddle for $\lambda >\sqrt{6}$ or $ \zeta >-\sqrt{6},$
        \item nonhyperbolic for  $\lambda =-\sqrt{6}$  or $ \zeta =\sqrt{6}.$
    \end{enumerate}
    \item $Z_6=(-\sqrt{6},0,1),$ with eigenvalues $\left\{6,-\sqrt{6} \zeta -6,3-\sqrt{\frac{3}{2}} \lambda \right\}$ and $q(Z_6)=2,$ it represents a stiff fluid solution. This point is a
    \begin{enumerate}
        \item source for $\lambda <\sqrt{6}$, $\zeta <-\sqrt{6},$
        \item saddle for $\lambda >\sqrt{6}$ or $\zeta >-\sqrt{6},$
        \item nonhyperbolic for $\lambda =\sqrt{6}$ or $\zeta =-\sqrt{6}.$
    \end{enumerate}
    \item $Z_7=(-\sqrt{6},0,-1),$ with eigenvalues $\left\{-6,6-\sqrt{6} \zeta ,-\sqrt{\frac{3}{2}} \lambda -3\right\}$ and $ q(Z_7)=2,$  it represents a stiff fluid solution. This point is 
    \begin{enumerate}
        \item sink for $\lambda >-\sqrt{6}$, $ \zeta >\sqrt{6},$
        \item saddle for $\lambda <-\sqrt{6}$ or $ \zeta <\sqrt{6},$
        \item nonhyperbolic for $\lambda =-\sqrt{6}$ or $ \zeta =\sqrt{6}.$
    \end{enumerate}
    \item $Z_{8}=(-\lambda, \sqrt{3-\frac{\lambda ^2}{2}},1),$ with eigenvalues $\left\{\lambda ^2,\frac{1}{2} \left(\lambda ^2-6\right),-\lambda  (\zeta +\lambda )\right\}$. This point exist for $-\sqrt{6}\leq \lambda \leq \sqrt{6}$ and is 
    \begin{enumerate}
        \item a saddle for
        \begin{enumerate}
            \item $-\sqrt{6}<\lambda <0, \zeta <-\lambda $ or 
            \item $0<\lambda <\sqrt{6}, \zeta >-\lambda $ or
            \item $-\sqrt{6}<\lambda <0, \zeta >-\lambda $ or
            \item $0<\lambda <\sqrt{6}, \zeta <-\lambda ,$
        \end{enumerate}
        \item nonhyperbolic for
        \begin{enumerate}
            \item $\lambda =0$ or
            \item $\zeta +\lambda =0$ or
            \item $\lambda =-\sqrt{6}$ or
            \item $ \lambda =\sqrt{6}.$
        \end{enumerate}
    \end{enumerate}
    As before, we calculate $q(Z_{8})=\frac{1}{2} \left(\lambda ^2-2\right)$ from where we infer that acceleration occurs for $\lambda^2<2$.
    \item $Z_{9}=(\lambda , \sqrt{3-\frac{\lambda ^2}{2}},-1),$ with eigenvalues $\left\{-\lambda ^2,-\frac{1}{2} \left(\lambda ^2-6\right),\lambda  (\zeta +\lambda )\right\}$. This point exist for $-\sqrt{6}\leq \lambda \leq \sqrt{6}$ and is
    \begin{enumerate}
        \item a saddle for
        \begin{enumerate}
            \item $-\sqrt{6}<\lambda <0, \zeta <-\lambda $ or
            \item $0<\lambda <\sqrt{6}, \zeta >-\lambda $ or
            \item $-\sqrt{6}<\lambda <0, \zeta >-\lambda $ or
            \item $0<\lambda <\sqrt{6}, \zeta <-\lambda ,$
        \end{enumerate}
        \item nonhyperbolic for
        \begin{enumerate}
            \item $\lambda =0$ or
            \item $\zeta +\lambda =0$ or
            \item $\lambda =-\sqrt{6}$ or
            \item $\lambda =\sqrt{6}.$
        \end{enumerate}
    \end{enumerate}
    Furthermore, for the asymptotic solution at the points, we derive $q(Z_{9})=\frac{1}{2} \left(\lambda ^2-2\right)$ from where we infer that acceleration occurs for $\lambda^2<2$.
     \item $Z_{10}=\left(x_{10},0,1\right)$, where 
  \begin{equation}
    \label{eqZ10} x_{10}=\frac{\frac{2^{2/3} \left(50-9 \zeta ^2\right)}{\sqrt[3]{9 \zeta  \left(\sqrt{2} \sqrt{r}-6 \zeta \right)+500}}+\sqrt[3]{2} \sqrt[3]{9 \zeta 
   \left(\sqrt{2} \sqrt{r}-6 \zeta \right)+500}+10}{3 \zeta },  
  \end{equation}
  where $r=9 \zeta ^4-132 \zeta ^2+500.$ This point exist for $\zeta \in \mathbb{R}$ but $ \zeta \neq 0.$ For $Z_{10}$ we have
   \begin{align*}
       \omega_{\phi}&=-\frac{ \sqrt[6]{2} \zeta  \left(9 \zeta  \left(\sqrt{2} \sqrt{r}-6 \zeta
   \right)+500\right)^{2/3} \left(3 \sqrt{2} \zeta +\sqrt{r}\right)}{\left(50-9 \zeta ^2\right)^2}\\ \nonumber &+\frac{250\
   2^{2/3} \left(9 \zeta  \left(\sqrt{2} \sqrt{r}-6 \zeta \right)+500\right)^{2/3}}{9\left(50-9 \zeta
   ^2\right)^2}\\ \nonumber &+\frac{1}{9}\Big(\sqrt[3]{2} \sqrt[3]{9 \zeta  \left(\sqrt{2} \sqrt{r}-6 \zeta \right)+500}+1\Big),
   \end{align*}
\begin{align*}
       q&=-\frac{3 \sqrt[6]{2} \zeta  \left(9 \zeta  \left(\sqrt{2} \sqrt{r}-6 \zeta \right)+500\right)^{2/3} \left(3
   \sqrt{2} \zeta +\sqrt{r}\right)}{2\left(50-9 \zeta ^2\right)^2}\\ \nonumber &
   +\frac{250\ 2^{2/3} \left(9 \zeta  \left(\sqrt{2}
   \sqrt{r}-6 \zeta \right)+500\right)^{2/3}}{6\left(50-9 \zeta ^2\right)^2}\\ \nonumber&+\frac{1}{6}\Big(\sqrt[3]{2} \sqrt[3]{9 \zeta 
   \left(\sqrt{2} \sqrt{r}-6 \zeta \right)+500}+4\Big).
   \end{align*}

  The eigenvalues of $Z_{12}$ are $\delta_i(\zeta,\lambda)$ for $i=1,2,3.$ Given the complexity of the expressions, we perform numerical analysis to conclude that this point is a source or saddle (see Fig. \ref{fig:7}).
  The physical parameters $\omega_{\phi}(Z_{10})$ and $q(Z_{10})$ are presented in  Fig. \ref{fig:6}. We observe that the equilibrium points can describe dust-like and radiation-like cosmological eras; however, $q(Z_{10})\geq 0$, the solution, cannot describe an accelerated universe. For large $\zeta$ we have that $q\rightarrow 0$ and $\omega_{\phi} \rightarrow -1/3$.
 \begin{figure}[ht!]
    \centering
    \includegraphics[scale=0.6]{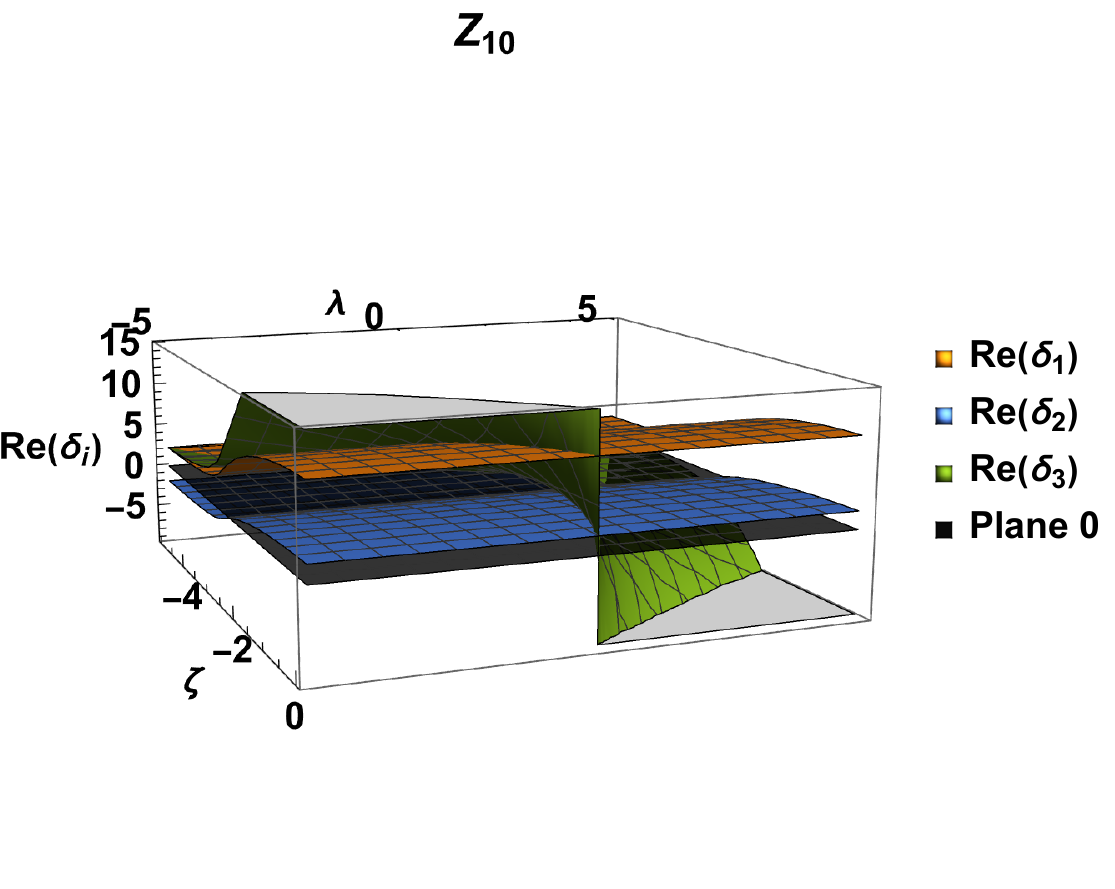}
    \includegraphics[scale=0.6]{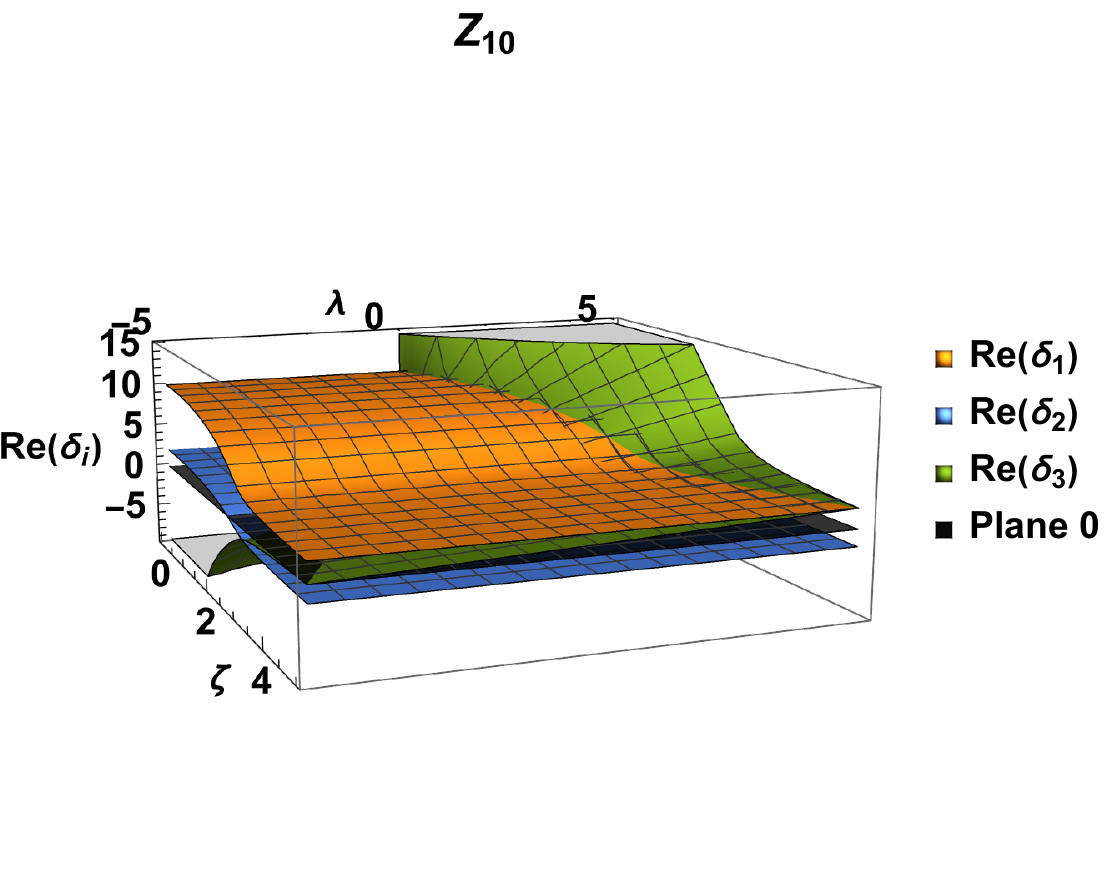}
    \caption{Real part of the eigenvalues of  $Z_{10}$ for  $\zeta<0 $ and $\zeta>0$.}
    \label{fig:7}
\end{figure}
\begin{figure}[ht!]
    \centering
    \includegraphics[scale=0.6]{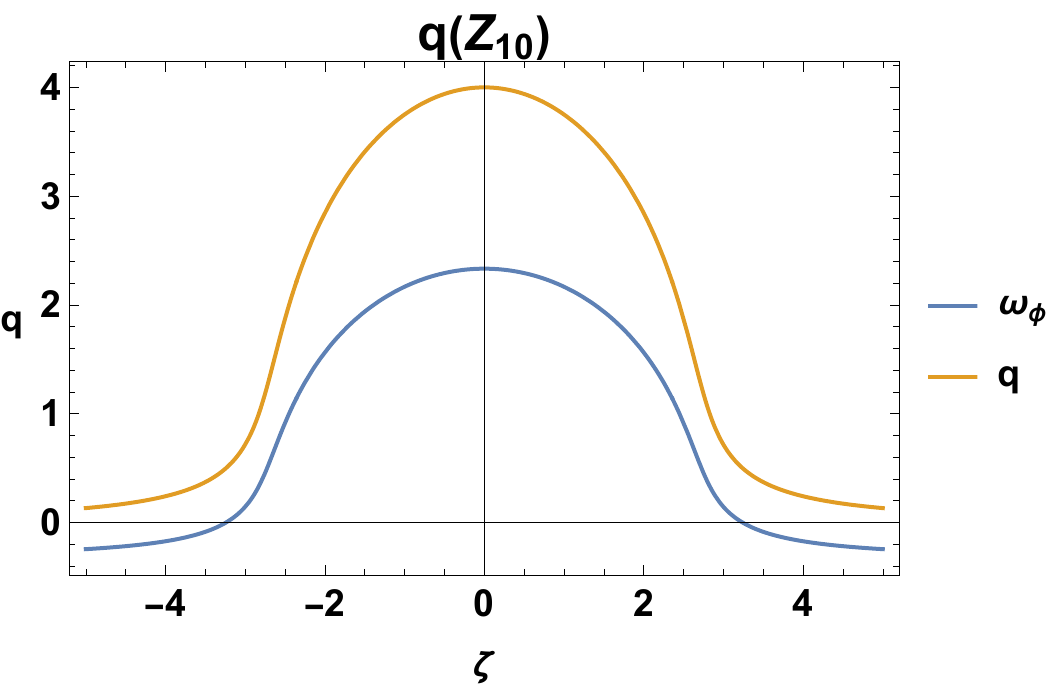}
    \caption{Plot of $q(Z_{10})$ and $\omega_{\phi}(Z_{10})$.}
    \label{fig:6}
\end{figure}

   \item $Z_{11}=\left(x_{11}, 0, -1\right)$, where 
   \begin{equation}
      \label{eqZ11} x_{11}=\frac{\frac{2^{2/3} \left(50-9 \zeta ^2\right)}{\sqrt[3]{9 \zeta  \left(6 \zeta +\sqrt{2} \sqrt{r}\right)-500}}+\sqrt[3]{2} \sqrt[3]{9 \zeta 
   \left(6 \zeta +\sqrt{2} \sqrt{r}\right)-500}-10}{3 \zeta },
   \end{equation} 
   where $r=9 \zeta ^4-132 \zeta ^2+500.$ This point exists for $\zeta<-\frac{5\sqrt{2}}{3}$ and $\zeta>\frac{5\sqrt{2}}{3}.$ For this point, we have 
   
   \begin{align*}
       \omega_{\phi}&=\frac{ \sqrt[6]{2} \zeta  \left(9 \zeta  \left(6 \zeta +\sqrt{2}
   \sqrt{r}\right)-500\right)^{2/3} \left(\sqrt{r}-3 \sqrt{2} \zeta \right)}{\left(50-9 \zeta
   ^2\right)^2}\\ \nonumber &+\frac{250\ 2^{2/3} \left(9 \zeta  \left(6 \zeta +\sqrt{2}
   \sqrt{r}\right)-500\right)^{2/3}}{9\left(50-9 \zeta ^2\right)^2}\\ \nonumber &+\frac{1}{9}\left(-\sqrt[3]{2} \sqrt[3]{9 \zeta  \left(6 \zeta
   +\sqrt{2} \sqrt{r}\right)-500}+1\right),
   \end{align*}
 \begin{align*}
     q&=\frac{3 \sqrt[6]{2} \zeta  \left(9 \zeta  \left(6 \zeta
   +\sqrt{2} \sqrt{r}\right)-500\right)^{2/3} \left(\sqrt{r}-3 \sqrt{2} \zeta \right)}{2\left(50-9 \zeta
   ^2\right)^2}\\ \nonumber &+\frac{250\ 2^{2/3} \left(9 \zeta  \left(6 \zeta +\sqrt{2}
   \sqrt{r}\right)-500\right)^{2/3}}{6\left(50-9 \zeta ^2\right)^2}\\ \nonumber &+\frac{1}{6}\left(-\sqrt[3]{2} \sqrt[3]{9 \zeta  \left(6 \zeta
   +\sqrt{2} \sqrt{r}\right)-500}+4\right).
 \end{align*}  
The eigenvalues of $Z_{11}$ are $\lambda_i(\zeta,\lambda)$ for $i=1,2,3.$ Given the complexity of the expressions, we perform numerical analysis to conclude that this point is a saddle (see Fig. \ref{fig:9}). We have presented plots for the case $\zeta >\frac{5\sqrt{2}}{3}$ because the other interval produces similar (symmetric) results. In Fig. \ref{fig:8} we give the evolution of the physical parameters $\omega_{\phi}(Z_{11})$ and $q(Z_{11})$ in terms of the free parameter $\zeta$.  Thus, the asymptotic solution describes ideal gas solutions, but an accelerated universe cannot be described. However, dust-like and radiation-like epochs are provided by the equilibrium points. For large $\zeta$ we have that $q\rightarrow 0$ and $\omega_{\phi} \rightarrow -1/3$.
 \begin{figure}[ht!]
    \centering
    \includegraphics[scale=0.7]{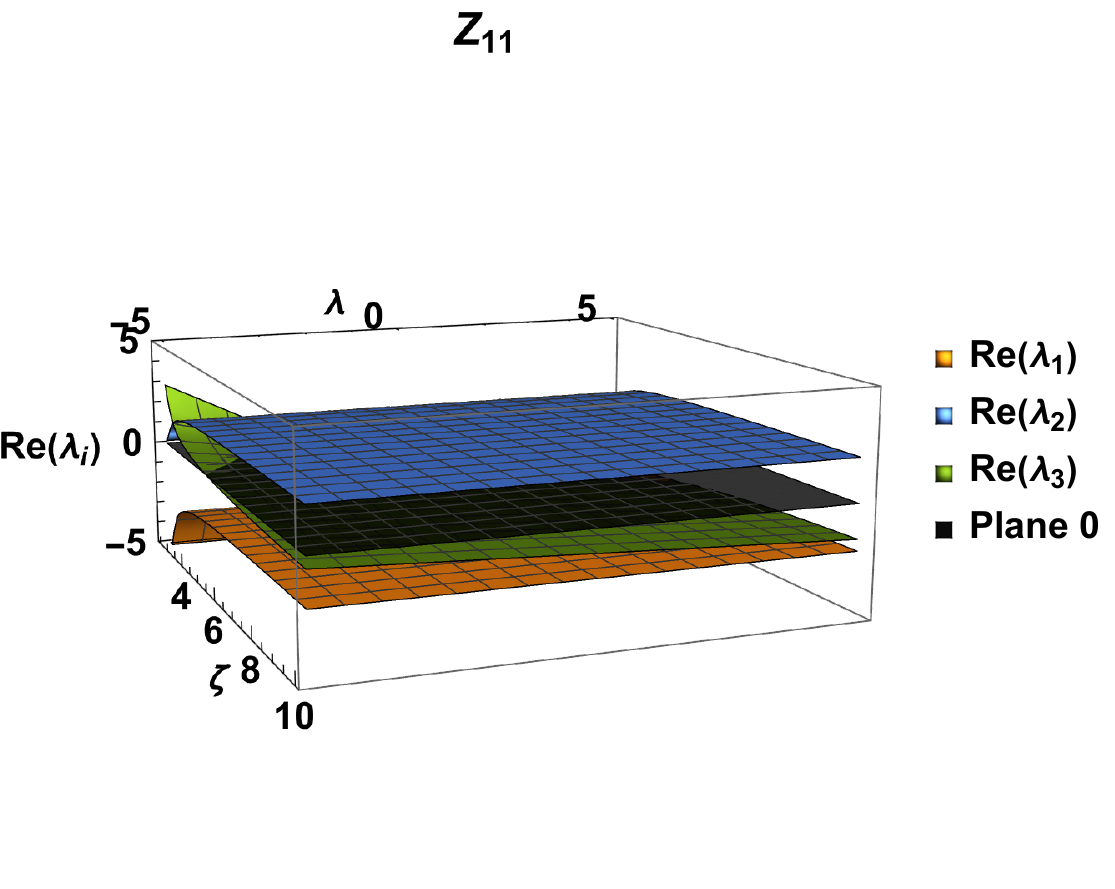}
    \caption{Real part of the eigenvalues of  $Z_{11}$.}
    \label{fig:9}
\end{figure}
\begin{figure}[ht!]
    \centering
    \includegraphics[scale=0.6]{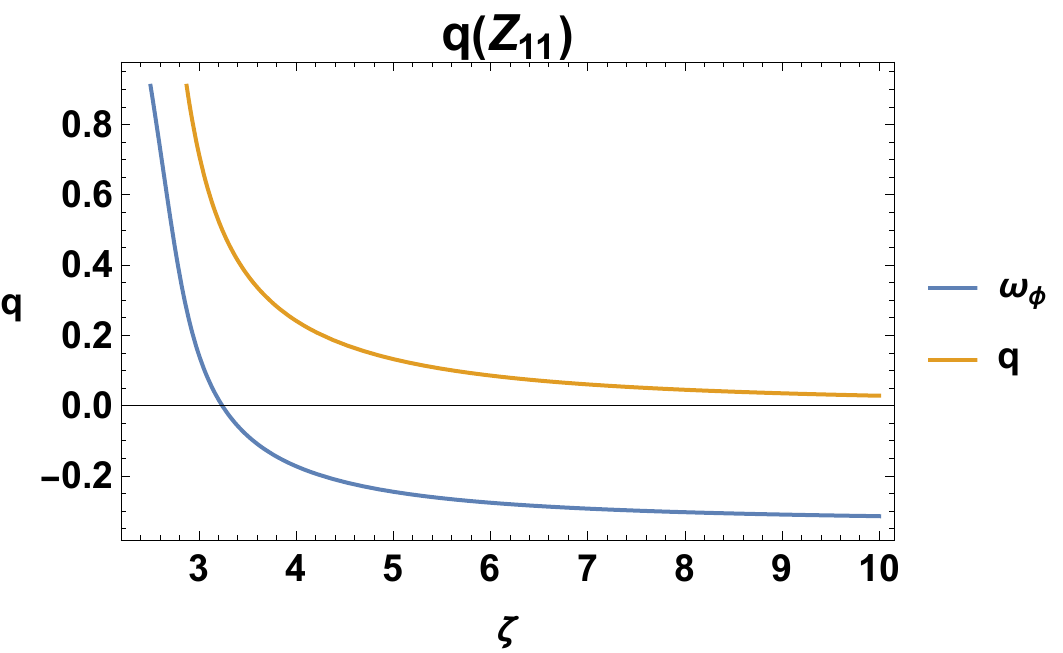}
    \caption{Plot of $q(Z_{11})$ and $\omega_{\phi}(Z_{11})$.}
    \label{fig:8}
\end{figure}
\item $Z_{12}=\left(x_{12},0,-1\right),$ where 
\begin{equation}
\label{eqZ12}
    x_{12}=\frac{\frac{4 \sqrt[3]{-1} 2^{2/3} \left(9 \zeta
   ^2-50\right)}{\sqrt[3]{54 \zeta ^2+9 \sqrt{2}
   \sqrt{\zeta ^2 r}-500}}+4 (-1)^{2/3}
   \sqrt[3]{2} \sqrt[3]{54 \zeta ^2+9 \sqrt{2}
   \sqrt{\zeta ^2 r}-500}-40}{12 \zeta },
\end{equation}
where $r=9 \zeta ^4-132 \zeta ^2+500.$  This point exists for $-\frac{5\sqrt{2}}{3}<\zeta<\frac{5\sqrt{2}}{3}.$ The eigenvalues for $Z_{12}$ are $\gamma_{i}(\zeta,\lambda),$ given the complexity of the expressions, we perform numerical analysis to conclude that this point is a sink or saddle (see Fig. \ref{fig:11}). For $Z_{14}$ we have $\omega_{\phi}=f_1(\zeta)$ and  $q=f_2(\zeta)$
given that these are long expressions we write them as $f_i(\zeta)$  but we verify that for $\zeta\rightarrow \pm \frac{5\sqrt{2}}{3},$ $\omega_{\phi}\approx 1.142$ and $q\approx 2.213$, see Fig. \ref{fig:10}.  
\begin{figure}[ht!]
    \centering
    \includegraphics[scale=0.7]{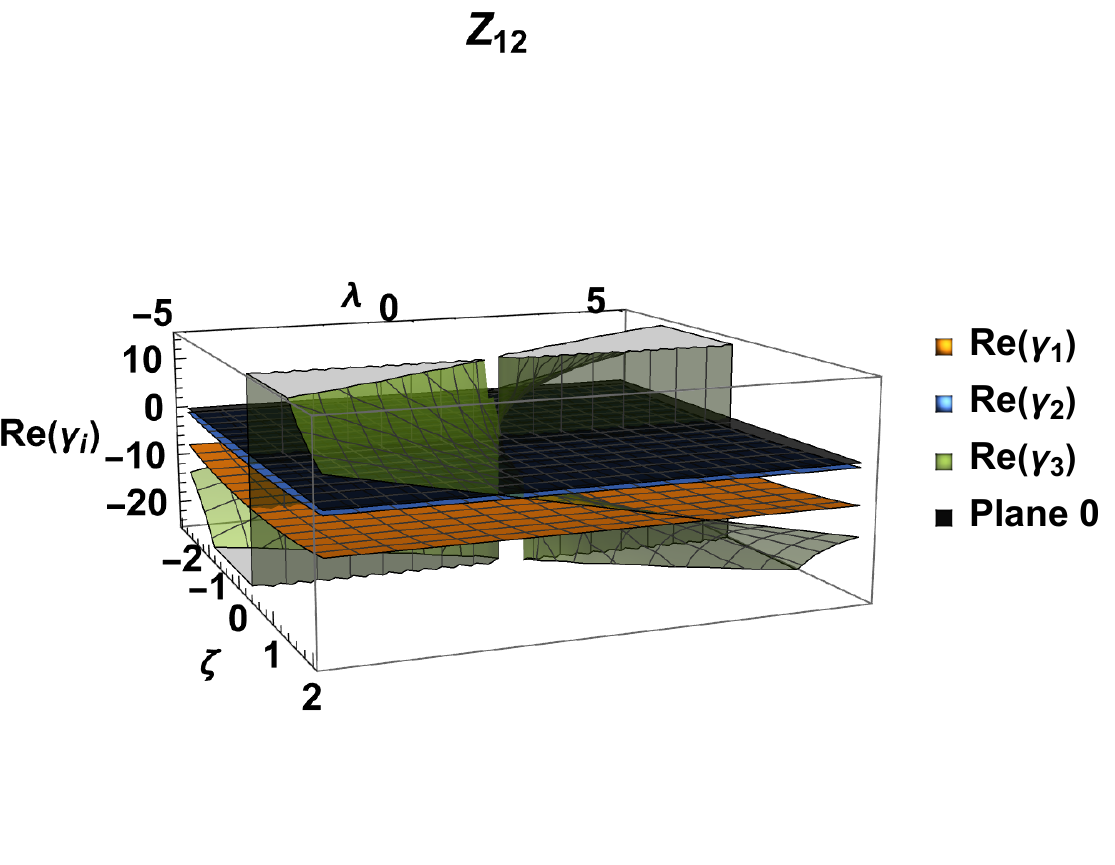}
    \caption{Real part of the eigenvalues of $Z_{12}$.}
    \label{fig:11}
\end{figure}
\begin{figure}[ht!]
    \centering
    \includegraphics[scale=0.6]{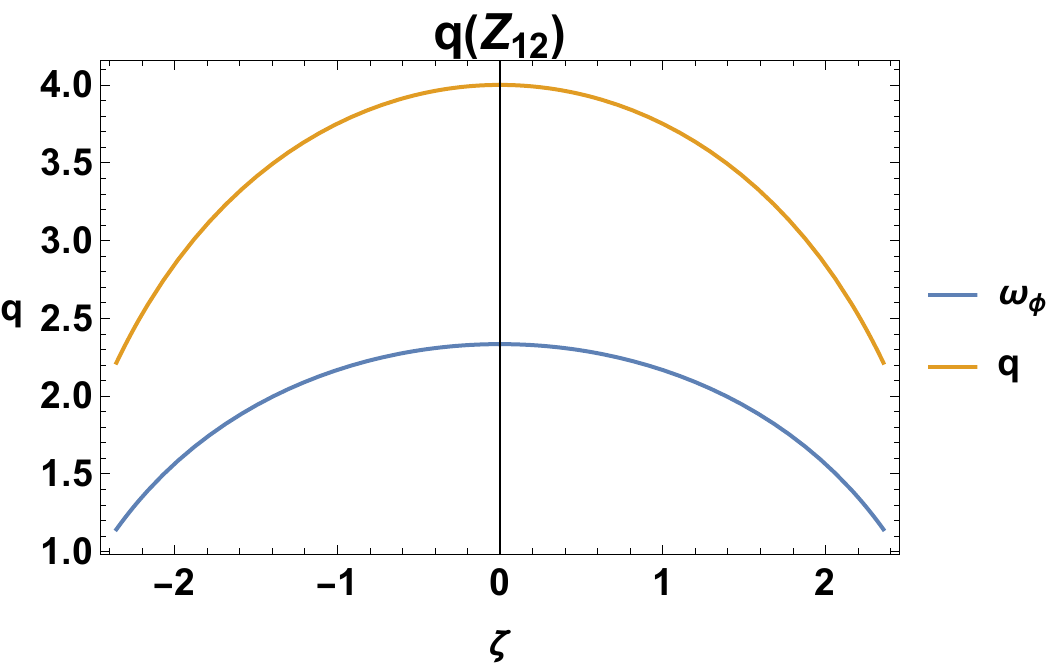}
    \caption{Plot of $q(Z_{12})$ and $\omega_{\phi}(Z_{12})$.}
    \label{fig:10}
\end{figure}

\end{enumerate}

\begin{figure}[ht!]
    \centering
    \includegraphics[scale=0.45]{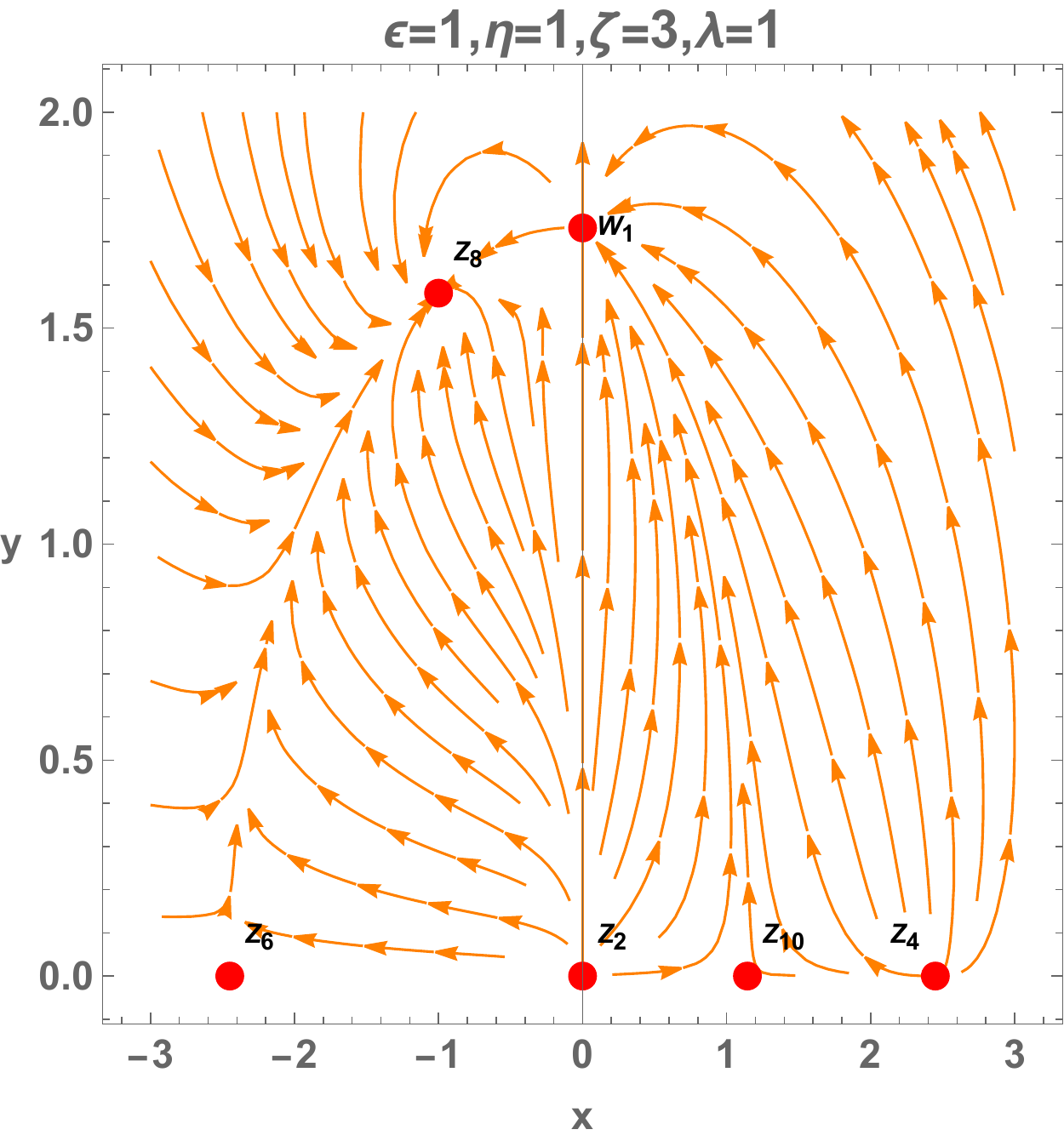}
    \includegraphics[scale=0.45]{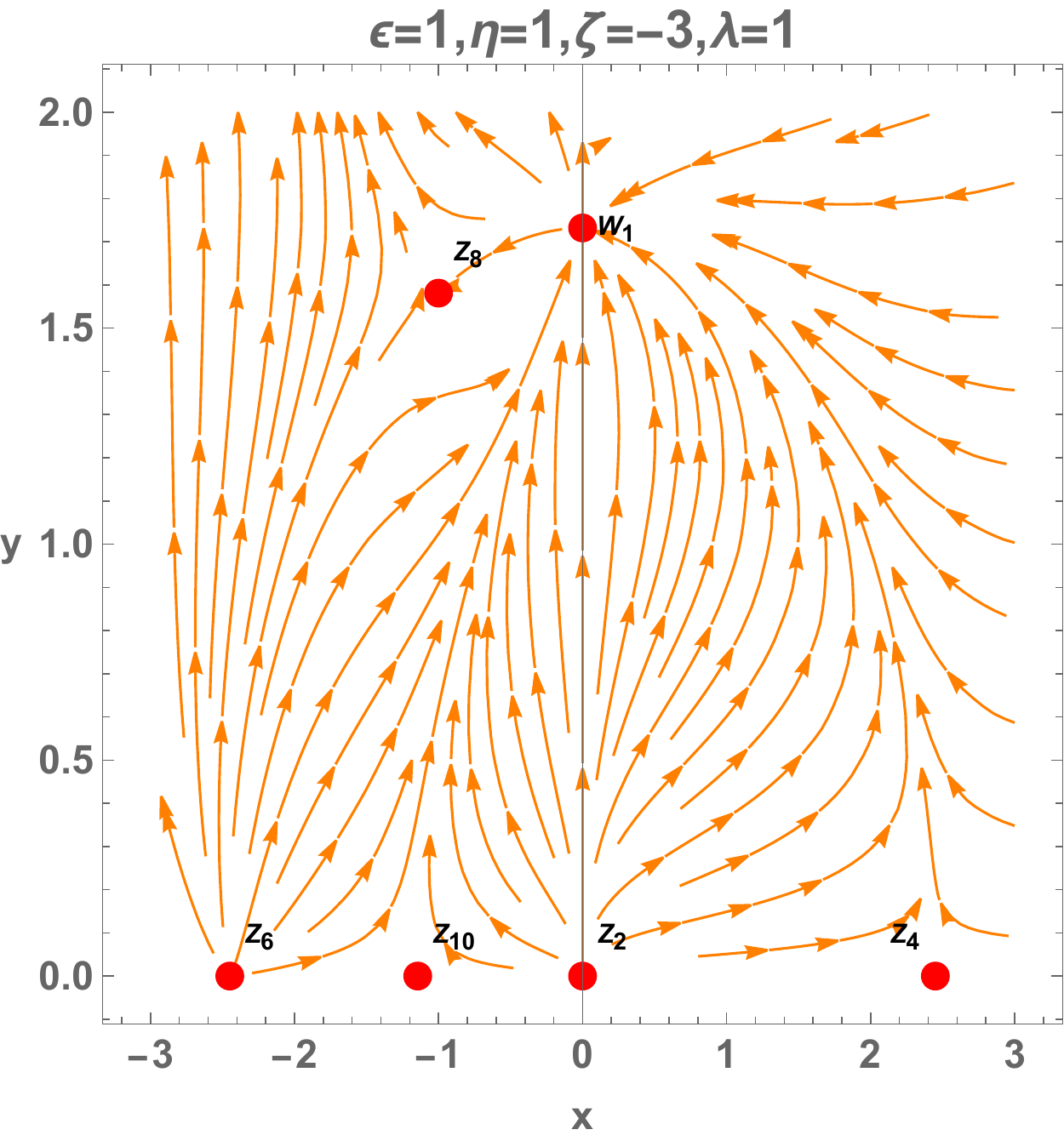}
    \includegraphics[scale=0.45]{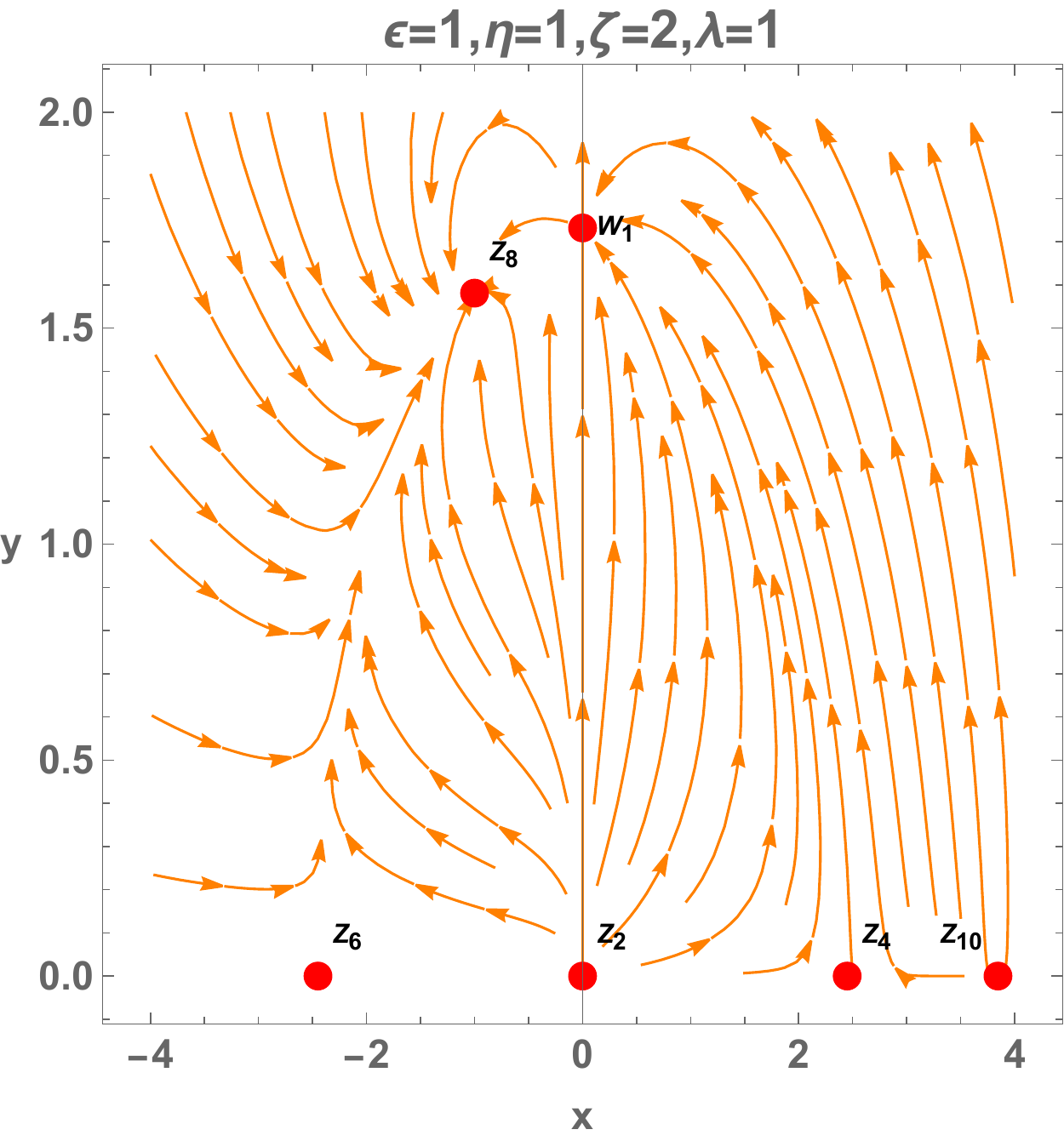}
    \includegraphics[scale=0.45]{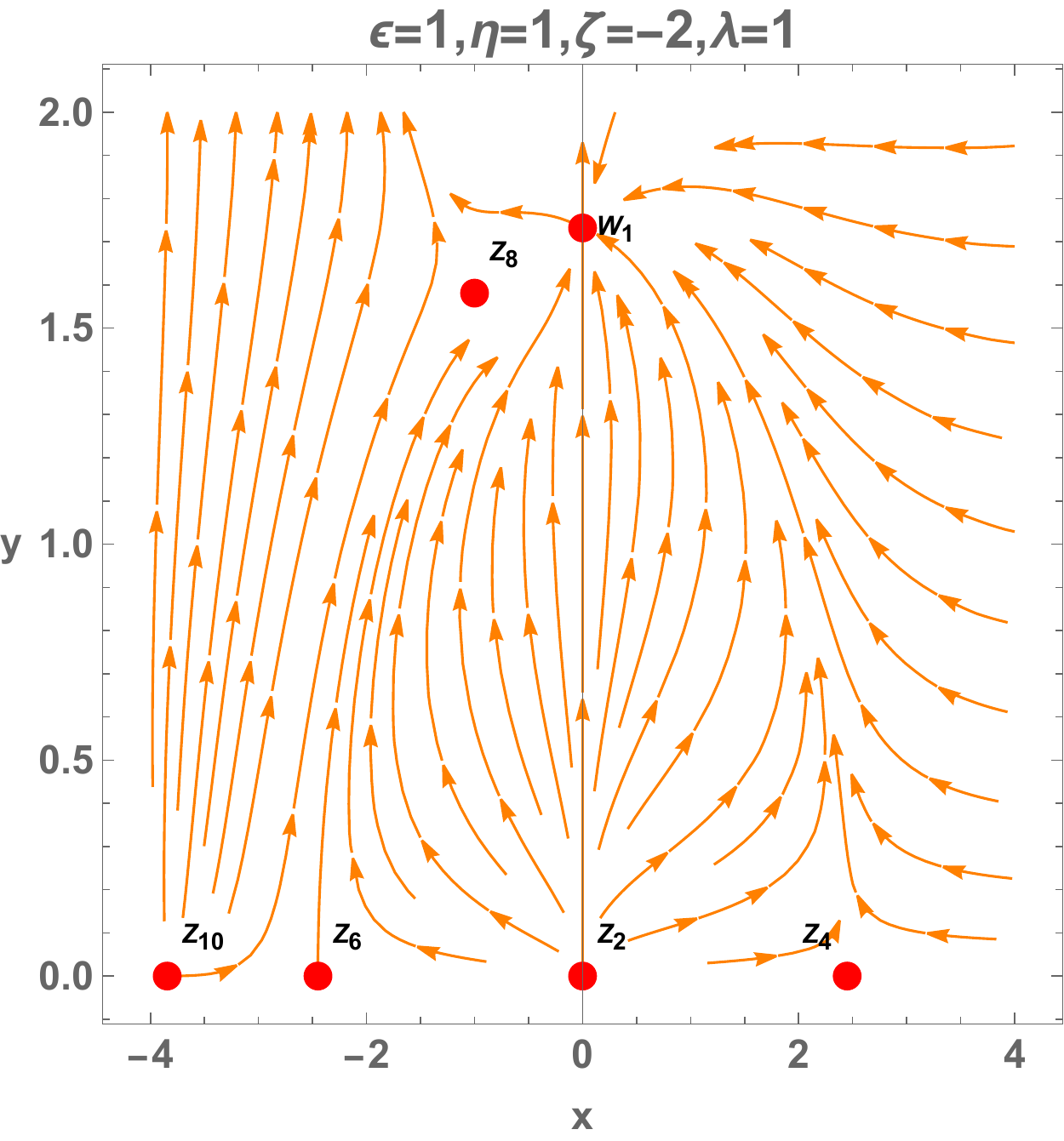}
    \caption{2D-Projection of system \eqref{exponentialf-1}, \eqref{exponentialf-2} and \eqref{exponentialf-3} setting $\epsilon=1, \lambda=1$ for $\eta=1$ with different values of $\zeta$. Here, the saddle point $W_1=(0,\sqrt{3})$ is a singularity in which both the numerator and denominator of the $y$ equation vanish.}
    \label{fig:12}
\end{figure}

\begin{figure}[h!]
    \centering
    \includegraphics[scale=0.45]{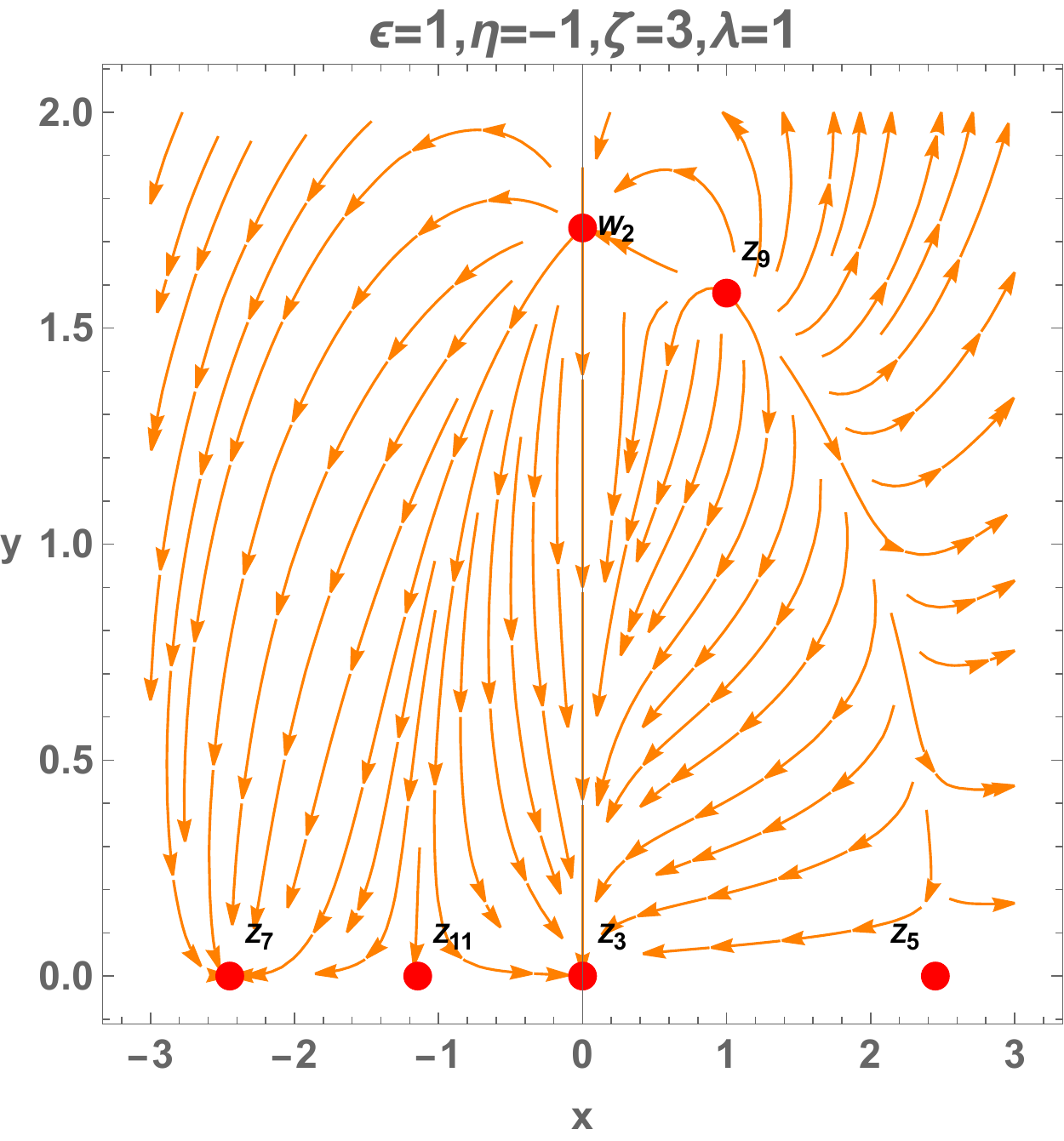}
    \includegraphics[scale=0.45]{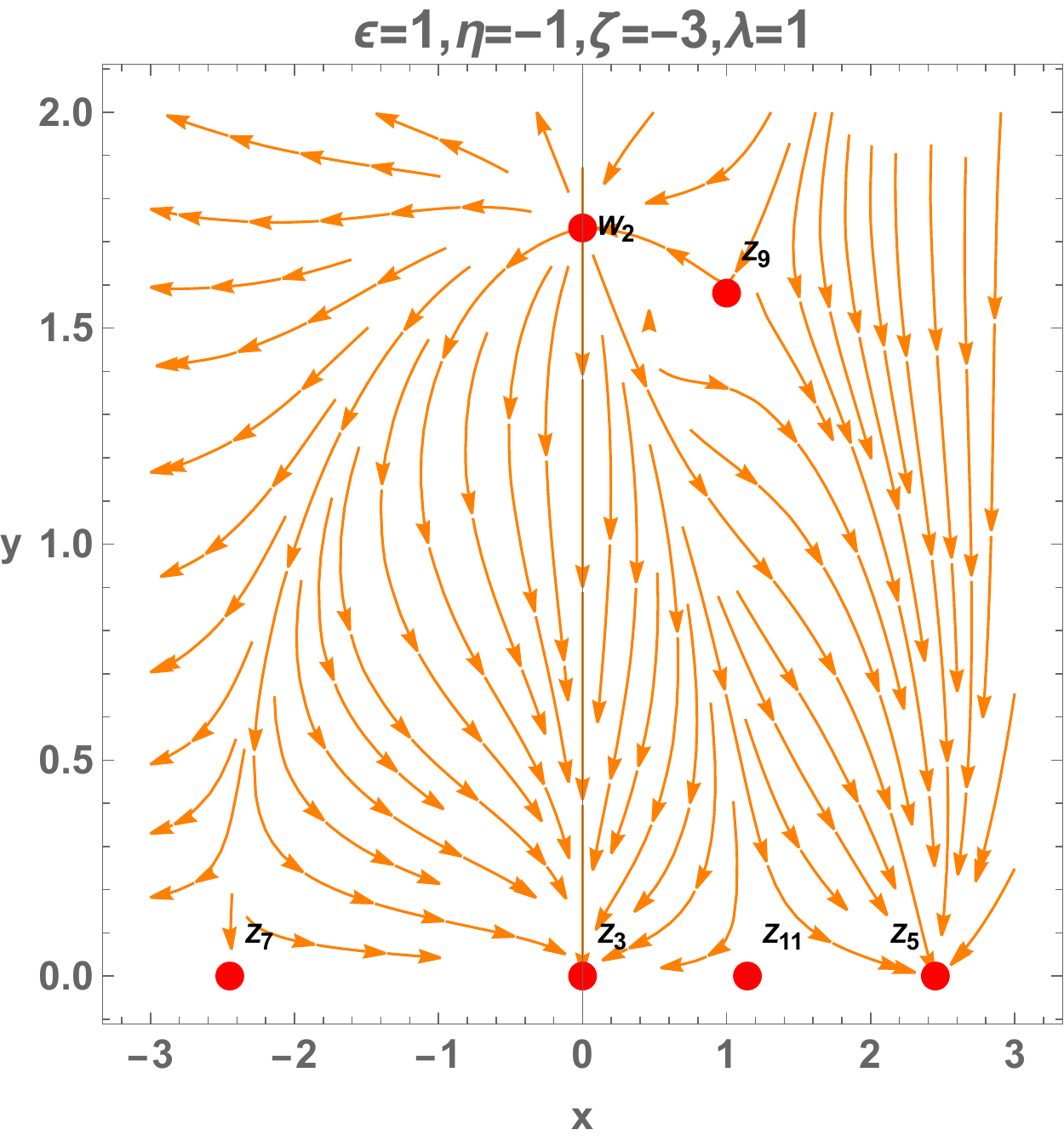}
    \includegraphics[scale=0.45]{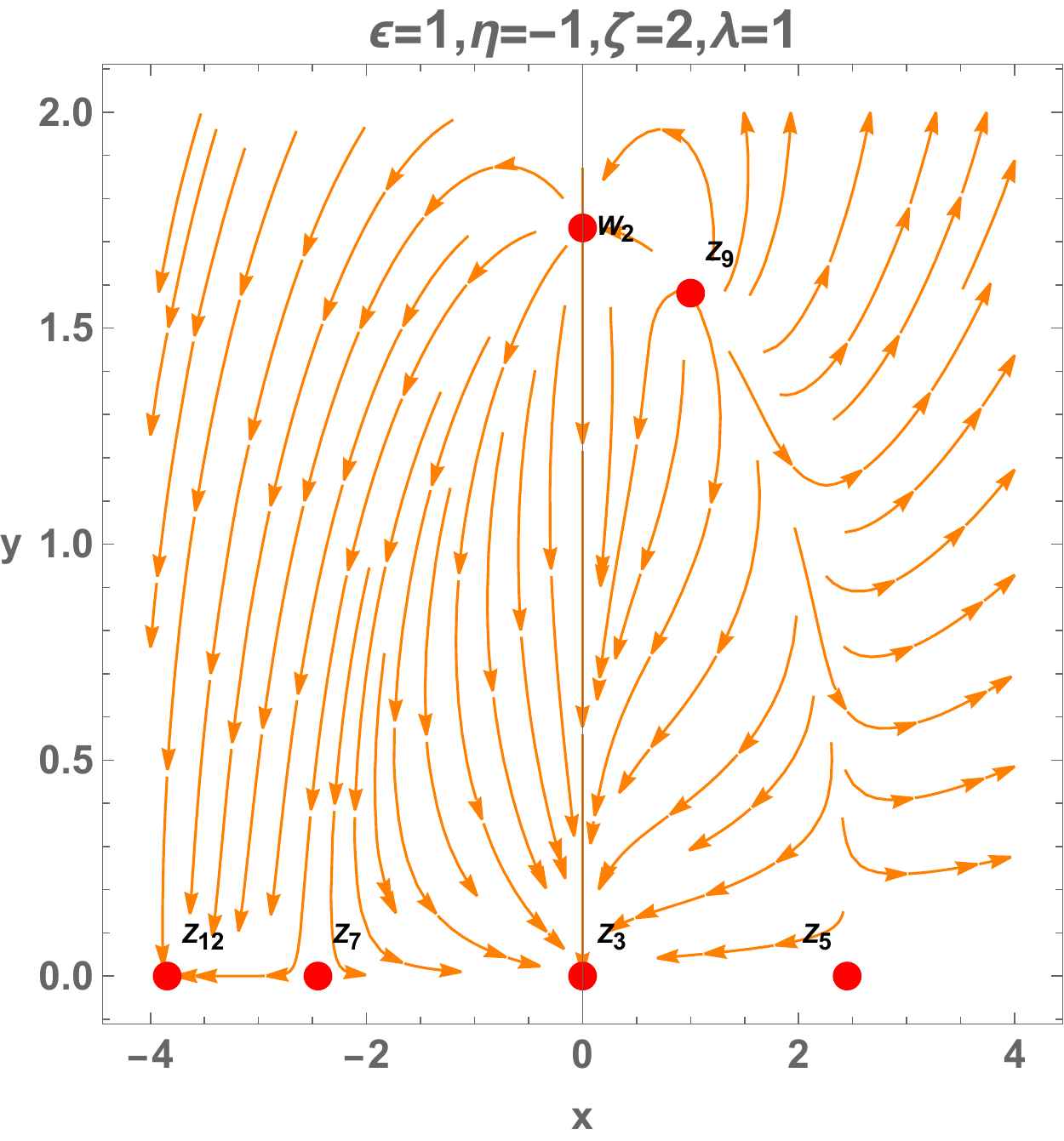}
    \includegraphics[scale=0.45]{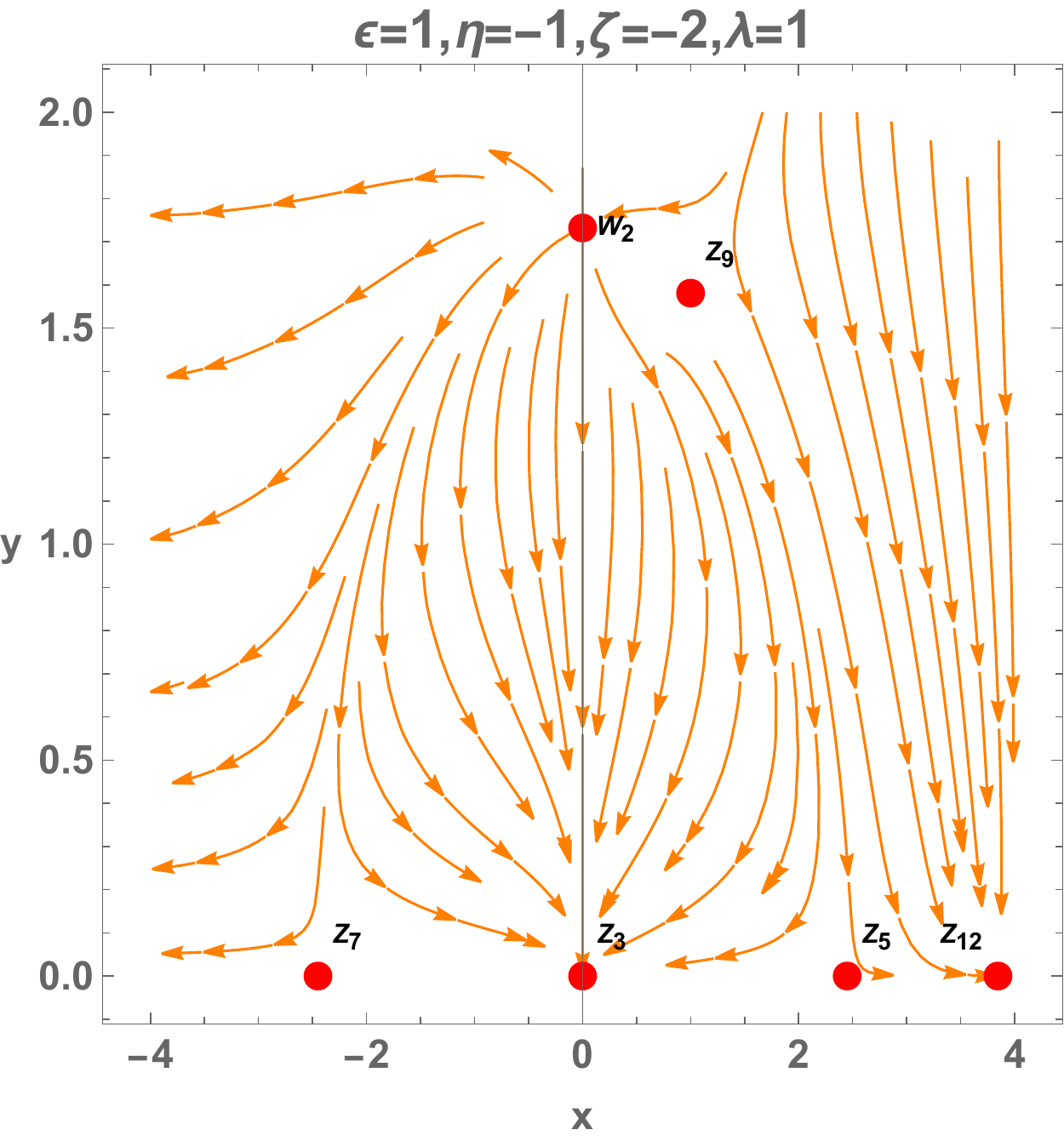}
    \caption{2D-Projection of system \eqref{exponentialf-1}, \eqref{exponentialf-2} and \eqref{exponentialf-3} setting $\epsilon=1, \lambda=1$ for $\eta=-1$ with different values of $\zeta$. Here, the saddle point $W_2=(0,\sqrt{3})$ is a singularity in which both the numerator and denominator of the $y$ equation vanish.}
    \label{fig:13}
\end{figure}

Phase-space diagrams for a 2D projection setting  $\epsilon=1, \eta=1, \lambda= 1$ and different values of $ \zeta$ are presented in Fig. \ref{fig:12}. We also present similar diagrams for the other 2D projection setting $\eta=-1$ in Fig. \ref{fig:13}. The stability analysis of the system is summarized in Table \ref{tab:4}. The existence of the points $Z_{10}, Z_{11}$ and $Z_{12}$ is discussed in appendix \ref{app}. Three-dimensional phase-space diagrams are presented in Fig. \ref{fig:13a} setting $\epsilon=1, \lambda=1$ and different values of $\zeta.$

\begin{figure}[ht!]
    \centering
    \includegraphics[scale=0.55]{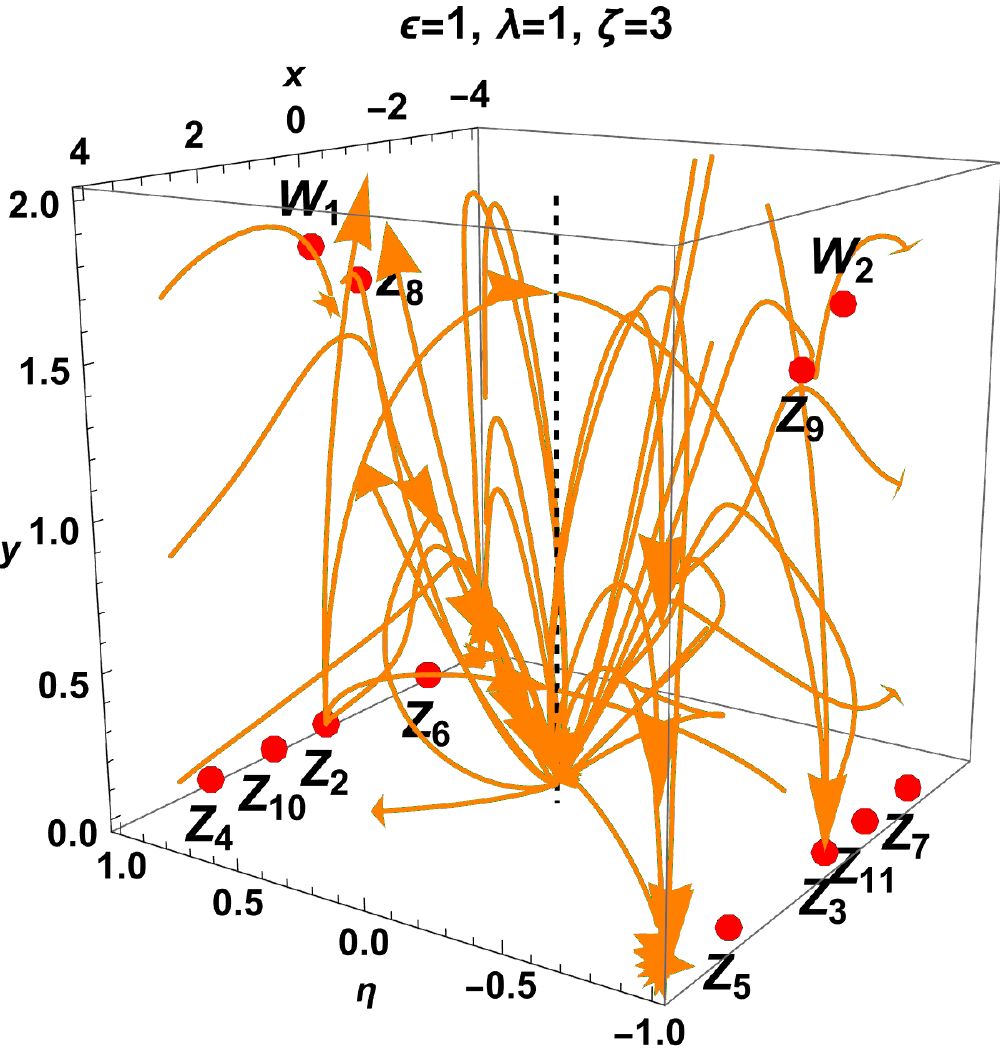}
    \includegraphics[scale=0.55]{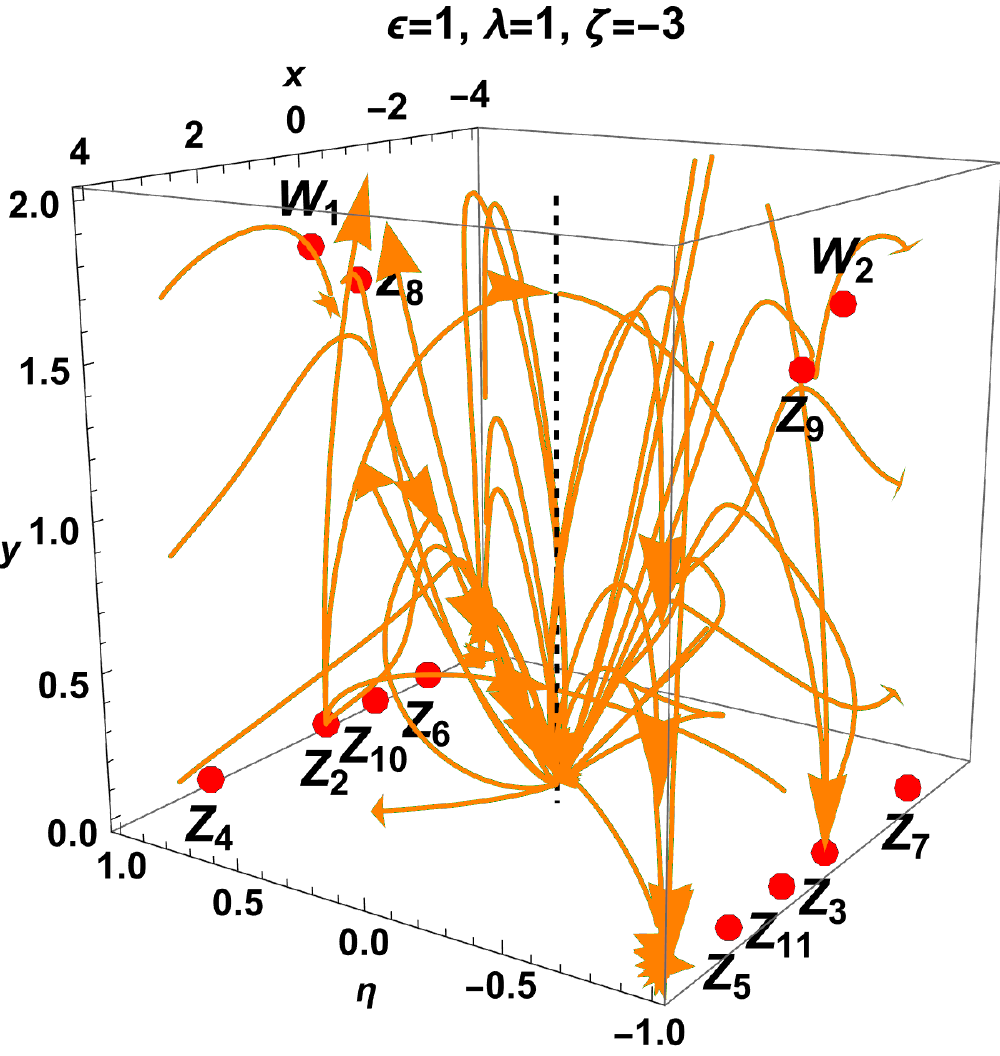}
    \includegraphics[scale=0.55]{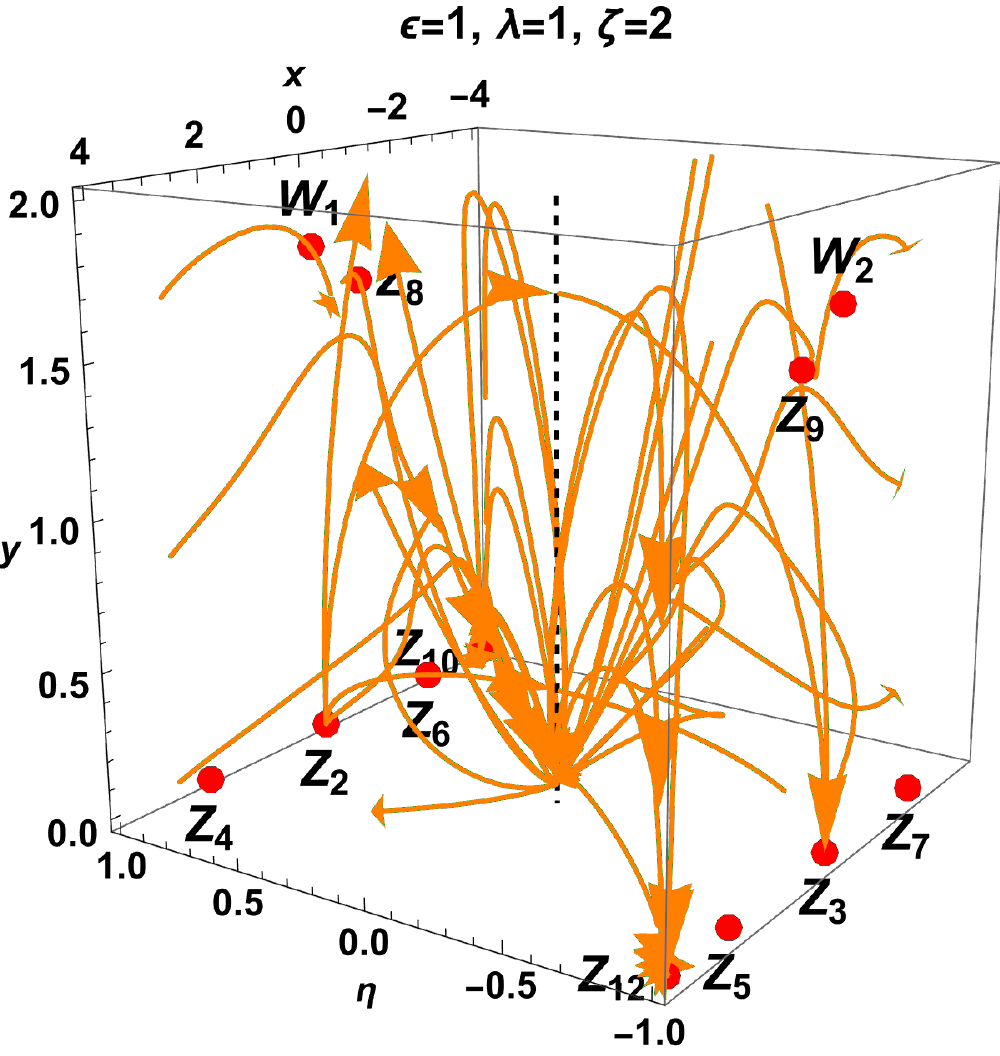}
    \includegraphics[scale=0.55]{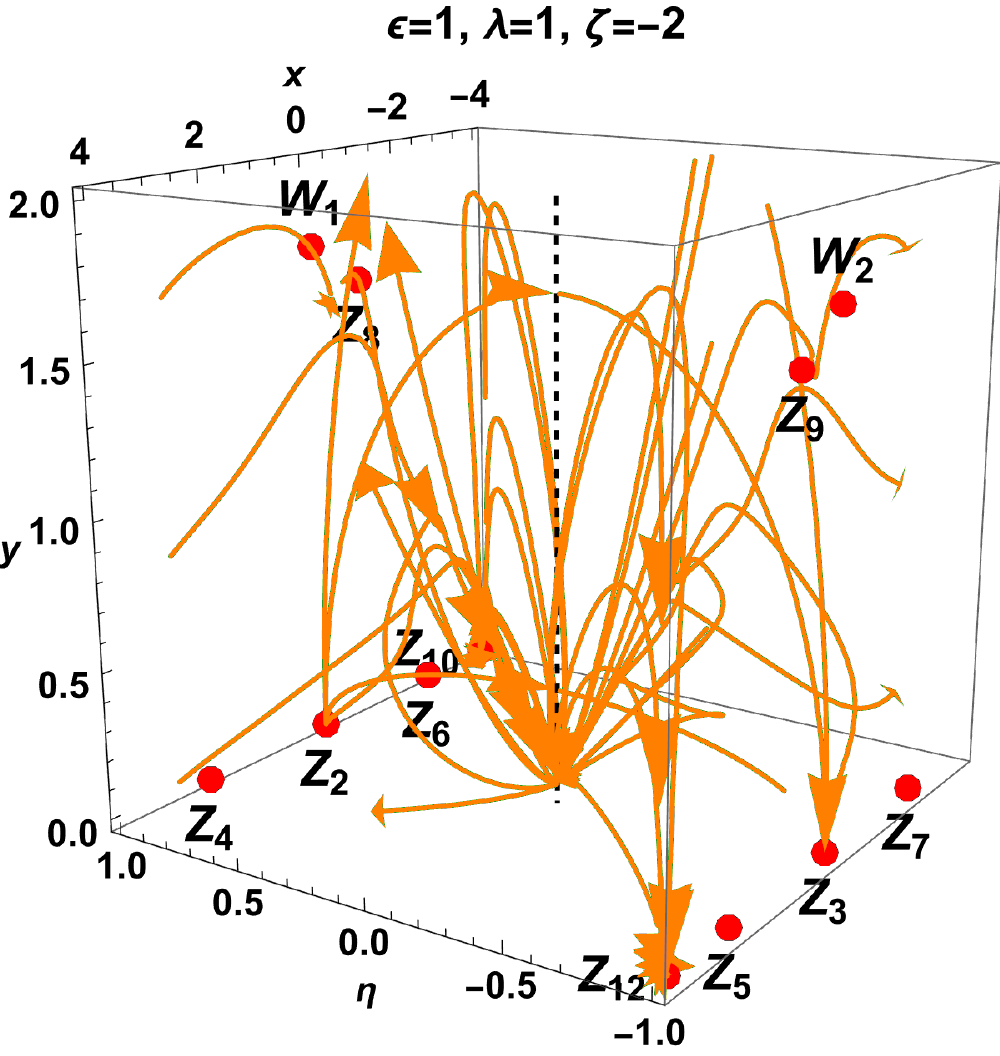}
    \caption{Three dimensional phase plot of of system \eqref{exponentialf-1}, \eqref{exponentialf-2} and \eqref{exponentialf-3} setting $\epsilon=1, \lambda=1$ with different values of $\zeta$. Here, the saddle points $W_1=(0,\sqrt{3},1)$ and $W_2=(0,\sqrt{3},-1)$ are singularities in which both the numerator and denominator of the $y$ equation vanish.}
    \label{fig:fig13a}
\end{figure}

\begin{table}[ht]
    \caption{Equilibrium points of system \eqref{exponentialf-1}, \eqref{exponentialf-2}, \eqref{exponentialf-3}  for $\epsilon=1$ with their stability conditions. It also includes the value of $\omega_{\phi}$ and $q.$ }
    \label{tab:4}
    \centering
\newcolumntype{C}{>{\centering\arraybackslash}X}
\centering
    \setlength{\tabcolsep}{1.6mm}
\begin{tabularx}{\textwidth}{cccccccc}
\toprule 
  \text{Label}  &  $x$ &  $y$ & $\eta$ & \text{Stability}& $\omega_{\phi}$ &  $q$\\\midrule  
  $Z_{1}$ &$0$ & $y$ & $0$  & nonhyperbolic & \text{indeterminate} & \text{indeterminate}\\   \midrule
   $Z_{2}$ &$0$ & $0$ & $1$ & \text{source} & $-1/3$ & $0$\\   \midrule
    $Z_{3}$ &$0$ & $0$ & $-1$  & \text{sink} & $-1/3$ & $0$\\   \midrule
    $Z_{4}$ &$\sqrt{6}$ & $0$ & $1$  &  source for $\lambda >-\sqrt{6}, \zeta >\sqrt{6}$ &&\\
    &&&& saddle for $\lambda <-\sqrt{6}$  or $ \zeta <\sqrt{6}$ &&\\
    &&&& nonhyperbolic for $\lambda =-\sqrt{6}$  or $ \zeta =\sqrt{6}$ & $1$ & $2$\\   \midrule
    $Z_{5}$ &$\sqrt{6}$ & $0$ & $-1$  & sink for $\lambda <\sqrt{6}, \zeta <-\sqrt{6}$ &&\\
    &&&& saddle for $\lambda >\sqrt{6}$ or $ \zeta >-\sqrt{6}$ &&\\
    &&&& nonhyperbolic for  $\lambda =-\sqrt{6}$  or $ \zeta =\sqrt{6}$ & $1$ & $2$\\   \midrule
    $Z_{6}$ &$-\sqrt{6}$ & $0$ & $1$  & source for $\lambda <\sqrt{6}, \zeta <-\sqrt{6}$&&\\
    &&&&  saddle for $\lambda >\sqrt{6}$ or $\zeta >-\sqrt{6}$&&\\
    &&&& nonhyperbolic for $\lambda =\sqrt{6}$ or $\zeta =-\sqrt{6}$ & $1$ & $2$\\   \midrule
    $Z_{7}$ &$-\sqrt{6}$ & $0$ & $-1$ & sink for $\lambda >-\sqrt{6}, \zeta >\sqrt{6}$&&\\
    &&&& saddle for $\lambda <-\sqrt{6}$ or $ \zeta <\sqrt{6}$&&\\
    &&&&  nonhyperbolic for $\lambda =-\sqrt{6}$ or $ \zeta =\sqrt{6}$ & $1$ & $2$\\   \midrule
     $Z_{8}$ & $-\lambda$ & $ \sqrt{3-\frac{\lambda ^2}{2}}$ & $1$ &  nonhyperbolic for  $\lambda\in\{0, -\zeta, -\sqrt{6}, \sqrt{6}\}$ & & \\
    &&&& saddle for & &\\
    &&&&$\begin{array}{cc}
            -\sqrt{6}<\lambda <0, \zeta <-\lambda & \text{or}\\ 
            0<\lambda <\sqrt{6}, \zeta >-\lambda & \text{or}\\
            -\sqrt{6}<\lambda <0, \zeta >-\lambda & \text{or}\\
            0<\lambda <\sqrt{6}, \zeta <-\lambda
\end{array}$ &$\frac{1}{3} \left(\lambda ^2-3\right)$&$\frac{1}{2} \left(\lambda ^2-2\right)$\\ \midrule
    $Z_{9}$& $\lambda$ & $ \sqrt{3-\frac{\lambda ^2}{2}}$ & $-1$  & nonhyperbolic for $\lambda\in\{0, -\zeta, -\sqrt{6}, \sqrt{6}\}$ & & \\
    &&&& saddle for & &\\
    &&&&$\begin{array}{cc}
      -\sqrt{6}<\lambda <0, \zeta <-\lambda  & \text{or}\\ 
       0<\lambda <\sqrt{6}, \zeta >-\lambda  & \text{or}\\ 
        -\sqrt{6}<\lambda <0, \zeta >-\lambda  & \text{or}\\ 
        0<\lambda <\sqrt{6}, \zeta <-\lambda
\end{array}$ &$\frac{1}{3} \left(\lambda ^2-3\right)$&$\frac{1}{2} \left(\lambda ^2-2\right)$\\ \midrule
 $Z_{10}$ & Eq. \eqref{eqZ10},& $0$ & $ 1$ &  source or saddle, Fig. \ref{fig:7} & $\geq -1/3$ & $\geq 0$ \\ \midrule
 $Z_{11}$ & Eq. \eqref{eqZ11} & $0$ & $- 1$ &  saddle, Fig. \ref{fig:9} & $\geq -1/3$ & $\geq 0$ \\ \midrule$Z_{12}$ & Eq. \eqref{eqZ12} & $0$ & $-1$& sink or saddle, Fig. \ref{fig:11} &$>1$&$>2$ \\ 
\bottomrule
    \end{tabularx}
\end{table}
\subsection{Dynamical system analysis of 3D system for $\epsilon=-1$}
The equilibrium points in the coordinates $(x,y,\eta)$ for system \eqref{exponentialf-1}, \eqref{exponentialf-2}, \eqref{exponentialf-3} and $\epsilon=-1$ are the following.
\begin{enumerate}
    \item $Z_1=(0,y,0),$ with eigenvalues $\{0,0,0\}.$ This is a nonhyperbolic set of points for $y>0$. The asymptotic solution at the point describes the Minkowski spacetime.
    \item $Z_2=(0,0,1),$ with eigenvalues $\{2,2,1\}.$ This point is a source and we verify that $q(Z_{2})=0$. The asymptotic solution describes a universe dominated by the Gauss-Bonnet term.
    \item $Z_3=(0,0,-1),$ with eigenvalues $\{-2,-2,-1\}.$ This point is a sink, and we also have that $q(Z_3)=0$. The asymptotic behaviour is similar to that of $Z_2.$
    \item $Z_{13}=(\lambda, \sqrt{3+\frac{\lambda^2}{2}},1),$ with eigenvalues $\left\{-\lambda ^2,-\frac{1}{2} \left(\lambda ^2+6\right),\lambda  (\zeta +\lambda )\right\}.$ For this point we have $q(Z_{13})=-\frac{1}{2}(\lambda ^2+2)$ this means that acceleration occurs for $\lambda \in \mathbb{R}$. The points are
    \begin{enumerate}
        \item sinks for 
        \begin{enumerate}
            \item $\lambda<0$ and $\zeta> - \lambda$ or
            \item $\lambda>0$ and $\zeta< - \lambda,$
 \end{enumerate}
 \item saddle for
 \begin{enumerate}
            \item $\lambda<0$ and $\zeta< - \lambda$ or
            \item $\lambda>0$ and $\zeta> - \lambda.$
 \end{enumerate}
 \item nonhyperbolic for $\lambda=0$ or $\zeta=-\lambda.$
    \end{enumerate}
    \item $Z_{14}=(-\lambda,\sqrt{3+\frac{\lambda^2}{2}},-1),$ with eigenvalues $\left\{\lambda ^2,\frac{1}{2} \left(\lambda ^2+6\right),-\lambda  (\zeta +\lambda )\right\}.$ For this points we have $q(Z_{14})=-\frac{1}{2}(\lambda ^2+2)$ this means that acceleration occurs for $\lambda \in \mathbb{R}$. The points are
    \begin{enumerate}
        \item sources for 
        \begin{enumerate}
            \item $\lambda<0$ and $\zeta> - \lambda$ or
            \item $\lambda>0$ and $\zeta< - \lambda,$
 \end{enumerate}
 \item saddle for
 \begin{enumerate}
            \item $\lambda<0$ and $\zeta< - \lambda$ or
            \item $\lambda>0$ and $\zeta> - \lambda.$
 \end{enumerate}
  \item nonhyperbolic for $\lambda=0$ or $\zeta=-\lambda.$
    \end{enumerate}
         \item $Z_{15}=\Big(x_{15},0,1\Big),$ where $\tilde{r}=9 \zeta ^4+132 \zeta ^2+500$ and 
    \begin{equation}
    \label{eqZ15}
x_{15}=\frac{\frac{2\ 2^{2/3} \left(9 \zeta ^2+50\right)}{\sqrt[3]{54 \zeta ^2+9 \sqrt{2} \sqrt{\zeta ^2 (-\tilde{r})}+500}}+2
   \sqrt[3]{2} \sqrt[3]{54 \zeta ^2+9 \sqrt{2} \sqrt{\zeta ^2 (-\tilde{r})}+500}+20}{6 \zeta }.        
    \end{equation}
    This point exists for $\zeta \in \mathbb{R}$ but $\zeta\neq 0.$ We verify that

    \begin{align*}
    \omega_{\phi}&=-\frac{\sqrt[6]{2} \sqrt{-\zeta ^2 \tilde{r}} \left(9 \sqrt{2} \sqrt{-\zeta ^2 \tilde{r}}+54 \zeta
   ^2+500\right)^{2/3}}{\left(9 \zeta ^2+50\right)^2}+\frac{2^{2/3} \left(9 \sqrt{2} \sqrt{-\zeta ^2 \tilde{r}}+54
   \zeta ^2+500\right)^{2/3}}{3 \left(9 \zeta ^2+50\right)} \\&+\frac{100\ 2^{2/3} \left(9 \sqrt{2} \sqrt{-\zeta ^2
   \tilde{r}}+54 \zeta ^2+500\right)^{2/3}}{9 \left(9 \zeta ^2+50\right)^2}+\frac{1}{9} \sqrt[3]{2} \sqrt[3]{9
   \sqrt{2} \sqrt{-\zeta ^2 \tilde{r}}+54 \zeta ^2+500}+\frac{1}{9},     
\end{align*}

 \begin{align*}
q&=-\frac{3 \sqrt{-\zeta ^2 \tilde{r}} \left(9
   \sqrt{2} \sqrt{-\zeta ^2 \tilde{r}}+54 \zeta ^2+500\right)^{2/3}}{2^{5/6} \left(9 \zeta
   ^2+50\right)^2}+\frac{\left(9 \sqrt{2} \sqrt{-\zeta ^2 \tilde{r}}+54 \zeta ^2+500\right)^{2/3}}{\sqrt[3]{2}
   \left(9 \zeta ^2+50\right)} \\&+\frac{50\ 2^{2/3} \left(9 \sqrt{2} \sqrt{-\zeta ^2 \tilde{r}}+54 \zeta
   ^2+500\right)^{2/3}}{3 \left(9 \zeta ^2+50\right)^2}+\frac{\sqrt[3]{9 \sqrt{2} \sqrt{-\zeta ^2 \tilde{r}}+54
   \zeta ^2+500}}{3\ 2^{2/3}}+\frac{2}{3}.
\end{align*}
 The eigenvalues for $Z_{15}$ are $\lambda_i(\zeta,\lambda),$ with $i=1,2,3$. Given the complexity of these expressions, we perform the analysis numerically, and we present it in Fig. \ref{fig:12b} we conclude that the point has a source or saddle behaviour.
 In Fig. \ref{fig:12a} we show that $q(Z_{15})$ and $\omega_{\phi}(Z_{15})$ are always positive and they go to infinity as $\zeta\rightarrow \pm \infty.$
\begin{figure}[ht!]
    \centering
    \includegraphics[scale=0.7]{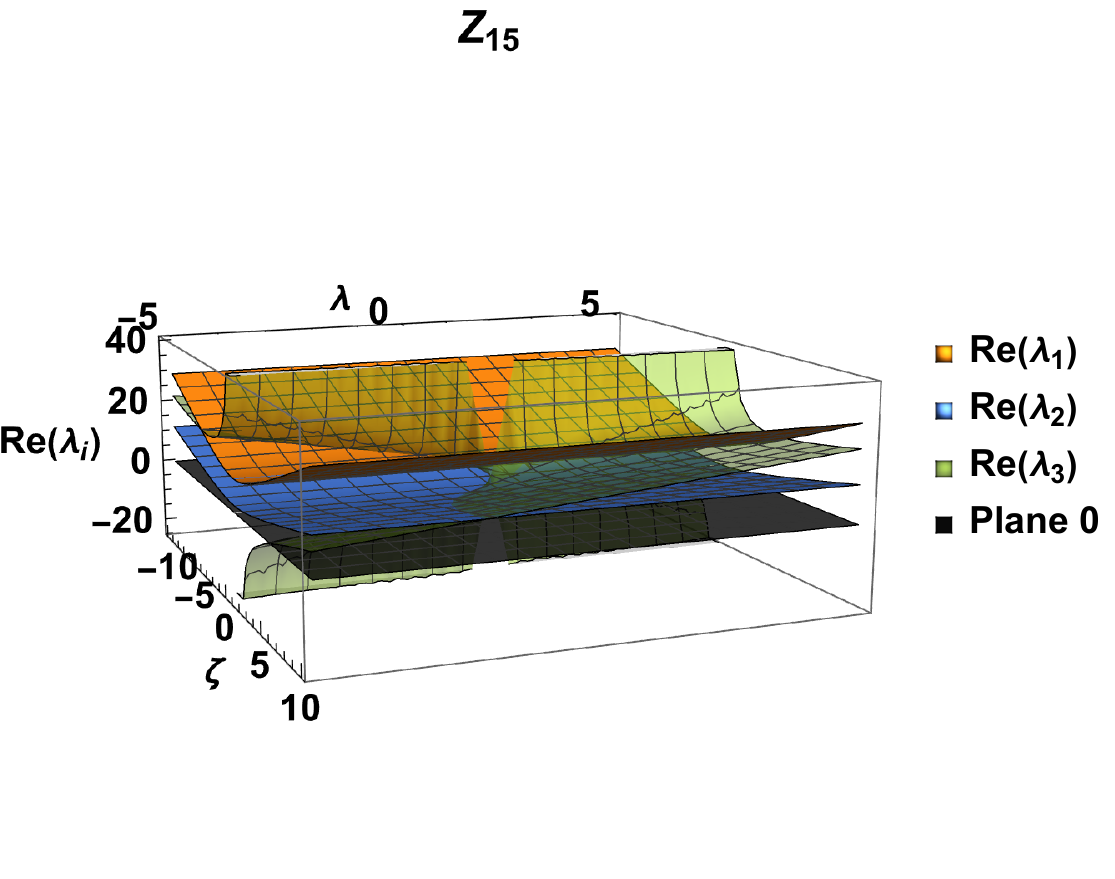}
    \caption{Real part of the eigenvalues of $Z_{15}.$ This point is a source or a saddle.}
    \label{fig:12b}
\end{figure}
\begin{figure}[ht!]
    \centering
    \includegraphics[scale=0.6]{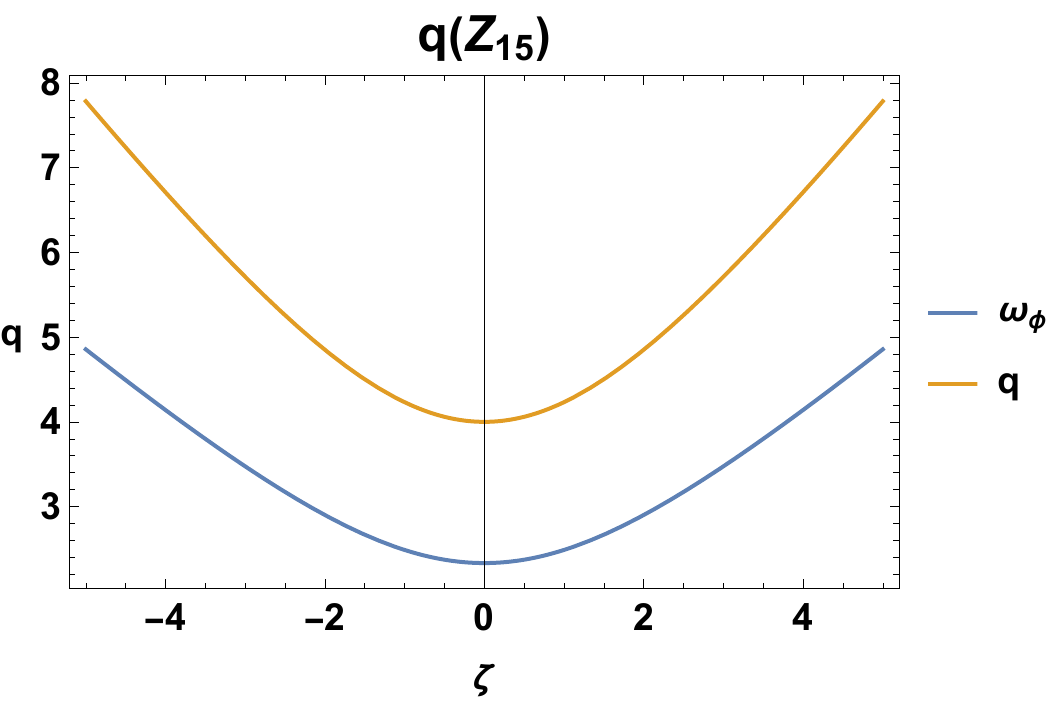}
    \caption{Plot of $q(Z_{15})$ and $\omega_{\phi}(Z_{15})$ they are both positive and go to $-\infty$ as $\zeta\rightarrow \pm \infty.$}
    \label{fig:12a}
\end{figure}
    \item $Z_{16}=\Big(x_{16},0,1\Big),$ where $\Tilde{r}=9 \zeta ^4+132 \zeta ^2+500$ and
    \begin{equation}
        \label{eqZ16}
        x_{16}=\frac{-\frac{4 \sqrt[3]{-1} 2^{2/3} \left(9 \zeta ^2+50\right)}{\sqrt[3]{9 \sqrt{2} \sqrt{-\zeta ^2 \tilde{r}}+54
   \zeta ^2+500}}+4 (-1)^{2/3} \sqrt[3]{2} \sqrt[3]{9 \sqrt{2} \sqrt{-\zeta ^2 \tilde{r}}+54 \zeta ^2+500}+40}{12
   \zeta }.
    \end{equation} 
    This point exists for $\zeta \in \mathbb{R}$ but $\zeta\neq 0.$ For $Z_{16}$ we have $\omega_{\phi}=f_1(\zeta)$ and  $q=f_2(\zeta)$.
The eigenvalues for $Z_{16}$ are $\lambda_i(\zeta,\lambda),$ with $i=4,5,6,$ given the complexity of the expressions, we perform numerical analysis to conclude that this point is a sink or saddle, see Fig. \ref{fig:13b}. Since the expressions for the EoS and deceleration parameters are lengthy and complicated, we write them as $f_i(\zeta)$. However, we verify that they are both negative and for $\zeta\rightarrow \pm \infty$, we have that $q\rightarrow -\infty$ and see $\omega_{\phi}\rightarrow -\infty$, see Fig. \ref{fig:13a}.
\begin{figure}[ht!]
    \centering
    \includegraphics[scale=0.7]{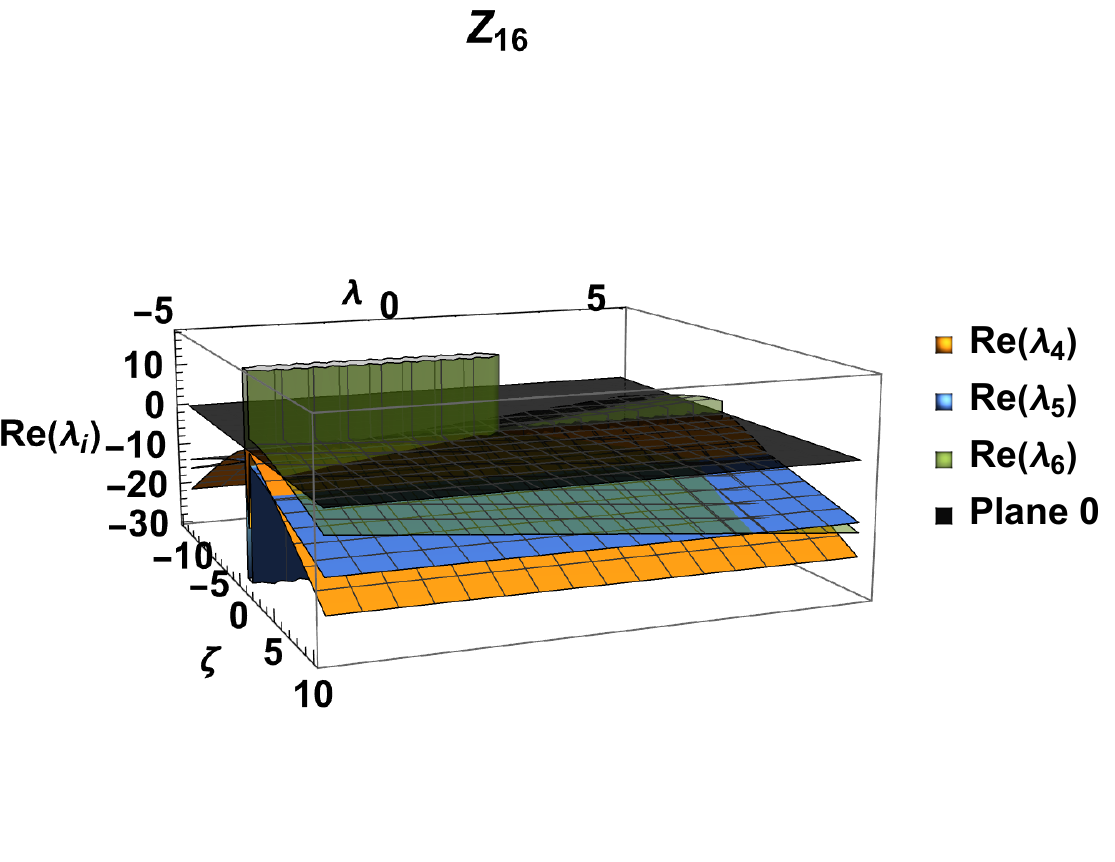}
    \caption{Real part of the eigenvalues of $Z_{16}.$ This point has sink or saddle behaviour.}
    \label{fig:13b}
\end{figure}
\begin{figure}[ht!]
    \centering
    \includegraphics[scale=0.6]{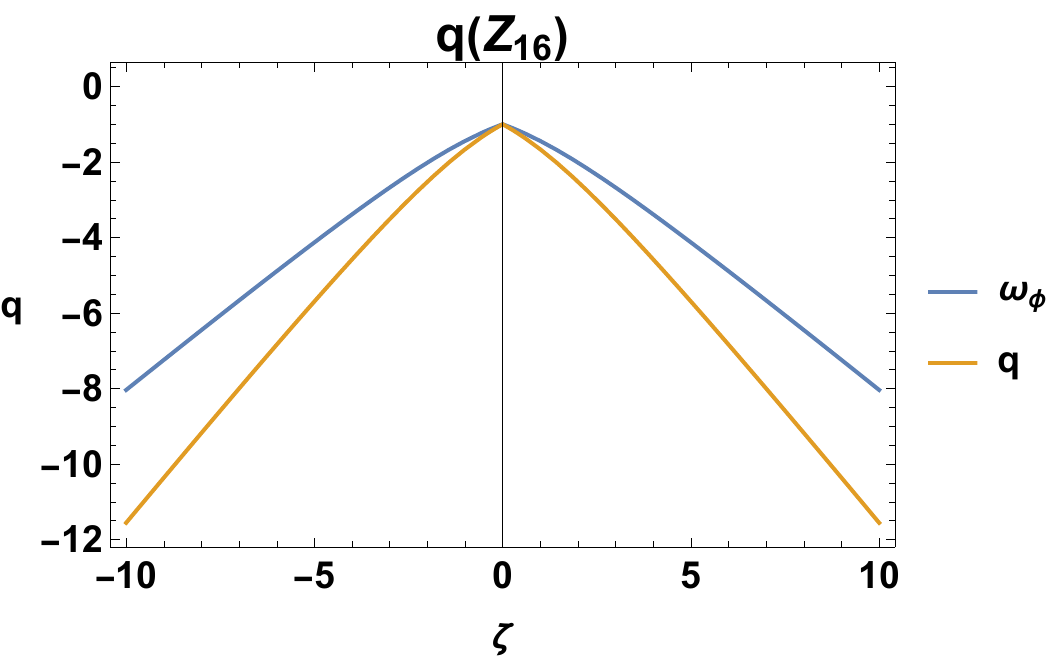}
    \caption{Plot of $q(Z_{16})$ and $\omega_{\phi}(Z_{16})$ they are both negative and go to $-\infty$ as $\zeta\rightarrow \pm \infty.$}
    \label{fig:13a}
\end{figure}
    \item $Z_{17}=\Big(x_{17},0,1\Big),$ where $\Tilde{r}=9 \zeta ^4+132 \zeta ^2+500$ and
    \begin{equation}
        \label{eqZ17}
        x_{17}=\frac{\frac{4 (-2)^{2/3} \left(9 \zeta ^2+50\right)}{\sqrt[3]{9 \sqrt{2} \sqrt{-\zeta ^2 \tilde{r}}+54 \zeta
   ^2+500}}-4 \sqrt[3]{-2} \sqrt[3]{9 \sqrt{2} \sqrt{-\zeta ^2 \tilde{r}}+54 \zeta ^2+500}+40}{12 \zeta }.
    \end{equation}
    This point exists for $\zeta \in \mathbb{R}$ but $\zeta\neq 0.$ The eigenvalues for this point are written symbolically as $\lambda_i(\zeta,\lambda)$ where $i=7,8,9.$ In Fig. \ref{fig:14b}, we see that the point has saddle behaviour. For $Z_{17}$ we have $\omega_{\phi}(Z_{17})=f_1(\zeta)$ and $q(Z_{17})=f_2(\zeta).$
    In Fig. \ref{fig:14a}, we show that the EoS and deceleration parameters are always negative and $\omega_{\phi}\rightarrow -\frac{1}{3},$ $q\rightarrow 0$ as $\zeta\rightarrow \pm \infty$.
    \begin{figure}[ht!]
    \centering
    \includegraphics[scale=0.7]{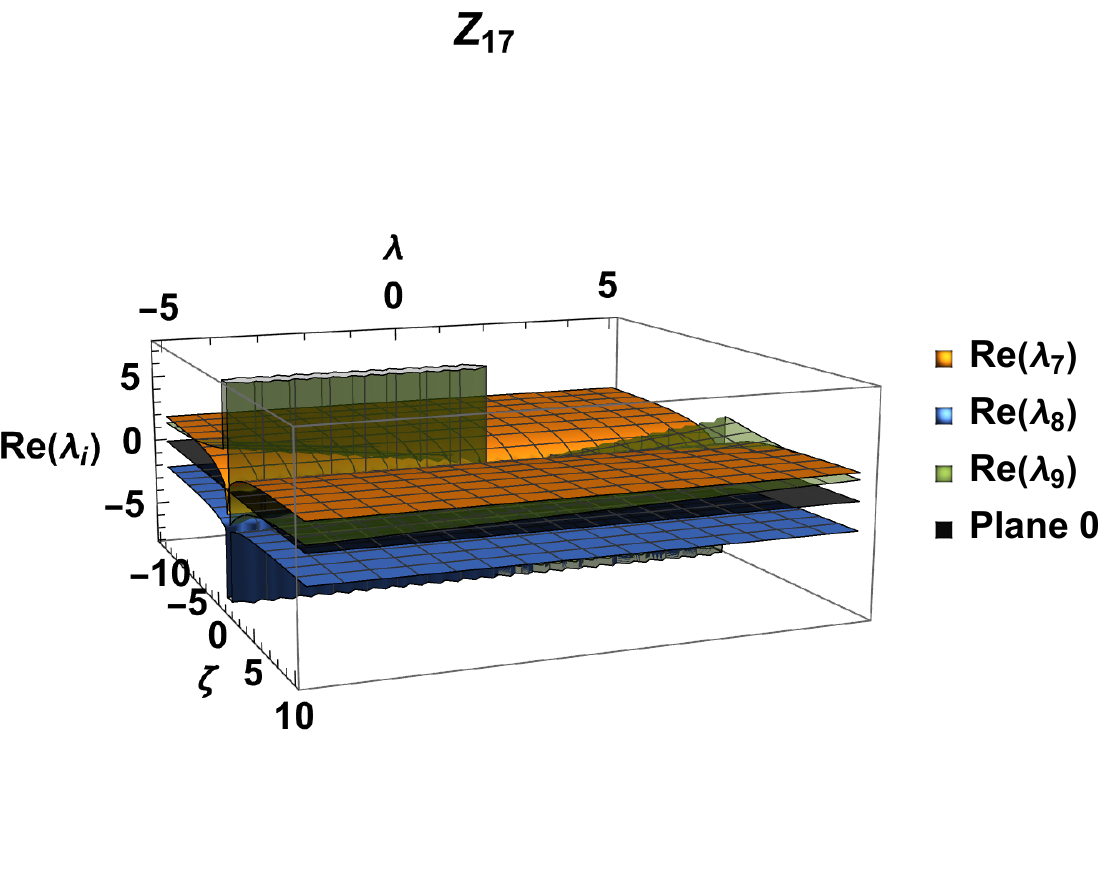}
    \caption{Real part of the eigenvalues of $Z_{17},$ we see that the point has saddle behaviour.}
    \label{fig:14b}
\end{figure}
\begin{figure}[ht!]
    \centering
    \includegraphics[scale=0.6]{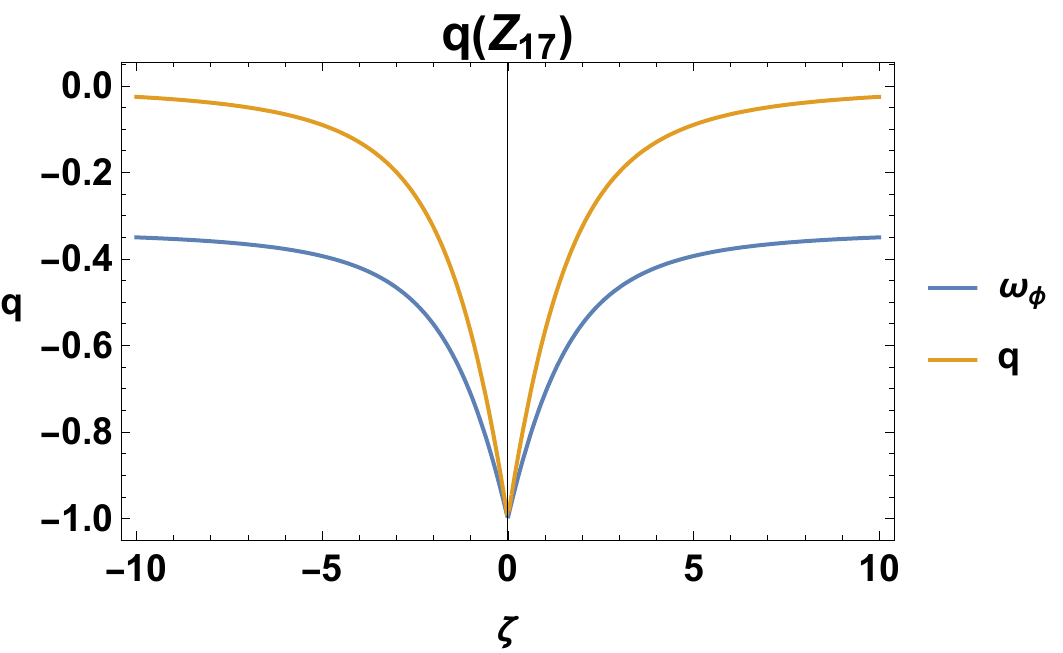}
    \caption{Plot of $q(Z_{17})$ and $\omega_{\phi}(Z_{17}).$ Here we see that $q\rightarrow 0$ and $\omega_{\phi}\rightarrow -\frac{1}{3}$ as $\zeta \rightarrow \pm  \infty$.}
    \label{fig:14a}
\end{figure}
    \item $Z_{18}=\Big(x_{18},0,-1\Big),$ where $\Tilde{r}=9 \zeta ^4+132 \zeta ^2+500$ and
    \begin{equation}
    \label{eqZ18}
        x_{18}=\frac{\frac{2\ 2^{2/3} \left(9 \zeta ^2+50\right)}{\sqrt[3]{9 \sqrt{2} \sqrt{-\zeta ^2 \tilde{r}}-54 \zeta ^2-500}}+2 \sqrt[3]{2} \sqrt[3]{9
   \sqrt{2} \sqrt{-\zeta ^2 \tilde{r}}-54 \zeta ^2-500}-20}{6 \zeta }.
    \end{equation}
     This point exists for $\zeta \in \mathbb{R}$ but $\zeta\neq 0.$ For this point, we have 
      \begin{align*}
        \omega_{\phi}&=\frac{\sqrt[6]{2} \sqrt{-\zeta ^2 \tilde{r}} \left(9 \sqrt{2} \sqrt{-\zeta ^2 \tilde{r}}-54 \zeta
   ^2-500\right)^{2/3}}{\left(9 \zeta ^2+50\right)^2}+\frac{2^{2/3} \left(9 \sqrt{2} \sqrt{-\zeta ^2 \tilde{r}}-54
   \zeta ^2-500\right)^{2/3}}{3 \left(9 \zeta ^2+50\right)}\nonumber \\ &+\frac{100\ 2^{2/3} \left(9 \sqrt{2} \sqrt{-\zeta ^2
   \tilde{r}}-54 \zeta ^2-500\right)^{2/3}}{9 \left(9 \zeta ^2+50\right)^2}-\frac{1}{9} \sqrt[3]{2} \sqrt[3]{9
   \sqrt{2} \sqrt{-\zeta ^2 \tilde{r}}-54 \zeta ^2-500}+\frac{1}{9},
    \end{align*}
    \begin{align*}
        q&=\frac{3 \sqrt{-\zeta ^2 \tilde{r}} \left(9
   \sqrt{2} \sqrt{-\zeta ^2 \tilde{r}}-54 \zeta ^2-500\right)^{2/3}}{2^{5/6} \left(9 \zeta
   ^2+50\right)^2}+\frac{\left(9 \sqrt{2} \sqrt{-\zeta ^2 \tilde{r}}-54 \zeta ^2-500\right)^{2/3}}{\sqrt[3]{2}
   \left(9 \zeta ^2+50\right)}\nonumber \\ &+\frac{50\ 2^{2/3} \left(9 \sqrt{2} \sqrt{-\zeta ^2 \tilde{r}}-54 \zeta
   ^2-500\right)^{2/3}}{3 \left(9 \zeta ^2+50\right)^2}-\frac{\sqrt[3]{9 \sqrt{2} \sqrt{-\zeta ^2 \tilde{r}}-54
   \zeta ^2-500}}{3\ 2^{2/3}}+\frac{2}{3}.
    \end{align*} and the eigenvalues are $\delta_i(\zeta,\lambda)$ for $i=1,2,3.$ In Fig. \ref{fig:15a}, we show the real part of the eigenvalues and conclude that $Z_{18}$ is a source. We also verified that the EoS and deceleration parameters are negative and go to minus infinity as $\zeta \rightarrow\pm \infty,$ see Fig. \ref{fig:15b}. 
    \begin{figure}[ht!]
    \centering
    \includegraphics[scale=0.7]{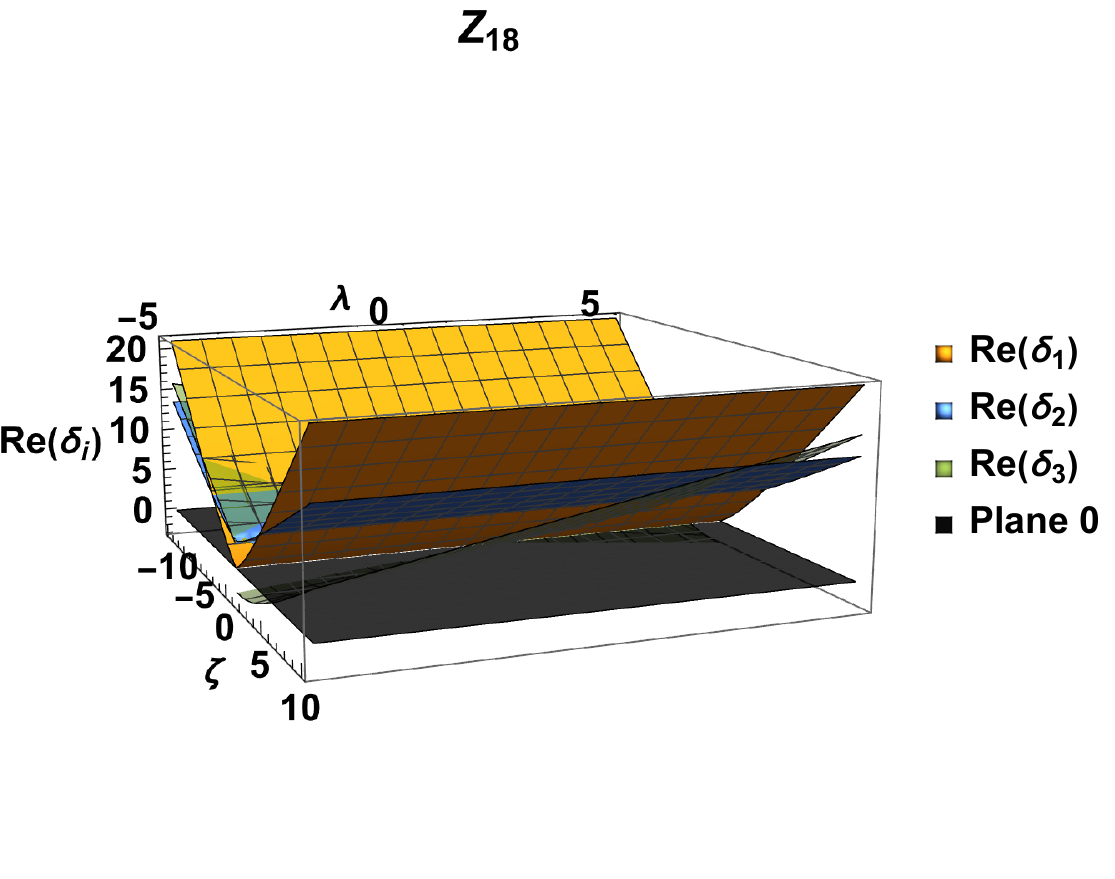}
    \caption{Real part of the eigenvalues of $Z_{18}$, we can see that the point is a source.}
    \label{fig:15a}
\end{figure}
    \begin{figure}[ht!]
    \centering
    \includegraphics[scale=0.6]{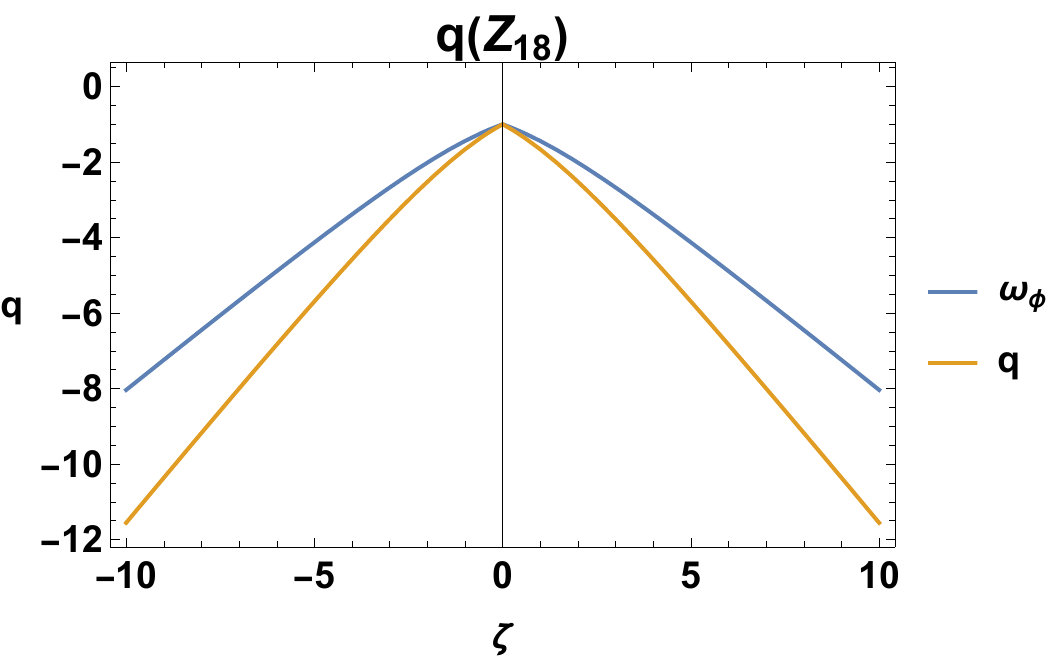}
    \caption{Plot of $q(Z_{18})$ and $\omega_{\phi}(Z_{18})$ we see that both are negative and go to $-\infty$ as $\zeta\rightarrow \pm \infty$.}
    \label{fig:15b}
\end{figure}
    \item $Z_{19}=\Big(x_{19},0,-1\Big),$ where $\Tilde{r}=9 \zeta ^4+132 \zeta ^2+500$ and

    \begin{equation}
        \label{eqZ19}
        x_{19}=\frac{-\frac{4 \sqrt[3]{-1} 2^{2/3} \left(9 \zeta ^2+50\right)}{\sqrt[3]{9 \sqrt{2} \sqrt{-\zeta ^2 \tilde{r}}-54 \zeta ^2-500}}+4 (-1)^{2/3}
   \sqrt[3]{2} \sqrt[3]{9 \sqrt{2} \sqrt{-\zeta ^2 \tilde{r}}-54 \zeta ^2-500}-40}{12 \zeta }.
    \end{equation}
     This point exists for $\zeta \in \mathbb{R}$ but $\zeta\neq 0.$
     The eigenvalues are $\delta_i(\zeta,\lambda)$ for $i=4,5,6.$ The stability analysis is performed numerically in Fig. \ref{fig:16a} where we see that $Z_{19}$ has sink or saddle behaviour.
     For $Z_{19}$ we have that $\omega_{\phi}(Z_{19})=f_{2}(\zeta,\lambda)$ and $q(Z_{19})=f_{2}(\zeta,\lambda)$ that is, are complicated expresions that depend on $\zeta$ and $\lambda$ therefore we study them in Fig. \ref{fig:16b} and see that they are always positive and go to $\infty$ as $\zeta \rightarrow \pm \infty.$
     \begin{figure}[ht!]
    \centering
    \includegraphics[scale=0.7]{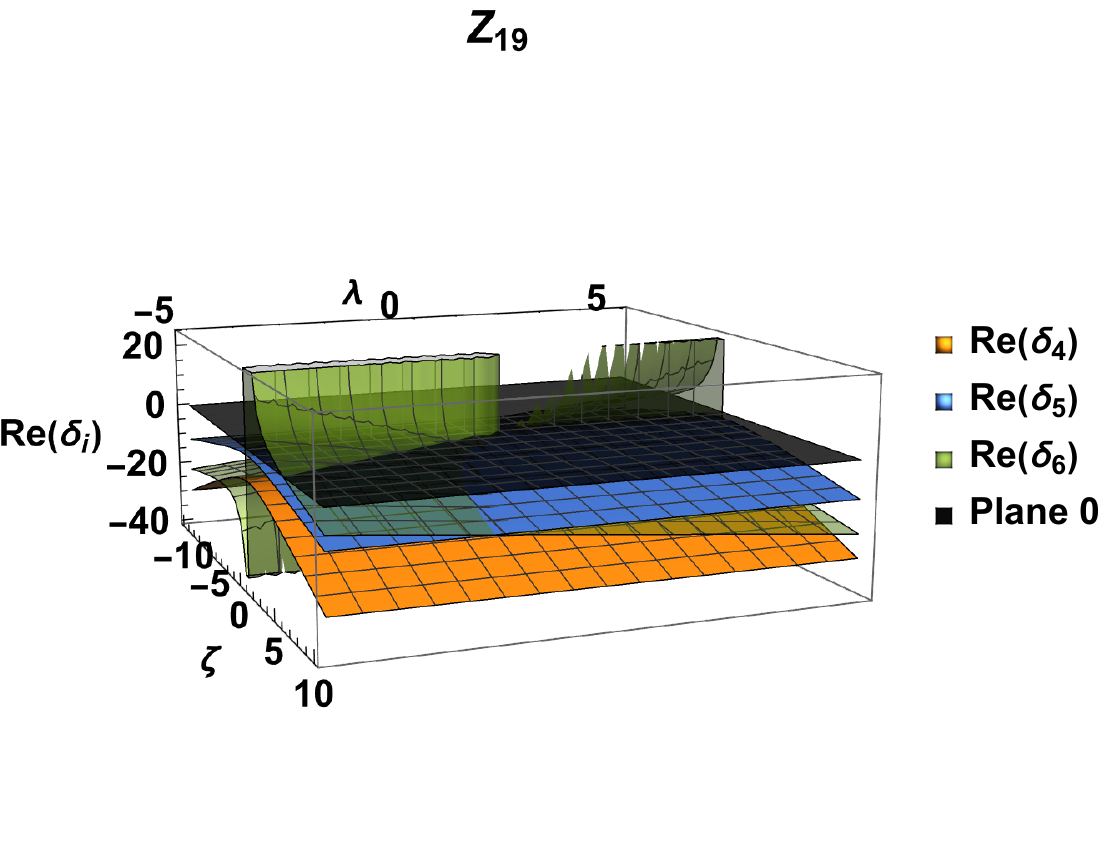}
    \caption{Real part of the eigenvalues of $Z_{19},$ we see that the point has sink or saddle behaviour.}
    \label{fig:16a}
\end{figure}
\begin{figure}[ht!]
    \centering
    \includegraphics[scale=0.6]{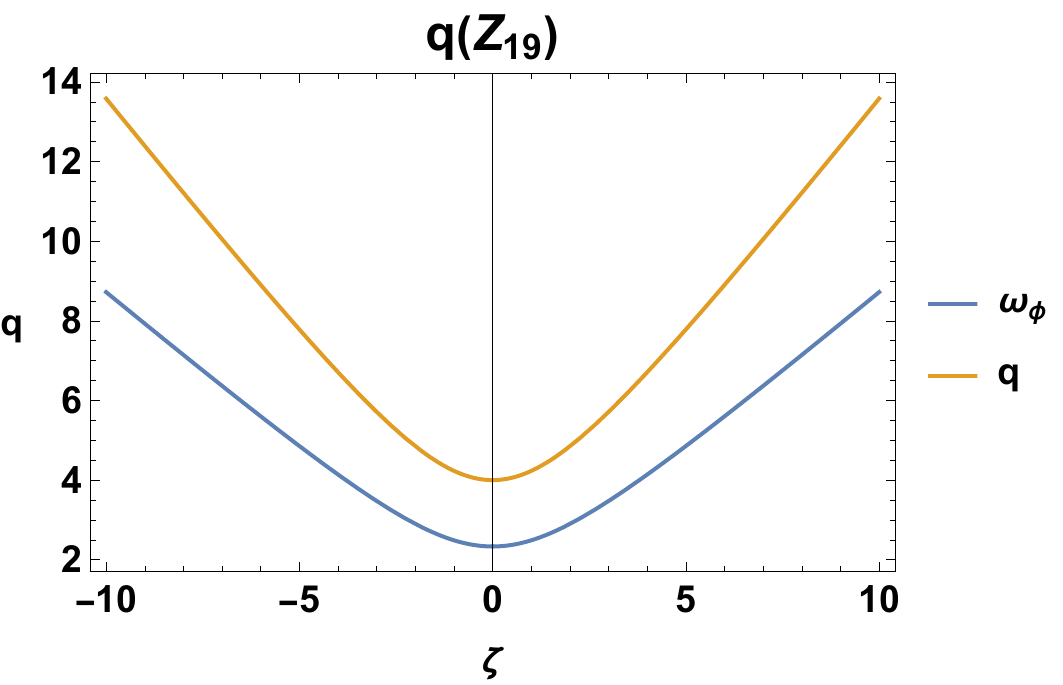}
    \caption{Plots for $q(Z_{19})$ and $\omega_{\phi}(Z_{19})$ we see that both are positive and go to $\infty$ as $\zeta\rightarrow \pm \infty$.}
    \label{fig:16b}
\end{figure}    

    \item $Z_{20}=\Big(x_{20},0,-1\Big),$ where $\Tilde{r}=9 \zeta ^4+132 \zeta ^2+500$ and
    \begin{equation}
        \label{eqZ20}
        x_{2}=\frac{\frac{4 (-2)^{2/3} \left(9 \zeta ^2+50\right)}{\sqrt[3]{9 \sqrt{2} \sqrt{-\zeta ^2 \tilde{r}}-54 \zeta ^2-500}}-4 \sqrt[3]{-2} \sqrt[3]{9
   \sqrt{2} \sqrt{-\zeta ^2 \tilde{r}}-54 \zeta ^2-500}-40}{12 \zeta }.
    \end{equation}
     This point exists for $\zeta \in \mathbb{R}$ but $\zeta\neq 0.$ The eigenvalues of $Z_{20}$ are $\delta_{i}(\zeta, \lambda)$ for $i=7,8,9$ and the stability behavior is saddle-like. For the EoS and deceleration parameters, they can be written as $\omega_{\phi}(Z_{20})=f_1(\zeta,\lambda)$, $q(Z_{20})=f_2(\zeta,\lambda)$ and we verify that they are both negative. Particularly if $\zeta \rightarrow \pm \infty$  we verify that $\omega_{\phi}(Z_{20}) \rightarrow -\frac{1}{3}$ and $q(Z_{20})\rightarrow 0.$
     \begin{figure}[ht!]
    \centering
    \includegraphics[scale=0.7]{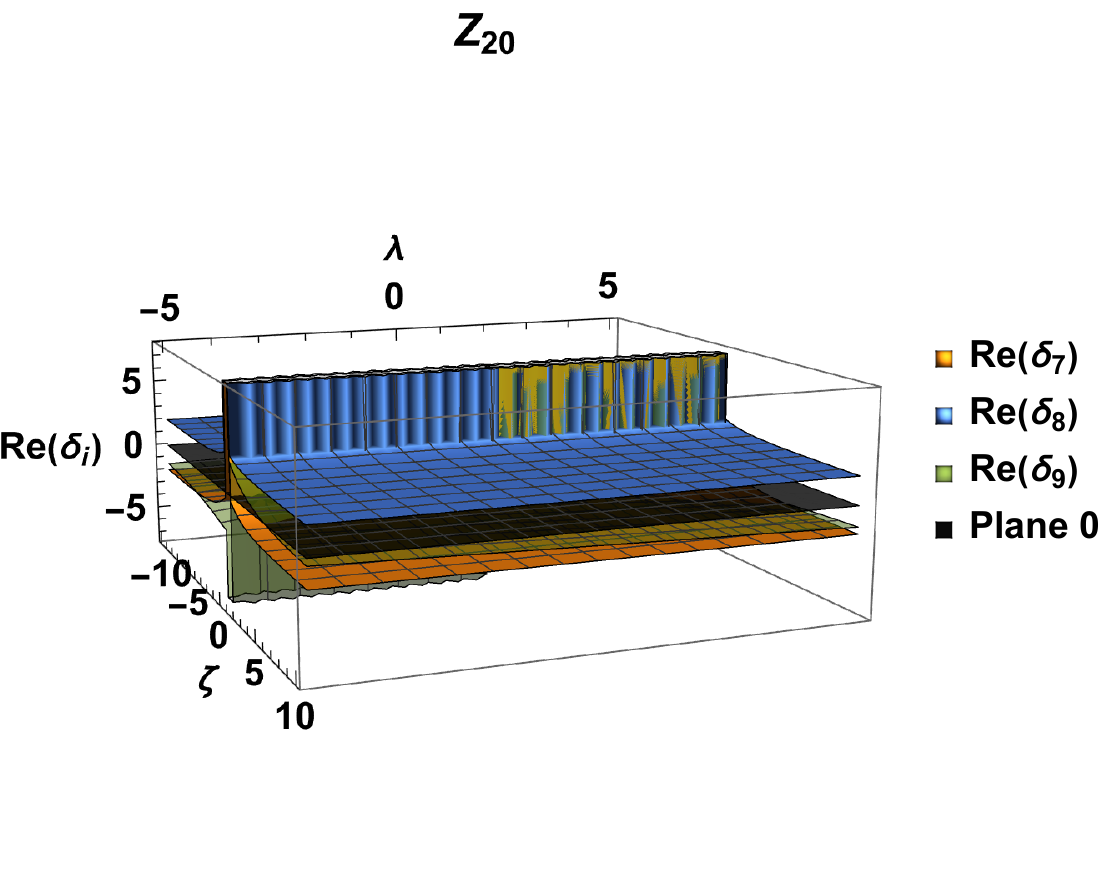}
    \caption{Real part of the eigenvalues of $Z_{20}$, we see that the point has saddle behaviour.}
    \label{fig:17a}
\end{figure}
\begin{figure}[ht!]
    \centering
    \includegraphics[scale=0.6]{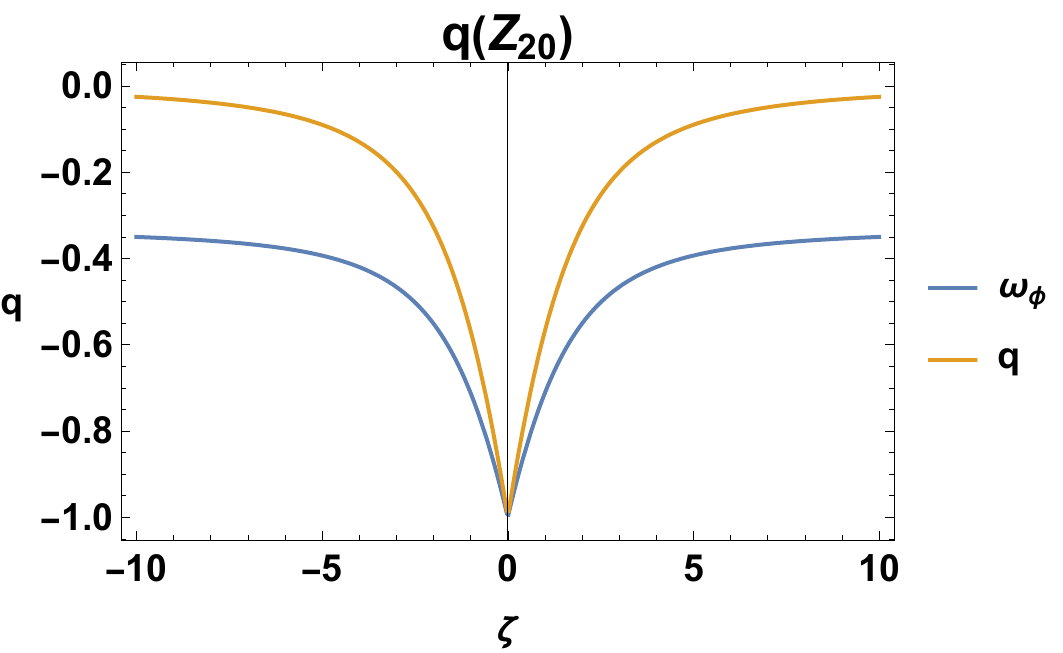}
    \caption{Plots of $q(Z_{20})$ and $\omega_{\phi}(Z_{20})$ we see that they are both negative but $q(Z_{20})\rightarrow 0$ and $\omega_{\phi}(Z_{20})\rightarrow -\frac{1}{3}.$}
    \label{fig:17b}
\end{figure}

\end{enumerate}

\begin{figure}[ht!]
    \centering
    \includegraphics[scale=0.45]{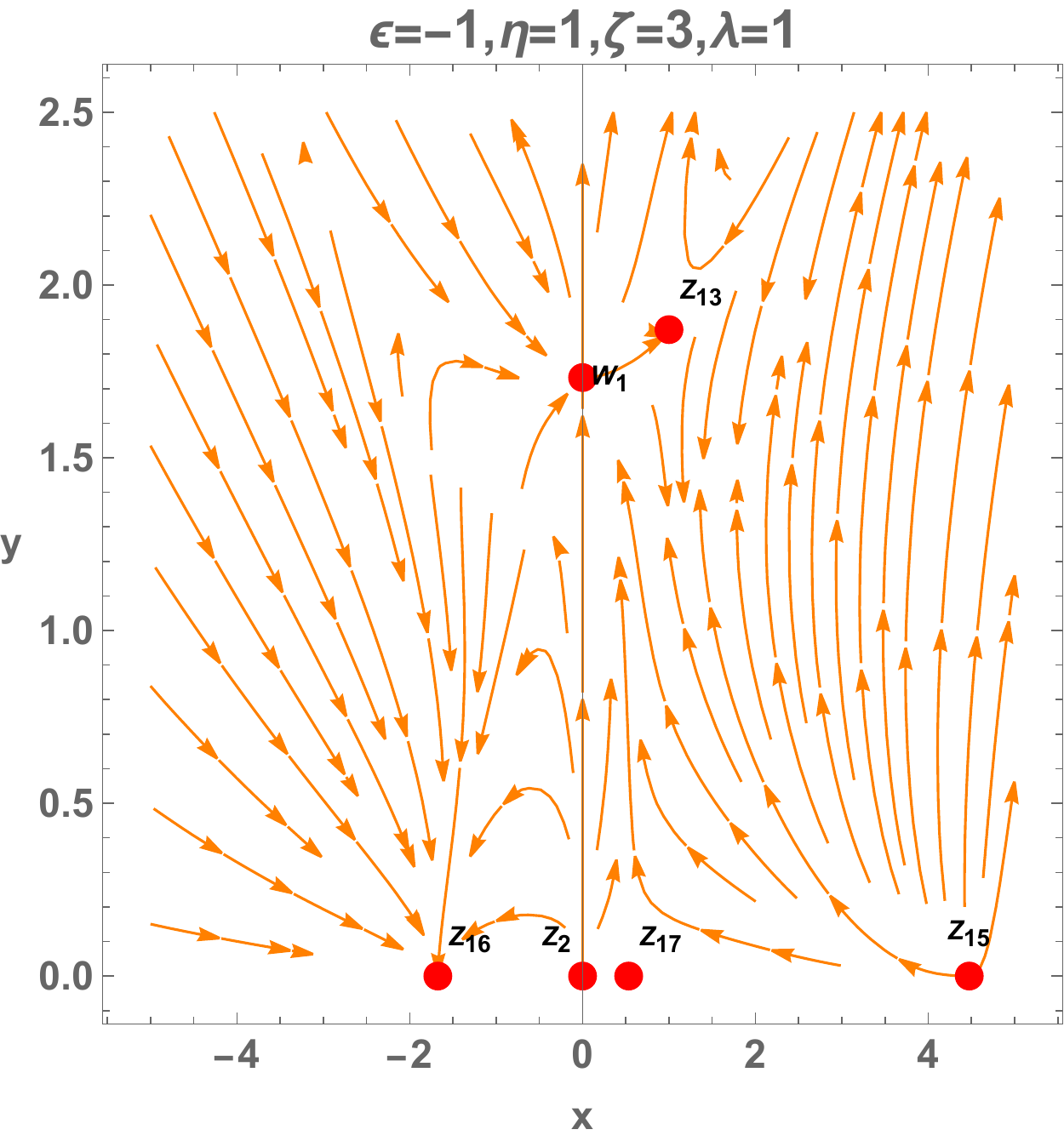}
    \includegraphics[scale=0.45]{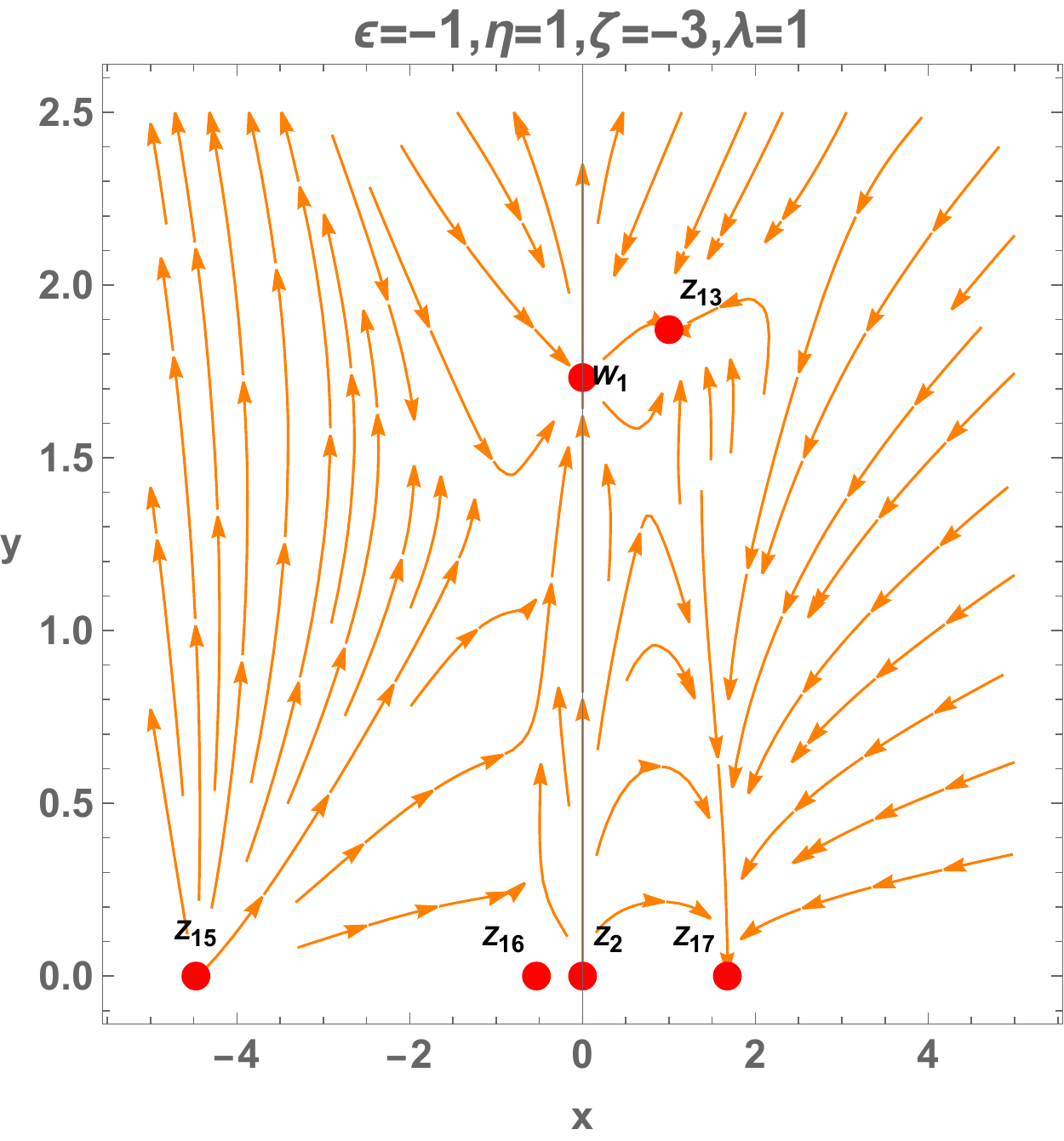}
    \includegraphics[scale=0.45]{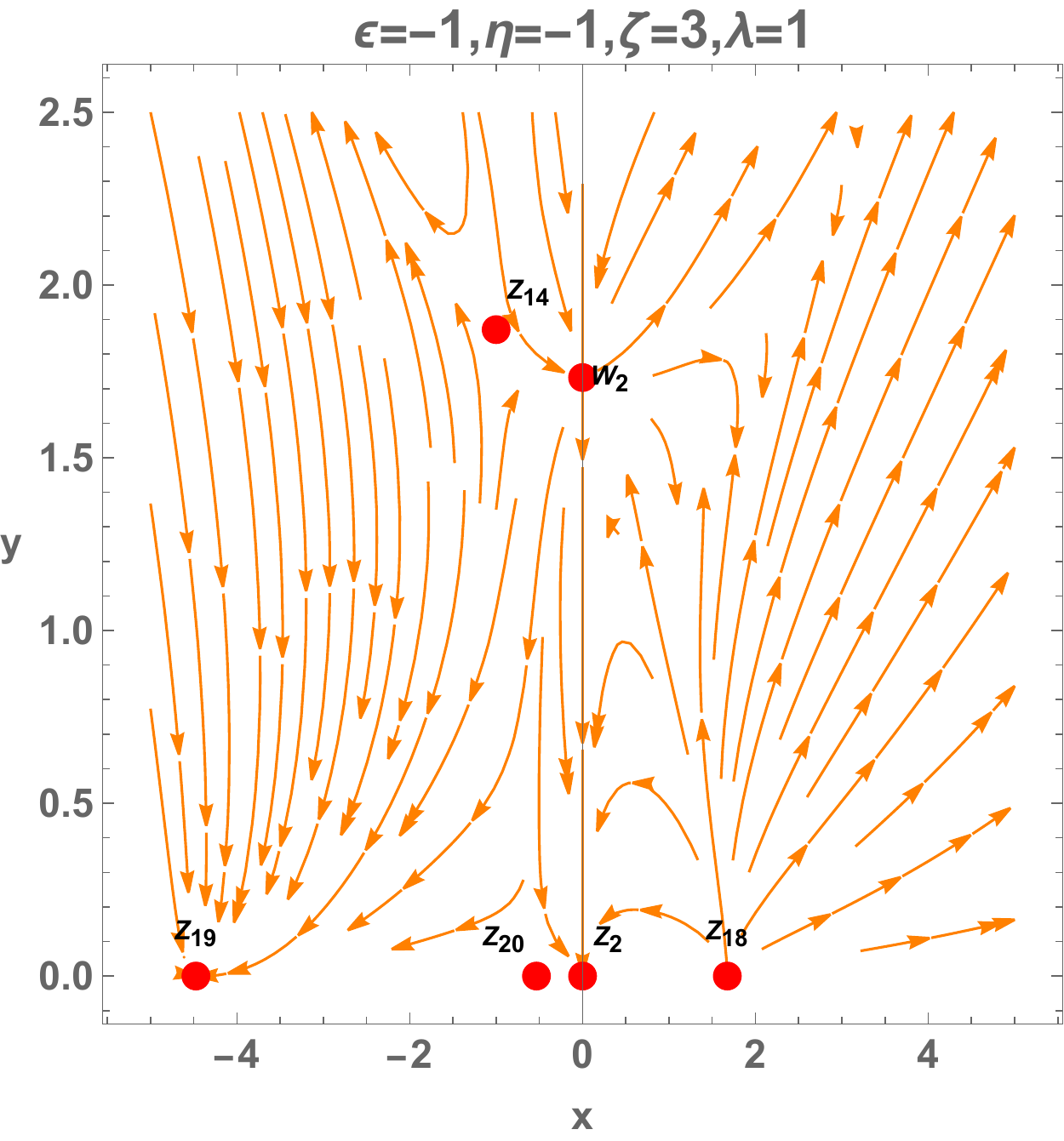}
    \includegraphics[scale=0.45]{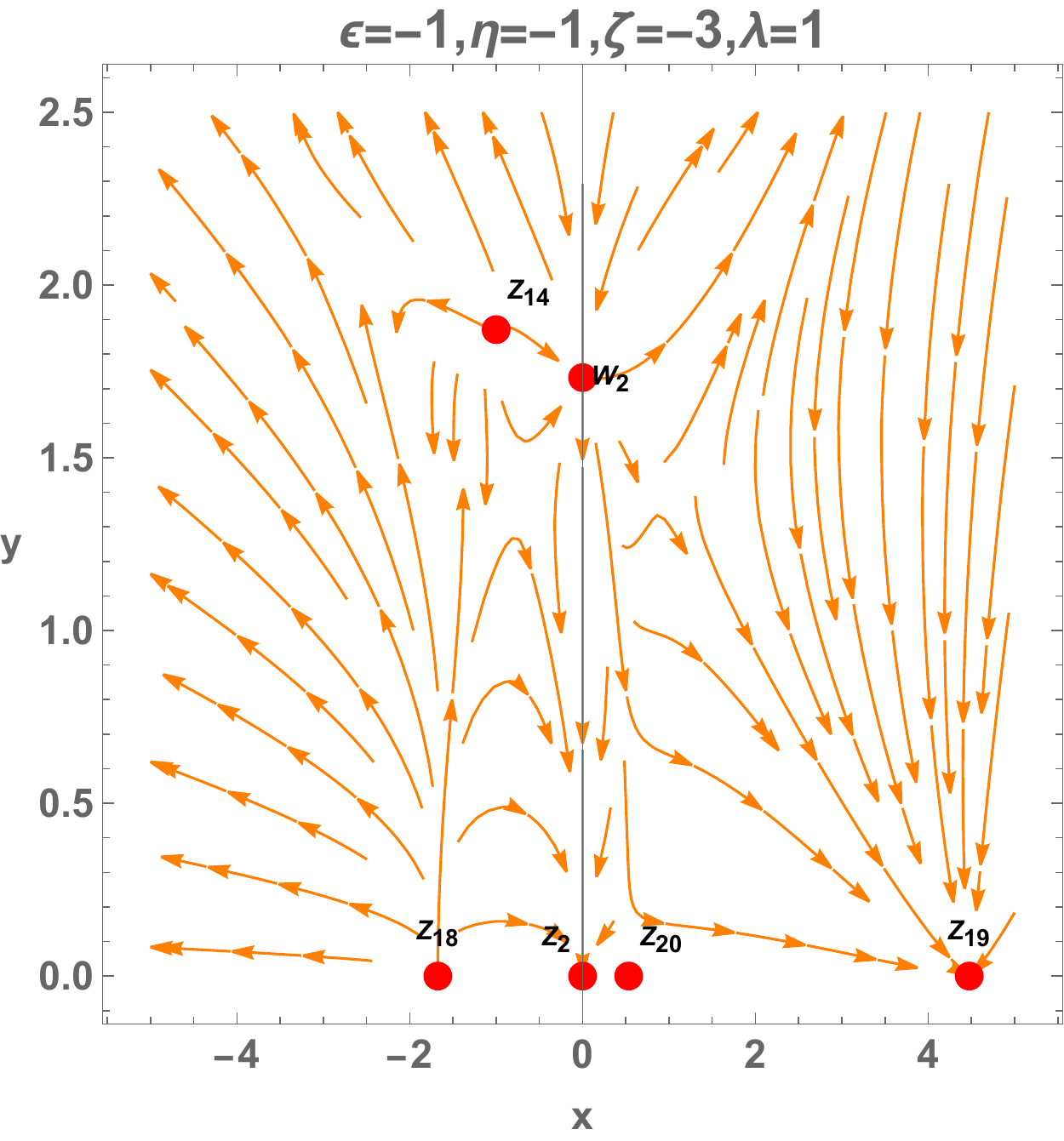}
    \caption{2D-Projection of system \eqref{exponentialf-1}, \eqref{exponentialf-2} and \eqref{exponentialf-3} setting $\epsilon=-1, \lambda=1$ for $\eta=-1$ with different values of $\zeta$. Here, the saddle points $W_1=(0,\sqrt{3})$ for $\eta=1$ and $W_2=(0,\sqrt{3})$ for $\eta=-1$ are singularities for which both the numerator and denominator of the $y$ equation vanish.}
    \label{fig:18}
\end{figure}

\begin{figure}
    \centering
    \includegraphics[scale=0.55]{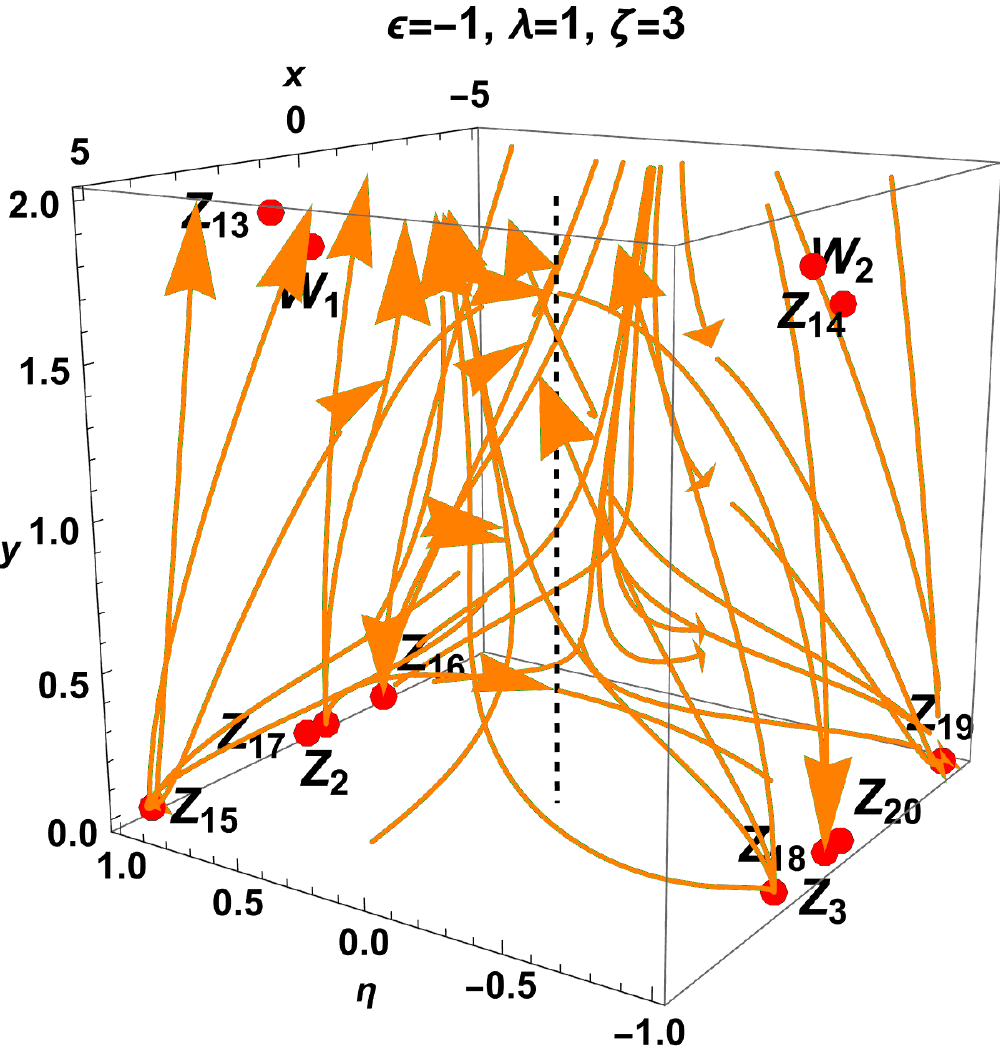}
    \includegraphics[scale=0.55]{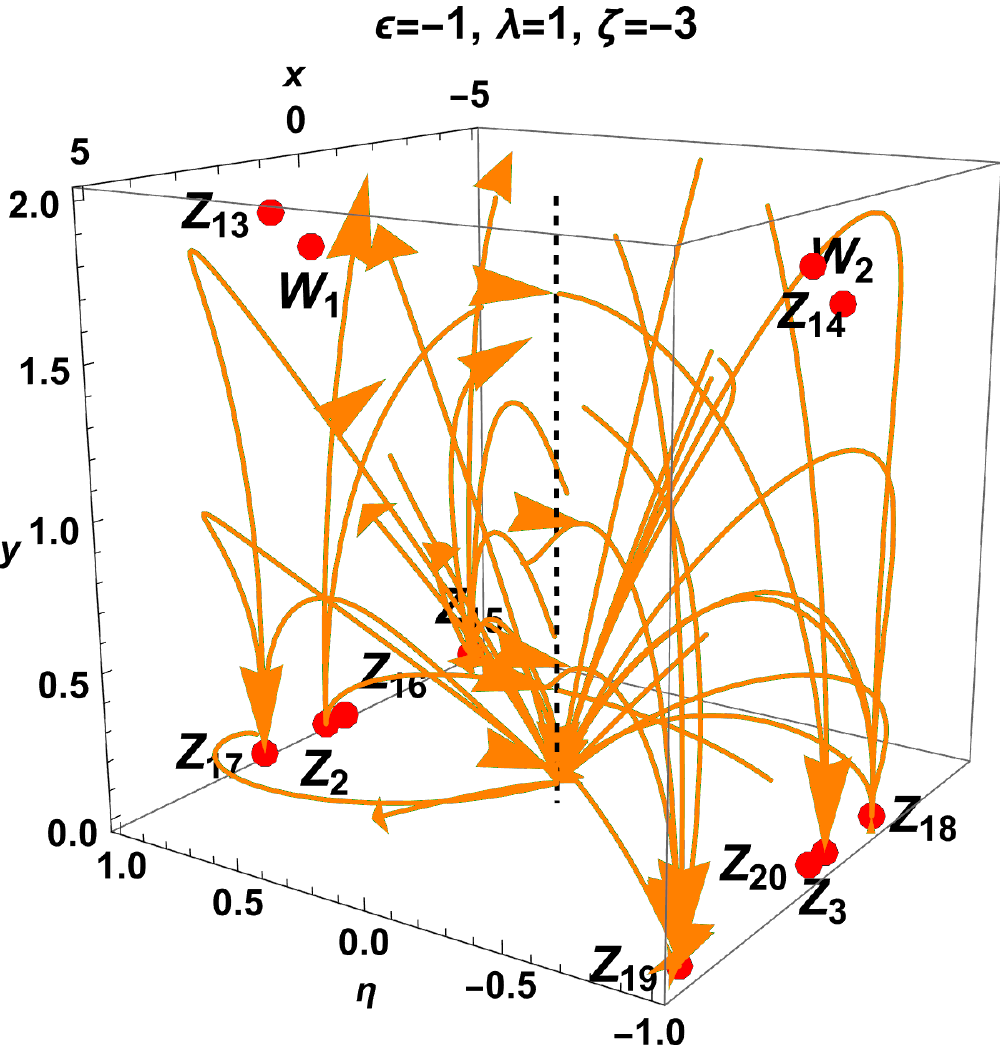}
    \caption{Three dimensional phase plot for \eqref{exponentialf-1}, \eqref{exponentialf-2} and \eqref{exponentialf-3} setting $\epsilon=-1, \lambda=1$ for different values of $\zeta.$ Here, the saddle points $W_1=(0,\sqrt{3},1)$ and $W_2=(0,\sqrt{3},-1)$ are singularities in which both the numerator and denominator of the $y$ equation vanish.}
    \label{fig:18a}
\end{figure}

In Fig. \ref{fig:18} we present phase-space diagrams for a 2D projection of system \eqref{exponentialf-1}, \eqref{exponentialf-2} and \eqref{exponentialf-3} setting $\epsilon=-1, \lambda=1$, $\eta=\pm 1$ and different values of $\zeta.$ Also three dimensional phase-space diagrams are presented in Fig. \ref{fig:18a} for $\epsilon=-1, \lambda=1$ and different values of $\zeta.$ The results of this section are summarized in Table \ref{tab:5}. The existence of the points $Z_{15}, Z_{16}, Z_{17}, Z_{18}, Z_{19}, Z_{20}$ is discussed in appendix \ref{app}.

\begin{table}[ht!]
\caption{Equilibrium points of system \eqref{exponentialf-1}, \eqref{exponentialf-2}, \eqref{exponentialf-3}  for $\epsilon=-1$ with their stability conditions. It also includes the value of $\omega_{\phi}$ and $q.$ }
    \label{tab:5}
    \centering
\newcolumntype{C}{>{\centering\arraybackslash}X}
\centering
    \setlength{\tabcolsep}{1.6mm}
\begin{tabularx}{\textwidth}{cccccccc}
\toprule 
    Label& $x$ & $y$ &$\eta$ &Stability &$\omega_{\phi}$ & $q$ \\ \midrule
    $Z_1$ & $0$ & $y$ & $0$ & nonhyperbolic & indeterminate & indeterminate \\ \midrule
    $Z_2$ & $0$ & $0$ & $1$ & source & $-\frac{1}{3}$ & $0$\\ \midrule
    $Z_3$ & $0$ & $0$ & $-1$ & sink & $-\frac{1}{3}$ & $0$\\ \midrule
    $Z_{13}$ & $\lambda$ & $ \sqrt{3+\frac{\lambda^2}{2}}$ & $1$ &   nonhyperbolic for  $\lambda=0$ or $\zeta=-\lambda$ & & \\
    &&&& sinks for & &\\
    &&&&$\begin{array}{cc}
            \lambda<0,\zeta >-\lambda& \text{or}\\ 
            \lambda>0,\zeta <-\lambda&\end{array}$
            & & \\
           &&&& saddle for & &\\
    &&&&
            $\begin{array}{cc}
            \lambda<0,\zeta <-\lambda& \text{or}\\ 
            \lambda>0,\zeta >-\lambda&\end{array}$& $-\frac{1}{3}(\lambda ^2+3)$ & $-\frac{1}{2}(\lambda ^2+2)$\\ \midrule
    $Z_{14}$ & $-\lambda$ & $ \sqrt{3+\frac{\lambda^2}{2}}$ & $-1$ &   nonhyperbolic for  $\lambda=0$ or $\zeta=-\lambda$ & & \\
    &&&& sources for & &\\
    &&&&$\begin{array}{cc}
            \lambda<0,\zeta >-\lambda& \text{or}\\ 
            \lambda>0,\zeta <-\lambda&\end{array}$
            & & \\
           &&&& saddle for & &\\
    &&&&
            $\begin{array}{cc}
            \lambda<0,\zeta <-\lambda& \text{or}\\ 
            \lambda>0,\zeta >-\lambda&\end{array}$& $-\frac{1}{3}(\lambda ^2+3)$ & $-\frac{1}{2}(\lambda ^2+2)$\\ \midrule
            $Z_{15}$ & Eq.\eqref{eqZ15} & $0$ & $1$ & source or saddle, see Fig. \ref{fig:12b} & $>0$, see Fig. \ref{fig:12a} &$>0$, see Fig. \ref{fig:12a}\\ \midrule
             $Z_{16}$ & Eq.\eqref{eqZ16} & $0$ & $1$ & sink or saddle, see Fig. \ref{fig:13b} & $<0$, see Fig. \ref{fig:13a} & $<0$, see Fig. \ref{fig:13a}\\ \midrule
              $Z_{17}$ & Eq.\eqref{eqZ17} & $0$ & $1$ & saddle, see Fig. \ref{fig:14b} & $\leq-\frac{1}{3}$, see Fig. \ref{fig:14a} &$\leq 0$, see  Fig. \ref{fig:14a}\\ \midrule
              $Z_{18}$ & Eq.\eqref{eqZ18} & $0$ & $-1$ & source, see  Fig. \ref{fig:15a} & $<0$, see Fig. \ref{fig:15b} & $<0$, see Fig. \ref{fig:15b}\\ \midrule
             $Z_{19}$ & Eq.\eqref{eqZ19} & $0$ & $-1$ & sink or saddle, see Fig. \ref{fig:16a} & $>0$, see Fig. \ref{fig:16b} & $>0$, see  Fig. \ref{fig:16b}\\ \midrule
              $Z_{20}$ & Eq.\eqref{eqZ20} & $0$ & $-1$ & saddle, see Fig. \ref{fig:17a} & $<0$, see Fig. \ref{fig:17b} &$<0$, see Fig. \ref{fig:17b}\\\midrule
\end{tabularx}
\end{table}

\subsection{Analysis of system \eqref{exponentialf-1}- \eqref{exponentialf-2} - \eqref{exponentialf-3} at infinity: Poincaré variables}
The numerical results presented in Figure  \ref{fig:fig13a} and \ref{fig:18a} suggest that there are non-trivial dynamics when $x \rightarrow \pm \infty$ and $y\rightarrow \infty$. For that reason, we introduce the Poincaré compactification variables along with the definition of $\eta$
\begin{equation}
\label{poinvar}
    x=\frac{\rho \cos \theta}{\sqrt{1-\rho^2}};\quad  y=\frac{\rho \sin \theta}{\sqrt{1-\rho^2}}; \quad \eta= \frac{H}{\sqrt{1+H^2}},
\end{equation}

we must find evolution equations for $(\rho, \theta, \eta)\in [0,1]\times[0, \pi]\times[-1, 1]$. Using eqs. \eqref{exponentialf-1}, \eqref{exponentialf-2}, \eqref{exponentialf-3} and \eqref{poinvar} we obtain the following system
\begin{small}
\begin{align}
\label{eqrho}
    \rho'&=-\frac{\rho(\rho^2-1)}{16L(\rho, \theta, \eta)}\Bigg\{4608 \eta ^7 (1-\rho^2)^{5/2}+768 \eta ^5 (1-\rho^2)^{3/2} \left(6 \cos ^2(\theta )+\rho ^2 ((3 \epsilon -1) \cos (2 \theta )+3 \epsilon
   -5)\right)\nonumber\\&-16 \eta ^3 \sqrt{1-\rho^2} \rho ^2 \left(\rho ^2 (39 \cos (4 \theta )+(48 \epsilon -4) \cos (2 \theta )+48
   \epsilon -75)-48 \cos ^2(\theta ) ((\epsilon +4) \cos (2 \theta )+\epsilon -4)\right)\nonumber\\&+48 \eta  \sqrt{1-\rho^2} \rho ^4
   \cos ^2(\theta ) (-36 \cos (2 \theta )+(8 \epsilon -1) \cos (4 \theta )-8 \epsilon -3)\nonumber\\&+384 \eta ^4 \rho  \left(\rho
   ^2-1\right) \cos (\theta ) \left(6 \zeta -3 \lambda +\cos (2 \theta ) \left(6 \zeta +3 \lambda +\rho ^2 (2 \zeta 
   (\epsilon -3)-\lambda )\right)+\rho ^2 (\lambda +2 \zeta  (\epsilon -3))\right)\nonumber\\&+32 \eta ^2 \rho ^3 \cos (\theta )
   \left(\rho ^2 (4 \cos (2 \theta ) (\zeta -4 \lambda -6 \zeta  \epsilon )+\cos (4 \theta ) (13 \zeta +4 \lambda -8
   \zeta  \epsilon -2 \lambda  \epsilon )-(\zeta  (16 \epsilon +9))+2 \lambda  (\epsilon +6))\right)\nonumber\\&+192 \eta ^2 \rho ^3 \cos (\theta )\left(((\epsilon -2) \cos (2
   \theta )+\epsilon +2) ((2 \zeta +\lambda ) \cos (2 \theta )+2 \zeta -\lambda )\right)\nonumber\\&+ (-\rho ^5(\cos (\theta )
   (\zeta  (40 \epsilon +94)+\lambda  (4 \epsilon +3)))+\rho ^5\cos (3 \theta ) (-18 \zeta +7 \lambda +8 \zeta  \epsilon -28
   \lambda  \epsilon )\nonumber\\&+3\rho ^5 \cos (5 \theta ) (-2 \zeta +7 \lambda +8 \zeta  \epsilon +4 \lambda  \epsilon )+\rho ^5(4 \epsilon
   -5) (2 \zeta +5 \lambda ) \cos (7 \theta ))\Bigg\},
   \end{align}
\end{small}
   \begin{small}
       \begin{align} 
    \label{eqtheta}
    \theta'&=-\frac{\sin (2 \theta )}{16L(\rho, \theta, \eta)}\Bigg\{-288 \eta ^4 \rho  \left(\rho ^2-1\right)^2 (2 \zeta +\lambda ) \cos (\theta )+576 \eta ^5 \left(1-\rho
   ^2\right)^{5/2}\nonumber\\&-48 \eta ^2 \rho ^3 \left(\rho ^2-1\right) (2 \zeta +\lambda ) \cos (\theta ) ((\epsilon -2) \cos (2
   \theta )+\epsilon +2)\nonumber\\&+\rho ^5 (2 \zeta +5 \lambda ) \cos (\theta ) (12 \cos (2 \theta )+(4 \epsilon -5) \cos (4
   \theta )-4 \epsilon -15)\nonumber\\&+96 \eta ^3 \rho ^2 \left(1-\rho ^2\right)^{3/2} ((\epsilon +4) \cos (2 \theta )+\epsilon
   -4)+6 \eta  \rho ^4 \sqrt{1-\rho ^2} (-36 \cos (2 \theta )+(8 \epsilon -1) \cos (4 \theta )-8 \epsilon -3)\Bigg\},
    \end{align}
   \end{small}
\begin{small}
\begin{align}
    \label{eqeta}
    \eta'&=\frac{\eta  \left(\eta ^2-1\right)}{L(\rho, \theta, \eta)}\Bigg\{36 \eta ^5 \left(1-\rho ^2\right)^{5/2}+6 \eta ^2 \rho ^3 \left(\rho ^2-1\right) \cos (\theta ) (-\lambda +\cos (2
   \theta ) (\lambda +\zeta  \epsilon )+\zeta  \epsilon )\nonumber\\&+2 \rho ^5 \cos (\theta ) \left(\zeta  \cos ^4(\theta )-2
   \lambda  \sin ^4(\theta )+\epsilon  (2 \zeta -\lambda ) \sin ^2(\theta ) \cos ^2(\theta )\right)+6 \eta ^3 \rho ^2
   \left(1-\rho ^2\right)^{3/2} ((3 \epsilon +2) \cos (2 \theta )+3 \epsilon -2)\nonumber\\&+\eta  \rho ^4 \sqrt{1-\rho ^2}
   \left(4 \sin ^4(\theta )+5 \cos ^4(\theta )-3 \epsilon  \sin ^2(2 \theta )\right)\Bigg\},
\end{align}
   \end{small}
defined on the phase-space
\begin{equation*}
    \{(\rho,\theta,\eta)\in \mathbb{R}^3: 0<\rho<1, -\pi\leq \theta \leq \pi, -1\leq \eta\leq 1\}.
\end{equation*}
Here, we used the notation
\begin{align*}
     L(\rho, \theta, \eta,\epsilon)&=288 \eta ^4 \left(\rho ^2-1\right)^2+48 \eta ^2 \rho ^2 \left(\rho ^2-1\right) ((\epsilon -2) \cos (2 \theta
   )+\epsilon +2)\\ &+\rho ^4 (4 \cos (2 \theta )+3 ((3-4 \epsilon ) \cos (4 \theta )+4 \epsilon +9)),
\end{align*} and defined a new time variable by

\begin{equation}
    f':=\sqrt{1-\rho^2}\frac{df}{d\tau}.
\end{equation}
\subsubsection{Analysis of system \eqref{eqrho}-\eqref{eqtheta}-\eqref{eqeta} for $\epsilon=1$}
\label{infinity1}
Note that the limit $\rho\rightarrow 1$ corresponds to $x,y\rightarrow \infty.$ The equilibrium points for system \eqref{eqrho}, \eqref{eqtheta} and \eqref{eqeta} with $\epsilon=1$ in the coordinates $(\rho, \theta, \eta)$ are the following.

\begin{enumerate}
    \item $T_1=(1,\frac{\pi}{2},\eta)$ with eigenvalues $\{0,0,0\}.$ This is a nonhyperbolic set of point with $q(T_1)=0.$
    \item $T_2=(1,0,0)$ with eigenvalues $\left\{-\frac{2 \zeta }{5},\frac{1}{10} (2 \zeta +5 \lambda ),\frac{2 \zeta }{5}\right\}.$ For this point we have that $\omega_{\phi}(T_2)$ and $q(T_2)$ blow up for $\eta=0,$ so we present the analysis in Fig. \ref{fig:20}. We see that for $\zeta=3, \lambda=1$ and negative values of $\eta$, $\omega_{\phi}(T_2)$ and $q(T_2)$ tend to minus infinity as $\rho\rightarrow 1$ but they tend to $-\frac{1}{3}$ and $0$ respectively as $\rho \rightarrow 0.$ For positive values of $\eta$ and $\zeta=3, \lambda=1$ the opposite occurs, $\omega_{\phi}(T_2)$ and $q(T_2)$ tend to infinity as $\rho\rightarrow 1$ but they tend to $-\frac{1}{3}$ and $0$ respectively as $\rho \rightarrow 0.$ If we consider negative values of $\zeta$ and $\lambda$ the behaviour is symmetric. We also see that this point is   
    \begin{enumerate}
    \item a saddle for $\lambda\in \mathbb{R}$, $\zeta\neq 0$, $\zeta\neq -\frac{5\lambda}{2}.$
    \item nonhyperbolic for 
    \begin{enumerate}
        \item $\zeta=0$ or
        \item $\zeta=-\frac{5\lambda}{2}$.
    \end{enumerate}
    \end{enumerate}
    \begin{figure}[ht!]
        \centering
        \includegraphics[scale=0.6]{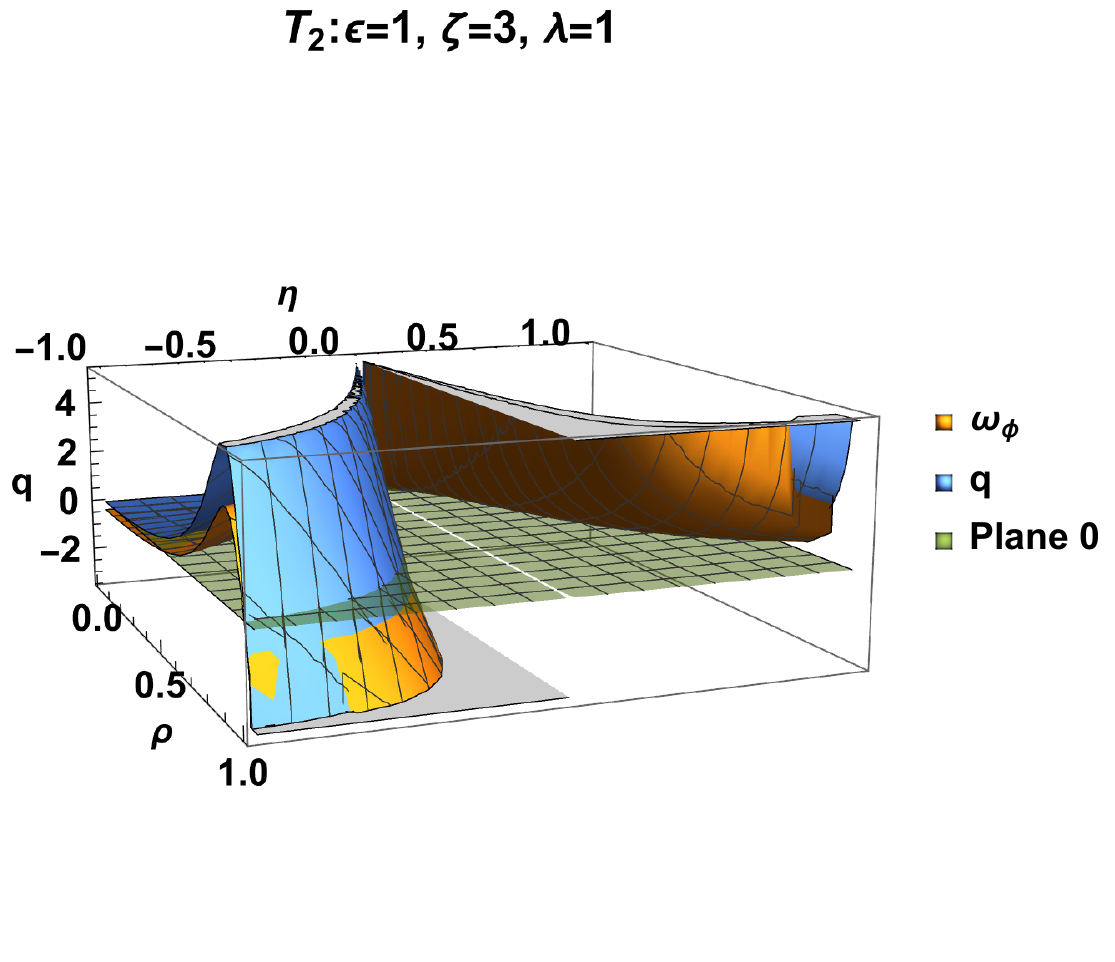}
        \includegraphics[scale=0.6]{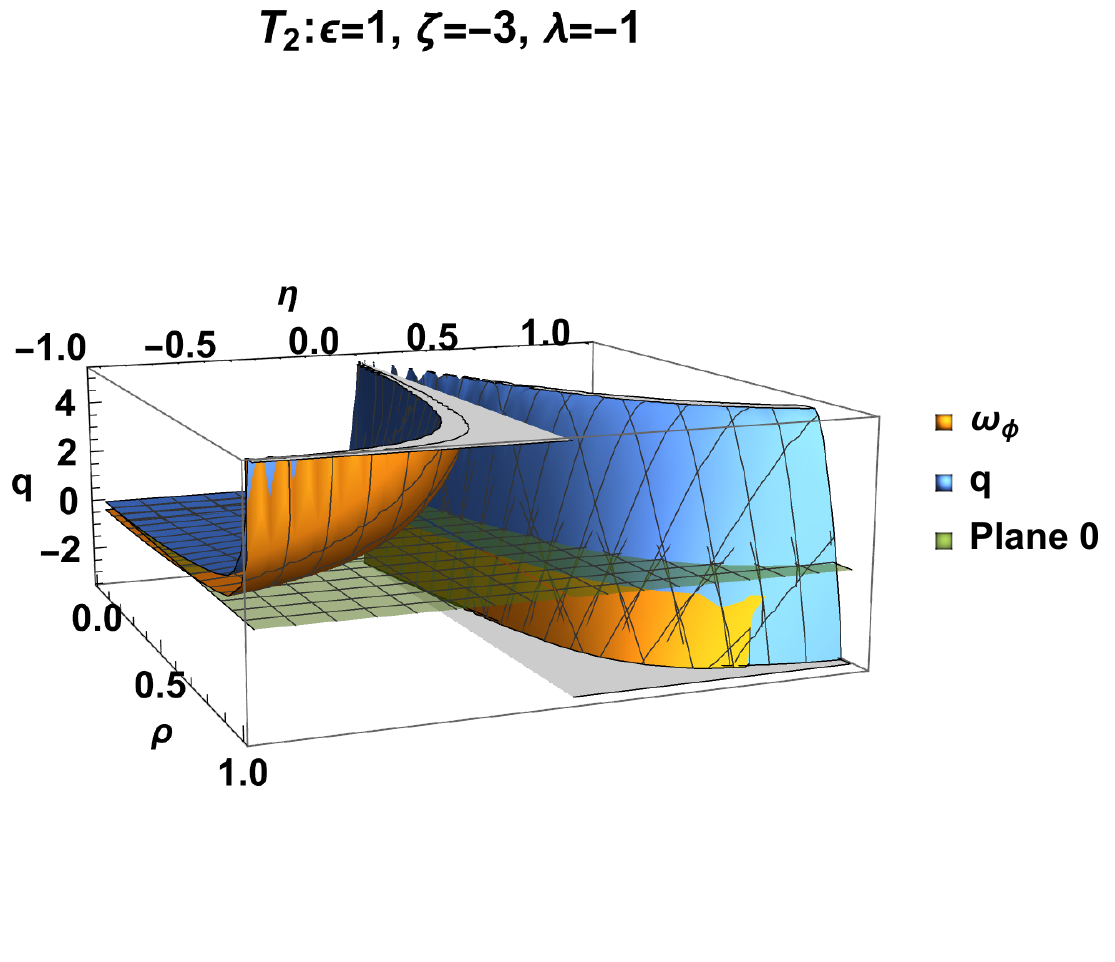}
        \caption{Plots of $\omega_{\phi}, q$ for $T_2$.}
        \label{fig:20}
    \end{figure}
    \item $T_3=(1,\pi,0),$ with eigenvalues $\left\{\frac{2 \zeta }{5},\frac{1}{10} (-2 \zeta -5 \lambda ),-\frac{2 \zeta }{5}\right\}.$ For this point the behaviour of $\omega_{\phi}(T_3)$ and $q(T_3)$ is similar that for $T_2,$ meaning that these parameters blow up as $\eta$ goes to $0$. See Fig. \ref{fig:21} in which we show that the behaviour is similar to the one described in Fig. \ref{fig:20} but symmetric to the sign change in $\zeta, \lambda.$ This point is also 
    \begin{enumerate}
    \item a saddle for $\lambda\in \mathbb{R}$, $\zeta\neq 0$, $\zeta\neq -\frac{5\lambda}{2}.$
    \item nonhyperbolic for 
    \begin{enumerate}
        \item $\zeta=0$ or
        \item $\zeta=-\frac{5\lambda}{2}$.
    \end{enumerate}
    \end{enumerate}
     \begin{figure}[ht!]
        \centering
        \includegraphics[scale=0.6]{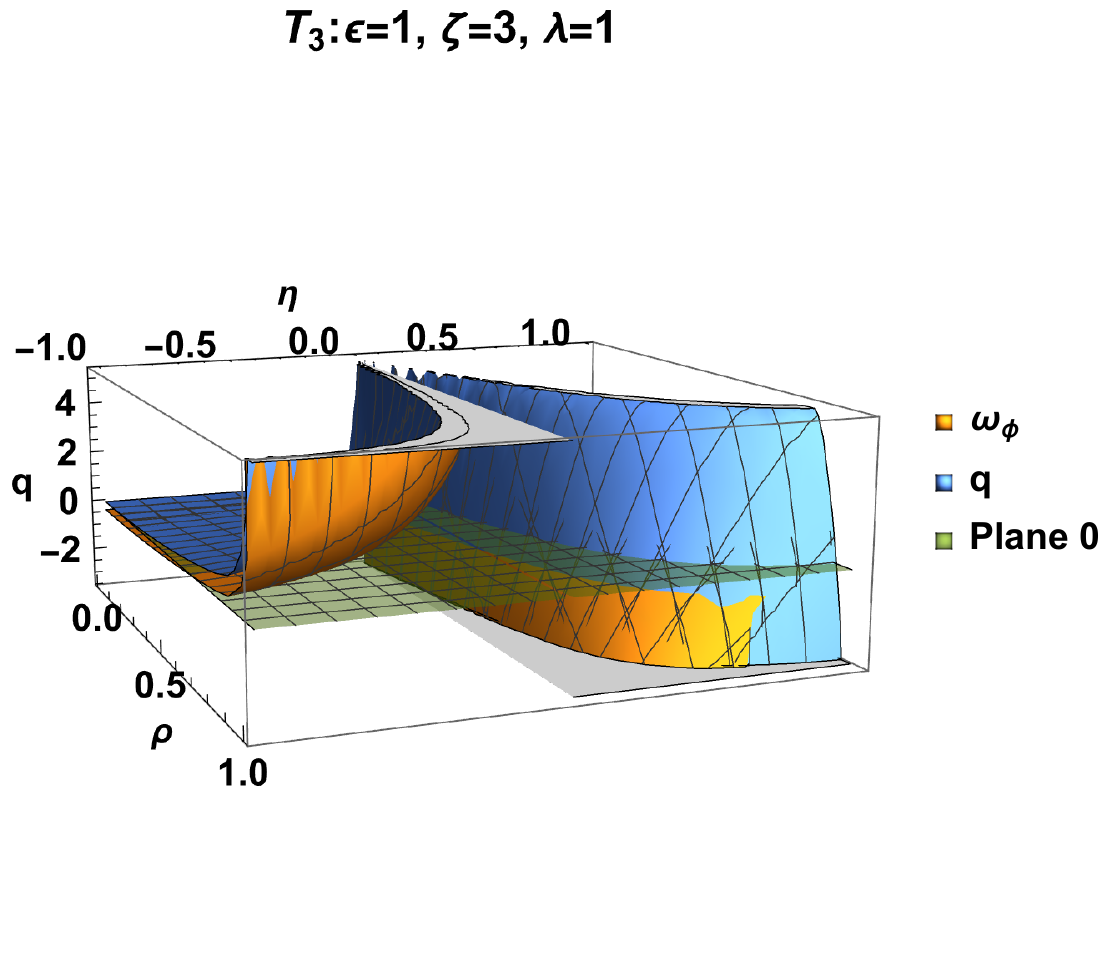}
        \includegraphics[scale=0.6]{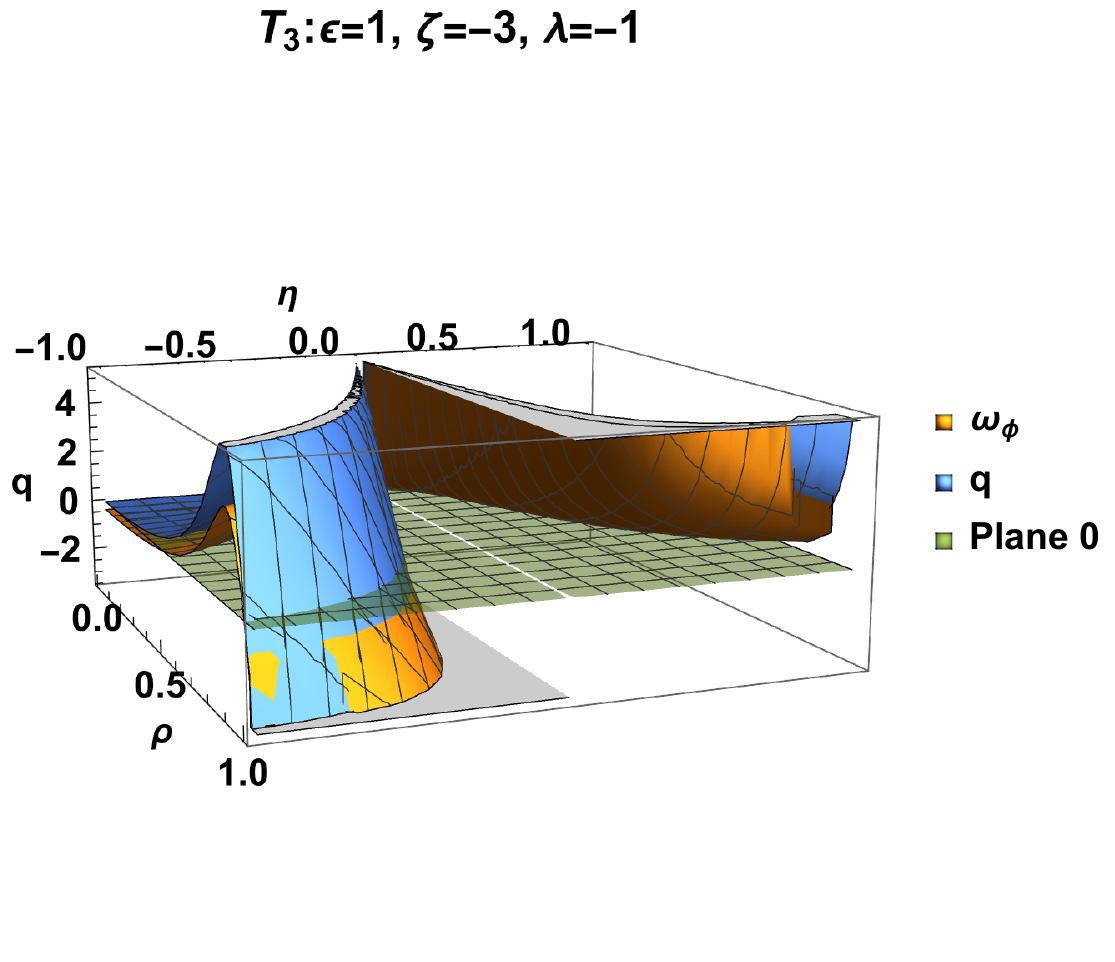}
        \caption{Plots of $\omega_{\phi}$ and $q$ for $T_3$.}
        \label{fig:21}
    \end{figure}
    \item $T_{4,5}=(1,0,\pm 1)$ with eigenvalues $\left\{-\frac{2 \zeta }{5},\frac{4 \zeta }{5},\frac{1}{10} (2 \zeta +5 \lambda )\right\}.$ For this points we have $\omega_{\phi}(T_{4,5})=f_1(\rho,\zeta)$ and $q(T_{4,5})=f_2(\rho,\zeta).$
we verify that $\lim_{\rho\rightarrow 1} (\omega_{\phi}(T_{4,5}))$ and $\lim_{\rho\rightarrow 1} (q(T_{4,5}))$ are directed infinities that depend on the sign of $-\zeta.$ However for $\rho \rightarrow 0$ we have that $\omega_{\phi}(T_{4,5})=-\frac{1}{3}$ and $q(T_{4,5})=0$, see Fig. \ref{fig:22}. Performing the stability analysis, we see that the points are
\begin{enumerate}
    \item saddle for $\lambda\in \mathbb{R}$, $\zeta\neq 0$, $\zeta\neq -\frac{5\lambda}{2}.$
    \item nonhyperbolic for 
    \begin{enumerate}
        \item $\zeta=0$ or
        \item $\zeta=-\frac{5\lambda}{2}$.
    \end{enumerate}
    \begin{figure}[ht!]
        \centering
        \includegraphics[scale=0.6]{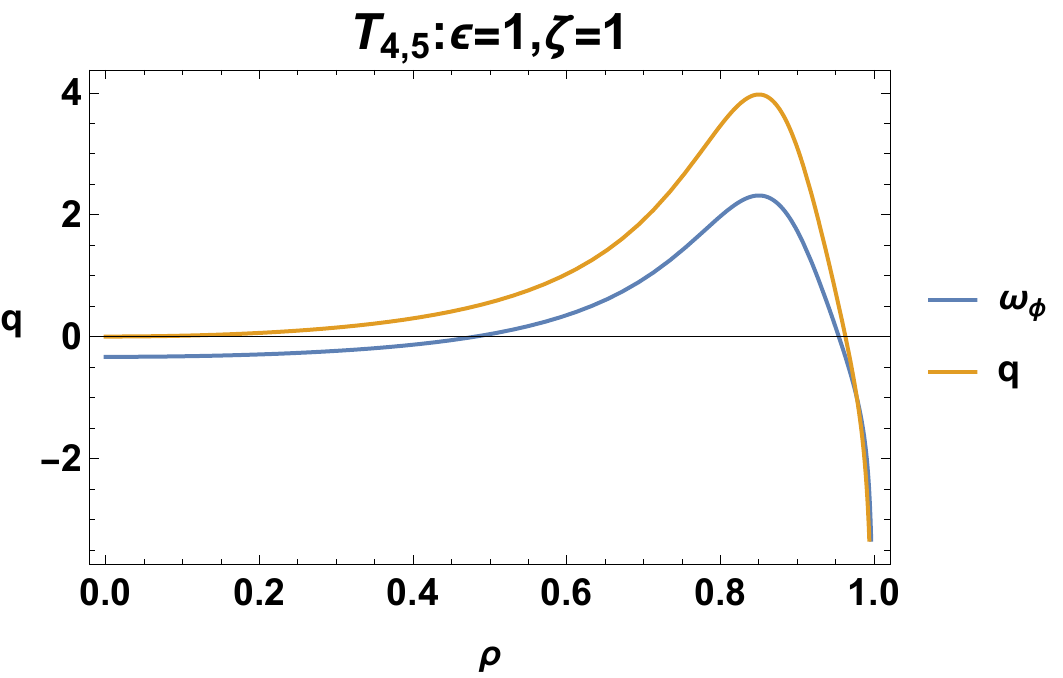}
        \includegraphics[scale=0.6]{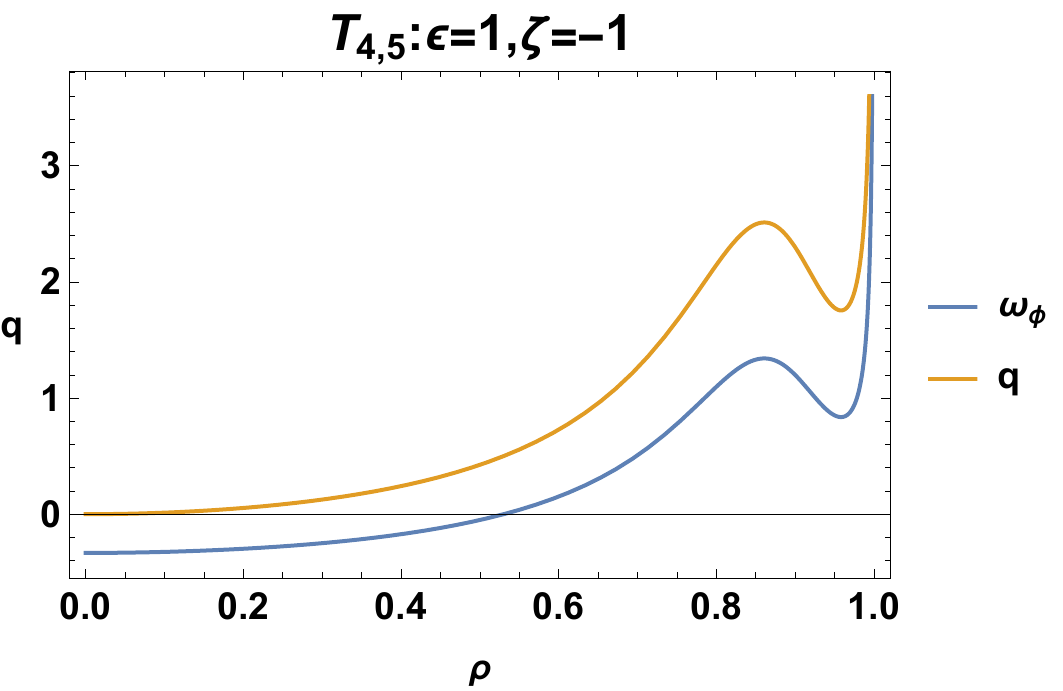}
        \caption{Plots of $\omega_{\phi}, q$ for $T_{4,5}.$}
        \label{fig:22}
    \end{figure}
\end{enumerate}
    \item $T_{6,7}=(1,\pi,\pm 1)$ with eigenvalues $ \left\{\frac{2 \zeta }{5},-\frac{4 \zeta }{5},\frac{1}{10} (-2 \zeta -5 \lambda )\right\}$. For this points we have $\omega_{\phi}(T_{6,7})=f_1(\rho,\zeta)$ and $q(T_{6,7})=f_2(\rho,\zeta).$
we verify that $\lim_{\rho\rightarrow 1} (\omega_{\phi}(T_{6,7}))$ and $\lim_{\rho\rightarrow 1} (q(T_{6,7}))$ are directed infinities that depend on the sign of $\zeta.$ However for $\rho \rightarrow 0$ we have that $\omega_{\phi}(T_{6,7})=-\frac{1}{3}$ and $q(T_{6,7})=0$, see Fig. \ref{fig:23}. By performing the stability analysis, we conclude that the points are
\begin{enumerate}
    \item saddle for $\lambda\in \mathbb{R}$, $\zeta\neq 0$, $\zeta\neq -\frac{5\lambda}{2}.$
    \item nonhyperbolic for 
    \begin{enumerate}
        \item $\zeta=0$ or
        \item $\zeta=-\frac{5\lambda}{2}$.
    \end{enumerate}
\end{enumerate}
    \begin{figure}[ht!]
        \centering
        \includegraphics[scale=0.6]{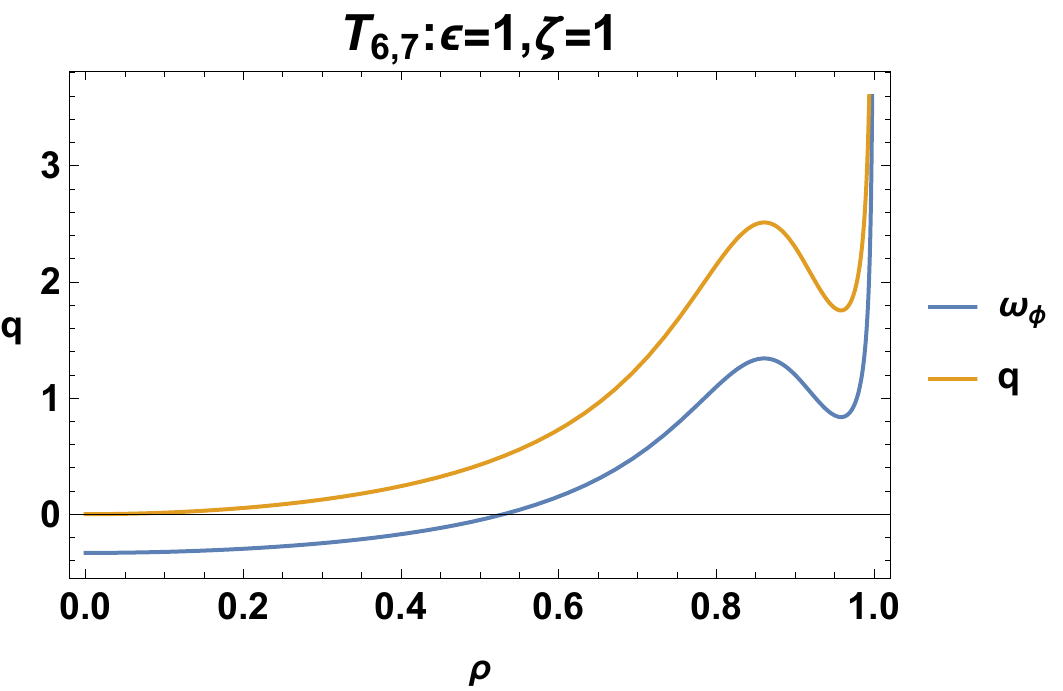}
        \includegraphics[scale=0.6]{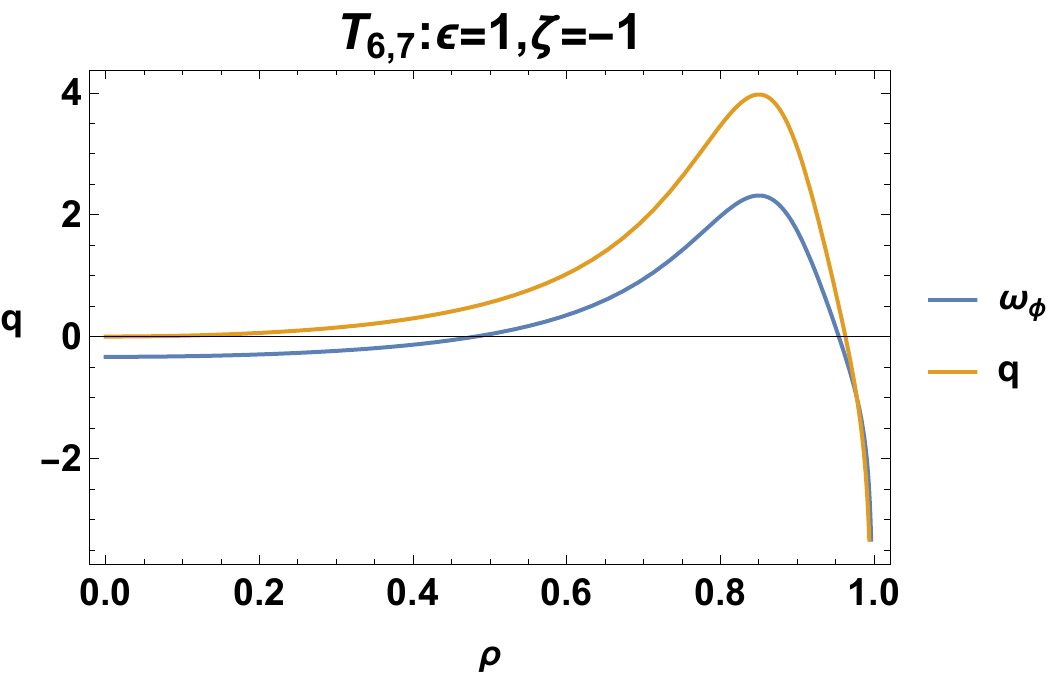}
        \caption{Plots of $\omega_{\phi}, q$ for $T_{6,7}.$}
        \label{fig:23}
    \end{figure}
        \item $S_1=(0,0,1),$ with eigenvalues $\{2,2,-1\}.$ This point is a saddle and has $\omega_{\phi}(S_1)=-\frac{1}{3}$ and $q(S_1)=0.$
    \item $S_2=(0,0,-1),$ with eigenvalues $\{-2,-2,1\}.$ This point is a saddle and has $\omega_{\phi}(S_2)=-\frac{1}{3}$ and $q(S_2)=0.$
     \item $S_3=(0,\frac{\pi}{2},1),$ with eigenvalues $\{2,2,1\}.$ This point is a source and has $\omega_{\phi}(S_3)=-\frac{1}{3}$ and $q(S_3)=0.$
    \item $S_4=(0,\frac{\pi}{2},-1),$ with eigenvalues $\{-2,-2,-1\}.$ This point is a sink and has $\omega_{\phi}(S_4)=-\frac{1}{3}$ and $q(S_4)=0.$
    \item $S_5=(0,\pi,1),$ with eigenvalues $\{2,2,-1\}.$ This point is a saddle and has $\omega_{\phi}(S_5)=-\frac{1}{3}$ and $q(S_5)=0.$
    \item $S_6=(0,\pi,-1),$ with eigenvalues $\{2,2,-1\}.$ This point is a saddle and has $\omega_{\phi}(S_6)=-\frac{1}{3}$ and $q(S_6)=0.$
    \item $S_7=(0,\theta,0)$ with eigenvalues $\{0,0,0\}$ is represented in Fig. \ref{fig:24} as a dashed red line. This set of points is nonhyperbolic with $\omega_{\phi}(S_7)=-\frac{1}{3}$ and $q(S_7)=0.$
\end{enumerate}
The points $S_i$ with $i=1,\ldots, 7$ have $\rho=0,$ this means that in the finite case, $x=y=0.$ Also, since $q(S_i)=0,$ the asymptotic solution for these points represents a universe dominated by the Gauss-Bonnet term. 
\begin{figure}[ht!]
    \centering
    \includegraphics[scale=0.6]{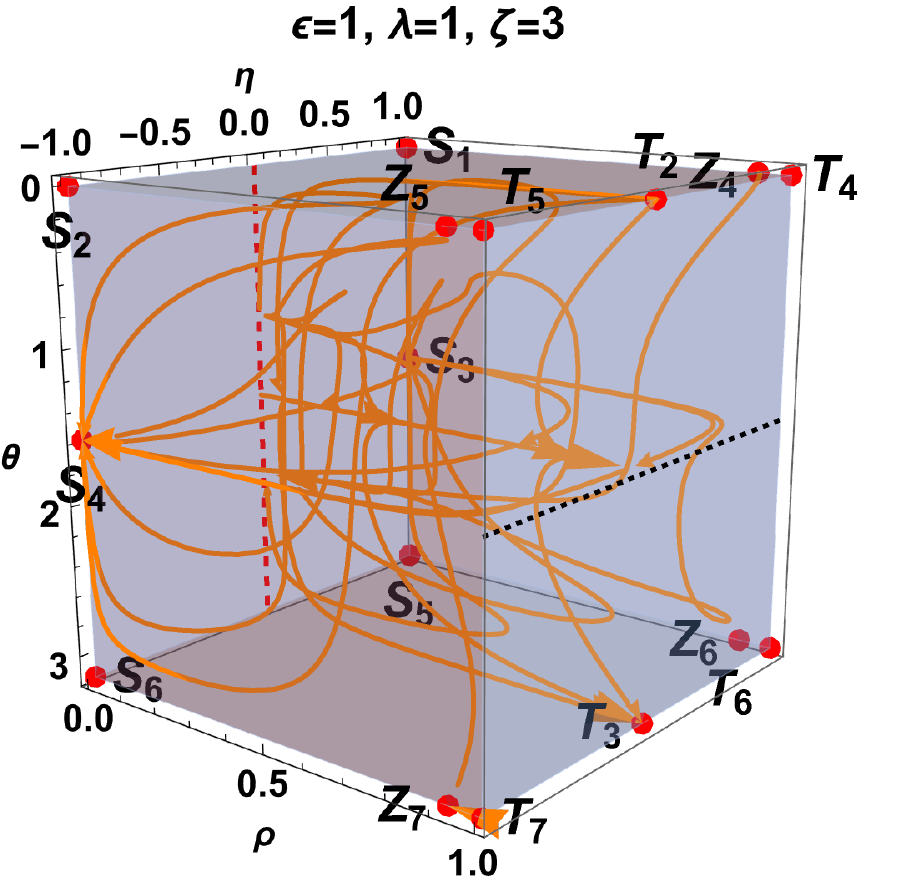}
    \includegraphics[scale=0.6]{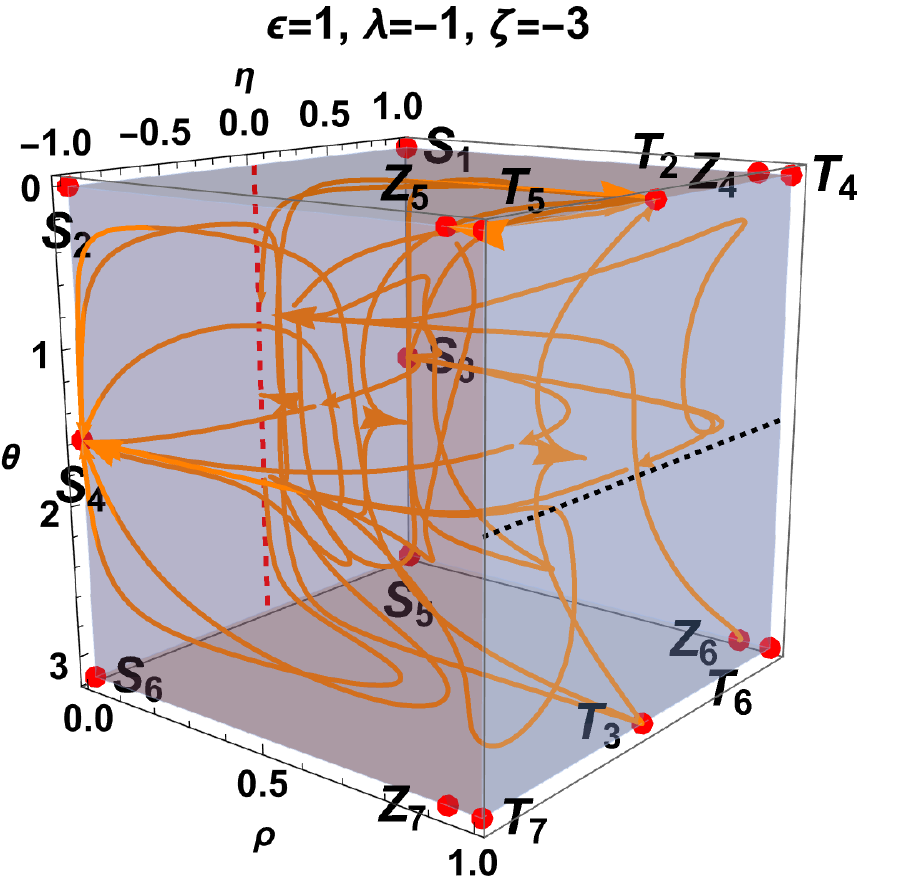}
    \caption{Three dimensional phase space for system \eqref{eqrho}, \eqref{eqtheta}, \eqref{eqeta} for different values of the parameters $\zeta,$ and $\lambda$. The dashed black line corresponds to $T_1$, and the dashed red line corresponds to $S_7.$}
    \label{fig:24}
\end{figure}

\begin{table}[ht!]
    \caption{Equilibrium points of system \eqref{eqrho}, \eqref{eqtheta}, \eqref{eqeta}  for $\epsilon=1$ with their stability conditions. It also includes the value of $\omega_{\phi}$ and $q.$}
    \label{tab:6}
 \centering
\newcolumntype{C}{>{\centering\arraybackslash}X}
\centering
    \setlength{\tabcolsep}{6.6mm}
\begin{tabularx}{\textwidth}{cccccccc}
\toprule 
        \text{Label}& $\rho$& $\theta$ & $\eta$ &\text{Stability}& $\omega_{\phi}$& $q$ \\\midrule
         $T_1$& $0$& $\frac{\pi}{2}$ & $0$ &nonhyperbolic& $-\frac{1}{3}$& $0$ \\\midrule$T_2$& $1$& $0$ & $0$ &saddle& see Fig. \ref{fig:20}& see Fig. \ref{fig:20} \\\midrule
         $T_3$& $1$& $\pi$ & $0$ &saddle& see Fig. \ref{fig:21}& see Fig. \ref{fig:21} \\\midrule
         $T_{4,5}$& $1$& $0$ & $\pm 1$ &saddle& see Fig. \ref{fig:22}& see Fig. \ref{fig:22} \\\midrule
         $T_{6,7}$& $1$& $\pi$ & $\pm 1$ &saddle& see Fig. \ref{fig:23}& see Fig. \ref{fig:23} \\\midrule
         $S_1$& $0$& $0$ & $1$ &saddle& $-\frac{1}{3}$&$0$ \\\midrule
          $S_2$& $0$& $0$ & $-1$ &saddle& $-\frac{1}{3}$&$0$ \\\midrule
           $S_3$& $0$& $\frac{\pi}{2}$ & $1$ &source& $-\frac{1}{3}$&$0$ \\\midrule
           $S_4$& $0$& $\frac{\pi}{2}$ & $-1$ &sink& $-\frac{1}{3}$&$0$ \\\midrule
            $S_5$& $0$& $\pi$ & $1$ &saddle& $-\frac{1}{3}$&$0$ \\\midrule
            $S_6$& $0$& $\pi$ & $-1$ &saddle& $-\frac{1}{3}$&$0$ \\\midrule
            $S_7$& $0$& $\theta$ & $0$ &nonhyperbolic& $-\frac{1}{3}$&$0$ \\\midrule
    \end{tabularx}
\end{table}

\subsubsection{Analysis of system \eqref{eqrho}- \eqref{eqtheta} - \eqref{eqeta} $\epsilon=-1$}
\label{infinity2}
The equilibrium points for system \eqref{eqrho}, \eqref{eqtheta} and \eqref{eqeta} for $\epsilon=-1$ are the same points as in section \ref{infinity1} that is, 
\begin{enumerate}
    \item $T_1=(1,\frac{\pi}{2},\eta),$ with eigenvalues $(0,0,0)$ this set of points is nonhyperbolic and is represented in Fig. \ref{fig:25} as a dashed black line. For this point we have $\omega_{\phi}(T_1)=-\frac{1}{3}$ and $q(T_1)=0.$
    \item $T_2=(1,0,0)$ with eigenvalues $\left\{-\frac{2 \zeta }{5},\frac{1}{10} (2 \zeta +5 \lambda ),\frac{2 \zeta }{5}\right\}$. The stability analysis is performed similarly to section \ref{infinity1}. For the study of $\omega_{\phi}(T_2)$ and $q(T_2)$ we see that these expressions blow up for $\eta=0,$ because of this we present Fig. \ref{fig:26}. For $\eta \rightarrow 0^+$ we verify that $\lim_{\rho\rightarrow 1}(\lim_{\eta \rightarrow 0^+}(\omega_{\phi}(T_2)))=\text{sgn}(\zeta) \infty$ and $\lim_{\rho\rightarrow 1}(\lim_{\eta \rightarrow 0^+}(q(T_2)))=\text{sgn}(\zeta) \infty$. On the other direction, that is $\eta \rightarrow 0^-$ we have $\lim_{\rho\rightarrow 1}(\lim_{\eta \rightarrow 0^-}(\omega_{\phi}(T_2)))=-\text{sgn}(\zeta) \infty$ and $\lim_{\rho\rightarrow 1}(\lim_{\eta \rightarrow 0^-}(q(T_2)))=-\text{sgn}(\zeta) \infty$. 

    \begin{figure}[ht!]
        \centering
\includegraphics[scale=0.6]{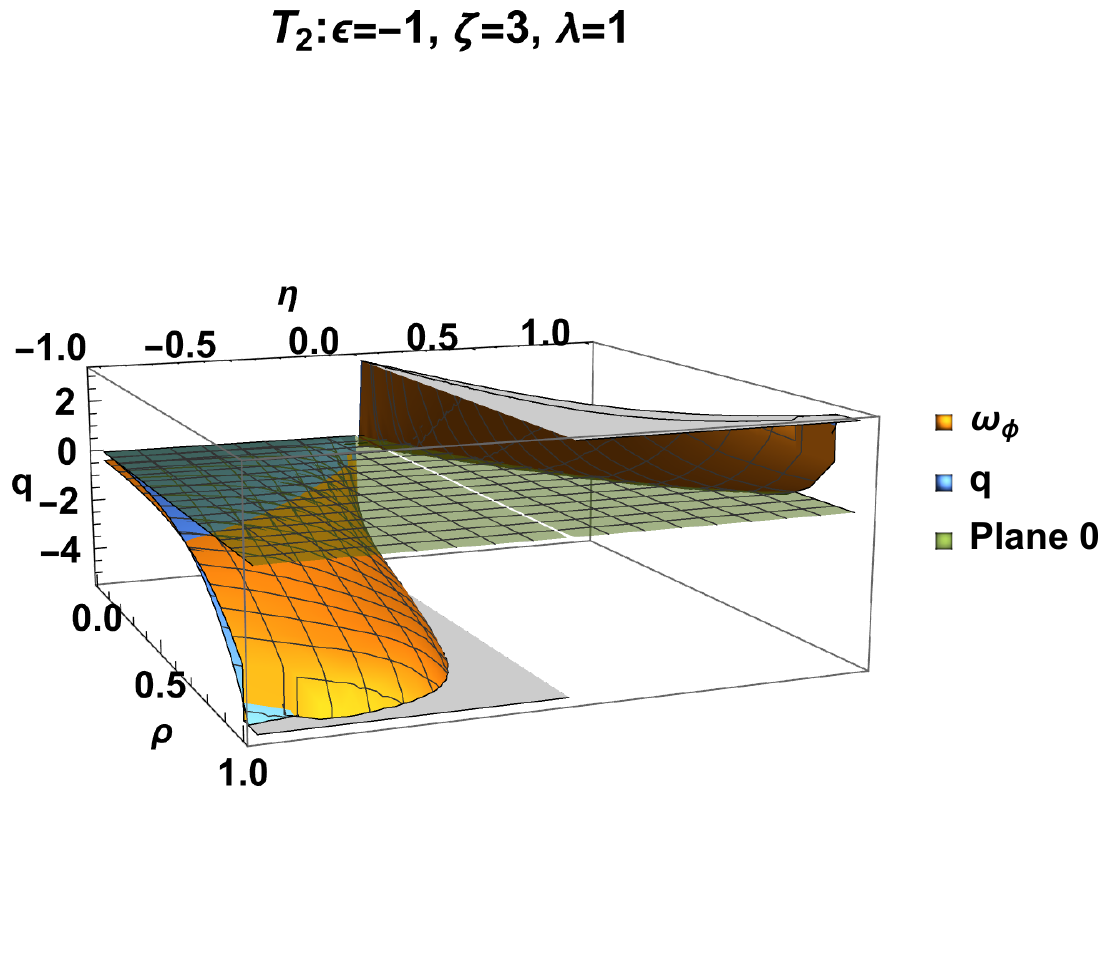}
\includegraphics[scale=0.6]{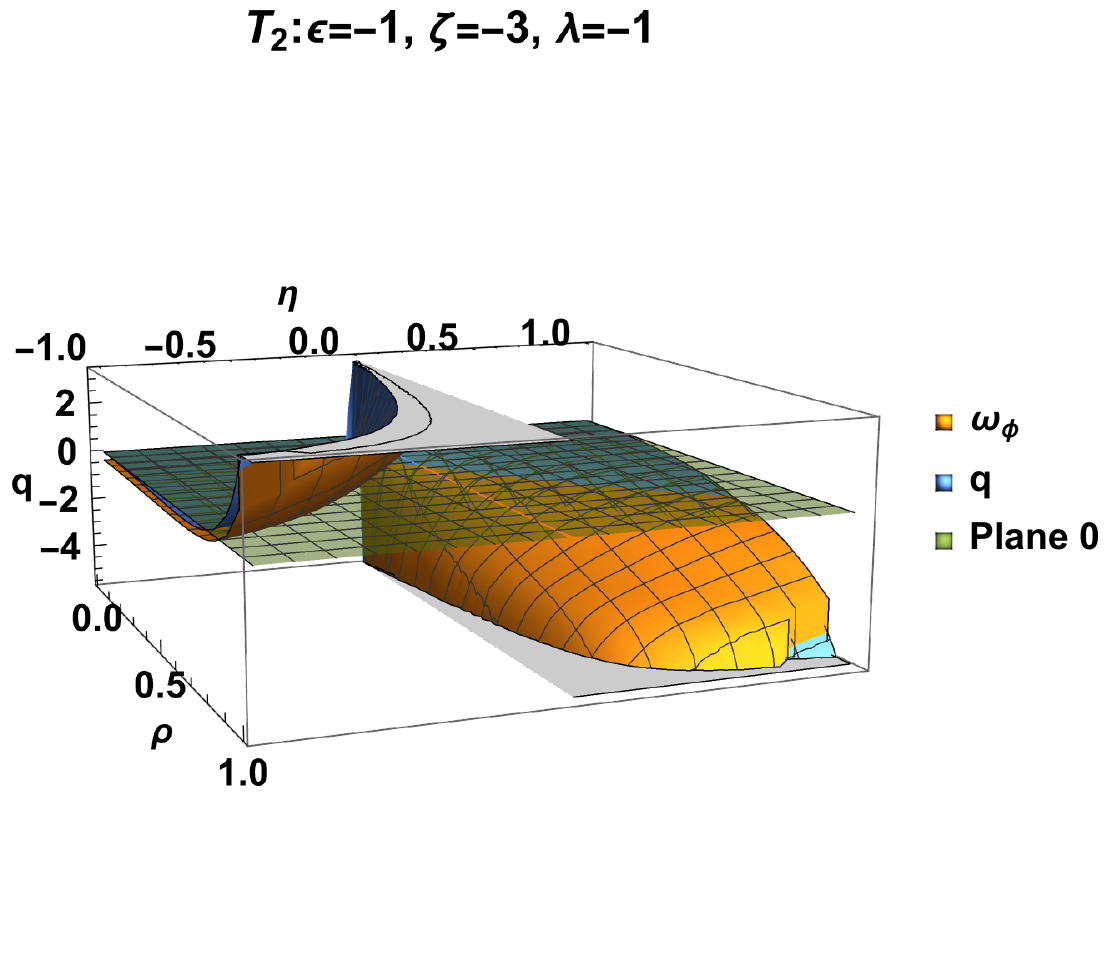}
        \caption{Plots of $\omega_{\phi}$ and $q$ for $T_2$.}
        \label{fig:26}
    \end{figure}
    
    \item $T_3=(1,\pi,0),$ with eigenvalues $\left\{\frac{2 \zeta }{5},\frac{1}{10} (-2 \zeta -5 \lambda ),-\frac{2 \zeta }{5}\right\}.$ The stability analysis is the same as in section \ref{infinity1}. Since the EoS and deceleration parameters blow up for $\eta=0,$ we study their behaviour in Fig. \ref{fig:27}. For $\eta \rightarrow 0^+$ we verify that $\lim_{\rho\rightarrow 1}(\lim_{\eta \rightarrow 0^+}(\omega_{\phi}(T_3)))=-\text{sgn}(\zeta) \infty$ and $\lim_{\rho\rightarrow 1}(\lim_{\eta \rightarrow 0^+}(q(T_3)))=-\text{sgn}(\zeta) \infty$. On the other direction, that is $\eta \rightarrow 0^-$ we have $\lim_{\rho\rightarrow 1}(\lim_{\eta \rightarrow 0^-}(\omega_{\phi}(T_3)))=\text{sgn}(\zeta) \infty$ and $\lim_{\rho\rightarrow 1}(\lim_{\eta \rightarrow 0^-}(q(T_3)))=\text{sgn}(\zeta) \infty$.
\begin{figure}[ht!]
    \centering
\includegraphics[scale=0.6]{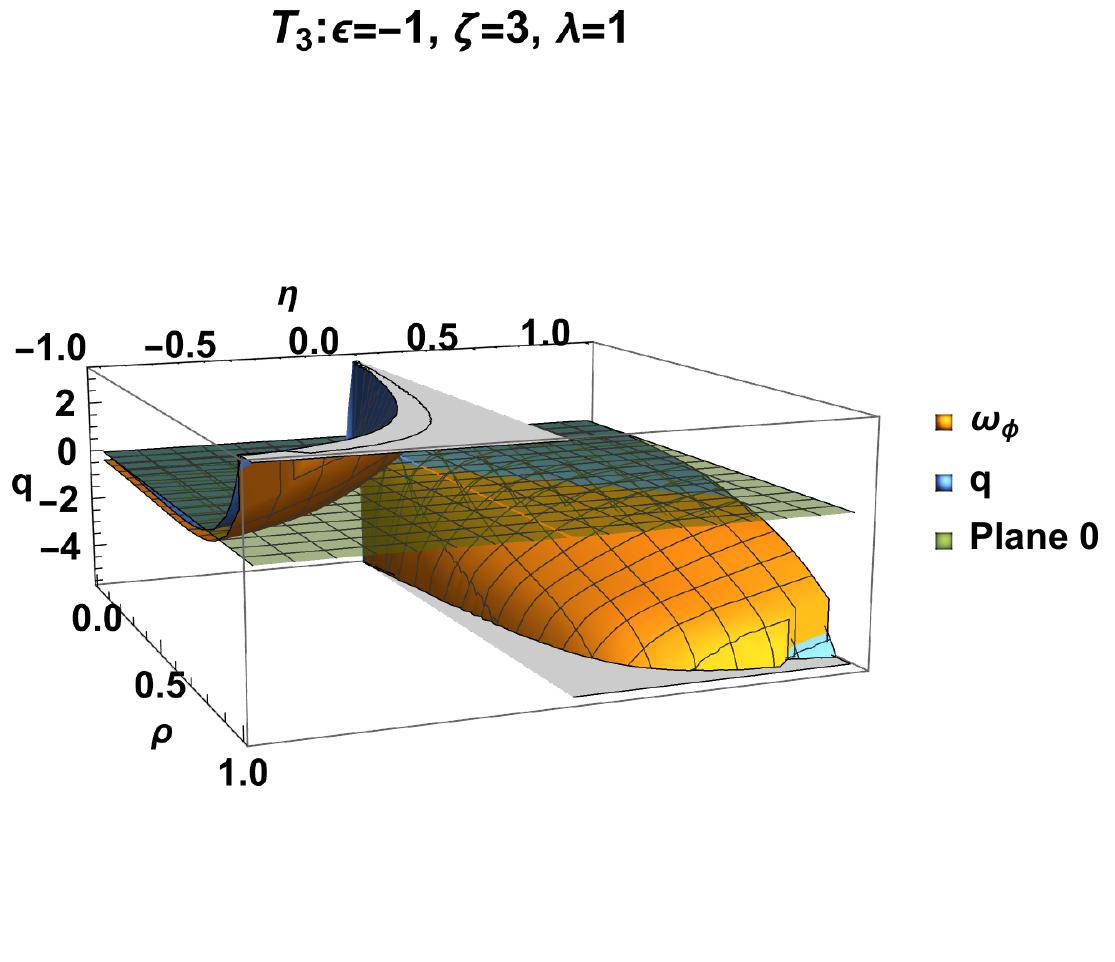}
\includegraphics[scale=0.6]{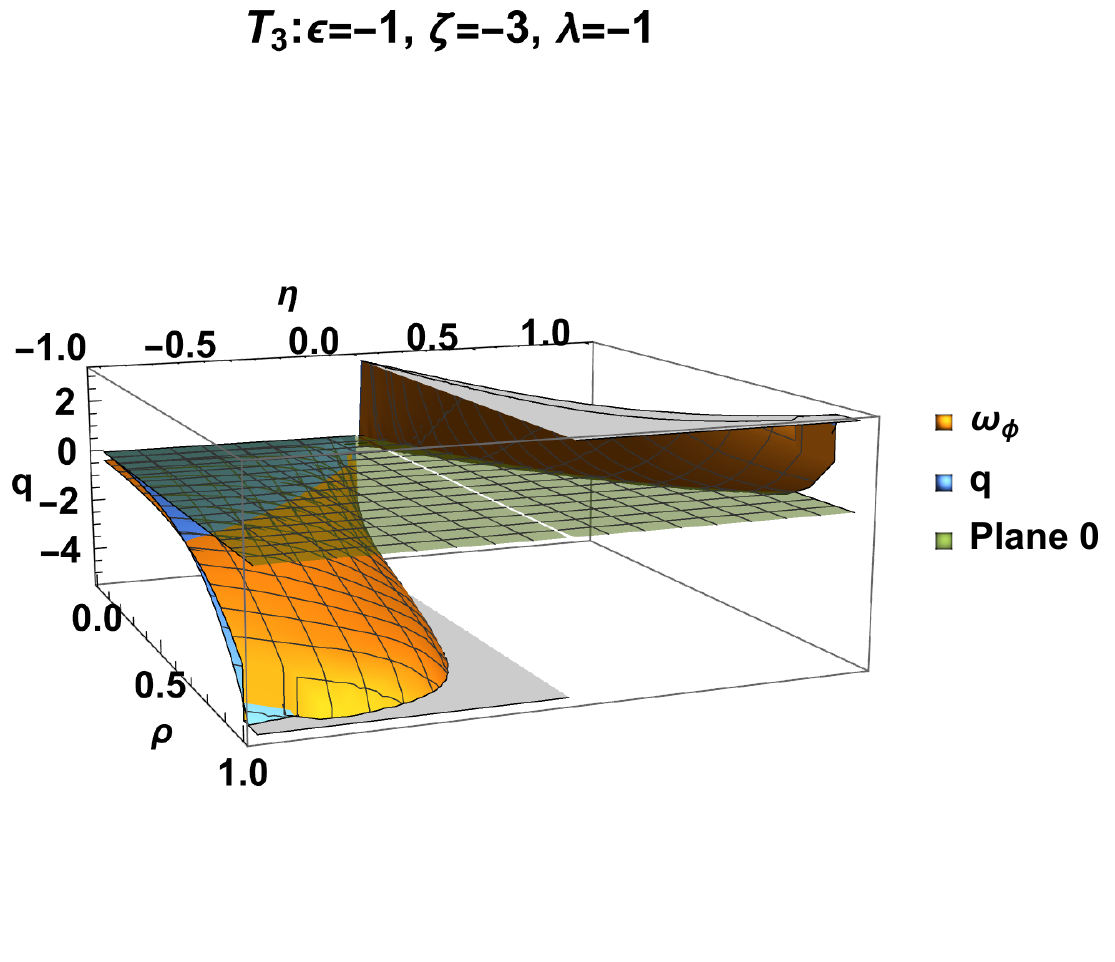}
    \caption{Plot of $\omega_{\phi}$ and $q$ for $T_3.$}
    \label{fig:27}
\end{figure}

    \item $T_{4,5}=(1,0, \pm 1),$ with eigenvalues $\left\{-\frac{2 \zeta }{5},\frac{4
   \zeta }{5},\frac{\zeta }{5}+\frac{\lambda }{2}\right\}.$ The stability analysis is the same as in section \ref{infinity1}. However, we verify that the limit as $\rho \rightarrow 1$ of the EoS and deceleration parameters are directed infinities that depend on the sign of $\zeta.$ We also see that $\lim_{\rho \rightarrow 0} (\omega_{\phi}(T_{4,5}))=-\frac{1}{3}$ and $\lim_{\rho \rightarrow 0} (q(T_{4,5}))=0$, see Fig. \ref{fig:28}.
   \begin{figure}[ht!]
       \centering
       \includegraphics[scale=0.6]{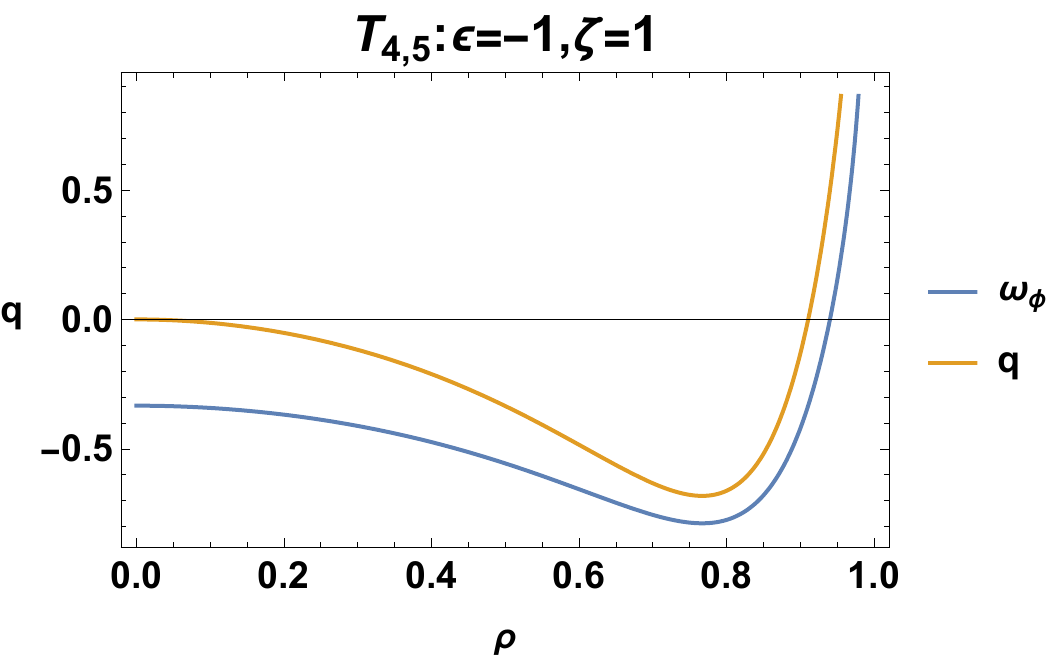}
       \includegraphics[scale=0.6]{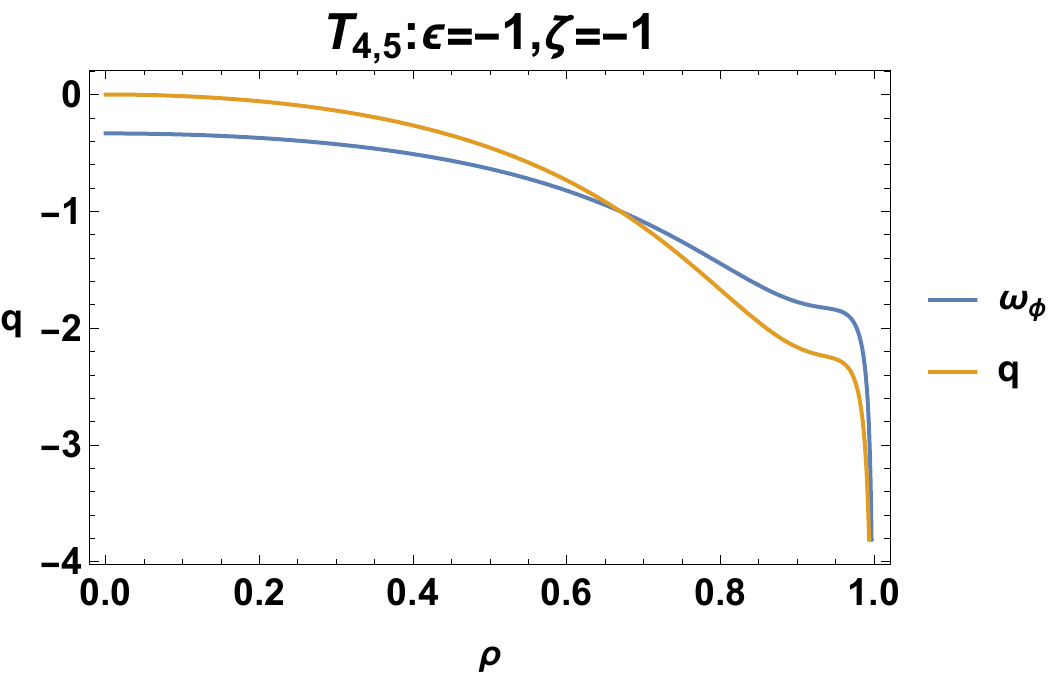}
       \caption{Plot of $\omega_{\phi}, q$ for $T_{4,5}.$}
       \label{fig:28}
   \end{figure}
   
    \item $T_{6,7}=(1,\pi,\pm 1),$ with eigenvalues $\left\{\frac{2 \zeta }{5},-\frac{4 \zeta }{5},-\frac{\zeta }{5}-\frac{\lambda }{2}\right\}.$ The stability analysis is the same as the one performed in section \ref{infinity1}. Something similar (to the previous two points) occurs to $\omega_{\phi}(T_{6,7})$ and $q(T_{6,7})$ that is, they have directed infinities, but in this case, they depend on the sign of $-\zeta$, see Fig. \ref{fig:29}. 
    \begin{figure}[ht!]
       \centering
    \includegraphics[scale=0.6]{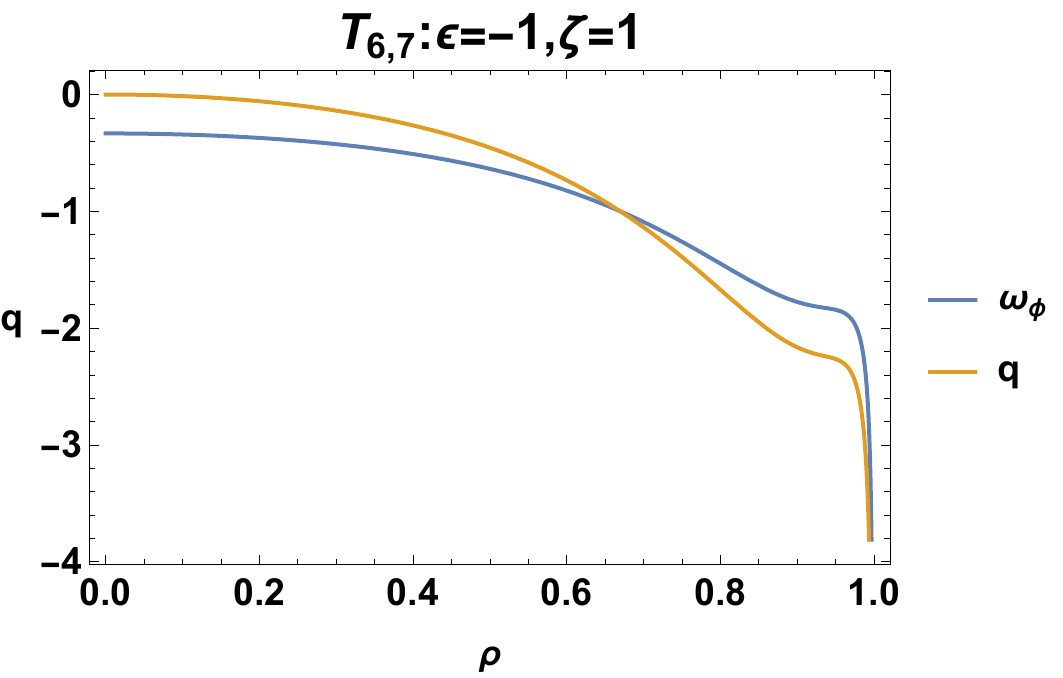}
    \includegraphics[scale=0.6]{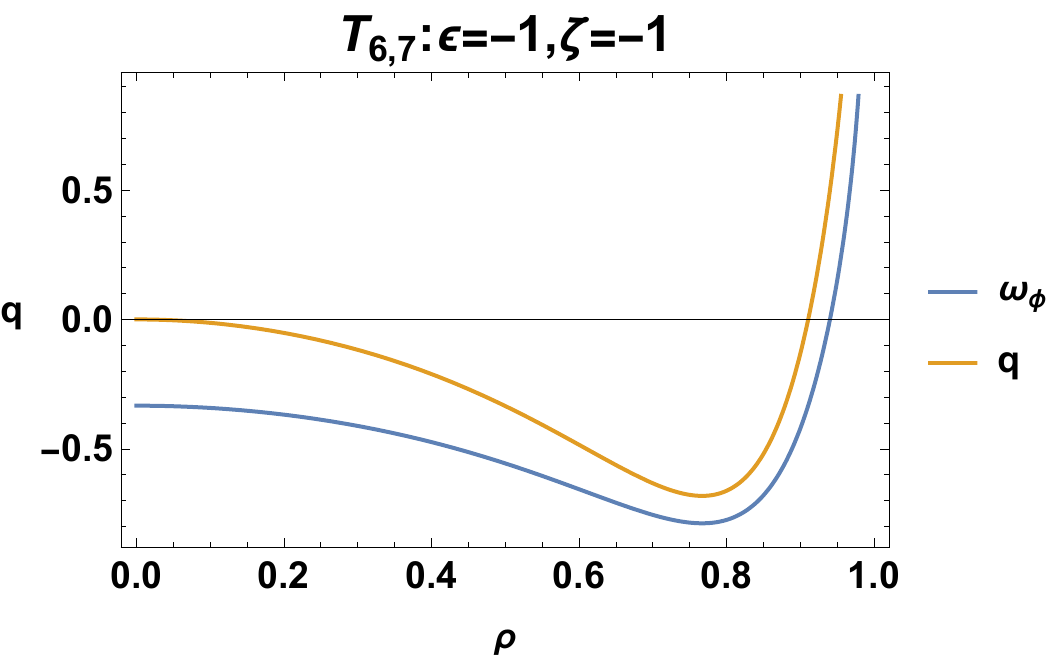}
       \caption{Plot of $\omega_{\phi}, q$ for $T_{6,7}.$}
       \label{fig:29}
   \end{figure}
   
    \item $S_1=(0,0,1),$ with eigenvalues $\{2,2,-1\}.$ This point is a saddle and has $\omega_{\phi}(S_1)=-\frac{1}{3}$ and $q(S_1)=0.$
    \item $S_2=(0,0,-1),$ with eigenvalues $\{-2,-2,1\}.$ This point is a saddle and has $\omega_{\phi}(S_2)=-\frac{1}{3}$ and $q(S_2)=0.$
     \item $S_3=(0,\frac{\pi}{2},1),$ with eigenvalues $\{2,2,1\}.$ This point is a source and has $\omega_{\phi}(S_3)=-\frac{1}{3}$ and $q(S_3)=0.$
    \item $S_4=(0,\frac{\pi}{2},-1),$ with eigenvalues $\{-2,-2,-1\}.$ This point is a sink and has $\omega_{\phi}(S_4)=-\frac{1}{3}$ and $q(S_4)=0.$
    \item $S_5=(0,\pi,1),$ with eigenvalues $\{2,2,-1\}.$ This point is a saddle and has $\omega_{\phi}(S_5)=-\frac{1}{3}$ and $q(S_5)=0.$
    \item $S_6=(0,\pi,-1),$ with eigenvalues $\{-2,-2,1\}.$ This point is a saddle and has $\omega_{\phi}(S_6)=-\frac{1}{3}$ and $q(S_6)=0.$
    \item $S_7=(0,\theta,0)$ with eigenvalues $\{0,0,0\}$ is represented in Fig. \ref{fig:25} as a dashed red line. This set of points is nonhyperbolic with $\omega_{\phi}(S_7)=-\frac{1}{3}$ and $q(S_7)=0.$
    
\end{enumerate}

Recall that the points $S_i$ with $i=1,\ldots, 7$ have $\rho=0,$ this means that in the finite case, $x=y=0.$ Also, since $q(S_i)=0,$ the asymptotic solution for these points represents a universe dominated by the Gauss-Bonnet term. 
We also have the following additional points where $\alpha=\arccot (\sqrt{2}).$ For these remaining points, we perform numerical analysis both on the real part of the eigenvalues and the behaviour of $\omega_{\phi}$ and $q.$

\begin{enumerate}
    \item $T_8=(1,\alpha,1).$
    The eigenvalues are $\lambda_i(\rho,\zeta,\lambda)$ for $i=1,2,3,$ the analysis is performed for some values of $\zeta$ and $\lambda$ in Fig. \ref{fig:31}, where we see that the point has a saddle or sink behaviour. However, since $\rho=1$, the point has source behaviour in the limit $\rho \rightarrow 1$. For this point, we verify that both $\omega_{\phi}(T_8)$ and $q(T_8)$ go to infinity  as $\rho \rightarrow 1$ therefore, the point cannot describe an accelerated universe regardless of the values of $\zeta$, and $\lambda$, see Fig. \ref{fig:30}, we also verify that they tend to $-\frac{1}{3}$ and $0$ respectively as $\rho \rightarrow 0.$   
        
        \begin{figure}[ht!]
        \centering
        \includegraphics[scale=0.6]{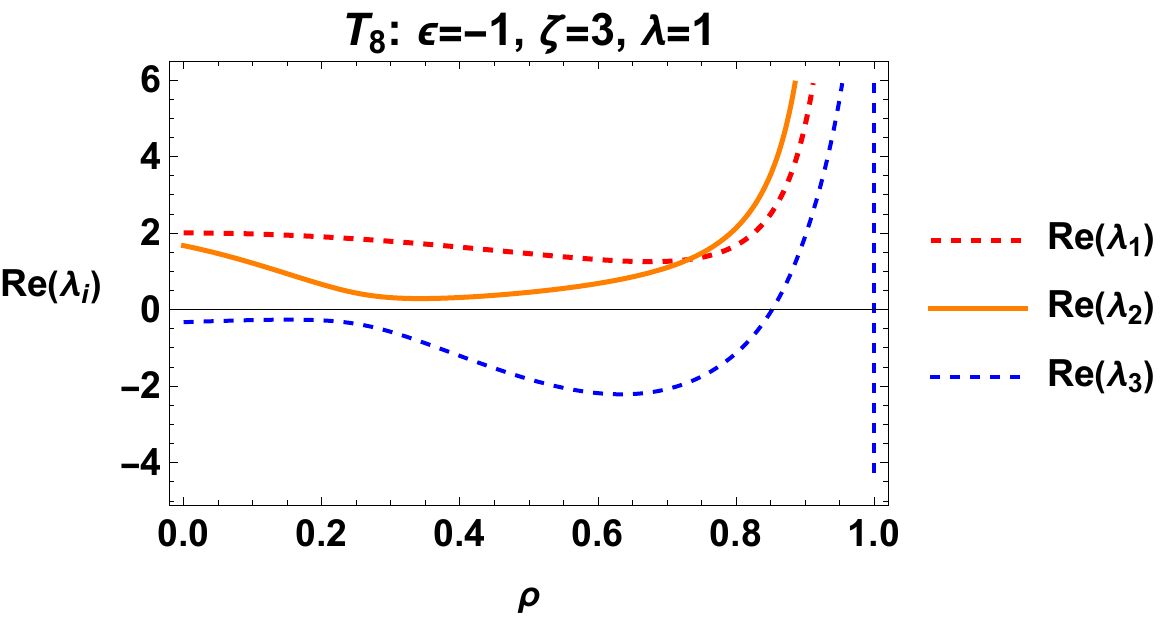}
        \includegraphics[scale=0.6]{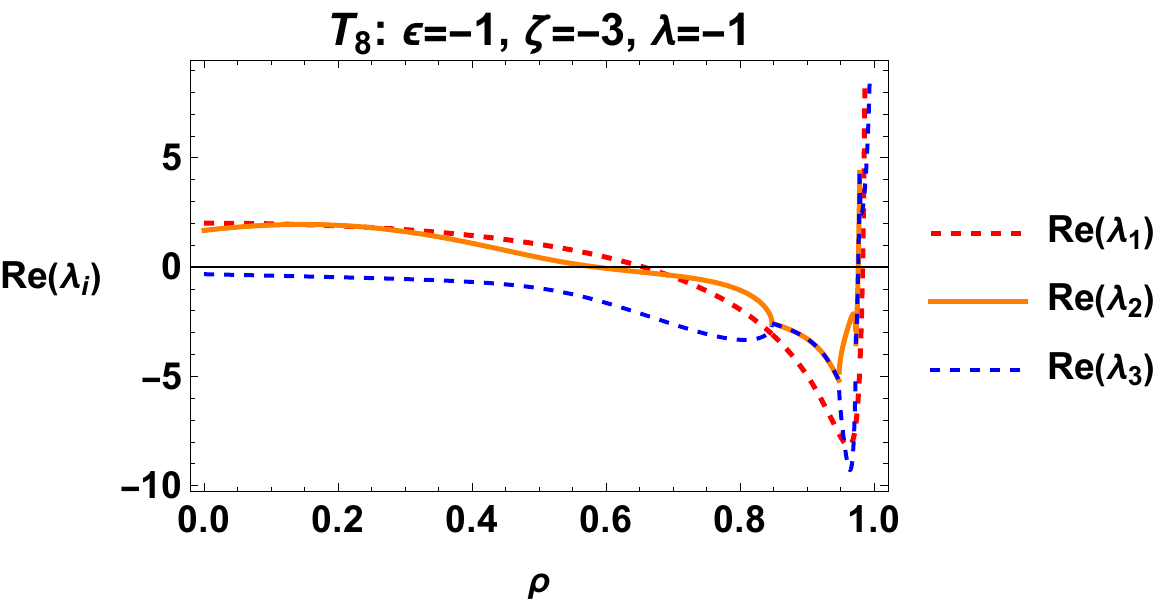}
        \caption{Real part of the eigenvalues of $T_8$ for different values of the parameters $\zeta$ and $\lambda$ with $0 \leq\rho\leq 1.$ This points exhibits source behaviour for $\rho \rightarrow 1$.}
        \label{fig:31}
    \end{figure}\begin{figure}[ht!]
        \centering
        \includegraphics[scale=0.6]{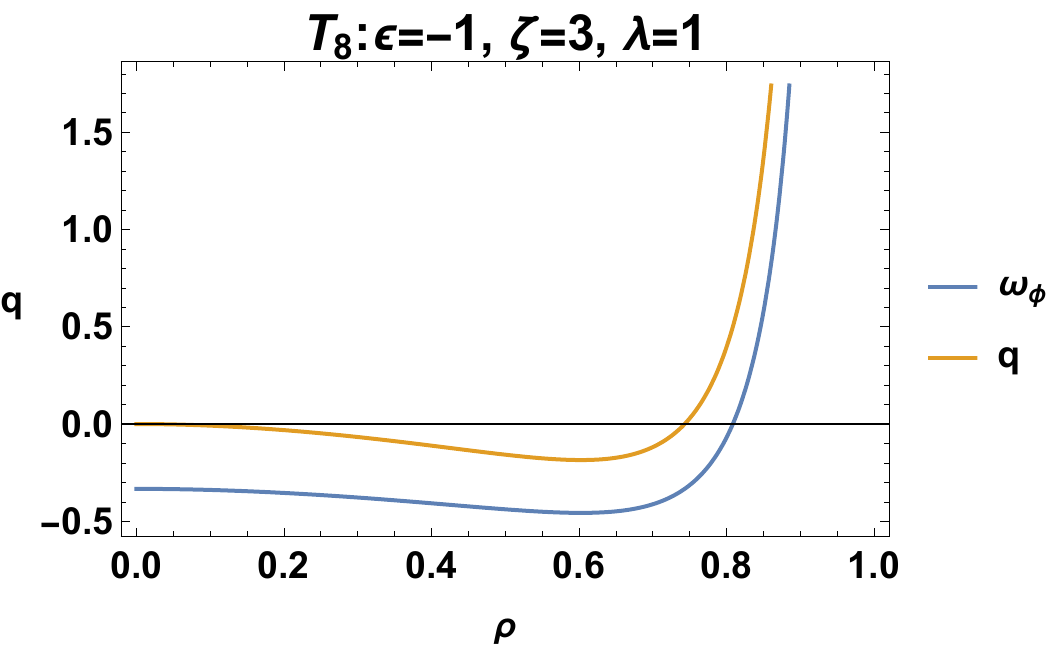}
        \includegraphics[scale=0.6]{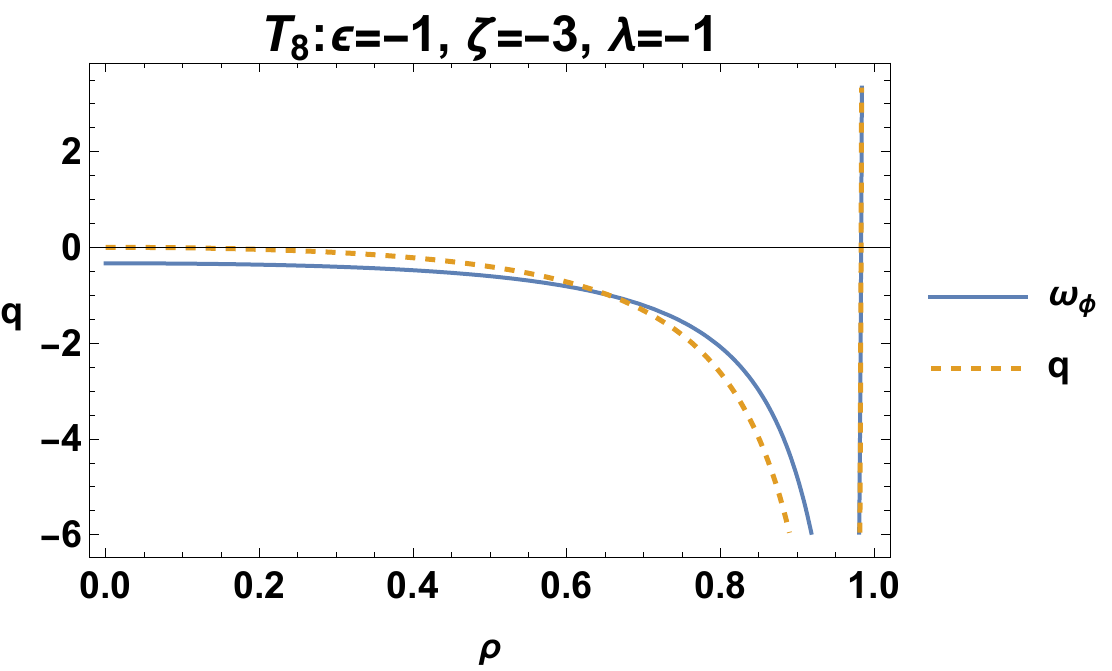}
        \caption{Plots of $\omega_{\phi}, q$ for $T_8.$}
        \label{fig:30}
    \end{figure}
    
    \item $T_9=(1,\alpha,-1).$ The eigenvalues are $\lambda_i(\rho,\zeta,\lambda)$ for $i=4,5,6,$ the analysis is performed for some values of $\zeta$ and $\lambda$ in Fig. \ref{fig:33}, the point has saddle or source behaviour. The point has sink behaviour as $\rho \rightarrow 1$. Also, we verify that both $\omega_{\phi}(T_9)$ and $q(T_9)$ go to infinity as $\rho \rightarrow 1$. That means that the point cannot describe an accelerated universe regardless of the values of $\zeta$ and $\lambda$, see Fig. \ref{fig:32}; we also verify that $\omega_{\phi}(T_9)$ and $q(T_9)$ tends to $-\frac{1}{3}$ and $0$ respectively as $\rho \rightarrow 0.$ 
        \begin{figure}[ht!]
        \centering
        \includegraphics[scale=0.6]{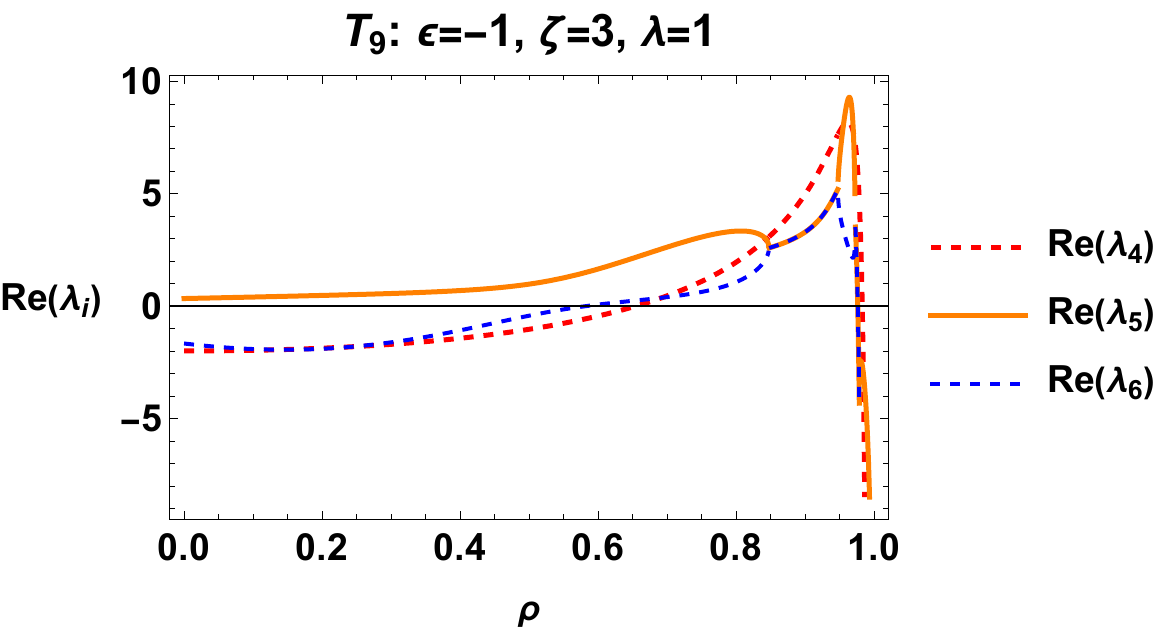}
        \includegraphics[scale=0.6]{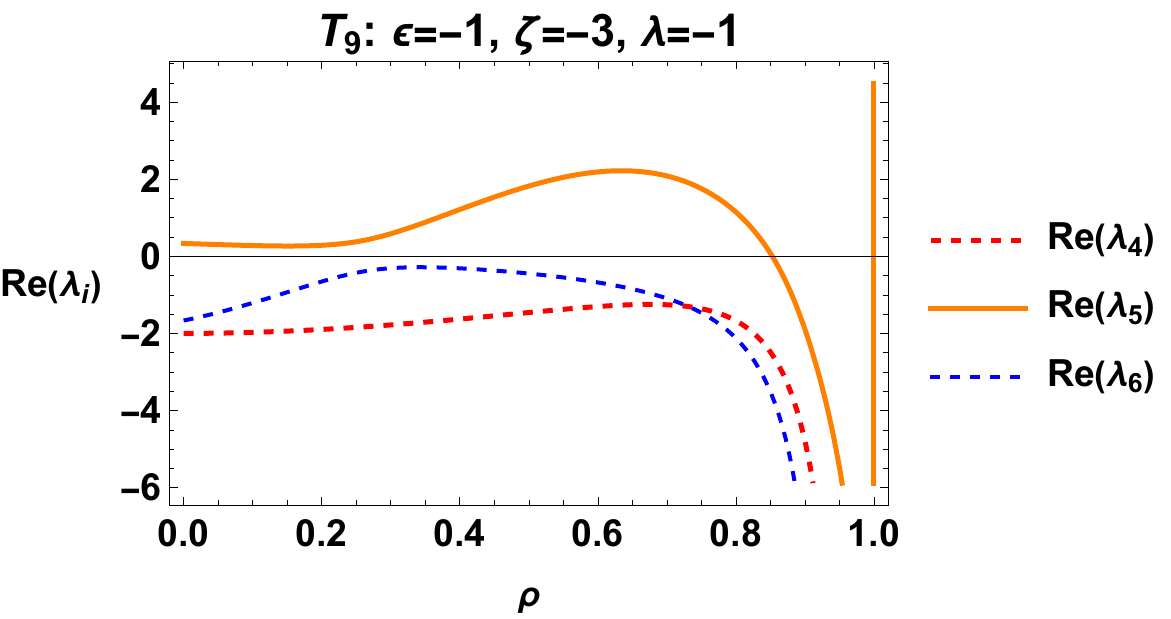}
        \caption{Real part of the eigenvalues of $T_9$ for different values of the parameters $\zeta$ and $\lambda$ with $0 \leq\rho\leq 1.$ This points exhibits saddle, source or sink behaviour.}
        \label{fig:33}
    \end{figure}
    
    \begin{figure}[ht!]
        \centering
        \includegraphics[scale=0.6]{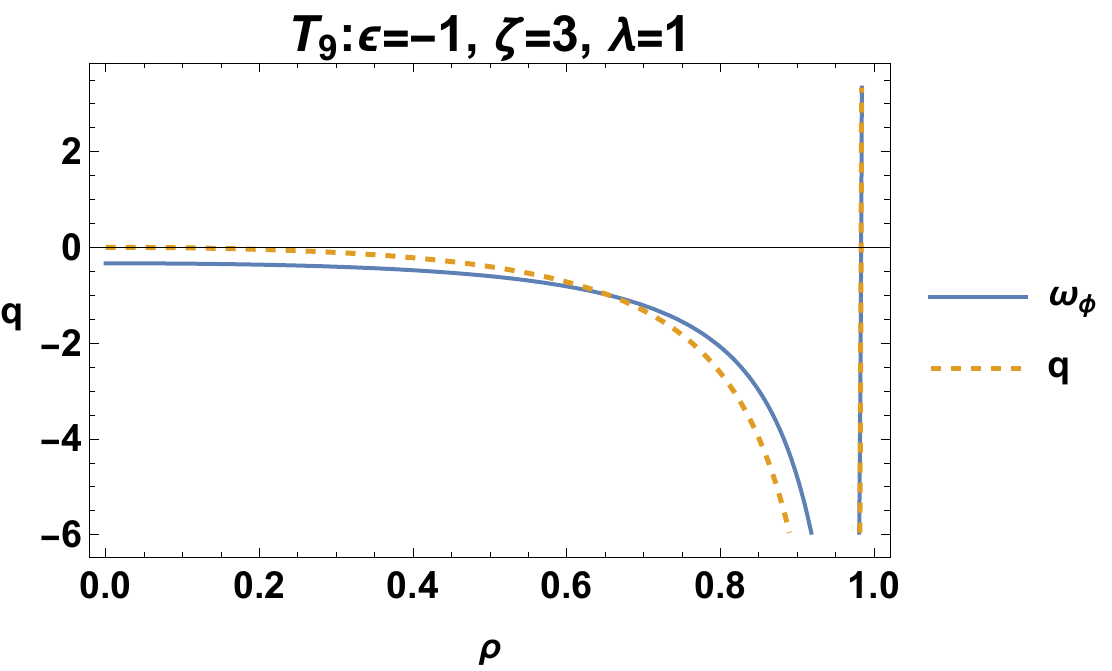}
        \includegraphics[scale=0.6]{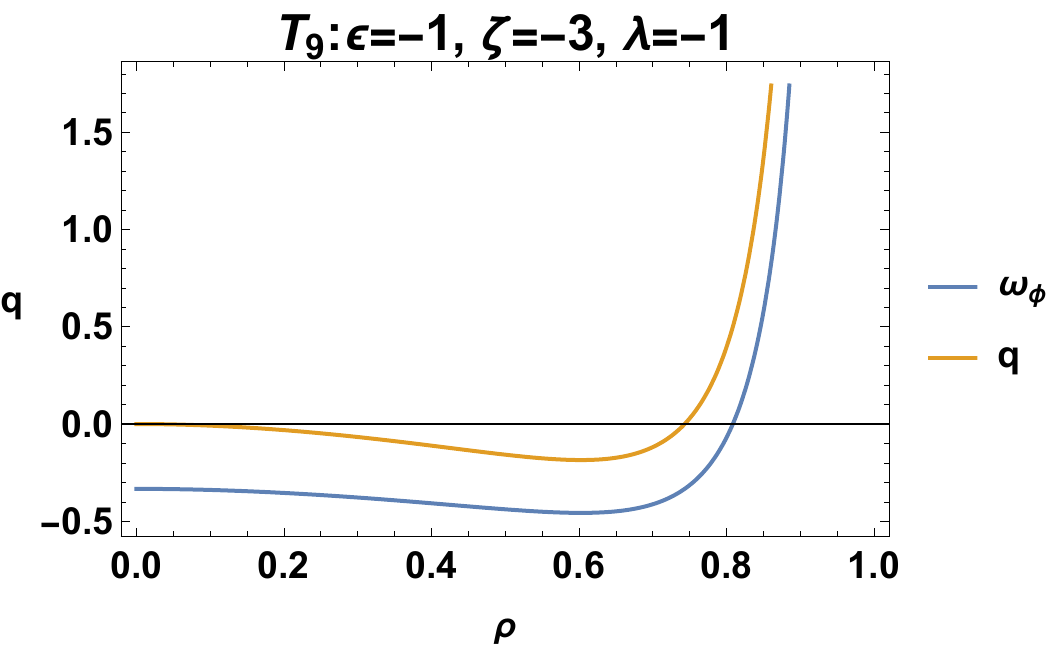}
        \caption{Plots of $\omega_{\phi}, q$ for $T_9.$}
        \label{fig:32}
    \end{figure}
    
    \item $T_{10}=(1,\pi-\alpha,1).$ The eigenvalues are $\delta_i(\rho,\zeta,\lambda)$ with $i=1,2,3$. The analysis is presented in Fig.  \ref{fig:35}, where we verify that the behaviour is symmetric as that of $T_8$ with respect to the signs of $\zeta$ and $\lambda.$ The interpretation of the physical parameters $\omega_{\phi}(T_{10}), q(T_{10})$ is similar as in $T_8$, see Fig. \ref{fig:34}.
           \begin{figure}[ht!]
        \centering
        \includegraphics[scale=0.6]{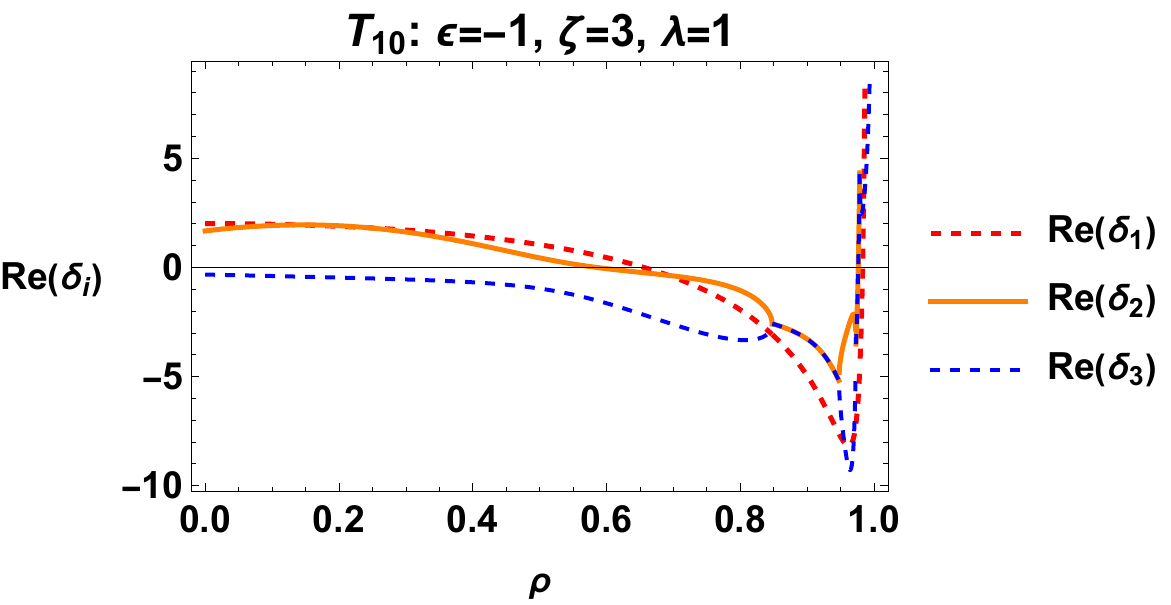}
        \includegraphics[scale=0.6]{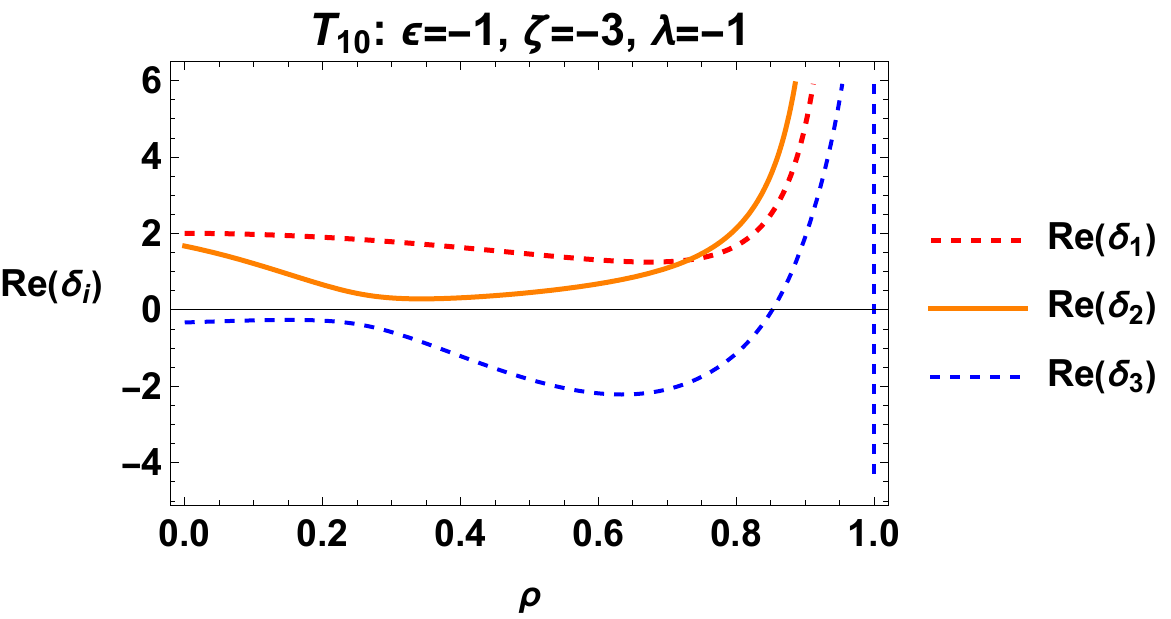}
        \caption{Real part of the eigenvalues of $T_{10}$ for different values of the parameters $\zeta$ and $\lambda$ with $0 \leq\rho\leq 1.$ This point exhibits source behaviour as $\rho \rightarrow 1$.}
        \label{fig:35}
    \end{figure}
    
        \begin{figure}[ht!]
        \centering
        \includegraphics[scale=0.6]{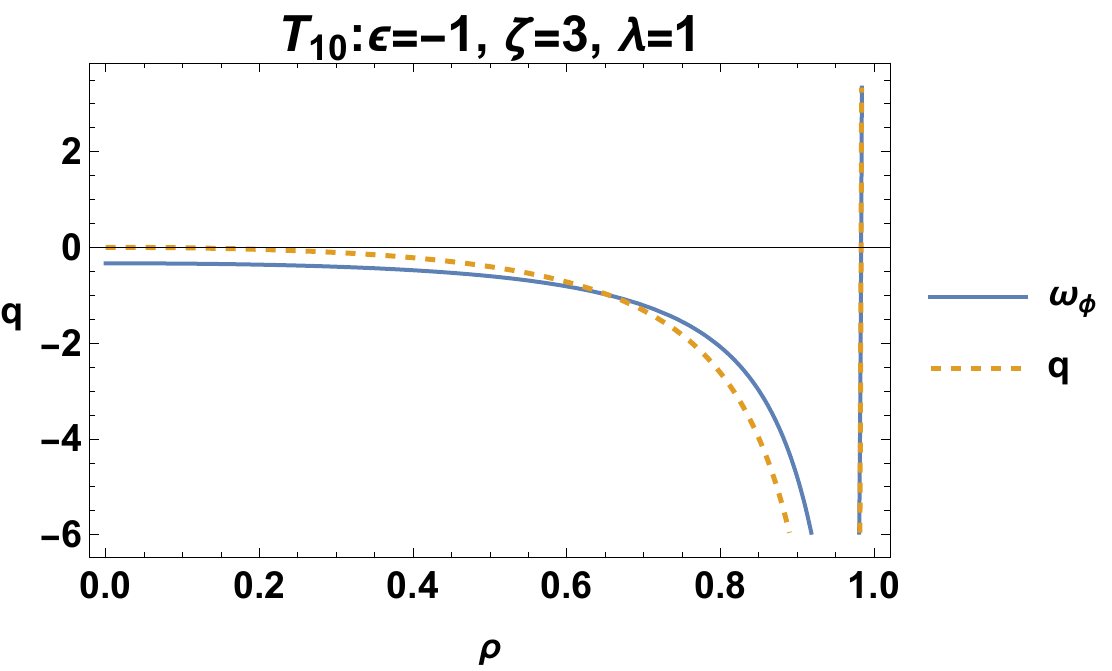}
        \includegraphics[scale=0.6]{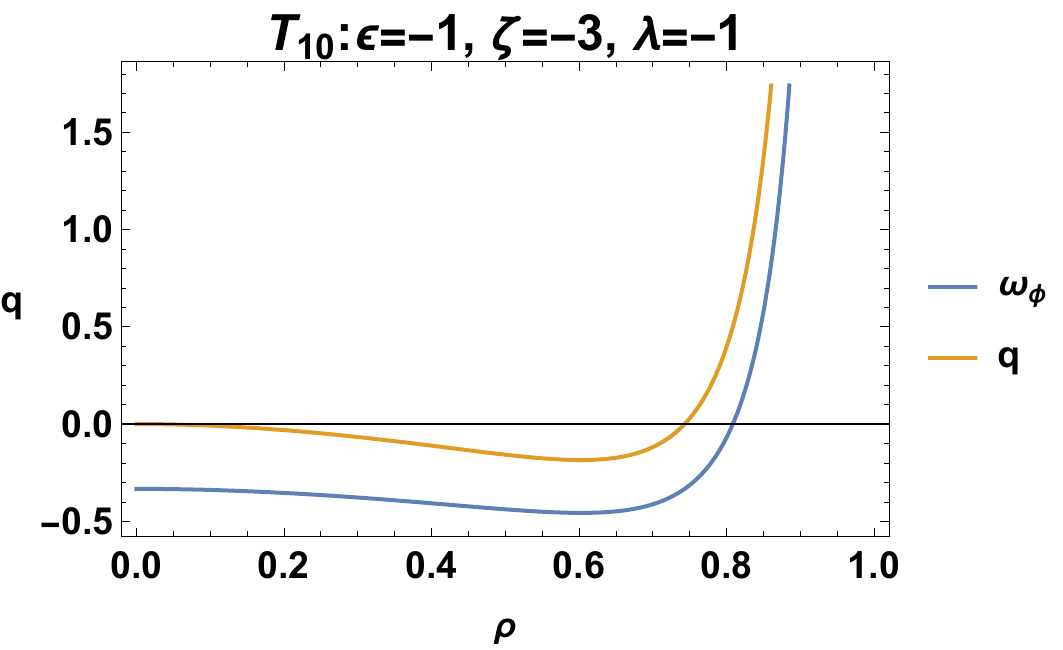}
        \caption{Plots of $\omega_{\phi}, q$ for $T_{10}.$}
        \label{fig:34}
    \end{figure}
    
    \item $T_{11}=(1,\pi-\alpha,-1).$  The eigenvalues are $\delta_i(\rho,\zeta,\lambda)$ with $i=4,5,6$. The analysis is presented in Fig.  \ref{fig:37}, where we verify that the behaviour is symmetric as that of $T_9$ with respect to the signs of $\zeta$ and $\lambda.$ The interpretation of the physical parameters $\omega_{\phi}(T_{11}), q(T_{11})$ is similar as in $T_9$, see Fig. \ref{fig:36}
        \begin{figure}[ht!]
        \centering
        \includegraphics[scale=0.6]{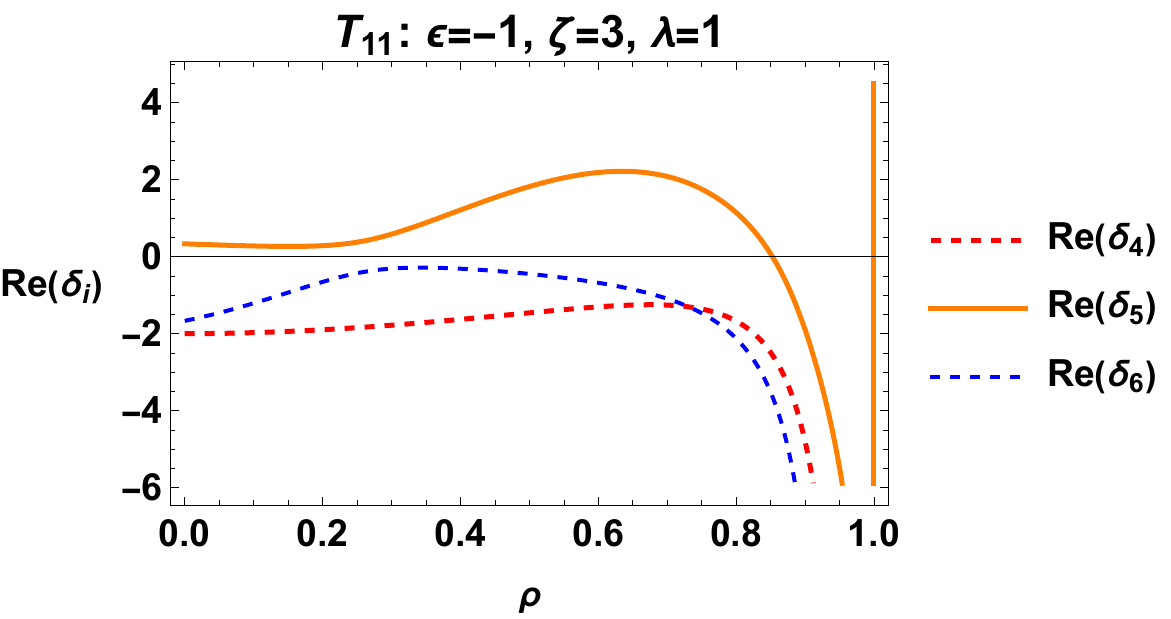}
        \includegraphics[scale=0.6]{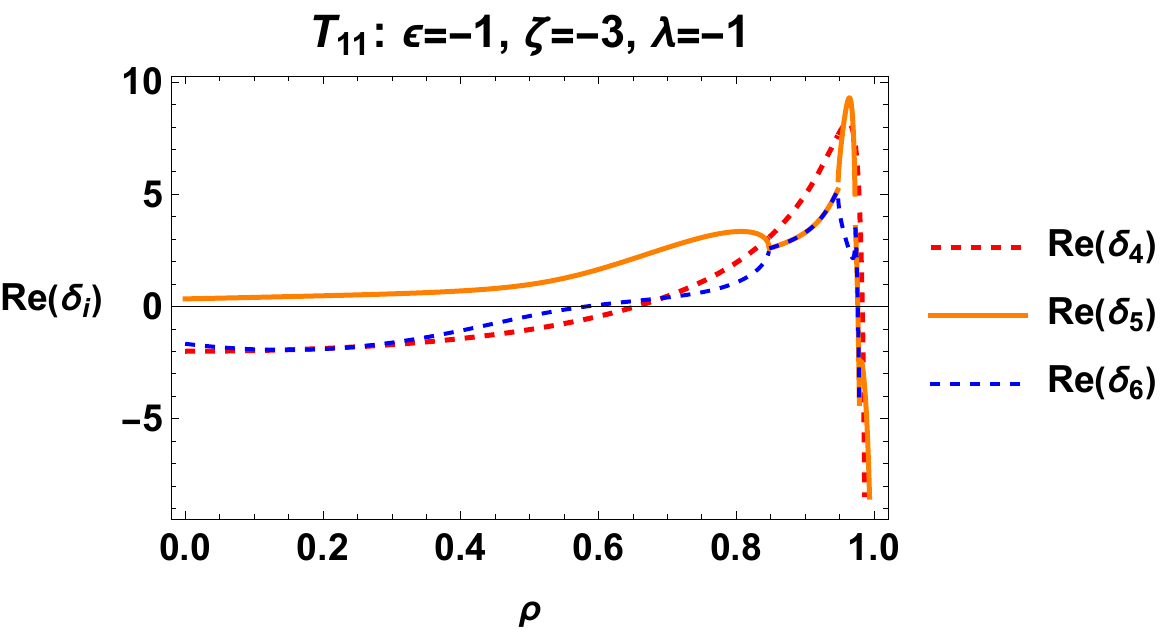}
        \caption{Real part of the eigenvalues of $T_{11}$ for different values of the parameters $\zeta$ and $\lambda$ with $0 \leq\rho\leq 1.$ This point exhibits saddle, source or sink behaviour.}
        \label{fig:37}
    \end{figure}
    
    \begin{figure}[ht!]
        \centering
        \includegraphics[scale=0.6]{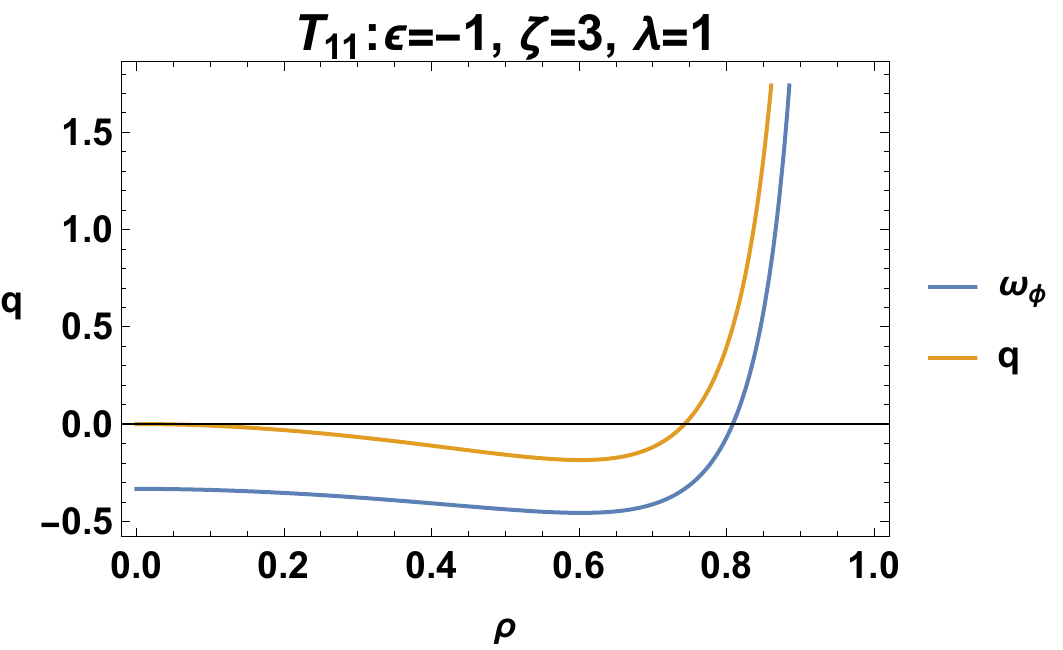}
        \includegraphics[scale=0.6]{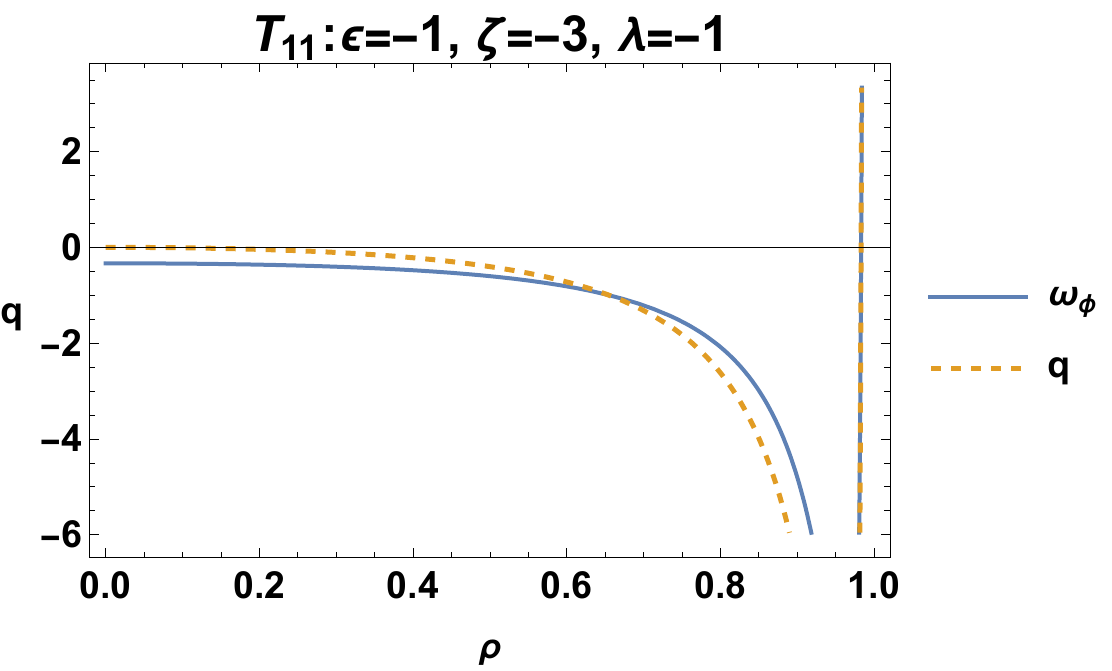}
        \caption{Plots of $\omega_{\phi}, q$ for $T_{11}.$}
        \label{fig:36}
    \end{figure}
    
    \item $T_{12}=(1,\alpha,0).$ The eigenvalues are $\gamma_i(\rho,\zeta,\lambda,\eta)$ with $i=1,2,3.$ The point has saddle behaviour for values of $\eta \neq 0,$ however since $\eta$ is zero for this point $T_{12}$ is nonhyperbolic, see Fig. \ref{fig:39}. The physical parameters $\omega_{\phi}(T_{12})$ and $q(T_{12})$ blow up for $\eta=0$, see Fig. \ref{fig:38}. In particular, we verify that $\lim_{\rho \rightarrow 1}(\lim_{\eta\rightarrow 0}(\omega_{\phi}(T_{12})))=\infty$ and $\lim_{\rho \rightarrow 1}(\lim_{\eta\rightarrow 0}(q(T_{12})))=\infty$ given this, the point cannot describe an accelerated universe.
        \begin{figure}[ht!]
        \centering
        \includegraphics[scale=0.6]{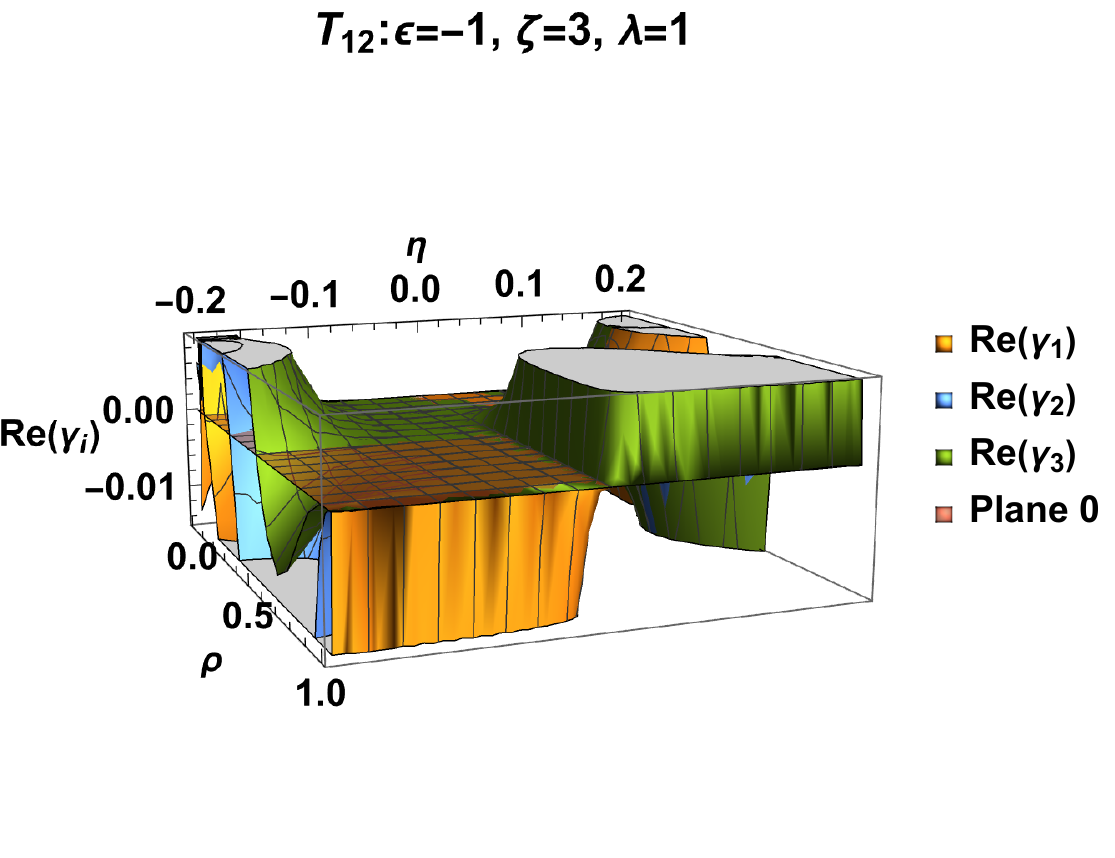}
        \includegraphics[scale=0.6]{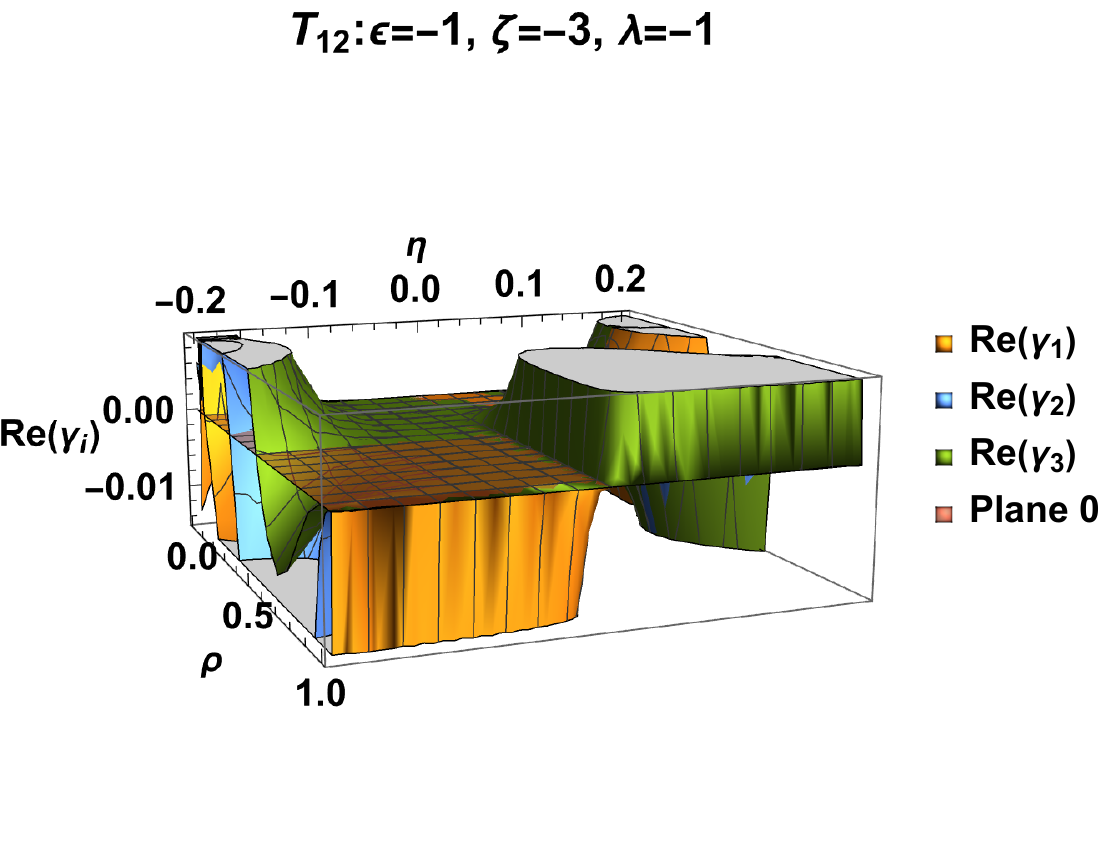}
        \caption{Real part of the eigenvalues of $T_{12}$ for different values of the parameters $\zeta$ and $\lambda$ with $0 \leq\rho\leq 1$ and $-1\leq \eta \leq 1.$ This point exhibits saddle or nonhyperbolic behaviour.}
        \label{fig:39}
    \end{figure}
     
        \begin{figure}[ht!]
        \centering
        \includegraphics[scale=0.6]{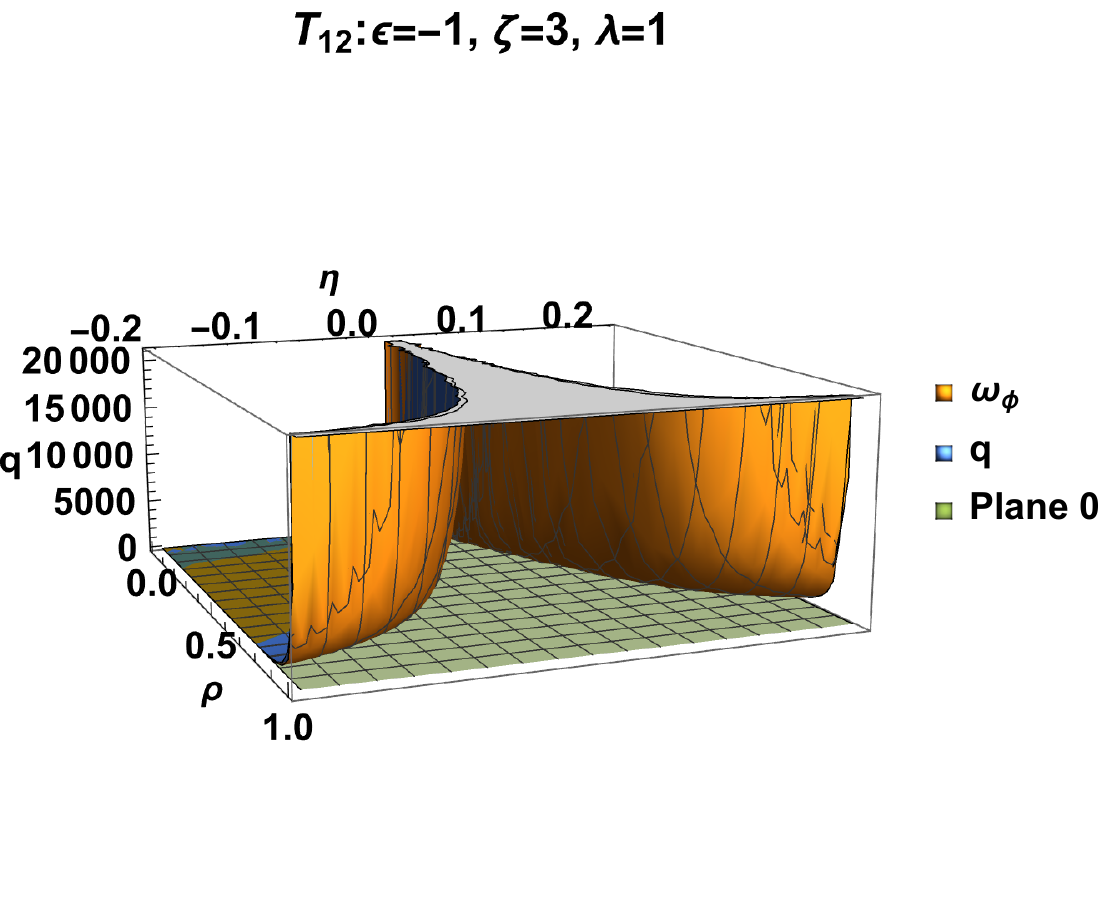}
        \includegraphics[scale=0.6]{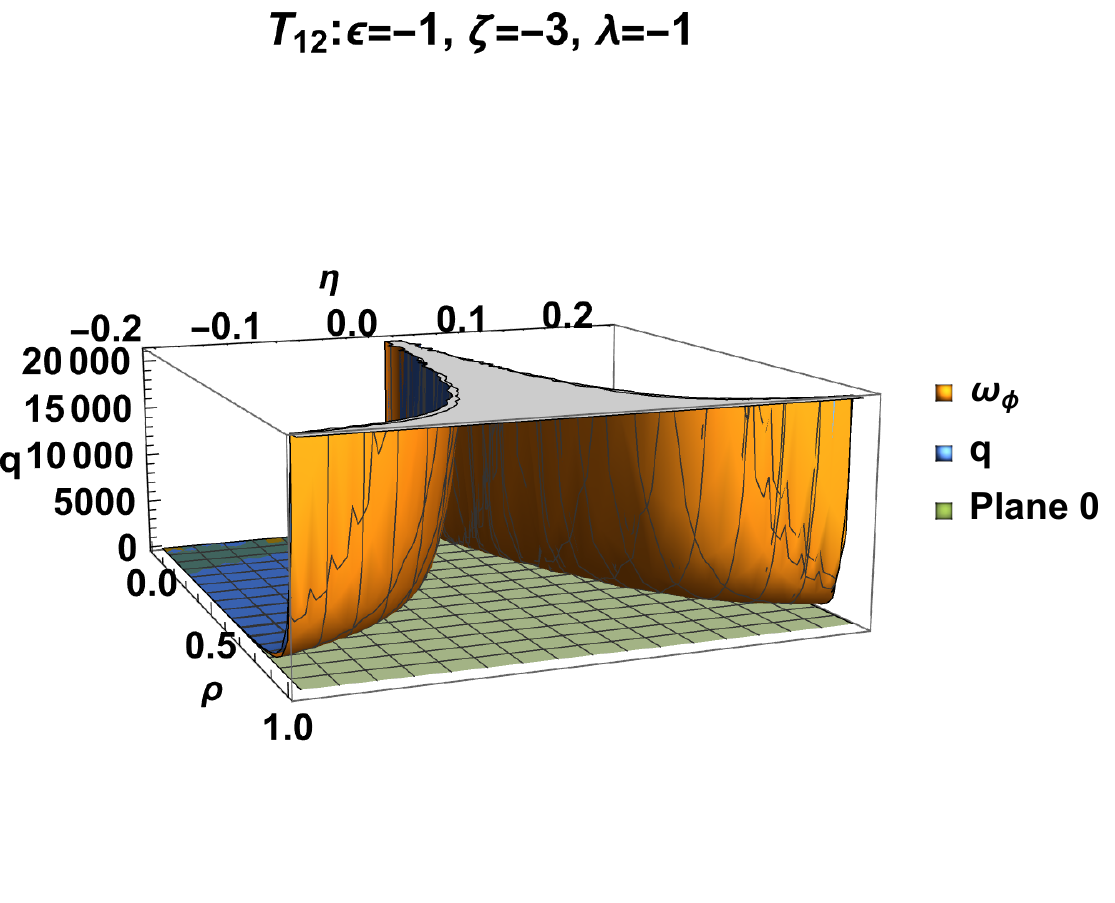}
        \caption{Plots of $\omega_{\phi}, q$ for $T_{12}.$}
        \label{fig:38}
    \end{figure}
    
    \item $T_{13}=(1,\pi-\alpha,0).$ The eigenvalues are $\gamma_i(\rho,\zeta,\lambda,\eta)$ with $i=4, 5, 6.$ The point has the same problems with $\eta=0$ but the stability analysis is similar to that of $T_{12}$, see Fig. \ref{fig:41} to compare. Once again we verify that $\lim_{\rho \rightarrow 1}(\lim_{\eta\rightarrow 0}(\omega_{\phi}(T_{13})))=\infty$ and $\lim_{\rho \rightarrow 1}(\lim_{\eta\rightarrow 0}(q(T_{13})))=\infty.$
        \begin{figure}[ht!]
        \centering
        \includegraphics[scale=0.6]{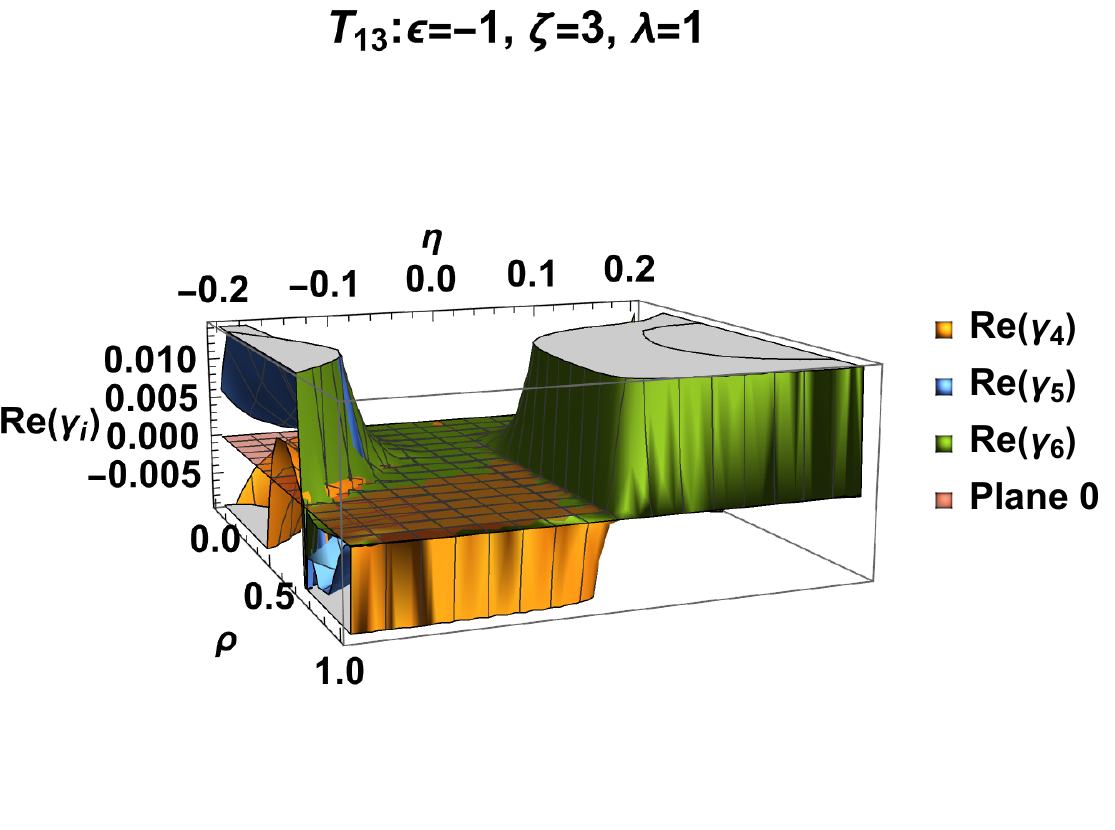}
        \includegraphics[scale=0.6]{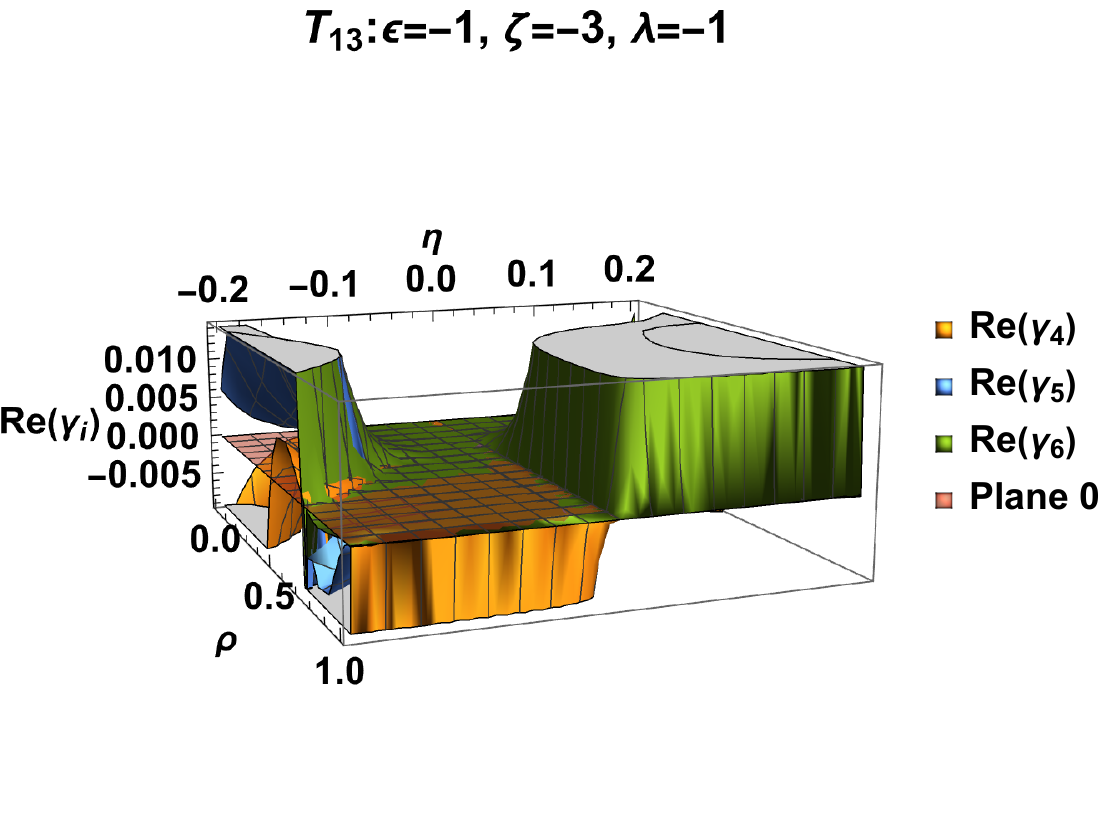}
        \caption{Real part of the eigenvalues of $T_{13}$ for different values of the parameters $\zeta$ and $\lambda$ with $0 \leq\rho\leq 1$ and $-1\leq \eta \leq 1.$ This point exhibits saddle and nonhyperbolic behaviour.}
        \label{fig:41}
    \end{figure}
     
     \begin{figure}[ht!]
        \centering
        \includegraphics[scale=0.6]{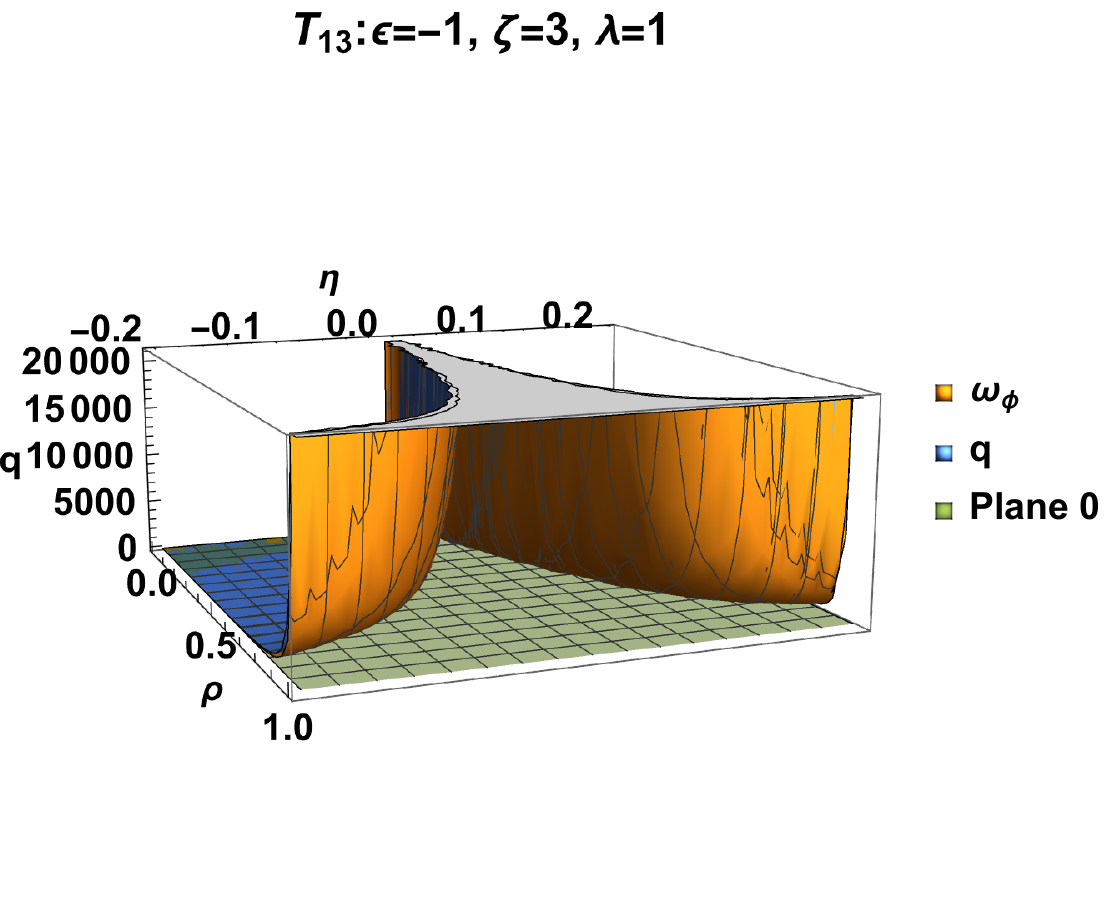}
        \includegraphics[scale=0.6]{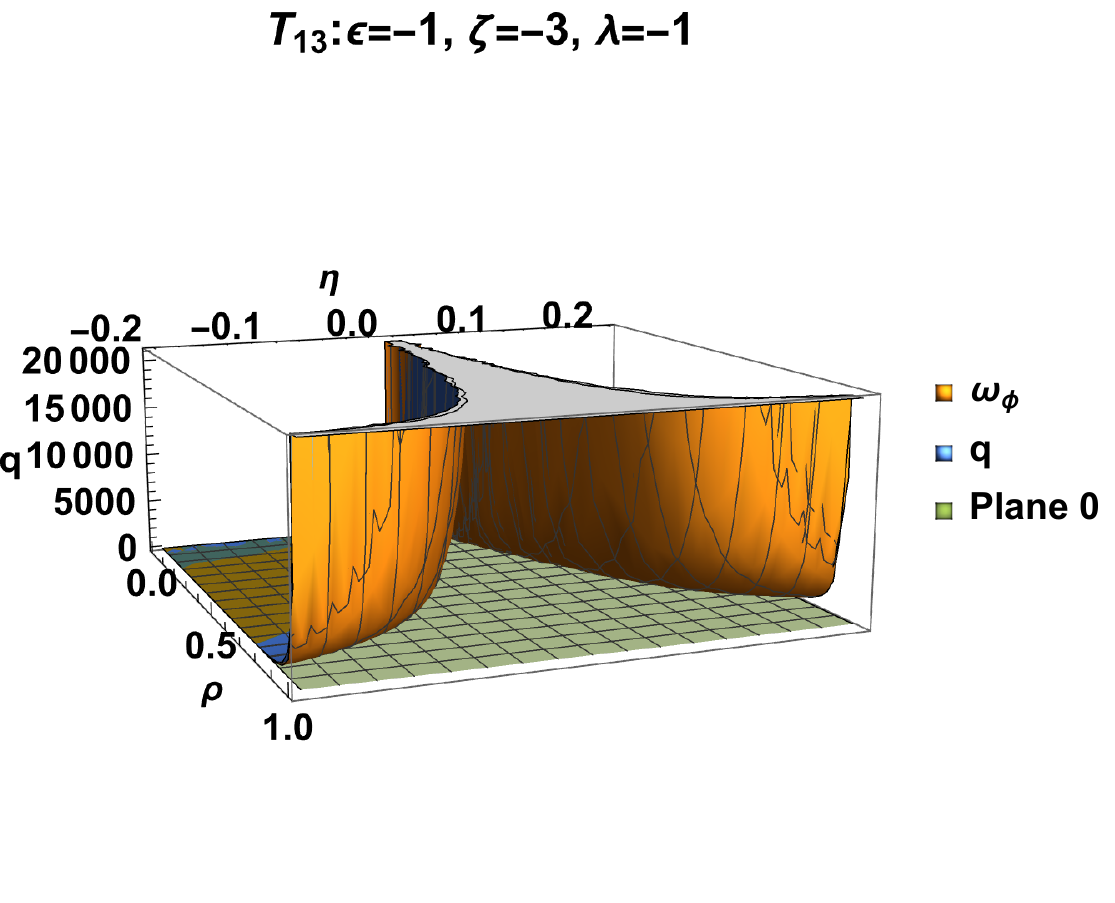}
        \caption{Plots of $\omega_{\phi}, q$ for $T_{13},$ we verify that $\lim_{\rho \rightarrow 1}(\lim_{\eta\rightarrow 0}(\omega_{\phi}(T_{13})))=\infty$ and $\lim_{\rho \rightarrow 1}(\lim_{\eta\rightarrow 0}(q(T_{13})))=\infty.$}
        \label{fig:40}
    \end{figure}
    
\end{enumerate}

In Fig. \ref{fig:25}, we present some three-dimensional phase-plot diagrams for $\epsilon=-1, \lambda=1$ and different values of $\zeta$. The results of this section are summarized in Table \ref{tab:7}. 
\begin{figure}[ht!]
    \centering
    \includegraphics[scale=0.6]{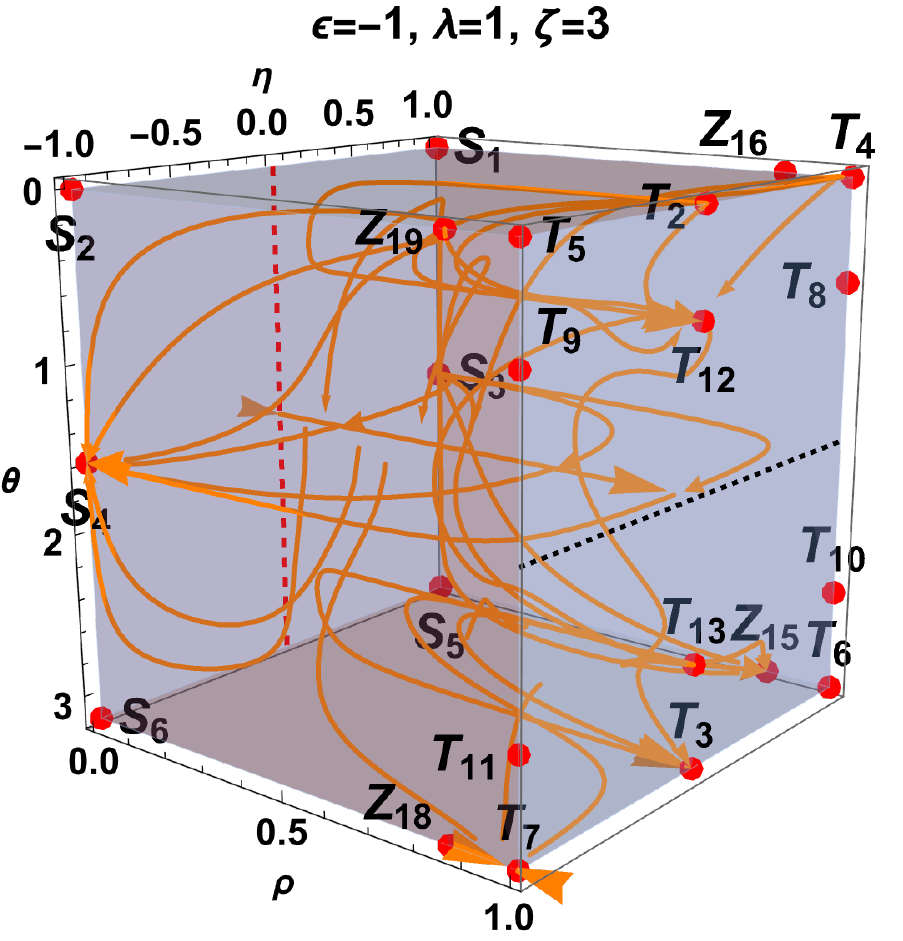}
    \includegraphics[scale=0.6]{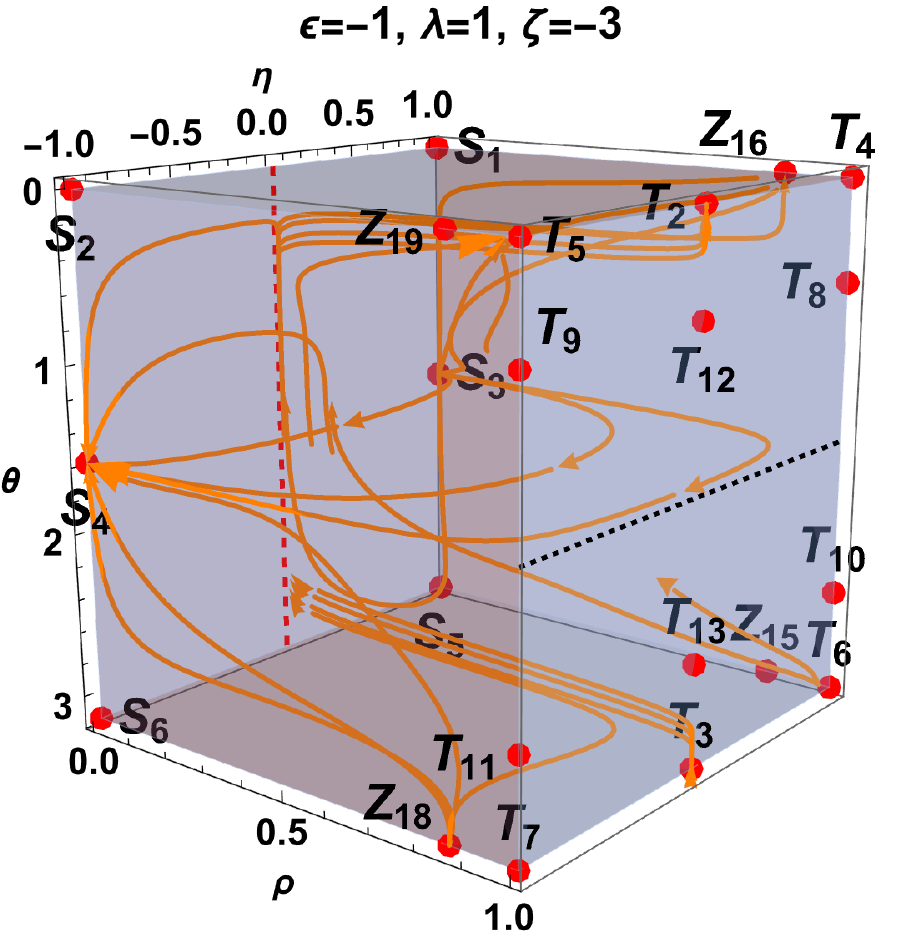}
    \caption{Three dimensional phase space for system \eqref{eqrho}, \eqref{eqtheta}, \eqref{eqeta} for different values of the parameters $\zeta,$ and $\lambda$. The dashed black line corresponds to $T_1$, and the dashed red line corresponds to $S_7.$}
    \label{fig:25}
\end{figure}

\begin{table}[ht!]
    \caption{Equilibrium points of system \eqref{eqrho}, \eqref{eqtheta}, \eqref{eqeta}  for $\epsilon=-1$ with their stability conditions. It also includes the value of $\omega_{\phi}$ and $q.$}
    \label{tab:7}
 \centering
\newcolumntype{C}{>{\centering\arraybackslash}X}
\centering
    \setlength{\tabcolsep}{6.6mm}
\begin{tabularx}{\textwidth}{cccccccc}
\toprule 
        Label& $\rho$& $\theta$ & $\eta$ &Stability& $\omega_{\phi}$& $q$ \\\midrule
         $T_1$& $0$& $\frac{\pi}{2}$ & $0$ &nonhyperbolic& $-\frac{1}{3}$& $0$ \\\midrule$T_2$& $1$& $0$ & $0$ &saddle& see Fig. \ref{fig:20}& see Fig. \ref{fig:20} \\\midrule
         $T_3$& $1$& $\pi$ & $0$ &saddle& see Fig. \ref{fig:21}& see Fig. \ref{fig:21} \\\midrule
         $T_{4,5}$& $1$& $0$ & $\pm 1$ &saddle& see Fig. \ref{fig:22}& see Fig. \ref{fig:22} \\\midrule
         $T_{6,7}$& $1$& $\pi$ & $\pm 1$ &saddle& see Fig. \ref{fig:23}& see Fig. \ref{fig:23} \\\midrule
         $S_1$& $0$& $0$ & $1$ &saddle& $-\frac{1}{3}$&$0$ \\\midrule
          $S_2$& $0$& $0$ & $-1$ &saddle& $-\frac{1}{3}$&$0$ \\\midrule
           $S_3$& $0$& $\frac{\pi}{2}$ & $1$ &source& $-\frac{1}{3}$&$0$ \\\midrule
           $S_4$& $0$& $\frac{\pi}{2}$ & $-1$ &sink& $-\frac{1}{3}$&$0$ \\\midrule
            $S_5$& $0$& $\pi$ & $1$ &saddle& $-\frac{1}{3}$&$0$ \\\midrule
            $S_6$& $0$& $\pi$ & $-1$ &saddle& $-\frac{1}{3}$&$0$ \\\midrule
            $S_7$& $0$& $\theta$ & $0$ &nonhyperbolic& $-\frac{1}{3}$&$0$ \\\midrule
            $T_8$& $1$& $\alpha$ & $1$ &see Fig. \ref{fig:31}& see Fig. \ref{fig:30}&see Fig. \ref{fig:30} \\\midrule
              $T_9$& $1$& $\alpha$ & $-1$ &see Fig. \ref{fig:33}& see Fig. \ref{fig:32}&see Fig. \ref{fig:32}  \\\midrule
                $T_{10}$& $1$& $\pi-\alpha$ & $1$ &see Fig. \ref{fig:35}& see Fig. \ref{fig:34}&see Fig. \ref{fig:34} \\\midrule
                $T_{11}$& $1$& $\pi-\alpha$ & $-1$ &see Fig. \ref{fig:37}& see Fig. \ref{fig:36}&see Fig. \ref{fig:36} \\\midrule
                $T_{12}$& $1$& $\alpha$ & $0$ &see Fig. \ref{fig:39}& see Fig. \ref{fig:38}&see Fig. \ref{fig:38}  \\\midrule
                $T_{13}$& $1$& $\pi-\alpha$ & $0$ &see Fig. \ref{fig:41}& see Fig. \ref{fig:40}&see Fig. \ref{fig:40} \\\midrule
    \end{tabularx}
\end{table}

\section{Conclusions}
\label{V}

The phase-space analysis of the gravitational field equations is a novel
mathematical approach to the model's asymptotic description and evolution of the
physical variables. In cosmological studies, we expect the
cosmological study model to provide the significant eras of cosmic history.
In this work, we considered a four-dimensional FLRW geometry and a second-order
modified gravitational theory with a scalar field coupled to the Gauss-Bonnet
scalar. In the limit where the scalar field is constant, the Gauss-Bonnet term
does not contribute to the gravitational Action Integral. The theory
reduces to General Relativity with a cosmological constant term. However, for
a dynamical scalar field, the physical properties of the present cosmological
model are distinct from that of the minimally coupled scalar field theory.

The gravitational\ Action Integral depends on two functions which are the
coupling function of the scalar field with the Gauss-Bonnet scalar and the
scalar field potential. For the coupling functions, we consider two functional
forms, the exponential function a power-law function, while the potential
we assume the exponential functional form. Moreover, a parameter
$\epsilon=\pm1$ has been introduced in the kinetic part of the scalar
field, such that the scalar field is a quintessence field, $\epsilon
=+1$, or a phantom field, $\epsilon=-1$.

In order to study the dynamical properties of the phase space and 
physical variables, we introduced dimensionless variables different from that
of the $H$-normalization. The latter is because, from the field equations, we
observed that it is possible in Einstein-Gauss-Bonnet scalar field theory that the Hubble function can cross sign during its evolution, which means that it can
vanish. Hence, the $H$-normalization, widely applied before, must be validated for global analysis and the complete reconstruction of the
cosmological history and epochs. Additionally, we observed that the dynamical
variables are not bounded in a finite regime, which means that to
perform a complete study of the phase-space, we assumed Poincare variables to
investigate the asymptotic behaviour of the model at infinity.

The two functional forms for the coupling function of the scalar field with
the Gauss-Bonnet scalar provide different cosmological evolution. Indeed, for
the exponential coupling function, only asymptotic solutions with deceleration
parameters $q=0$ and $q=-1$ exist. However, scaling solutions with
a deceleration parameter, $q\neq0$, exist only for the power-law coupling
function. Last but not least, the stability properties of the asymptotic
solutions were investigated.

This study extends and completes previous results in the literature in
Einstein-Gauss-Bonnet scalar field cosmology \cite{dn1,dn2}. The analysis
indicates that the theory can explain the main eras of cosmological
history. In future work, we plan to extend the further analysis by introducing
matter source components and new functional forms for the
scalar field potential and the coupling function.

\section*{Acknowledgments}

Alfredo David  Millano was supported was supported by Agencia Nacional de Investigación y Desarrollo (ANID) Subdirección de Capital Humano/Doctorado Nacional/año 2020 folio 21200837, Gastos operacionales Proyecto de tesis/2022 folio 242220121, and by Vicerrectoría de Investigación y Desarrollo Tecnológico (VRIDT) at Universidad Católica del Norte. GL was funded through Concurso De Pasantías De Investigación Año 2022, Resolución VRIDT No. 040/2022 and Resolución VRIDT No. 054/2022. He also thanks the support of Núcleo de Investigación Geometría Diferencial y Aplicaciones, Resolución VRIDT N°096/2022, and Andronikos Paliathanasis acknowledges VRIDT-UCN through Concurso de Estadías de Investigación, Resolución VRIDT N°098/2022.

\appendix
\section{Existence of special equilibrium points}\label{app}
The special points are $Z_{12}, Z_{13}, Z_{14}, Z_{19}, Z_{20}, Z_{21}, Z_{22}, Z_{23}, Z_{24}.$ We show that these points exist by analyzing  the $x-$equation from system \eqref{exponentialf-1}, \eqref{exponentialf-2} and \eqref{exponentialf-3} while setting $\epsilon=1, y=0, \eta=1.$ The equation reads

\begin{equation}
    \frac{dx}{d\tau}=\frac{x \left(x^2-6\right) \left(\zeta  x^3-10 x^2+6 \zeta  x-12\right)}{5 x^4-12 x^2+36}.
\end{equation}
By analyzing  the numerator we know that $x=0, x=\pm \sqrt{6}$ are equilibrium points for the $x-$equation. We need to examine the following polynomial of third degree, 
\begin{equation*}
    \zeta  x^3-10 x^2+6 \zeta x-12=0,
\end{equation*}
which can be rewritten as  
\begin{equation}
\label{poly}
    X^3-\Big(\frac{100}{3 \zeta ^2}-6 \Big)X-\frac{2000}{27 \zeta ^3}+\frac{8}{\zeta }=0
\end{equation}
where, we have divided by $\zeta$ and used the change of variable $X=x-\frac{10}{3 \zeta }.$ Now the polynomial \eqref{poly} has the form \begin{equation}
\label{polystandard}
    X^3+p X+q=0,
\end{equation}

The sign of the determinant $\Delta$ determines the nature of the roots. For $\Delta>0$, the polynomial has three real roots; for $\Delta<0$, it has one real root. For polynomials in the form \eqref{polystandard}, the discriminant is
$\Delta=-4 p^3-27 q^2,$ in our case we have

\begin{equation}
    \Delta=-\frac{96 \left(9 \zeta ^4-132 \zeta ^2+500\right)}{\zeta ^4},
\end{equation}
that is always negative for $\zeta \in \mathbb{R}$ which means there is only one real root, and it is $Z_{12}.$

We must do the same for the other values of $\eta$ and $\epsilon.$ Setting $\epsilon=1$ and $\eta=-1$ we have the other projection of system \eqref{exponentialf-1}, \eqref{exponentialf-2} and \eqref{exponentialf-3}. The $x-$equation reads

\begin{equation}
    \frac{dx}{d\tau}=\frac{x \left(x^2-6\right) \left(\zeta  x^3+10 x^2+6 \zeta  x+12\right)}{5 x^4-12 x^2+36}.
\end{equation}

With this, we can write a polynomial as before

\begin{equation}
    \label{poly2}
    X^3-\Big(\frac{100}{3 \zeta ^2}-6\Big)X+\frac{2000}{27 \zeta ^3}-\frac{8}{\zeta }=0.
\end{equation}
The discriminant is 
\begin{equation*}
    \Delta=-\frac{96 \left(9 \zeta ^4-132 \zeta ^2+500\right)}{\zeta ^4},
\end{equation*}
which is also negative for all values of $\zeta.$ Once again, there is only one real root. This root is $Z_{13}$ if $\zeta<-\frac{5\sqrt{2}}{3}$ and  $\zeta>\frac{5\sqrt{2}}{3}$ or $Z_{14}$ if $-\frac{5\sqrt{2}}{3}<\zeta<\frac{5\sqrt{2}}{3}.$  

For the case $\epsilon=-1$ we take a similar approach, setting $\eta=1, y=0$ in \eqref{exponentialf-1}, \eqref{exponentialf-2} and \eqref{exponentialf-3} gives the first projection, and the $x-$equation is

\begin{equation}
    \frac{dx}{d\tau}=\frac{x \left(x^2+6\right) \left(\zeta  x^3-10 x^2-6 \zeta  x+12\right)}{5 x^4+12 x^2+36}.
\end{equation}

Once again, we write the following polynomial as
\begin{equation}
    \label{poly3}
    X^3+\Big(-\frac{100}{3 \zeta ^2}-6\Big) X-\frac{2000}{27 \zeta ^3}-\frac{8}{\zeta }=0,
\end{equation}
with discriminant

\begin{equation*}
    \Delta=\frac{96 \left(9 \zeta ^4+132 \zeta ^2+500\right)}{\zeta ^4}.
\end{equation*}
This discriminant is always positive for $\zeta \neq 0,$ this means 3 real roots which are $Z_{19}, Z_{21}$ and $Z_{22}.$

Finally we study the final projection for $\epsilon=-1,$ that is we set $\eta=-1, y=0$ and we write the $x-$equation as 
\begin{equation*}
    \frac{dx}{d\tau}=\frac{x \left(x^2+6\right) \left(\zeta  x^3+10 x^2-6 \zeta  x-12\right)}{5 x^4+12 x^2+36}.
\end{equation*}

The polynomial for this case is
\begin{equation}
    \label{poly4}
    X^3+\Big(-\frac{100}{3 \zeta ^2}-6\Big)+\frac{2000}{27 \zeta ^3}+\frac{8}{\zeta }=0.
\end{equation}
Now the discriminant is
\begin{equation*}
    \Delta=\frac{96 \left(9 \zeta ^4+132 \zeta ^2+500\right)}{\zeta ^4},
\end{equation*}
once again $\Delta>0$ for $\zeta\neq 0.$ The three real roots are $Z_{22}, Z_{23}$ and $Z_{24}.$


\begin{thebibliography}{99}                                                                                               %
\bibitem {Aref1}A. Linde, Phys. Lett. B 108, 389 (1982)

\bibitem {guth}A. Guth, Phys. Rev. D 23, 347 (1981)

\bibitem {f1}K. Sato, MNRAS 195, 467 (1981)

\bibitem {f2}J.D\ Barrow and A. Ottewill, J. Phys. A\ 16, 2757 (1983)

\bibitem {ref1a}I. P. Neupane, Class. Quantum Grav. 25, 125013 (2008)

\bibitem {ref1}A.D. Linde, Phys. Lett. B 129, 177 (1983)

\bibitem {ref2}A.R. Liddle, Phys. Lett. B 220, 502 (1989)

\bibitem {ref3}T. Charters, J.P. Mimoso and A. Nunes, Phys. Lett. B 472, 21 (2000)

\bibitem {newinf}J.D. Barrow, Phys.\ Rev. D 48, 1585 (1993)

\bibitem {star}A.A. Starobinsky, Phys. Lett. B 91, 99 (1980)

\bibitem {ib1}E.O. Pozdeeva and S. Yu. Vernov, arXiv:2211.10988 (2022)

\bibitem {ib2}D.Y. Cheong, H.M. Lee and S.C. Park, Phys. Lett. B 805, 135453 (2020)

\bibitem {sup1}A.G. Riess et al. Astron. J., 116, 1009 (1998)

\bibitem {Clifton1}T. Clifton, P.G. Ferreira, A. Padilla and C. Skordis, Phys.
Rept. {513}, 1 (2012)

\bibitem {Nojiri:2017ncd}S.~Nojiri, S.~D.~Odintsov and V.~K.~Oikonomou,
Phys.\ Rept.\ {692}, 1 (2017)

\bibitem {sf8}E. Di\ Valentino, O. Mena, S. Pan, L. Visinelli, W. Yang, A.
Melchiorri, D.F. Mota, A.G. Riess and J. Silk, Class. Quantum\ Grav. 38,
153001 (2021)

\bibitem {lvl}D. Lovelock, J.\ Math. Phys. 13, 874 (1972)

\bibitem {lvl2}D. Lovelock, J. Math. Phys. 12, 498 (1971)

\bibitem {lvl3}A. Mardones and J. Zanelli, Class. Quantum Grav. 8, 1545 (1991)

\bibitem {gb1}F. Canfora, A.\ Giacomini and S.A. Pavluchenko, Gen. Relativ.
Grav. 46, 1805 (2014)

\bibitem {gb2}S.D. Maharaj, B. Chiambwe and S. Hansraj, Phys. Rev. D 91,
084049 (2015)

\bibitem {gb3}G. Papallo and H.S. Reall, JHEP 11, 109 (2015)

\bibitem {gb4}Y. Brihaye and L. Ducobu, Int. J. Mod. Phys. D 25, 1650084 (2016)

\bibitem {gb6}S.K. Marya, A. Banerjee, A. Pradhan and D. Yadav, EPJC 82, 552 (2022)

\bibitem {gb7}M. Minamitsuji and S.\ Tsujikawa, Phys. Rev. D 106, 064008 (2022)

\bibitem {gb8}S.D. Odintsov, D. Saez-Chillion Gomez and G.S. Sharov, Phys.
Dark Univ. 37, 101100 (2022)

\bibitem {gb9}F. Gomez, S. Lepe, V.C. Orozco and P.\ Salgado, EPJC 82, 906 (2022)

\bibitem {gb10}S. Hasraj, D. Krupanandn, A. Banerjee and C. Hasraj, Annals.
Phys. 445, 169070 (2022)

\bibitem {gb11}D.J.Gross and J.H.Sloan, Nucl. Phys. B 291, 41 (1987)

\bibitem {hor}H. Lu and Y. Pang, Phys. Lett. B 809, 135717 (2020)

\bibitem {gbm01}B. Li, J.D. Barrow and D.F.\ Mota, Phys. Rev. D 76, 044027 (2007)

\bibitem {gbm02}N.M. Garcia, T. Harko,\ F.S.N. Lobo and J.P. Mimoso, J. Phys.
Conf. Ser. 314, 012060 (2011)

\bibitem {gbm03}S. Nojiri, S.D. Odintsov, V.K.\ Oikonomou and A.V. Popov,
Nuclear Phys. B 973, 115617 (2021)

\bibitem {bb1}I. Fomin, EPJC 80, 1145 (2020)

\bibitem {bb2}R.A. Konoplya, T. Pappas and A. Zhidenko, Phys. Rev. D 101,
044054 (2020)

\bibitem {bb2b}F. Atamurotov, S. Shaymatov, P.\ Sheoran and S. Siwach, JCAP 08,
045 (2021)

\bibitem {bb3}H. Witek, L. Gualtieri and P. Pani, Phys. Rev. D 101, 124055 (2020)

\bibitem {bb4}H.S Vieira, V.B. Bezerra, C.R. Muniz and M.S Cunha, EPJC 82, 669 (2022)

\bibitem {bb5}Z. Luy, N. Jiang and K. Yagi, Phys.\ Rev. D 105, 064001 (2022)

\bibitem {in1}S. Chakraborty, T. Paul and S. SenGupta, Phys. Rev. D 98, 083539 (2018)

\bibitem {in2}I.V. Fomin, Physics of Particles and Nuclei 49, 525 (2018)

\bibitem {in3}S.A. Venekoudis and F.P. Fronimos, Eur. Phys. J. Plus 136, 308 (2021)

\bibitem {in4}S.D. Odintsov, V.K. Oikonomou and F.P. Fronimos, Annals of
Physics 420, 168250 (2020)

\bibitem {in5}S.D. Odintsov, V.K. Oikonomou and F.P. Fronimos, Nuclear Physics
B 958, 115135 (2020)

\bibitem {in6}P. Kanti, R. Gannouji and N. Dadhich, Phys. Rev. D 92, 083524 (2015)

\bibitem {dn1}N. Chatzarakis and V.K. Oikonomou, Annals of Physics 419, 168216 (2020)

\bibitem {dn2}K.F.\ Dialektopoulos, J.L.\ Said and Z. Oikonomopoulou,
arXiv:2211.06076 (2022)
\end{thebibliography}
\end{document}